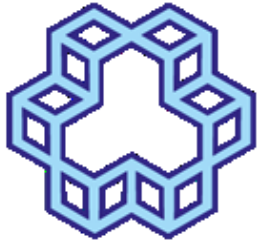

# Improved Forward Problem Modeling in Magnetic Induction Tomography for Biomedical Applications

Thesis Submitted in Partial Fulfillment of the
Requirements for the Degree of
Philosophy of Doctorate in
Electrical Engineering, Bioelectric

By:
**Hassan Yazdanian**

Supervisors:
**Dr. Reza Jafari**
**Dr. Hamid Abrishami Moghaddam**

September 2021

<div dir="rtl">
به نام خداوند جان و خرد

چه دانم های بسیار است لیکن من نمی دانم...

«مولانا»
</div>

*I know that I know nothing.*

"*Socrates*"

*To my wife and our daughter*

The following individuals have examined the dissertation entitled "Improved forward problem modeling in magnetic induction tomography for biomedical applications" presented by Hassan Yazdanian in partial fulfillment of the requirements for the degree of Doctor of Philosophy in Electrical Engineering- Bioelectric, and they certify that it is fully adequate, in scope and quality, and graded as very good.

**Examining Committee:**

1- Dr. Reza Jafari, K. N. Toosi University of Technology
   Supervisor

2- Dr. Hamid Abrishami Moghaddam, K. N. Toosi University of Technology
   Co-supervisor

3- Dr. Reza Faraji-Dana, University of Tehran
   Examiner

4- Dr. Hamid Soltanian-Zadeh, University of Tehran
   Examiner

5- Dr. Somayyeh Chamaani, K. N. Toosi University of Technology
   Examiner

6- Dr. Ali Khadem, K. N. Toosi University of Technology
   Examiner

7- Dr. Farhad Akbari Boroumand
   Chairperson at defense

# Declaration

**Thesis title:** Improved forward problem modeling in magnetic induction tomography for biomedical applications

**Supervisors:** Dr. Reza Jafari and Dr. Hamid Abrishami Moghaddam

**Student Name:** Hassan Yazdanian

**Student ID:** 9323646

I, Hassan Yazdanian confirm that the work presented in this thesis is my own. Where information has been derived from other sources, I confirm that this has been indicated in the thesis. No portion of the work referred to in the thesis has been submitted in support of an application for another degree or qualification of this or any other university or other institute of learning.

Signed……………………………………….…….. (Hassan Yazdanian)

Date: January 23, 2022 (3 Bahman 1400)

# Copyright Statement

1- The copyright of this thesis rests with the author. Copies of this thesis, either in full or in extracts and whether in hard or electronic copy, is allowed only with the consent of the author or the library of the Faculty of Electrical Engineering of K. N. Toosi University of Technology.
   This page must form part of any such copies made.

2- Intellectual Property of this thesis and any reproductions of copyright works in the thesis, for example graphs and tables, which may be described in this thesis belong to K. N. Toosi University of Technology and cannot be transferred to a third without the prior written permission of the university.
   Also, in the case of using thesis results, it must be cited properly.

# Acknowledgements

First and foremost, I would like to express gratitude to my main supervisor, Dr. Reza Jafari, for his guidance, kindness, and patience throughout my PhD studies. He truly taught me how to stand on my own in research. Secondly, I deeply indebted to my co-supervisor, Dr. Hamid Abrishami Moghddam, for his inspiring support and advice during the course of my PhD thesis. His insight and management had a noticeable effect on the progress rate of this research. It was a great pleasure working with these two brilliant scientists.

Besides my supervisors, I would like to thank the rest of my thesis committee: Dr. Reza Faraji-Dana, Dr. Hamid Soltanian-Zadeh, Dr. Somayyeh Chamaani, and Dr. Ali Khadem, for their insightful comments and hard questions to improve this thesis.

My sincere thanks also goes to Dr. Kim Knudsen for kindly hosting me at the Department of Applied Mathematics and Computer Science at Denmark Technical University (DTU) during my PhD visit.

I owe many thanks to all my friends and fellow colleagues for their help throughout this journey. I would particularly like to thank Dr. Mohammad Reza Yousefi for sharing his experiences in magnetic induction tomography with me, Dr. Maedeh Hadinia for helping me about element free Galerkin method and about the analytical solution of 2D MIT, and Mrs. Nasireh Dayarian for her supports and helpful discussions.

At last, but not least, my wonderful wife, our beautiful daughter, my parents, my wife's parents and my homeland deserve special thanks for everything they have done for me through all these years.

# Abstract

The main contribution of this thesis is to investigate the incorporating skin and proximity effects in magnetic induction tomography (MIT) coils and present an improved model for the two-dimensional (2D) forward problem. To evaluate the performance of the improved method in possible biomedical applications, a 16 coils 2D MIT system is modeled, and the finite element method (FEM) is employed to solve the forward problem. Results show the difference between the induced voltages obtained from the early and improved method falls into the meaningful range in terms of achievable conductivity contrast from the improved one. This finding revealed the importance and necessity of using the improved forward method for modeling 2D MIT coils for biomedical applications.

Another contribution of this thesis is to present the hybrid finite element–element free Galerkin (FE-EFG) method for solving the improved forward model. Previous studies on the biomedical MIT have employed mesh-based methods like FEM for solving the forward problem. These techniques have meshing task problems in complex geometries like the human head model. On the other hand, the element free Galerkin (EFG) method is computationally expensive. In this thesis, for the first time, the hybrid FE-EFG method is employed to take both advantages of FE and EFG methods. Comparison of FE and FE-EFG methods through solving the forward problem for a 16 coils 2D MIT system with a synthetic circular phantom revealed that the latter was more accurate compared to the former for the same run-time.

To see the importance of the improved forward method, it is required to apply it for conductivity image reconstruction. Thus, another contribution of this thesis is to present an improved technique for solving the MIT inverse problem by considering the skin and proximity effects in coils. To this end, the regularized Gauss-Newton algorithm is adapted to the improved forward method. The regularization parameter is chosen by an adaptive method. A new Jacobian matrix calculation is presented based on a standard technique that is compatible with the improved forward method. To compare the early and improved forward methods, a 2D 8-coil MIT system is modeled, and image reconstruction is performed for synthetic phantoms. Results show that it is crucial to use the improved forward method for the reconstruction of the absolute conductivity values.

The final contribution of this thesis focuses on comparison and improvements of techniques for conductivity image reconstruction in MIT. In previous studies, generally, the real part (in-phase component) and the imaginary part (quadrature component) of the induced voltages have been used in low- and high-conductivity imaging, respectively. However, techniques for image reconstruction in the mid-range conductivity have not been well-established in the MIT literature. In this thesis, the necessity of using both the real and imaginary parts of the induced voltages in MIT is shown, and a technique that uses both the real and imaginary parts of the induced voltages for conductivity image reconstruction is presented. This technique includes conductivity image reconstruction in the mid-range of conductivity. Results indicate the efficiency of the proposed technique compared to the previous one, especially in the mid-range conductivity applications.

**Keywords:** Complex-valued induced voltages, conductivity imaging, element free Galerkin method, finite element method, forward problem, Gauss-Newton algorithm, Jacobian matrix, magnetic induction tomography, skin and proximity effects.

# Contents

# List of figures:

# List of tables:

# List of abbreviations:

| | |
|---|---|
| 1D | One Dimensional |
| 2D | Two Dimensional |
| 3D | Three Dimensional |
| AE | Absolute Error |
| BEM | Boundary Element Method |
| CC | Conductivity Contrast |
| CSF | Cerebrospinal Fluid |
| CT | Computed Tomography |
| DEM | Diffusion Element Method |
| ECT | Electrical Capacitance Tomography |
| eFEM | edge Finite Element Method |
| EFG | Element Free Galerkin |
| EIT | Electrical Impedance Tomography |
| EMATs | Electro-Magnetic Acoustic Transducers |
| FDM | Finite Difference Method |
| FE | Finite Element |
| FE-EFG | Finite Element-Element Free Galerkin |
| FEM | Finite Element Method |
| FIT | Finite Integration Technique |
| GM | Gray Matter |
| GN | Gauss Newton |
| ICH | Intracranial hemorrhage |
| m° | milli-degree |
| MIT | Magnetic Induction Tomography |
| MLS | Moving Least Square |
| MQS | Magneto-quasi-statics |
| MRI | Magnetic Resonance Imaging |
| MVP | Magnetic Vector Potential |
| nFEM | nodal Finite Element Method |
| PE | Position Error |
| PPs | Performance Parameters |
| PSF | Point Spread Function |
| RE | Relative Error |
| RES | Resolution |
| SD | Shape Deformation |
| SNR | Signal to Noise Ratio |
| SVD | Singular Value Decomposition |
| TCD | Total Current Density |
| TSVD | Truncated Singular Value Decomposition |
| VIE | Volume Integral Equation |
| WM | White Matter |

# List of symbols:

| Symbol | Description |
|---|---|
| $\mathcal{A}$ | Column matrix containing all FE node potentials in FEM |
| $\mathcal{A}^{EFG}$ | Column matrix containing the EFG nodal parameters in FE-EFG method |
| $\mathcal{A}^{FE}$ | Column matrix containing all FE node potentials in FE-EFG method |
| $\vec{A}$ | Magnetic vector potential |
| $\vec{A}_0$ | Primary magnetic vector potential |
| $\vec{A}_r$ | Reduced magnetic vector potential |
| $A_z$ | $z$-component of magnetic vector potential |
| $A(\mathbf{x})$ | Matrix used in construction of MLS basis function |
| $a(\mathbf{x})$ | Column matrices containing position-dependent coefficient in MLS approximation |
| $B(\mathbf{x})$ | Matrix used in construction of MLS basis function |
| $\vec{B}$ | Magnetic flux density vector |
| $B_0$ | Magnitude of primary magnetic field |
| $B_1$ | Magnitude of total magnetic field |
| $\vec{D}$ | Electric flux density vector |
| $\vec{E}$ | Electric field intensity vector |
| $\vec{H}$ | Magnetic field intensity vector |
| $f$ | Frequency |
| $\mathcal{F}$ | Right hand side matrix in system of equations after assembly over all nodes in FEM |
| $\mathcal{F}^{EFG}$ | Right hand side matrix in system of equation of EFG method in FE-EFG method |
| $\mathcal{F}^{FE}$ | Right hand side matrix in system of equation of FE method in FE-EFG method |
| $\mathbf{H}$ | Hessian matrix |
| $\mathbf{H}_p$ | Hessian matrix for proposed technique |
| $I$ | Electric current |
| $\vec{J}$ | Total current density vector |
| $\vec{J}_e$ | Eddy current density vector |
| $\vec{J}_s$ | Source current density vector |
| $J_z$ | $z$-component of total current density vector |
| $\mathbf{J}$ | Jacobian matrix |
| $\mathbf{J}_p$ | Jacobian matrix for proposed technique |
| $\mathcal{K}$ | Left hand side matrix in system of equations after assembly over all nodes in FEM |
| $\mathcal{K}^{EFG}$ | Left hand side matrix in system of equation of EFG method in FE-EFG method |
| $\mathcal{K}^{FE}$ | Left hand side matrix in system of equation of FE method in FE-EFG method |
| $m$ | Number of independent measurements |
| $m_{po}$ | Maximum order of monomial basis functions |
| $\vec{m}$ | Magnetic moment vector |
| $n$ | Number of image pixel |
| $n_{sn}$ | Number of support nods |
| $\vec{n}$ | Outward normal unit vectors |
| $N^{EFG}$ | Number of EFG nodes in FE-EFG method |
| $N^{FE}$ | Number of FE nodes in FE-EFG method |
| $\mathbf{p}(\mathrm{x})$ | Column matrices containing monomial basis function in MLS approximation |

| | | |
|---|---|---|
| $V_0$ | | Voltage induced by the primary magnetic field |
| $V_1$ | | Voltage induced by the total magnetic field |
| $V_1^E$ | | Voltage induced by the total magnetic field, obtained by the early forward method |
| $V_1^I$ | | Voltage induced by the total magnetic field, obtained by the improved forward method |
| $V_{1,GT}^I$ | | Voltage induced by the total magnetic field, obtained by the improved forward method on an extremely fine mesh (ground-truth) |
| $V_1^{Normal}$ | | Voltage induced by the total magnetic field for normal head |
| $V_1^{ICH}$ | | Voltage induced by the total magnetic field for ICH |
| $\mathbf{V}_F(\boldsymbol{\sigma})$ | | Real-valued column matrix obtained from the forward solver |
| $\overline{\mathbf{V}}_F(\boldsymbol{\sigma})$ | | Complex-valued column matrix obtained from the forward solver |
| $\mathbf{V}_{pF}(\boldsymbol{\sigma})$ | | Real-valued column matrix obtained from the forward solver used in the proposed technique |
| $\mathbf{V}_M$ | | Real-valued column matrix containing real part of measured voltages |
| $\overline{\mathbf{V}}_M$ | | Complex-valued column matrix containing measured voltages |
| $\mathbf{V}_{pM}$ | | Real-valued column matrix containing measured voltages used in the proposed technique |
| $\mathbf{R}$ | | Regularization matrix |
| $\mathcal{S}$ | | Local FE coefficient matrix |
| $\mathcal{T}$ | | Local FE coefficient matrix |
| $W(r)$ | | Weight function for MLS approximation |
| $\Delta B$ | | Magnitude of the secondary magnetic field |
| $\Delta V$ | | Voltage induced by the secondary magnetic field |
| $\Delta V^E$ | | Voltage induced by the secondary magnetic field, obtained by the early forward method |
| $\Delta V^I$ | | Voltage induced by the secondary magnetic field, obtained by the improved forward method |
| $\Delta V^{Normal}$ | | Voltage induced by the secondary magnetic field for normal head |
| $\Delta V^{ICH}$ | | Voltage induced by the secondary magnetic field for ICH |
| $\varepsilon$ | | Permittivity |
| $\varepsilon_r$ | | Relative permittivity |
| $\lambda$ | | Regularization parameter |
| $\mu$ | | Permeability |
| $\mu_0$ | | Free space permeability |
| $\mu_r$ | | Relative permeability |
| $\rho$ | | Gain factor |
| $\tau$ | | Coefficient used in adaptive regularization method |
| $\Phi$ | | Scalar potential |
| $\phi_i(x)$ | | MLS basis function |
| $\phi(\sigma)$ | | Least-squares objective function |
| $\sigma$ | | Conductivity |
| $\boldsymbol{\sigma}$ | | Conductivity column matrix |
| $\sigma_h$ | | Homogeneous conductivity value |
| $\omega$ | | Angular frequency |
| $\Omega_C$ | | Conducting region |
| $\Omega_N$ | | Non-conducting region |
| $\Omega_S$ | | Source region |
| $\Omega^{EFG}$ | | Domain solved by EFG in FE-EFG method |

| | |
|---|---|
| $\Omega^{FE}$ | Domain solved by FE in FE-EFG method |

*All electromagnetic field quantities used in this thesis are in the frequency domain (Phasor representation, $e^{-j\omega t}$ convention is used through this thesis).

# Chapter 1:   Introduction

## 1.1   Electrical Tomography

Tomography methods can be classified into two general groups: 1) Hard-field tomography such as X-ray imaging and magnetic resonance imaging (MRI), and 2) Soft field tomography such as electrical tomography [1]. In the former, the excitation field lines almost propagate as straight lines, whereas in the latter, they diffuse in the medium. Due to this feature, image reconstruction is much more difficult for soft-field tomography. Electrical tomography, as a soft-field modality, includes three subclasses: electrical impedance tomography (EIT) [2], electrical capacitance tomography (ECT) [3], and magnetic induction tomography (MIT) [4]. Electrical tomography can image passive electromagnetic properties (conductivity, permittivity, and permeability) inside an object in medical and industrial applications. All subclasses are non-invasive, non-destructive, non-radiative, low-cost, and portable. So far, commercial EIT systems have been provided to image lung ventilation. Salvia ELISA 800$^{VIT}$ [5] and PulmoVista 500 [6] are examples of these systems.

## 1.2   Magnetic Induction Tomography (MIT)

MIT is the youngest member of the electrical tomography family and is in the progress stage. So far, no commercial MIT system is available. In a typical MIT system, an array of coils is used as both the exciter and sensor placed in the boundary of imaging space. Most of the MIT systems use an alternating magnetic field for exciting the imaging region by one coil, which is called primary magnetic field. This generates another magnetic field inside the conductive objects in the imaging region through the induced eddy currents, this field is called secondary magnetic field. The total magnetic field, which is the sum of the primary and secondary fields, induces voltages in the MIT coils. The measured voltages can be fed to reconstruction algorithms to obtain an image of the conductivity distribution inside an object.

MIT is believed to be more advantageous than other electrical tomography methods in some aspects. First, it is difficult and in some cases impossible to attach EIT and ECT electrodes to samples under measurement [7]. Second, MIT operates through an air gap, consequently, the difficulties with contact electrodes such as their unknown or variable impedance and position for EIT electrode [8] are avoided completely which yields to more accurate image reconstruction. Finally, unlike electrical field, magnetic field can penetrate high resistivity materials (e.g., skull and fat). These advantages make MIT more attractive, especially for applications in which the imaging region is inside a non-conductive shell. It is noteworthy that MIT is capable of imaging all passive electromagnetic properties of materials; i.e., conductivity, permittivity, and permeability [9].

## 1.3 Biomedical Applications of MIT

Generally, MIT applications for conductivity imaging can be divided into two categories: high- and low-conductivity imaging [1]. Since the conductivity distribution to be imaged is vastly different in these two categories, software and hardware components are mainly different for them. The biomedical applications of MIT belong to the latter category. The following is a list of biomedical applications which have been proposed for MIT in the literature:
1- Intracranial hemorrhage (ICH) imaging [10]–[19],
2- Imaging of lung ventilation [20], [21],
3- Other applications including inductive measurement of wound conductivity [1], imaging of spinal cord conductivity [22], imaging of heart conductivity [23], and brain cryosurgery monitoring [24].

In all of these applications, conductivities to be imaged are less than 2 S/m. Therefore, the secondary magnetic field is generally much weaker than the primary magnetic field. This results in the weak voltages induced by the secondary magnetic field compared to the high conductivity applications. As can be seen from the list, ICH imaging is a more attractive application for biomedical MIT. Therefore, we explain more about the importance of this application in the following.

### 1.3.1 Intracranial hemorrhage (ICH) imaging

Based on the world stroke organization report in 2019, there are over 13.7 million new strokes each year, and one in four people over age 25 will have a stroke in their life around the world [25]. Over the last several decades, the burden of stroke in the world has shifted from developed to developing countries [26].

Stroke is a medical emergency condition in which disturbance in the blood supply to the brain causes the rapidly developing loss of brain function. Stroke generally can be divided to ischemia and hemorrhage. The former is due to interruption of the blood supply, and the latter is due to the rupture or leakage of a blood vessel within the skull. Therefore, ischemic stroke and hemorrhagic stroke are opposites [27]. If a stroke is suspected, it is imperative to determine the type of stroke quickly and accurately. For ischemia, thrombolytic therapy may be administered to dissolve blood clots in a process called thrombolysis. In this way, the damage caused by the blockage of blood vessels can be limited. Based on the guidelines, thrombolysis can be beneficial provided that it is administered within 3 to 4.5 hours of stroke onset (golden time), and the possibility of a hemorrhage is completely ruled out [28]. It makes sense that thrombolytic therapy is contraindicated in the case of a hemorrhagic stroke because they make the situation worse by prolonging bleeding within the skull space. In consequence, a rapid distinguishing between ischemia and hemorrhage is vital.

Computed tomography (CT) and MRI, besides the physical examinations, are the most important and the most common methods for diagnosing stroke [28]. These imaging modalities can distinguish between ischemia and hemorrhage for physicians. However, they are expensive and non-portable with some side effects. In addition, they are sometimes inaccessible within the golden time, especially in small towns and rural areas.

Since there is a significant difference between blood conductivity and most the tissues, MIT can potentially be employed as a modality to distinguish between ischemia and hemorrhage [29]. It should be noted that MIT probably never reach the spatial resolution of MRI or CT, but it can be employed to rapid initial diagnosis as a low-cost and portable system.

## 1.4 Statements of the problems

The main challenge in the biomedical MIT is that induced voltages are inherently weak, and their detection not only requires the improvement of hardware component but also necessitates a more elaborate computational part in order to acquire more reliable imaging data. Image reconstruction in MIT generally involves the minimization of error between the measured data and modeled data. The forward problem generates the modeled data. To achieve a reconstructed image with a desired degree of accuracy, it is crucial to have a highly accurate forward model which can be fitted as much as possible to the measured data in the reconstruction process; since minor errors in the forward model can causes considerable errors when it is entered in the process of image reconstruction. In addition, accurate forward computation is vital to the process of MIT hardware design.

A straightforward possibility to strengthen the magnitude of induced voltages in the biomedical MIT would be to increase the operating frequency. However, by increasing frequency, skin and proximity effects become more noticeable in conductors exposed by alternating magnetic fields and cannot be ignored [30]. It means, unlike static cases, the current densities in coils are no longer constant or independent of position within conductors and position of coils relative to each other. Previous studies on MIT forward problem have used simplified Maxwell's equations which assume a constant and position-independent total current density (TCD) inside the coils (ignoring skin effect). Moreover, they assume that TCD is independent of the relative position of the coils (ignoring proximity effect). These assumptions can cause considerable errors in the solution of the biomedical MIT forward problem in which the frequency of several MHz is often employed.

Furthermore, reviewing the studies conducted on the biomedical MIT forward problem revealed that almost all of them have employed mesh-based approaches like the finite element method (FEM) to solve the forward problem. Some applications, like ICH imaging, involve the complex geometry in which generation of a mesh is a time-consuming and challenging task [31]. In addition, in some other applications like imaging of lung ventilation by EIT [32], the shape and size of the boundary and internal regions of the object under study are subject to frequent deformations. These applications in which mesh distortion is inevitable are susceptible to producing errors in solution of the forward problem.

Moreover, by solving the MIT forward problem in previous studies, it was found that the conductivity distribution is proportional to the real and the imaginary part of the induced voltages in the low- and high-conductivity applications, respectively. [21]. There are other potential applications like imaging of regions containing biomedical conductive polymers [33] in which the conductivity is classified in the mid-range. However, reconstruction techniques for mid-range conductivity applications have not been well-established in the MIT literature.

## 1.5 Aims and objectives

The aims and objectives of this thesis can be summarized as follows:
1- The first aim of this thesis is to present an improved MIT forward model by incorporating skin and proximity effects in the exciter and sensor coils. Consideration of the skin and proximity effects requires the use of a position-dependent TCD in Maxwell's equations inside the coil domain. The performance of this improved forward model should be evaluated in comparison with the early model for predicting induced voltages in biomedical MIT applications.

2- The second aim of this thesis is to evaluate the biomedical MIT conductivity reconstruction when the improved forward problem is employed compared to when the early one is used. To this end, the reconstruction algorithm should be adapted to the improved forward model. Then, the reconstruction results by using the early and improved forward model should be compared quantitatively.

3- The third aim of this thesis is to improve the accuracy of the forward problem solution by employing a mesh-free numerical method. Therefore, a mesh-free method should be adapted to the improved forward model, and its accuracy should be evaluated in comparison with FEM, as the most common mesh-based method, for predicting induced voltages in the biomedical MIT applications.

4- The final aim of this thesis is to develop a reconstruction technique for the mid-range conductivity applications. For this purpose, first, the relation between the mid-range conductivity values and the real and imaginary parts of induced voltages should be clarified. Then, a reconstruction technique should be adapted based on this relation.

## 1.6 Thesis organization

This thesis comprises seven chapters and four appendices. Each chapter starts with a brief introduction on the subject of the chapter and continues by the main body:

- In Chapter 2, the MIT principles are explained, and the fundamental components of a typical biomedical MIT system are presented. A comprehensive review of the studies on the forward and inverse problem of biomedical MIT is provided then.

- In Chapter 3, ignoring skin and proximity effects in MIT coils is addressed by introducing the improved forward model. First, the formulation for the early and improved forward methods is extracted. The chapter continues to present the procedure of computation of the induced voltage. Then, numerical implementation of the improved method is validated with a simple analytical test problem. Next, through modeling a 2D 16 coils MIT system and using a synthetic phantom and a realistic head model, the solution of the forward problem based on the improved method will be investigated and compared to that of the early method.

- In Chapter 4, it is tried to solve the improved MIT forward problem by the element-free Galerkin method (EFG). First, the fundamental of the EFG method is presented. A hybrid method of FEM and EFG method (FE-EFG method) is next formulated to solve the biomedical MIT forward problem. Then, FE and FE-EFG methods are compared through solving the improved forward problem for a 16 coils 2D MIT system with a synthetic circular phantom.

- In Chapter 5, the regularized Gauss-Newton (GN) algorithm is adapted to the improved forward problem model for MIT conductivity image reconstruction. The methods of calculation of the Jacobian matrix are next explained in detail. Then, an adaptive method for choosing the regularization parameter is presented. Next, through modeling a 2D 8-coil MIT system and using synthetic phantoms, the conductivity image reconstruction based on the improved forward model will be investigated and compared to that of the early one.

- In chapter 6, the dependency of induced voltages on the conductivity is discussed in various MIT applications. Then, a new technique is proposed which works for all ranges of conductivity values. The GN algorithm and the Jacobian matrix are derived for this technique. Through five 2D numerical examples, the proposed technique is compared with the early one, and results are evaluated quantitatively.

- In Chapter 7, conclusions about the research and the results obtained in the thesis are discussed. Furthermore, some suggestions are provided to approach as future work.

At the end of this thesis, appendices are placed, followed by a bibliography.

# Chapter 2: State of the art in biomedical MIT

The chapter aims to outline the biomedical MIT imaging technique. First, the MIT principles are explained through an illustrative example. Next, the fundamental components of a typical biomedical MIT system are presented. Finally, a comprehensive review of the studies conducted on the forward and inverse problems of biomedical MIT is provided.

## 2.1 Biomedical MIT Principles

To explain the principles of MIT, we start with a simple example. Then, the relation between the induced voltage and conductivity to be imaged is clarified based on the result obtained from the simple example.

### 2.1.1 A single channel MIT system

Figure 2.1 shows a simple model of two coaxial coils with a target domain which is known as a single channel MIT system in the literature [34]. By passing an alternating current from the excitation coil placed around the target object, an alternating magnetic field is generated which is called *primary magnetic field*, $B_0$ (the magnitude of the field). This generates another magnetic field inside the conductive target object by inducing eddy currents, this field is called *secondary magnetic field*, $\Delta B$. The *total magnetic field*, $B_1$, which is the sum of the primary and secondary fields, can be measured by the sensing coil. Figure 2.2 shows the phasor diagram for $B_0$, $\Delta B$, and $B_1$ which is redrawn from [35].

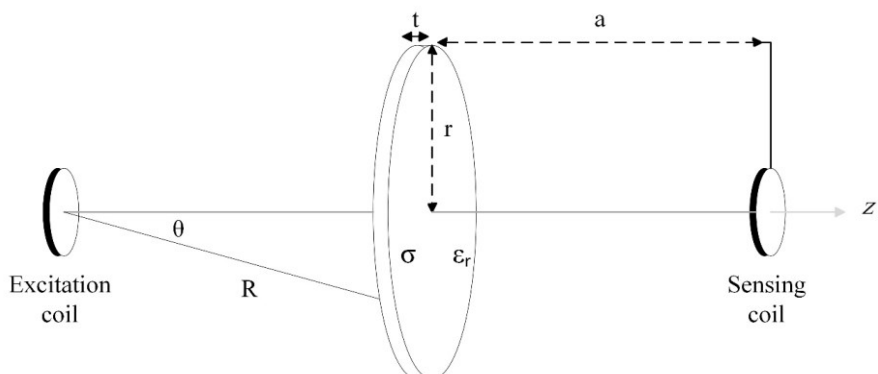

Figure 2.1. A single channel MIT system. The magnetic field generated by the excitation coil has been shown in a conductive disk at the spherical coordinate. The radius and thickness of the disk are $r$ and $t$, respectively. It is coaxially placed between the excitation and sensing coils. The distance between the centers of coils is $2a$. The conductivity and relative permittivity of disk is $\sigma$ and $\varepsilon_r$, respectively, and it is assumed to be non-magnetic ($\mu_r = 1$). It is presumed that the penetration depth of the excitation field is greater than $t$. Redrawn from [34].

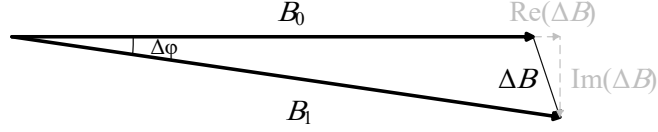

Figure 2.2. Phasor diagram of the magnetic field in biomedical MIT [35]. The magnitude of the primary, secondary, and total magnetic field are indicated by $B_0$, $\Delta B$, and $B_1$ respectively.

Because biological tissues are non-magnetic, it is assumed that $\mu_r=1$ in most biomedical applications, except for applications aimed at measuring hepatic iron stores [36], [37]. Consequently, in most studies on the biomedical MIT, reconstruction of the conductivity or both conductivity and permittivity distribution inside tissues has been considered.

To explore the relation between the electrical properties of the tissue and the induced field in the sensing coil, let's consider Figure 2.1. In this figure, a disk (as a biological tissue) with a radius of $r$ and thickness of $t$ is coaxially placed between two small coils (considered as magnetic dipoles). It is assumed that $t \ll r$. If the conductivity and relative permittivity of the disk is $\sigma$ and $\varepsilon_r$, respectively, and it is non-magnetic ($\mu_r = 1$), the relative change in the induced magnetic field and consequently in the induced voltage in the sensing coil due to the presence of the disk between coils can be expressed as follows [34] (see Appendix 1):

$$\frac{\Delta V}{V_0} = \frac{\Delta B}{B_0} = Q\mu_0\omega(\omega\varepsilon_0\varepsilon_r - j\sigma) \qquad (2.1)$$

where $V_0$, $\Delta V$, and $V_1 = V_0 + \Delta V$ are proportional to $B_0$, $\Delta B$, and $B_1$, respectively, and $\omega$ is the angular frequency. In (2.1), Q is a constant coefficient related to the problem geometry. When the imaging region is empty, the voltage $V_0$ is induced in the sensing coil by the primary magnetic field $B_0$. When the disk is placed between coils, the induced voltage changes by $\Delta V$ and $V_1$ is induced in the sensing coil. To obtain (2.1), it is assumed that the penetration depth of the excitation field, $\delta$, is much greater than the thickness of the disk, $t$, which is known as the weakly coupled field approximation. Thus, the effect of conductive disk on the excitation field is negligible. This approximation is acceptable for the conductivity and frequency ranges applied in the biomedical MIT [34].

### 2.1.2 Relation between conductivity and induced voltage

In this section, we clarify the relationship between the conductivity to be imaged and the induced voltage.

It is assumed that the excitation coil in the single-channel MIT system is derived by the input current $i_s(t) = I_0\cos(\omega t)$. The induced voltages in the sensing coil in the absence and presence of the disk are denoted by $v_0(t) = A_0\cos(\omega t + \theta)$ and $v_1(t) = A_1\cos(\omega t + (\theta - \Delta\theta))$, respectively, where $A_0$ and $A_1$ are the amplitude of the voltages and $\theta$ and $(\theta - \Delta\theta)$ are the phase shifts of the voltages with respect to the excitation current. Phasor representation of the voltages can be written as follows:

$$\begin{aligned} V_0 &= A_0 e^{j\theta} = Re\{V_0\} + jIm\{V_0\} \\ V_1 &= A_1 e^{j(\theta - \Delta\theta)} = Re\{V_1\} + jIm\{V_1\} \end{aligned} \qquad (2.2)$$

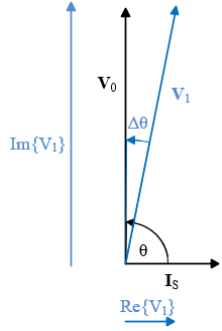

Figure 2.3 Phasor representation diagram of excitation current ($I_s$) and induced voltages. $V_0$ is the induced voltage when the imaging space of MIT system is empty (background voltage). $V_1$ is the induced voltage when an inhomogeneity is placed in the imaging space.

where $Re\{\cdot\}$ and $Im\{\cdot\}$ indicate the real and imaginary parts of the voltages, respectively. Figure 2.3 shows the phasor diagram of the excitation current ($I_s$) and induced voltages ($V_0$ and $V_1$). Simplifying (2.1) yields the following equation:

$$V_1 = V_0[(1 + Q\omega^2\mu_0\varepsilon) - jQ\omega\mu_0\sigma] \qquad (2.3)$$

The phase angle of $V_0$ will be $\theta = \pi/2$ for MIT systems with lossless coils. Therefore, it can be written as $V_0 = j|V_0|$. Substituting $V_0$ into (2.3) yields:

$$V_1 = |V_0|[Q\omega\mu_0\sigma + j(1 + Q\omega^2\mu_0\varepsilon)] \qquad (2.4)$$

The phase change of induced voltage, $\Delta\theta$ in Figure 2.3, is very small in biomedical applications of MIT [3, 17] so that one can assume $Im\{V_1\} = |V_0|(1 + Q\omega^2\mu_0\varepsilon) \cong Im\{V_0\} = |V_0|$ and, therefore, (2.4) can be rewritten as

$$V_1 \cong |V_0|(Q\omega\mu_0\sigma + j1) \qquad (2.5)$$

It is worth noting that $\angle V_1 = \tan^{-1}(1/Q\omega\mu_0\sigma) = \pi/2 - \tan^{-1}(Q\omega\mu_0\sigma)$ and $\angle V_1 - \angle V_0 = -\Delta\theta$, then, phase change $\Delta\theta$ can be obtained as:

$$\Delta\theta = \tan^{-1}(Q\omega\mu_0\sigma) \qquad (2.6)$$

Using the approximation $\tan(\alpha) \cong \alpha$ for small $\alpha$, (2.6) will be simplified as follows:

$$\Delta\theta \cong Q\omega\mu_0\sigma \qquad (2.7)$$

In conclusion, the following results can be inferred from (2.1), (2.4) and (2.7):
   A- $Im\{\Delta V/V_0\}$ is proportional to the conductivity as shown by (2.1).
   B- $Re\{V_1\}$ is proportional to the conductivity as shown by (2.4).
   C- $\Delta\theta$ is proportional to the conductivity as shown by (2.7).
   D- $Im\{\Delta V/V_0\}$, $Re\{V_1\}$, and $\Delta\theta$ are proportional to the angular frequency $\omega$.

Based on this conclusion in some studies $Im\{\Delta V/V_0\}$ [38]–[40], in some other studies $Re\{V_1\}$ [41]–[45], and in some other studies $\Delta\theta$ [14]–[17], [29], [46], [47] have been used for conductivity imaging in biomedical MIT applications. Since the conductivity of biological tissue is low (less than 2 S/m), $\Delta V$ is inherently weak. Consequently, the frequency range of several MHz is used in biomedical MIT to increase the magnitude of the induced signals.

It noteworthy that the above conclusion is only valid for low-conductivity MIT applications, including biomedical applications. For high-conductivity applications, which is out of the scope of this thesis, the imaginary part of induced voltages has been employed for conductivity reconstruction [48]. There are other potential applications like imaging of regions containing biomedical conductive polymers [33] in which the conductivity is classified in the mid-range. However, techniques for conductivity reconstruction in the mid-range conductivity have not been well-established in the MIT literature.

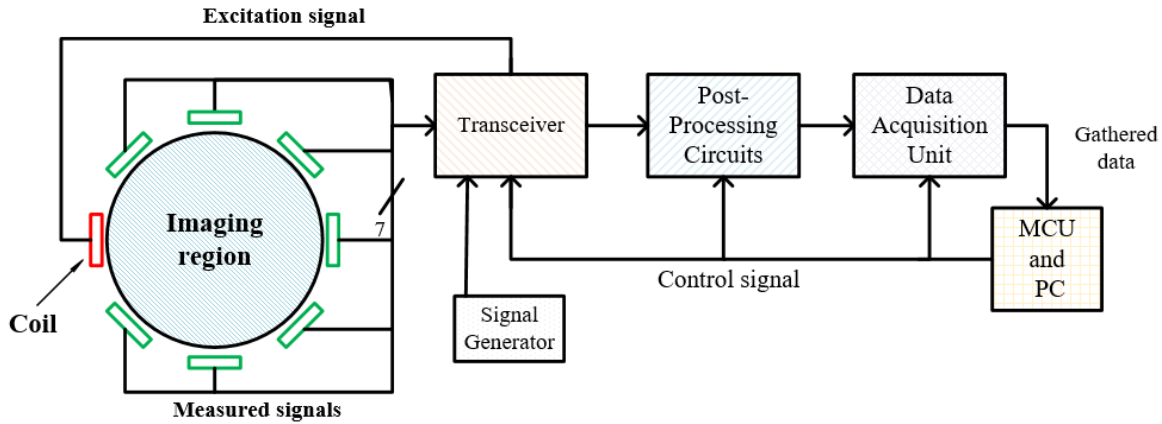

Figure 2.4 General block diagram of a typical MIT system (redrawn from [1]).

## 2.2 MIT Hardware

Figure 2.4 shows a general block diagram of a typical MIT system. Regardless of the application, MIT systems mainly include the following components:
1- An array of excitation and sensing coils placed around the imaging space.
2- Electronic circuitry including a signal generator, power amplifier, receiving buffers, multiplexers, and filters.
3- Data acquisition unit which demodulates, digitizes, and sends data to PC.
4- A computer to reconstruct an image from the logged data by running an algorithm.

The system acts as follows: Signal generator first produces a sinusoidal voltage, and then a power amplifier boosts the current and voltage amplitude of the initial signal. The resultant signal is fed into an excitation coil, and an alternating magnetic field is generated. After interacting the primary field with the target object, the resultant secondary magnetic field is measured by the sensing coils. Excitation and sensing coils are sequentially selected by multiplexers. The induced voltages in the sensing coils are buffered and then fed to a data acquisition to be demodulated and digitized. Finally, the logged data is transferred to a computer to reconstruct an image of the conductivity distribution. In Appendix 4, a review of the biomedical MIT hardware has been presented.

## 2.3 MIT software

The software module in MIT uses the acquired data and elaborated reconstruction algorithms to generate an image of the conductivity distribution. The reconstruction algorithm generally involves solutions to the forward and inverse problems. As shown in Figure 2.5(a), the forward problem employs an applied current pattern and a given conductivity distribution to generate simulated data. In biomedical MIT, forward models are generally stated based on the differential form of Maxwell's equations and then solved by numerical methods such as FEM. As shown in Figure 2.5(b), the inverse problem compares the simulated and measured data to estimate a conductivity distribution. The inverse problem algorithms are generally based on the steepest descent or Newton's method in biomedical MIT. In the following sections, we review the methods used for solving the biomedical MIT forward and inverse problems.

## 2.4 Review of the methods for solving forward problem in biomedical MIT

Forward problem is a well-defined problem based on Maxwell's equation with a unique solution. To achieve a reconstructed conductivity with a desired degree of accuracy, it is crucial to provide an accurate solution of the forward problem; since small errors in the forward model can generate considerable errors in the solution of the inverse problem. In addition, accurate forward computation is vital to the process of MIT hardware design. On the other hand, the method for solving the forward problem should be computationally efficient. It is usually not possible to simultaneously achieve an accurate and fast computing method, and a trade-off would be required. The methods that have been reported so far to solve the biomedical MIT forward problem can generally be classified into: analytical method, FEM, and finite difference method (FDM). Besides these methods, some other methods have been used for solving the biomedical MIT forward problem. In this section, we briefly review all these methods.

It is noteworthy that in all studies are reviewed here, the simplified Maxwell's equations have been used which assume there is no eddy current neither in the excitation coil region nor in the sensing coil region. It means ignoring skin and proximity effects inside the coils.

### 2.4.1 Analytical method

Analytical methods can be applied to problems with simple and symmetric geometry. The main advantage of this approach is its high computational speed. However, obtaining the analytical solution even for simple geometries involves complex algebraic operations. This approach is mainly applied for the validation of numerical methods. By knowing the analytical solution for a given problem, one can evaluate the accuracy and precision of the numerical methods for simple geometries.

In some studies on biomedical MIT, an analytical method has been used to solve the forward problem. In 2008, Dekdouk et al. [49] have presented an analytical solution for a six-layer spherical model to study the feasibility of detecting brain edema using MIT. This analytical solution was obtained for a single channel MIT system, and the displacement current was ignored. In this study, the Helmholtz equation for the electric field was used in the spherica

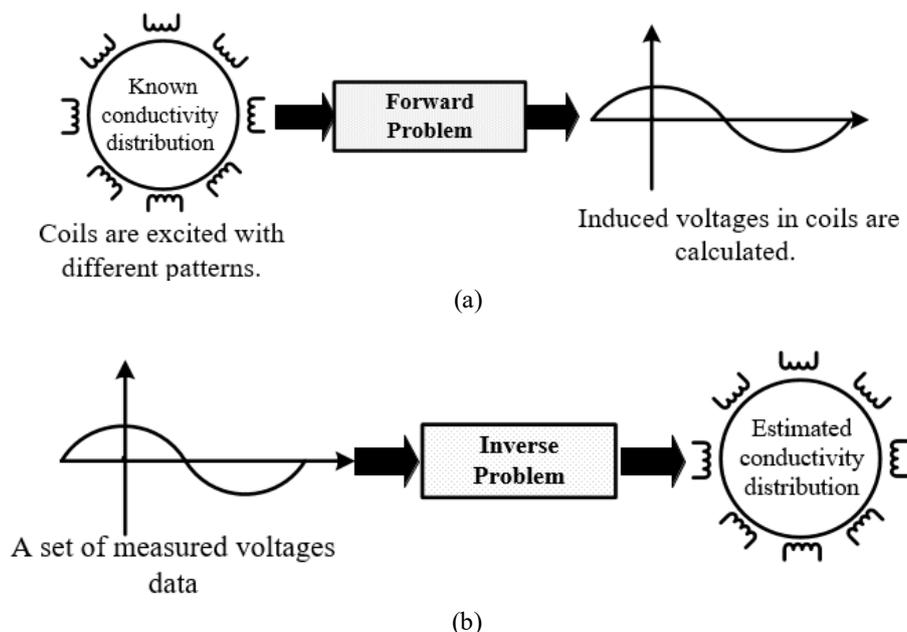

(a)

(b)

Figure 2.5 (a) A schematic view for the forward problem and (b) A schematic view for the inverse problem (redrawn from [1]).

l coordinates. The result showed that brain edema with conductivity and radius of 0.8624 S/m and 27 mm, respectively, located at the center of the head could be detected by an MIT system with a 40 dB signal-to-noise ratio (SNR) and operating frequency of 10 MHz.

In another study in 2010, Xu et al. [50] have presented an analytical solution with a four-layer model and one excitation coil. In this study, the forward problem of MIT was constructed by the Helmholtz equation for magnetic vector potential (MVP) in the spherical coordinate and then solved by the method of separation of variables at 1 MHz. The permittivity of tissues was considered besides their conductivities in this work. Based on the authors' claim, this analytical solution can be beneficial to the fast calculation of the MIT sensitivity matrix for the problem with regular geometry.

Since biological tissues have irregular shapes and complex structures, there is no analytical solution for them. Then, numerical methods are used to solve the forward problem in realistic geometries. In the following, we briefly introduce the numerical methods used to solve the forward problem of biomedical MIT.

### 2.4.2 Finite element method

FEM is a technique to find an approximate solution of the partial differential equations. This method has widely been used for the solution of the MIT forward problem. In FEM, the problem domain is meshed into small elements. Usually, triangular and tetrahedral elements are used for two- and three-dimensional problems, respectively. The unknown scalar or vector function is approximated by a simple function which is called the shape function. The combination of shape functions for an element makes a basis function for that element. Then, the unknown function is stated over all elements based on the linear combination of the basis functions. In this way, the partial differential equation governed on the problem is turned into a system of linear equations.

In MIT, the imaging space and the object inside it, coils array, and enough air space outside of the coils array to consider the distribution of the fields (if there is no electromagnetic shield) should be meshed. Depending on the unknowns are scalar or vector functions, the nodal finite element method (nFEM) or edge finite element method (eFEM) usually have been used, respectively [51]. In nFEM and eFEM, the shape functions are scalar and vector functions, respectively. If nFEM is used to approximate vector functions, it must be guaranteed that the normal and tangential components of the unknown vector function are continuous [51]. In consequence, the gauged version of the forward problem formulation is used. On the other hand, if eFEM is used to approximate vector functions, only the continuity of the tangential component must be guaranteed, and there is no need to impose the gauge to the forward problem formulation [51]. Furthermore, applying boundary conditions for vector functions is more complicated by nFEM.

In many studies, FEM has been used to solve the biomedical MIT forward problem. In 2004, Merwa et al. [52] and Hollaus et al. [53] used $\vec{A}_r - \vec{A}_r, V$ formulation to set up the forward problem and solved it by 3D eFEM in the human head domain where $\vec{A}_r$ and V are the reduced magnetic vector potential and electric scalar potential, respectively (see Appendix 2 for various MIT forward problem formulations). In this study, an almost realist head model was employed to assess the sensitivity of MIT to brain edema. The model consisted of four layers: cerebrospinal fluid (CSF), white matter, gray matter, and edema. It was meshed by about 50000 second-order edge elements. Both conductivity and permittivity of the tissues were considered. Results showed that brain edema with a contrast of 2 S/m to the background and a diameter of 40 mm located at the center of the head could be detected by an MIT system with 1 $nV/\sqrt{Hz}$ noise and operating frequency of 100 kHz. In 2006, Scharfetter et al. [54] also employed the same approach to solve the forward problem in a single-step 3D image reconstruction algorithm for biomedical MIT.

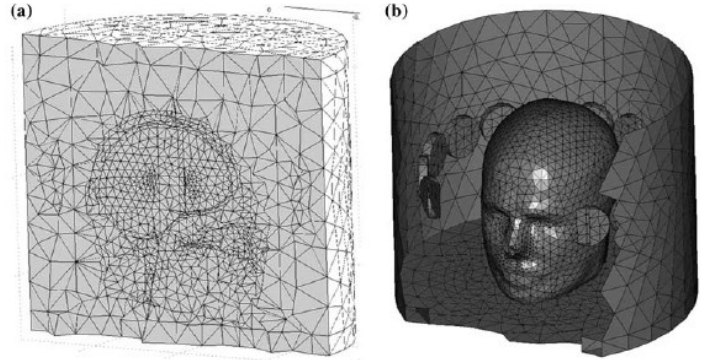

Figure 2.6 FEM mesh used in [9] for simulating MIT. (a) Tetrahedral elements in a cross section and (b) Triangular boundary elements showing the location of the coils relative to the head.

In 2006, Soleimani and Lionheart [41] formulated the forward problem by $\vec{A} - \vec{A}, V$ method where $\vec{A}$ is the MVP. Then, they solved the forward problem by 3D eFEM for biomedical applications of MIT. In this study, the forward problem domain was meshed by 208000 tetrahedral elements. The operating frequency was 10 MHz.

In 2009, Zolgharni et al. [29], [46] evaluated the sensitivity of MIT to detect ICH by modeling the forward problem. They solved the forward problem by 3D eFEM in the commercial package, Comsol Multiphysics, for a realistic head model comprising 12 different tissues. In this study, $\vec{A} - \vec{A}$ formulation was used. Figure 2.6 shows the mesh used in this work which consists of 82852 second-order tetrahedral elements. The total runtime for solving the forward problem was 65 hours. The relative error between the proposed solution and an analytical solution in a test problem ranged from 3.76 to 0.25% for different meshes. Results revealed that for a peripheral stork with a volume of 50 cm$^3$, 27% of induced voltages are above the noise level of a practical MIT system with 20 m° (millidegree) phase noise and operating frequency of 10 MHz. They mentioned that to detect the same percentage of the voltages due to a centrally located ICH with a volume of 7.7 cm$^3$, a reduction in phase noise to 1 m° is necessary.

In 2010, Chen et al. [12], in a similar simulation study, evaluated the detectability of ICH by an MIT system with the operating frequency of 10 MHz. In this study, a realistic head model comprising 18 different tissues was used. This model was meshed to 213709 tetrahedral elements. The time-harmonic vector wave equation for the electric field was employed to describe the forward problem and solved it by the 3D eFEM.

In 2019, Xiao et al. [16] solved the forward problem by FEM to evaluate the sensitivity of a cambered MIT compared to cylindrical MIT in imaging of cerebral hemorrhage. The forward problem was formulated by $\vec{A} - \vec{A}$ method and solved in Comsol Multiphysics. A realistic human head model comprising of six tissue types with different hemorrhage volumes and positions was employed for simulation. The model was meshed to 828154 tetrahedral elements. Figure 2.6 shows the head model and the relative position of cylindrical and cambered MIT coil arrays used in this study. The simulation accuracy of the forward problem was validated with different tetrahedral mesh sizes. The results manifested that the cambered coil arrangement improves the MIT sensitivity to near-subsurface locations. In 2021, a similar numerical study was done by Lv and Luo [18]. They combined the coils suitable for transcranial magnetic stimulation and MIT setup aiming to solve the detection and focusing of deep brain tissue. The forward problem was formulated by $\vec{A} - \vec{A}$ method and solved in Comsol Multiphysics.

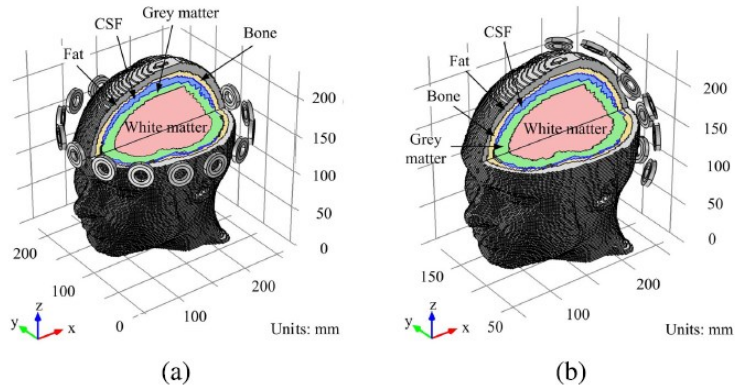

Figure 2.7 MIT coil array with realistic head model used in [16]. (a) Cylindrical coil array. (b) Cambered coil array.

The nodal finite element method has usually been used for studies in which the forward problem is formulated by $\vec{A}_0 - \vec{A}_0, V$ method where $\vec{A}_0$ is the MVP of the primary magnetic field. For example, in 1999, Gencer et al. [55] used $\vec{A}_0 - \vec{A}_0, V$ formulation to define the forward problem. They then solved it by nFEM in a simulation study aimed at imaging the tissue conductivity via contactless measurements. They used the second-order hexahedral elements to mesh the geometry. Results manifested that inclusion with a volume of 1 cm$^3$ at 2 cm depth in the medium conductivity can be detected. In 2003, this research group also presented the practical results in their following study [56].

In 2009, Gursoy and Scharfetter [20] employed $\vec{A}_0 - \vec{A}_0, V$ formulation and 3D nFEM solver for the forward problem in order to study the feasibility of lung imaging by MIT. The problem domain was discretized by 100000 elements. Two different operating frequencies, 100 kHz and 600 kHz used in this study. Results revealed that lung imaging is feasible by MIT; however, some software and hardware improvements are vital before clinical usage.

In several studies, the 2D finite element method has been used to solve the biomedical MIT forward problem. For instance, in 2007, Wang et al. [57] solved the forward problem using 2D nFEM, a three-layer circular model, and $\vec{A} - \vec{A}$ formulation. Their purpose was to explore the possibility of using MIT for human head imaging. In a similar study in 2014, Liu et al. [13] studied the feasibility of imaging the low-contrast and small volume inclusions inside the human brain using MIT. They solved the forward problem by nFEM and considering $\vec{A} - \vec{A}$ formulation. The problem domain was meshed into 800 triangular elements. The experimental results obtained from a three-layer phantom showed that inclusion with the minimum contrast 0.03 S/m and volume of 3.4 cm$^3$ is detectable.

In 2016, Han et al. [45], with the aim of comparison of three reconstruction algorithms, solved the MIT forward problem by 2D nFEM for biomedical applications. They employed $\vec{A} - \vec{A}$ formulation to demonstrate the forward problem. They modeled an 8-coil 2D MIT system at 10 MHz and used a circular phantom meshed to 1152 elements. The results from solving the forward problem by nFEM used as the modeled measured voltage data for the conductivity image reconstruction.

In 2017, Xiao et al. [47] employed 2D nFEM for solving MIT forward problem in imaging of ICH application to investigate the effect of inter-tissue inductive coupling. They used $\vec{A} - \vec{A}$ formulation to define the forward problem. Then, they modeled a 2D 16-coil MIT system and solved the forward problem by Comsol Multiphysics. A 2D head model comprising of seven tissue types was used. The results revealed that the introduction of inter-tissue inductive coupling could reduce the errors of multi-frequency imaging technique.

In 2019, Chen et al. [15] also used the 2D nFEM to solve the MIT forward problem for the location monitoring of the anomaly in the biological tissues. In this study, the $\vec{A} - \vec{A}$ formulation in Comsol Multiphysics was employed to define the forward problem. Based on this forward problem, they proposed a new reconstruction algorithm that improved anomaly reconstruction accuracy and reduced the cerebral hemorrhage prediction time. It is noteworthy that in both [15] and in [47], researchers have considered the 2D forward problem, whereas the circular coils were assumed.

As illustrated, FEM has widely been employed in numerous studies to solve the biomedical MIT forward problem. Along with all the benefits of this method, its computational costs depending on the mesh used to discretize the problem domain can be considerably high. In addition, the accuracy of FEM usually depends on the mesh size. In other words, the finer mesh is used, the more accurate result is obtained. It is noteworthy that although the commercial software packages are widely available for FEM, but they are more beneficial to solve the electromagnetic fields problems and not optimized for solving the forward problem. Since for each activation of an excitation coil, it is required to rerun the package which leads to repetition of some procedures and consequently increasing the computation time. In addition, to solve the inverse problem, pieces of information at the mesh level from the source code are required, which is not accessible or hard to extract from commercially available FEM packages. Furthermore, for some inverse algorithms, it is necessary to solve the forward problem in each iteration. Thus, using the software packages makes the connection between inverse and forward problems difficult. Finally, the commercial packages usually are expensive, whereas the low-cost feature of MIT makes it attractive.

### 2.4.3 Finite difference method

In the finite difference method, the problem geometry is discretized by Cartesian meshes (rectangular grid in 2D and cubic grid in 3D), and the partial differential equations in the frequency domain are approximated by finite difference equations. Mesh generation is easier and computation time is faster for FDM compared to FEM. But, in contrast to FEM, it is hard to model complex geometries by FDM due to the nature of the underlying mesh. In several biomedical MIT studies, the forward problem has been solved by FDM. In most of these studies, $\vec{A}_0 - \vec{A}_0, V$ formulation has been employed to describe the forward problem. In the following, we review some of these studies.

In 2001, Morris et al. [58] developed a finite difference model to simulate MIT measurements from biological tissues. This model had three steps. In the first step, the current distribution induced by the primary magnetic field within the target object was computed. Both conductivity and permittivity of the target object were considered. In the second step, the induced voltage in the sensing coil was calculated. In the last step, amplitude and phase changes of the induced voltage in sensing coil due to presence of the target object inside the imaging region was calculated. The simulation results obtained from this study were in a good agreement with the practical measurements obtained from a single channel MIT system with a 10 MHz operating frequency.

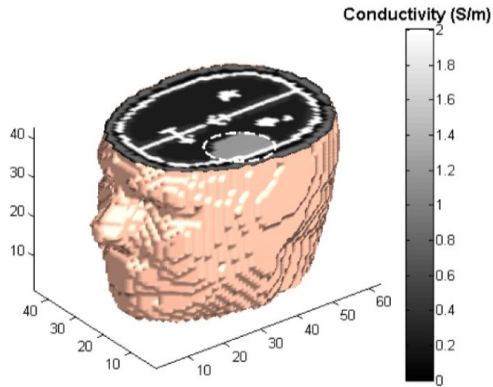

Figure 2.8 Head model used in [59] with coronal cutaway plane showing conductivity distribution. Gray circle shows the hemorrhage.

In 2010, Zolgharni et al. [11] studied the detectability of ICH by MIT through numerical modeling with FDM. They used the frequency difference technique (at the frequency of 1 MHz and 10 MHz) to benefit from the frequency dependency of the conductivity of the biological tissues. In this study, the simulation was done for 56 coils positioned on a hemisphere in the shape of a helmet. A realistic human head model consisting of 12 tissues was employed, and it was discretized to 474897 cubes with the size of 2 mm. The results revealed that an MIT system with 3 m° phase noise level could detect a peripheral hemorrhagic stroke with a volume of 49 ml.

In the same year, Dekdouk et al. [59] employed the finite difference method, called the impedance method, to solve the MIT forward problem for biomedical applications. In the impedance method, after discretization of the problem domain to the regular-sized cubic cell, each cell is modeled as a network of lumped impedances and sources. In consequence, the problem is reduced to an equivalent electrical network. Figure 2.8 shows the model used in this study which was meshed to 175770 elements with a volume of 3.5 mm$^3$. The results manifested that if the conductivities to be imaged are less than 0.5 S/m, then the impedance method reaches an error of less than 0.4% in comparison with FEM (commercial FE solver, Comsol Multiphysics).

In 2011, Ramos and Wolff [60] also used the impedance method to solve the biomedical MIT forward problem. In this study, the distribution of the magnetic field and eddy current was calculated numerically in a sphere exposed by the field of a coil, and the result was compared to an analytical model. The simulation result was also compared to the practical measurements. It was reported that the impedance method has a good agreement with experimental data in the frequency range from 100 kHz to 20 MHz and for low conductivity (< 10 S/m) objects.

In 2017, Korjenevsky and Sapetsky [61] employed the time-domain finite difference to approximate the time-domain Maxwell's equations and simulate an MIT measuring system for biomedical applications. They simulated a 16-coil MIT system operating at 20 MHz and compared the numerical results with experimental ones. The experimental result was in agreement with numerical results. The results were used for imaging of low contrast inclusions in a conductive background.

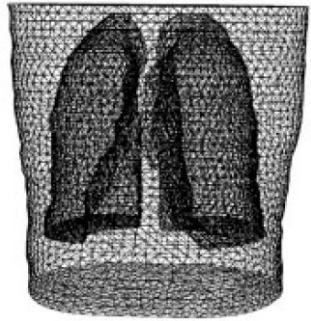

Figure 2.9 Mesh of the thorax and lungs used in [63]

It is noteworthy that $\vec{A}_0 - \vec{A}_0, V$ formulation is obtained under the weakly coupled field approximation. Thus, using the FDM method based on $\vec{A}_0 - \vec{A}_0, V$ formulation in case of high conductivity objects and/or high excitation frequencies cause considerable errors.

### 2.4.4 Other methods

In this section, we review the studies which have employed other methods to solve the MIT forward problem in biomedical applications.

In 2012, Caeiros and Martins [62] used the finite integration technique (FIT) to develop an optimized forward problem solver for the complete characterization of the electromagnetic properties of biological tissues in MIT. The basic idea behind FIT resides in transforming the integral form of Maxwell's equations into a set of matrix equations defined over a dual staggered grid. In this study, the forward problem was defined based on $\vec{A} - \vec{A}, V$ formulation and solved by FIT. The obtained results were compared with analytical solutions, which revealed the proposed method could give a low error solution even with relatively coarse grids. Thus, it reduced the overall system size and consequently the solution time.

In 2011, Engleder and Steinbach [63] formulated the MIT forward problem in terms of boundary integral equations. In this study, two models were presented: the eddy current model and a reduced simplified model. The eddy current model was defined based on the time-harmonic inhomogeneous wave equation for the electric field. Since this model was computationally expensive, the reduced simplified model was proposed to be used in iterative inverse algorithms. The reduced model was defined based on $\vec{A}_0 - \vec{A}_0, V$ formulation and solved by the boundary element method (BEM). Briefly, in BEM, the boundaries of domains are discretized, and field equations are solved on the boundaries. By determining the field on the boundary of a domain, one can obtain the field at any point inside the domain by the Green's function. The result obtained from the reduced model compared to the results for the eddy current model (implemented by FEM) revealed that the reduced model is a good approximation for biomedical applications. Figure 2.9 shows the mesh used to discretize the boundary of the thorax and lungs, which consists of 13076 boundary elements and 7548 boundary nodes.

In 2021, Tillieux and Goussard [64], with the aim of reducing the computational complexity of the biomedical MIT forward problem, proposed a method based on the volume integral equation (VIE) technique. Generally, methods like FEM require the discretization of a domain much larger than the actual object to avoid numerical artifacts at the interfaces, which increases the computational cost. The VIE-based method proposed in [64] only requires the discretization of the volume of interest, which reduces considerably the number of unknowns and results in a lower computational cost. The forward model was validated with a scattering theory. Results showed that the complexity of the proposed model was comparable to FEM, but since the number of unknowns was greatly reduced, the total computational cost was improved

Table 2.1 Comparison between the main methods used for solving the biomedical MIT forward problem

| Method | Advantages | Disadvantages |
|---|---|---|
| Analytical | ✓ Very fast<br>✓ Exact solution | ✗ Hard to obtain<br>✗ Only applicable to simple and symmetric geometries |
| FEM | ✓ Allows the modeling of complex geometries<br>✓ Often more accurate than FDM<br>✓ The most popular method for solving biomedical MIT forward problem | ✗ Approximates solution<br>✗ Requires domain mesh<br>✗ Higher computational costs<br>✗ Generation of mesh for complex geometry is time-consuming |
| FDM | ✓ Easy to implement<br>✓ Lower computational costs<br>✓ Easy and fast mesh generation | ✗ Approximates solution<br>✗ Requires domain mesh<br>✗ Not appropriate for unstructured mesh<br>✗ Less accurate compared to FEM |

significantly.

At the end of this section, we summarize the advantages and disadvantages of the main methods used for solving the biomedical MIT forward problem in Table 2.1.

## 2.5   Review of the methods for solving inverse problem in biomedical MIT

Image reconstruction in MIT is performed via the solution of an inverse problem to estimate conductivity distribution inside an object. The MIT inverse problem usually is nonlinear, ill-posed, ill-conditioned, and underdetermined.

The *nonlinearity* in the problem arises from the fact that the sensitivity of the measurements to the conductivity distribution is a function of position in the imaging space. The sensitivity is high in the periphery of imaging space but dramatically decreases by going towards the center. Furthermore, a local change in conductivity distribution affects all measurement channels.

The *ill-posedness* is explained based on the Hadamard definition, which says a problem is well-posed if it has all of three conditions:
a)  Existence: a solution must exist for the problem.
b)  Uniqueness: the solution must be unique.
c)  Stability: the solution depends in a continuous (stable) way on the observed data.

If the problem violates any of these requirements, it is ill-posed. Often for inverse problems, including MIT, the first two criteria could be met; however the third condition is problematic. Since small changes in measured data can often leads to a considerable change in reconstructed conductivity.

Another difficulty arises when the MIT inverse problem is discretized. This is quantified by the condition number of the Jacobian matrix, which appears in the discretized inverse problem. The condition number represents the ratio of the maximal to the minimal singular value of the Jacobian matrix. For the MIT inverse problem, the singular values of the Jacobian show a significant decay towards zero giving a large condition number. A problem with the large condition number said to be *ill-conditioned*.

In MIT, usually, the number of unknown conductivity elements is much greater than the number of measurements. This causes the MIT inverse problem to be *underdetermined*.

Algorithms used to solve the MIT inverse problem can generally be grouped into linear single-step and iterative algorithms. In the following, we review algorithms used for solving the biomedical MIT inverse problem.

## 2.5.1 Linear Single-step algorithms

Solution of nonlinear inverse problems involves employing nonlinear reconstruction algorithms, which require high computational expenses. To resolve this difficulty in the electrical tomography field, the inverse problem is linearized. Thus, the nonlinear problem is turned into a linear approximation which can be solved by the matrix calculation. These algorithms are based on the assumption of a slight difference between the true conductivity distribution and the given initial conductivity distribution.

Single-step linear algorithms are fast and suitable for real-time imaging, but they are often capable of producing qualitative images. They are generally used in time- or state-difference imaging to detect perturbations within a conductive background. The linear back-projection method, Tikhonov regularization method, and truncated singular value decomposition (TSVD) method are examples of the single-step linear algorithms used in the biomedical MIT.

### 2.5.1.1 Linear back-projection

Linear back-projection is a fast and straightforward method that is generally used for hard-field tomography modalities. Applying this method to soft field tomography often causes blurred images. This method is based on the linearization of the forward problem map between the conductivity distribution and the induced voltages. The linearized map is obtained by calculating the Jacobian matrix, which converts the conductivity distribution to an approximation of the induced voltages. To estimate the conductivity distribution, it is only required to calculate the inverse of the Jacobin matrix. However, the Jacobin matrix in MIT is ill-conditioned, and the system of equations is also underdetermined. Thus, this matrix cannot be directly inverted. In the linear back-projection, the transpose of the Jacobian matrix is considered as the inverse mapping from the induced voltages to the conductivity distribution [3].

Al-Zeibak and Sunders [65], who developed the first biomedical MIT system in 1993, used the back-projection algorithm to solve the MIT inverse problem. They aimed to determine human body composition non-invasively and in-vivo. In 2000, Korjenevsky et al. [66] employed the back-projection algorithm to reconstruct the conductivity distribution for biomedical applications. Figure 2.10 shows an example of reconstructed conductivities in this study. In 2017, the same research group studied the feasibility of using the back-projection method for the reconstruction of low contrast perturbations in a conducting background in MIT [61]. The results proved the applicability of back projection in biomedical MIT.

Simplicity and high speed are the main advantages of the back-projection algorithm. However, the quality of the reconstructed images by the algorithm is low due to large approximation errors, which cause artifacts in the images.

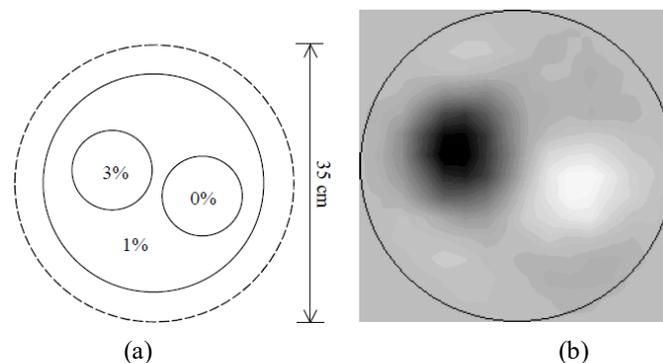

(a)  (b)

Figure 2.10 Example of the reconstructed conductivity by the linear back projection method in [66]. (a) True conductivity distribution (saline concentration by weight is shown), (b) Reconstructed conductivity.

### 2.5.1.2 Tikhonov regularization method

The regularization tools have been developed to solve the ill-posed inverse problems. Regularization methods enforce prior information to the problem leading to more stable approximate solutions. Tikhonov method is one of the most commonly used regularization methods for solving ill-posed inverse problems. As mentioned earlier, the Jacobian matrix in MIT cannot be directly inverted. Thus, Tikhonov regularization method is applied to the problem and obtained an inverse for the Jacobian matrix. This method is called the Tikhonov regularization reconstruction algorithm. The performance of this method is highly dependent on the choice of a regularization parameter. A small regularization parameter yields a more accurate solution; however, a very small parameter can result in an unstable problem. On the other hand, a large parameter increases the solution error and causes over-smoothed images.

In 2005, Merwa et al. [67] used the Tikhonov regularization algorithm to reconstruct the changes in the complex conductivity distribution inside biological tissues. In 2006, Scharfetter et al. [54] employed this reconstruction algorithm to study theoretical limits of spatial resolution and contrast to noise ratio of biomedical MIT. Two years later, Vauhkonen et al. [68] developed an MIT system to reconstruct the conductivity of biological tissues in which the Tikhonov regularization method was used as the inverse problem solver.

In 2012, Wei and Soleimani [69] developed an MIT system for biomedical applications in which they employed the Tikhonov regularization algorithm for solving the inverse problem. In 2016, Han et al. [45] also used the Tikhonov regularization algorithm to reconstruct the conductivity for biomedical application and compared the result with other algorithms.

In 2019, Xiao et al. [16] solved the inverse problem by the Tikhonov regularization algorithm to image ICH through a cambered MIT. In 2021, Chen et al. [17] also used this algorithm for local detection of ICH in a combined planar MIT system. Tikhonov regularization algorithm has been used in numerous studies for detecting cerebral hemorrhage [11], [12], [14], [47], [70], [71], as well. The serious disadvantage of the Tikhonov regularization algorithm is over-smoothing of the reconstructed image, which results in blurred boundaries between the target object and the background. This phenomenon can be resolved to a certain degree by setting a threshold value on the reconstructed images as proposed in [45].

### 2.5.1.3 Truncated singular value decomposition (TSVD) method

Singular value decomposition (SVD) is a powerful tool in linear algebra to study matrices. SVD of the MIT Jacobian matrix can reveal the ill-posedness of the inverse problem and ill-conditioning and rank-deficiency of the matrix. An almost linear decay of the singular values of the Jacobian matrix shows the problem is very ill-posed. Usually, there is a distinguishable gap between large and small singular values, which means the Jacobian matrix is rank-deficient and the matrix columns are linearly dependent. It is also caused to a large condition number for the Jacobian matrix which shows its ill-conditioning nature.

Truncated singular value decomposition (TSVD) is a form of regularization which is popular in biomedical MIT [21], [43], [54], [67], [72]–[74]. By means of this method, the ill-conditioning of the Jacobian matrix is treated by explicitly truncating the singular values at the level at which the erroneous singular values settle or at the noise level of the measurement data. As an example, Gursoy and Scharfetter used this method to investigate the effect of anisotropic muscle tissues that cover the rib cage on MIT image of the chest [74], to evaluate the effect of receiver coil orientations on the imaging performance of biomedical MIT [43], and to continuously monitor lung function using a planar coil MIT system [21].

### 2.5.2 Iterative algorithms

The nonlinear nature of the MIT inverse problem causes the low-quality images reconstructed by the linear single-step algorithms. Solving the inverse problem by the iterative algorithm improves the quality of images. For this reason, iterative reconstruction algorithms are often employed in biomedical MIT systems. These algorithms usually act based on the minimization of a functional of the squared difference between measured and simulated voltages. The simulated voltages are obtained from solving the forward problem fed by the last estimation of the conductivity distribution. Iterative algorithms are generally computational demanding and time-consuming which are recognized as the main drawback of these methods. One can classify these algorithms into linear and nonlinear groups. Generally, linear methods are more suitable for difference imaging (time-, state-, or frequency-difference), while the nonlinear ones are proper for absolute imaging.

#### 2.5.2.1 Landweber algorithm

Landweber algorithm is a first-order linear iterative method which has been used to solve ill-posed inverse problems. This algorithm is a special case of the gradient descent method in which the opposite direction to the gradient of the forward map at the current point is chosen as the search direction. The Landweber algorithm employs only the first-order derivative of the forward map specified by the Jacobian matrix. On the other hand, the Jacobian matrix is not updated during the iterations, and thus, it is classified as a linear iterative algorithm. These features reduce the computational demands for the Landweber algorithm. This method is often characterized by slow convergence. Furthermore, due to the linear nature of this algorithm, it might not be proper for highly nonlinear problems as in those including complex distributions of conductivity.

As an example of using the Landweber algorithm in biomedical MIT, in 2009, Zolgharni et al. [46] employed this algorithm to image cerebral hemorrhage. They compared the results for the Landweber algorithm to the Tikhonov regularization method. Figure 2.11 shows an example of reconstructed conductivity distributions for different phase noise in this study. As can be seen, the Landweber algorithm shows superior performance over the Tikhonov algorithm.

#### 2.5.2.2 Regularized Gauss-Newton algorithm

Regularized GN algorithm is a second-order nonlinear iterative algorithm which is widely used in the biomedical MIT. This algorithm, unlike the Landweber method, employs both first-and second-order derivatives of the forward map. In nonlinear algorithms, the forward model is updated, and the Jacobian matrix is recalculated in each iteration. This offers improvements over equivalent linear reconstruction algorithms at the cost of increasing computational expenses. This algorithm is also known as the "modified Newton-Raphson method" [75] or the "damping Gauss-Newton method" [76] in the literature.

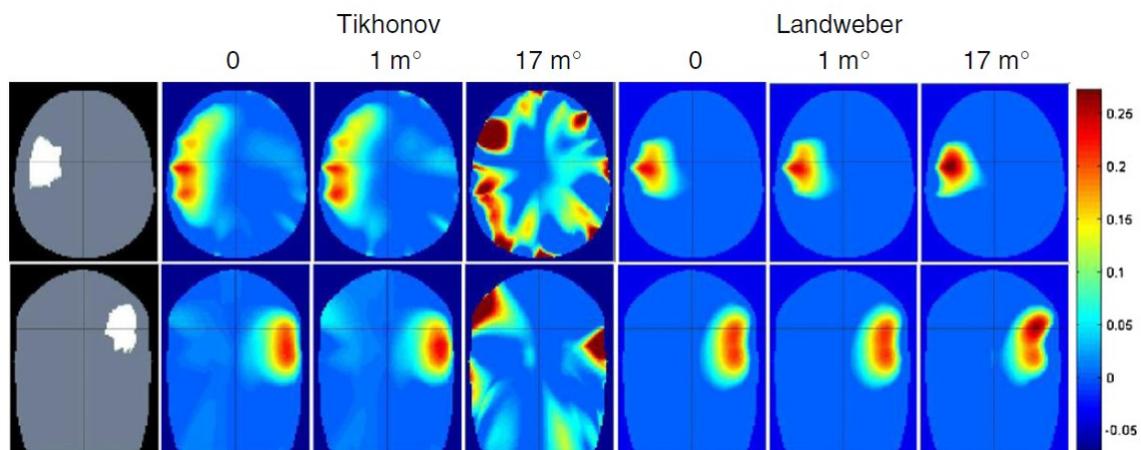

Figure 2.11 The reconstructed conductivity images from a large peripheral ICH [46]. The first column shows the true location of the simulated ICH. Other columns show the reconstructed conductivity distribution by Tikhonov regularization and Landweber algorithms for various noise levels.

In 2006, Soleimani and Lionheart [41] employed the regularized GN algorithm to reconstruct conductivity ranges encountered in biomedical applications. The results demonstrated the advantages of the nonlinear reconstruction algorithm both in terms of locating inclusions and reconstruction of the absolute conductivity values. In 2010, Dekdouk et al. [76] also used the regularized GN algorithm to consider the possibility of using absolute MIT voltage measurements for monitoring the progress of a peripheral ICH in a human brain. For the regularization of the algorithm, they proposed a regularization operator based on prior structural information of the head tissues. They numerically demonstrated that the hemorrhage region could be recovered provided that the prior information containing the distribution of the head tissues is accurate.

In 2014, Liu et al. [13] used the regularized GN algorithm to study the feasibility of imaging the low-contrast perturbation and small volume object in human brains. They evaluated the algorithm by a practical three-layer brain phantom. Then, they demonstrated that all of the conductivity perturbation objects in the phantom could be clearly distinguished in the reconstructed images. In 2020, the same research group employed the regularized GN algorithm to verify the feasibility of in vivo MIT [77]. They conducted an animal study with abdominal subcutaneous injection of conducting solution and showed the feasibility of in vivo application for MIT. In 2016, Han et al. [45] also employed the regularized GN algorithm in comparison with two other reconstruction algorithms for biomedical MIT.

In the regularized GN algorithm, it is required to choose a regularization parameter which usually imposes a difficulty to the user. In 2016, Dekdouk et el. [78] presented an adaptive version of the GN algorithm in which an adaptive process was applied to tune this parameter during iterations with the intent to improve the algorithm convergence. The process was automated and low time-consuming. In this study, the performance of two other nonlinear algorithms named the Levenberg-Marquardt and Powell Dog Leg method was evaluated as well for biomedical MIT.

In 2020, Tillieux and Goussard [64] employed an algorithm called quasi-Newton method to reconstruct conductivity images in biomedical MIT. This method is an alternative to Newton's method and is applicable where the Jacobian or Hessian matrix is unavailable or is too expensive to compute at every iteration. This method has lower computational costs per iteration, like the gradient-based algorithm, and also has a good convergence rate, like the Newton-based algorithm.

Figure 2.12 shows a general block diagram for a nonlinear iterative reconstruction algorithm in MIT. As can be seen, the forward problem is an integral part of the conductivity reconstruction by MIT. Concerning difficulties of the MIT inverse problem such as ill-posedness and ill-conditioning, it is crucial to employ a highly accurate forward problem model to produce simulated voltages which can be fitted as well as possible to the measured data.

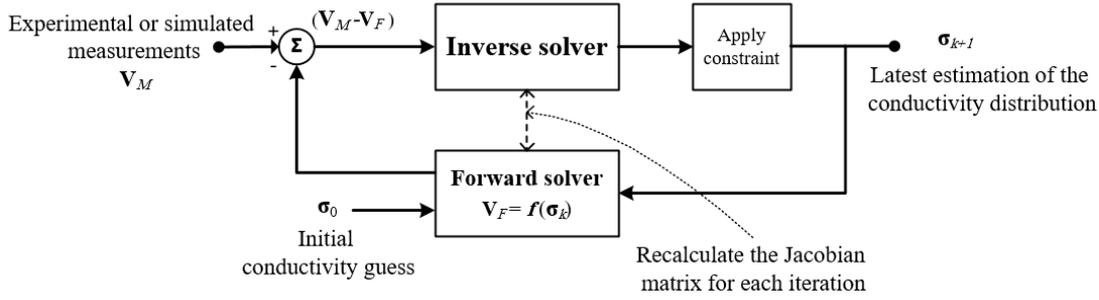

Figure 2.12 A typical block diagram for a non-linear iterative algorithm. The algorithm is initiated by a guess for the conductivity distribution ($\sigma_0$). With regard to prior information about the problem, it is possible to apply some constraints on the reconstructed conductivity. New estimated conductivity ($\sigma_{k+1}$) is fed to the forward problem ($V_F = f(\sigma_k)$). The Jacobian matrix is recalculated at each iteration by the forward solver and used in the inverse problem at the same iteration. The output of the forward problem ($V_F$) is subtracted from true measurements ($V_M$) and the result is fed to the input of the inverse solver. Then, the next update for the conductivity is obtained. This iterative process continue to reach a predefined error level or number of iterations.

### 2.5.3 Other methods

So far, we have introduced the classical algorithm for conductivity reconstruction in biomedical MIT. In recent years, in order to improve the reconstruction accuracy and achieve fast imaging in biomedical applications, artificial intelligence algorithms have been employed in some studies. In 2019, Chen et al. [15] proposed a novel reconstruction algorithm for biological tissue imaging by MIT. This fast algorithm was based on a stacked auto-encoder neural network composed of a multi-layer automatic encoder. The superiority of the algorithm was verified through comparison with the back-projection algorithm. In 2020, the same group employed four deep learning networks, including restricted Boltzmann machine, deep belief network, stacked autoencoder, and denoising autoencoder to solve the nonlinear reconstruction problem of biomedical MIT [79]. Results showed that in comparison with the back-projection algorithm, the deep learning networks improve the accuracy and the speed of conductivity reconstruction.

In 2021, a complex convolutional neural network and generative adversarial network-based algorithm was proposed to predict the object location and shape simultaneously in biomedical MIT [80]. The algorithm was compared to the Newton-Raphson method, optimized fully-connected network, and stacked autoencoder. The experimental results on practical data showed the proposed algorithm had superior performance over other methods. A comparison with traditional reconstruction algorithms indicated that these new methods could significantly improve the spatial and temporal resolution of MIT by fast and autonomous learning the non-linear relationship between input and output data.

Table 2.2 summarizes the advantages and disadvantages of the main algorithms used for solving the biomedical MIT inverse problem.

## 2.6 MIT studies in Iran

The only study conducted in Iran to build an MIT system was done by Yousefi et al. [81]–[83] at K. N. Toosi University of Technology. This system was named MIT-KNTU1 and was a 2D magnetic induction tomography. The imaging space diameter was 10 cm, and its height was 20 cm. In this system, 16 dual coils were employed for excitation and sensing. In excitation mode, dual coils were in series, and in sensing mode, dual coils acted as a gradiometer. They have employed a dual-frequency phase-sensitive demodulation technique to improve voltages measurement by sensing coil [81]. The excitation frequency was 10 kHz which made the system suitable for industrial MIT applications. From the coil shape point of view, MIT-KNTU1 was unique in its kind. In other words, this system was only 2D system which employed rectangular-shaped coils with long height. Figure 2.13 shows this system. The forward problem in MIT-KNTU1 has been solved by both FEM and combined wavelet-based mesh free-finite element method [82]. In this research, for the first time, a mesh-free method has been employed for solving MIT forward problem. Furthermore, the inverse problem has been solved by regularized GN algorithm.

Table 2.2 Comparison between the main algorithms used for solving the biomedical MIT inverse problem

| Algorithm type | Example | Advantages | Disadvantages |
|---|---|---|---|
| Linear single-step | • Linear back projection<br>• Tikhonov regularization<br>• Truncated SVD | ✓ Fast and suitable for real-time imaging<br>✓ Simple to implement | ✗ Inappropriate for absolute value conductivity reconstruction<br>✗ Blurred and low quality images<br>✗ Large approximation errors |
| Linear iterative | • Landweber | ✓ Better image quality than single-step algorithms<br>✓ Lower computational cost compared to nonlinear algorithms | ✗ More suitable for difference imaging<br>✗ Inappropriate for highly nonlinear problems |
| Nonlinear iterative | • Regularized GN | ✓ Highest image quality<br>✓ Appropriate for absolute imaging | ✗ Computational demanding and time consuming |

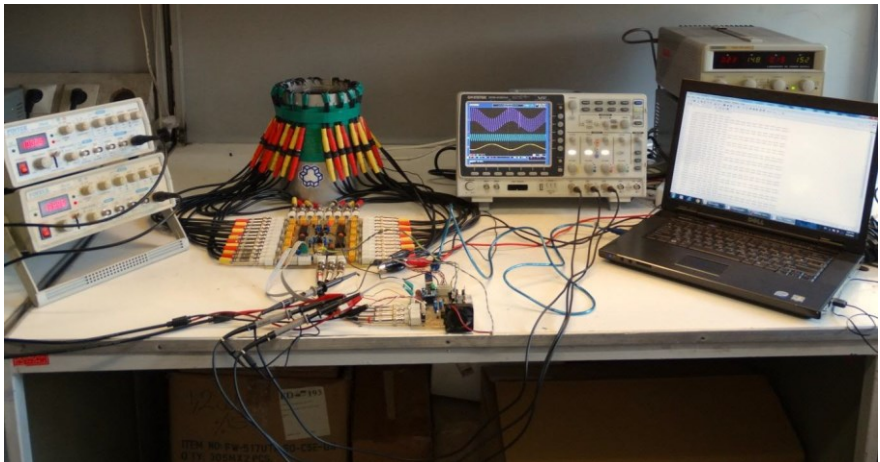

Figure 2.13 MIT-KNTU1 system [83].

## 2.7 Conclusions

In this chapter, the biomedical MIT principles were illustrated, the relation between the induced voltage and the conductivity to be imaged was clarified, and the hardware and software parts for a typical biomedical MIT system were presented. Then, we systematically reviewed the studies on the MIT forward problem for biomedical applications. As mentioned earlier, in all of these studies, the simplified Maxwell's equations have been used, which ignore skin and proximity effects inside the coils regions. In this thesis, we aim to employ a new formulation for the forward problem in which skin and proximity effects are considered. Since FEM is the most popular method in biomedical MIT; we solve the new forward problem by this numerical method.

Almost in all previous studies on the biomedical MIT, the forward problem has been solved by mesh-based methods. The results obtained from these methods highly depend on the used mesh. Only in one study in our group, a wavelet-based mesh-free method has been employed for solving the industrial MIT forward problem [82]. In this thesis, we focus on the biomedical MIT and use another mesh-free method to solve the new forward problem.

We also reviewed the studies on the biomedical MIT inverse problem in this chapter. As seen, the regularized GN algorithm is the most popular method to reconstruct absolute conductivity values. Therefore, we employ this method to evaluate the new forward problem model compared to the previous one for conductivity image reconstruction. Furthermore, we adapt the regularized GN algorithm to develop a technique for the mid-range conductivity image reconstruction.

# Chapter 3: MIT Forward Problem I

The low conductivity MIT, including biomedical MIT, requires the improvement of both software and hardware components in order to acquire more reliable imaging data. As shown in Chapter 2, the forward problem solution is an integral part of the inverse problem solution and image reconstruction. To achieve a reconstructed image with a desired degree of accuracy, it is crucial to provide an accurate solution to the forward problem; since small errors in the forward model can generate considerable errors in the solution of the inverse problem. Furthermore, accurate forward computation is vital to the process of MIT hardware design.

As pointed in Chapter 2, the secondary magnetic field induced by the eddy currents in biomedical MIT is generally much smaller than the primary magnetic field. A straightforward possibility to strengthen the magnitude of the secondary magnetic field would be to increase the operating frequency (see eq. (2.1)). A wide range of frequencies has been used in the biomedical MIT, which includes 20 kHz [24], 100 kHz [53], 1 MHz [11], 10 MHz [64], and 13.56 MHz [77]. However, by increasing excitation frequency, skin and proximity effects become more noticeable in conductors exposed by alternative magnetic fields and cannot be ignored [30]. It means, unlike static cases, the current densities in coils are no longer constant or independent of position within conductors and position of coils relative to each other. Considering skin and proximity effects in MIT coils becomes much more important when the parallel excitation mode as in [84] is used and coils as in [16] become closer to each other.

In general, three formulations have been used for solving eddy current and skin effect problems in the literature [30]. In the first formulation, the TCD in a current-carrying conductor equals the total current divided by conductor cross-sectional area in which a constant space function is considered for TCD and skin and proximity effects are ignored. The second formulation is based on an incomplete equation for source current density. The third formulation, which is based on Maxwell's equations, considers a complete equation for the source current density and a position-dependent function for TCD. Using the third formulation, skin and proximity effects are accurately considered in eddy current and skin effect problems.

Previous studies in MIT forward problem used the first formulation for solving the governing equations [16], [59], [62], [85]. In [85], it has been assumed there is no eddy current neither in the excitation coil region nor in the sensing coil region. Here, we call this method as *early method*. In this thesis, for the first time, the third formulation has been used for modeling of TCD in the MIT coils in an effort to solve more accurately the forward problem. This method is referred to as *improved method* in this thesis. The improved method has been adopted previously for modeling electromagnetic acoustic transducers (EMATs) [86], [87] employed in non-destructive testing. However, the situation is entirely different in MIT; because it uses an array of coils aimed at imaging the medium conductivity.

We implement the improved method for 2D MIT forward problem. Ideally, the forward problem in MIT must be solved in 3D; however, it requires extensive computational and storage resources. Obviously, if one aims at implementing the improved method and considering skin and proximity effects in 3D, the situation will be more challenging [62]. This is the reason why it would be reasonable to use specific object geometry and coil configuration to be able to solve a 3D problem as a 2D one. Figure 3.1 shows the specific object geometry and coil configuration. In fact, we study on a cross section of a 3D problem which has infinite length perpendicular to this cross section. Some researchers have employed analytical solutions to solve the 2D MIT forward problem and ignored skin and proximity effects in coils [88], [89]. Additionally, some other researchers have solved 2D MIT forward problem by using the FEM and considering the early method [14], [78].

This chapter is organized as follows: In Section 3.1, the MIT forward problem formulation is introduced, and equations for the early and improved methods are presented. The procedure of computation of the induced voltage in the biomedical MIT is next explained. In Section 3.2, the numerical implementation of the improved method is first validated with a simple analytical test problem. Then, in Section 3.2.2, through modeling a 2D 16-coil MIT system and using a circular phantom, the solution of the forward problem based on the improved method will be investigated and compared to that of the early method. After that, the results obtained for the circular phantom are discussed and analyzed. Then, through using a 2D head phantom, the solution of the forward problem based on the early and improved forward methods will be inspected in a biomedical application. Finally, Section 3.3 presents the conclusion.

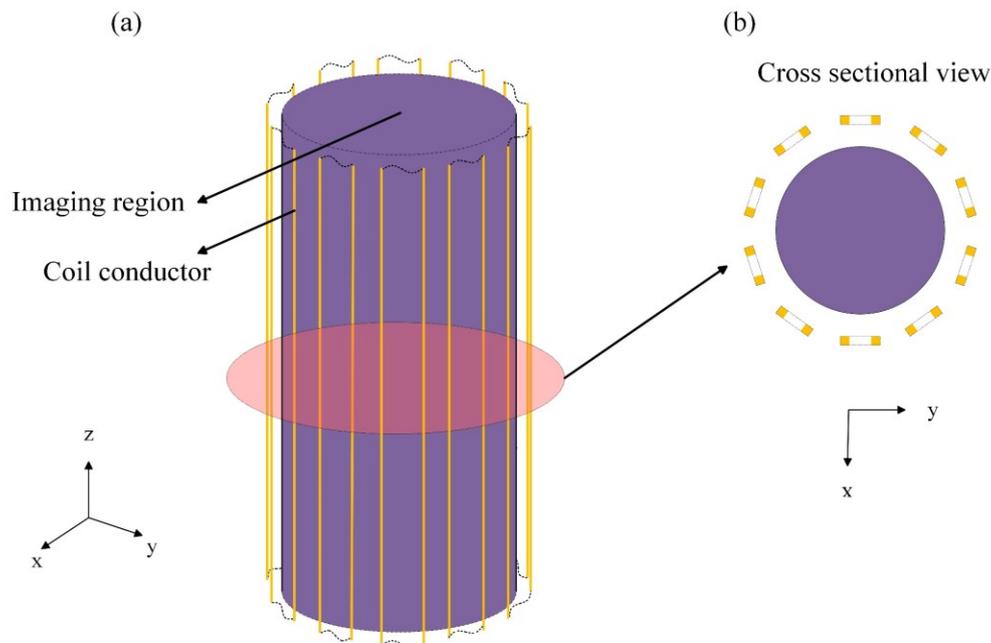

Figure 3.1 (a) 3D representation of an MIT system which can be considered as a 2D problem. The conductivity distribution is invariant along the $z$-axis. The current passes through coils' conductor in $z$ direction. (b) The cross sectional view of the 3D problem.

## 3.1 2D MIT Forward Problem Formulation

The forward problem in MIT includes a classical eddy current problem. A wide range of industrial and medical applications of MIT use the magneto-quasi-static (MQS) approximation [90] in solving eddy current problems. Under MQS approximation, for a given conductivity $\sigma$, permittivity $\varepsilon$, and angular frequency $\omega$, it is assumed that $\omega\varepsilon/\sigma \ll 1$; that is the displacement current can be neglected. To obtain MIT forward problem formulation, we start from time-harmonic Maxwell's equations under magneto-quasi-static approximation:

$$\nabla \times \vec{E} = -j\omega\vec{B} \quad (3.1)$$

$$\nabla \times \vec{H} = \vec{J} \quad (3.2)$$

$$\nabla \cdot \vec{D} = 0 \quad (3.3)$$

$$\nabla \cdot \vec{B} = 0 \quad (3.4)$$

where $\vec{E}$ is the electric field intensity, $\vec{B}$ is the magnetic flux density, $\vec{H}$ is the magnetic field intensity, $\vec{J}$ is the TCD of current-carrying conductors, $\vec{D}$ is the electric flux density, and $j = \sqrt{-1}$. Assuming that materials have linear and isotropic electrical and magnetic properties, then $\vec{B} = \mu\vec{H}$, $\vec{D} = \varepsilon\vec{E}$, and $\vec{J} = \sigma\vec{E}$ where $\mu$ is the magnetic permeability. As mentioned in Chapter II, it is assumed that $\mu_r = 1$ in most biomedical applications. Thus, we consider $\mu = \mu_0$ throughout this thesis except otherwise specified.

Let the MVP be defined as follow:

$$\vec{B} = \nabla \times \vec{A} \quad (3.5)$$

Using the constitutive relations $\vec{B} = \mu\vec{H}$ and substituting (3.5) into (3.2) gives:

$$\frac{1}{\mu_0}\nabla \times (\nabla \times \vec{A}) = \vec{J} \quad (3.6)$$

Substituting (3.5) into (3.1) yields the following relationship between the electric field intensity $\vec{E}$ and the vector potential $\vec{A}$,

$$\vec{E} = -j\omega\vec{A} - \nabla\Phi \quad (3.7)$$

where $\Phi$ is a scalar potential function. The total current density can be considered as

$$\vec{J} = \vec{J}_s + \vec{J}_e \quad (3.8)$$

where $\vec{J}_s$ is named the source current density and $\vec{J}_e$ is the eddy current density vector and they are related to potentials as follows:

$$\vec{J}_s = -\sigma\nabla\Phi \quad (3.9a)$$

$$\vec{J}_e = -j\omega\sigma\vec{A} \quad (3.9b)$$

It is worthwhile noting that neither $\vec{J}_s$ nor $\vec{J}_e$ is not evaluated physically; they are mathematical quantities. Only the total current of the coil, $i(t)$, is a measurable quantity [91].

In 2D MIT modeling, it is assumed that electric currents flow only in the $z$ direction. Consequently, only the $z$-component of $\vec{A}$ and $\vec{J}$ are considered. Assuming $\vec{A} = A_z(x,y)\,\vec{a}_z$ and $\vec{J} = J_z(x,y)\,\vec{a}_z$, and using a vector identity[1] and Coulomb gauge $\nabla \cdot \vec{A} = 0$, (3.6) is simplified

---

[1] $\nabla \times (\nabla \times \vec{A}) = \nabla(\nabla \cdot \vec{A}) - \nabla^2\vec{A}$

to:

$$\frac{1}{\mu_0} \nabla^2 A_z(x,y) = -J_{zk}(x,y) \tag{3.10}$$

where $J_{zk}$ is the TCD of the $k$-th conductor and is related to the total current of $k$-th conductor by [30]:

$$\iint_{R_k} J_{zk}(x,y)\,\mathrm{d}s = I_k \tag{3.11}$$

where $R_k$ and $I_k$ denote cross-section region and phasor current of $k$-th current-carrying conductor, respectively. Furthermore, (3.8) and (3.9) gives:

$$J_{zk}(x,y) = -\mathrm{j}\omega\sigma\, A_z + J_{sk}. \tag{3.12}$$

Before moving to the next section, we summarized approximations used to define the forward problem in this thesis:
a- MQS approximation [90] which is obtained by neglecting the displacement current term in the Ampere-Maxwell's law and resulting Ampere's law, (3.2).
b- 2D approximation which is valid for a specific object geometry and coil configuration (Figure 3.1).

### 3.1.1 Early forward method

In previous studies, the MIT forward problem region $\Omega$, as shown in Figure 3.2(a), has been divided into two sub-regions: conducting or eddy current region $\Omega_C$ and non-conducting region $\Omega_N$ such that $\Omega = \Omega_C \cup \Omega_N$ [29]. In this model, $\Omega_C$ is the imaging region of interest and $\Omega_N$ includes the MIT coils and air. The field in $\Omega_N$ is treated as a static problem; i.e., no eddy current is considered in coils (sources), $J_e = \mathrm{j}\omega\sigma A_z = 0$ in $\Omega_N$, and $J_{zk}$ in (3.10) is constant inside the current-carrying conductor. Thus, $J_{zk}$ can be obtained from (3.11) as

$$J_{zk} = \frac{I_k}{S_k} \tag{3.13}$$

where $S_k$ is the cross-sectional area of the $k$-th current-carrying conductor. By substituting (3.13) into (3.10), the equation for $\Omega_N$ becomes:

$$\frac{1}{\mu_0} \nabla^2 A_z(x,y) = -J_{zk} = -\frac{I_k}{S_k} \quad \text{in} \quad \Omega_N. \tag{3.14}$$

The equation for $\Omega_C$ is obtained by substituting (3.12) into (3.10) and knowing that $J_{sk} = 0$ in $\Omega_C$:

$$\frac{1}{\mu_0} \nabla^2 A_z(x,y) - \mathrm{j}\omega\sigma\, A_z(x,y) = 0 \quad \text{in} \quad \Omega_C. \tag{3.15}$$

Equations (3.14) and (3.15) form the forward problem formulation in 2D MIT based on the early method which has been used in the literature [14], [45].
The discretized FE equivalents of (3.14) and (3.15) are obtained from the following equation [19]:

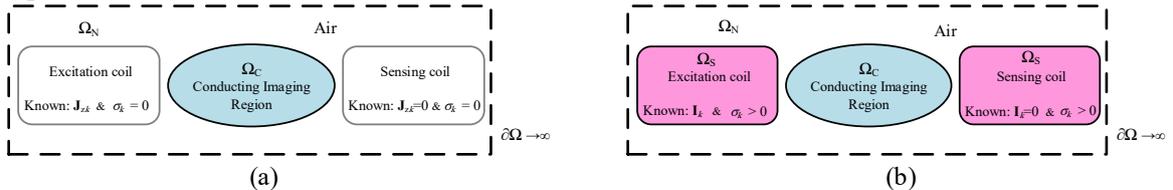

Figure 3.2. Schematic view of the forward problem (a) based on the early method [29] and (b) based on the improved method; $\Omega_C$: conducting region; $\Omega_N$: non-conducting region; $\Omega_S$: source region; $\partial\Omega$: truncation boundary; $J_{zk}$: phasor TCD; $I_k$: phasor current; $\sigma_k$: conductivity of the $k$-th current-carrying conductor.

$$\left(\frac{1}{\mu_0}\mathcal{S} + j\omega\sigma\kappa\mathcal{T}\right)\mathcal{A} = \mathcal{T}\mathcal{J}. \tag{3.16}$$

Here, the column matrices $\mathcal{A}_{r\times 1}$ and $\mathcal{J}_{r\times 1}$ are the phasors of FE node potentials and node total current density values, respectively, and $r$ is the number of interpolation nodes. The square matrices $\mathcal{S}_{r\times r}$ and $\mathcal{T}_{r\times r}$ are the usual FE coefficient matrices. Note that constant $\kappa$ in (3.16) is equal to zero for the FE elements in $\Omega_N$ and is unity for the FE elements in $\Omega_C$ in Figure 3.2(a).

The equivalent circuit model for coils surrounded by air in the early forward method has been shown in Figure 3.4(a). The inductance $L_0$ is that caused by the magnetic flux outside each conductor and $R_0$ is the DC resistance of each conductor.

### 3.1.2 Improved forward method

In the forward problem based on the improved method, the region $\Omega$ as shown in Figure 3.2(b) is partitioned into three sub-regions: source or coil region $\Omega_S$, conducting region $\Omega_C$, and non-conducting region $\Omega_N$; i.e., $\Omega = \Omega_S \cup \Omega_C \cup \Omega_N$. The improved method considers eddy currents and consequently skin and proximity effects in coils region. It means that $J_e(x,y) \neq 0$ in $\Omega_S$. Furthermore, in $\Omega_S$, we have the source current density $J_s$ which is constant over every cross-sectional surface of current-carrying conductors at each frequency [92]. In the following, it is shown that $J_s$ is constant for any given surface normal to the direction of current flow.

The electric scalar potential distribution $\Phi(z)$ along a straight conductor of arbitrary cross section
and of length L (see Figure 3.3) must satisfy the Laplace's equation [92]:

$$\frac{d}{dz}\sigma\frac{d}{dz}\Phi(z) = 0 \tag{3.17}$$

The solution is given in terms of the potential values at the end surface of the conductor at $z = 0$ and $z = L$, respectively. It result in [92]:

$$\Phi(z) = \frac{\Phi(L) - \Phi(0)}{L}z + \Phi(0) \tag{3.18}$$

From (3.9a) and (3.18), one obtain $J_s$ for the 2D problem as follows:

$$J_s = \sigma E_s = -\sigma\frac{d}{dz}\Phi(z) = \sigma\frac{\Phi(0) - \Phi(L)}{L} \tag{3.19}$$

Clearly, $J_s$ is constant. However, the magnitude of $J_s$ is not known since $\Phi(0)$ and $\Phi(L)$ are not given.

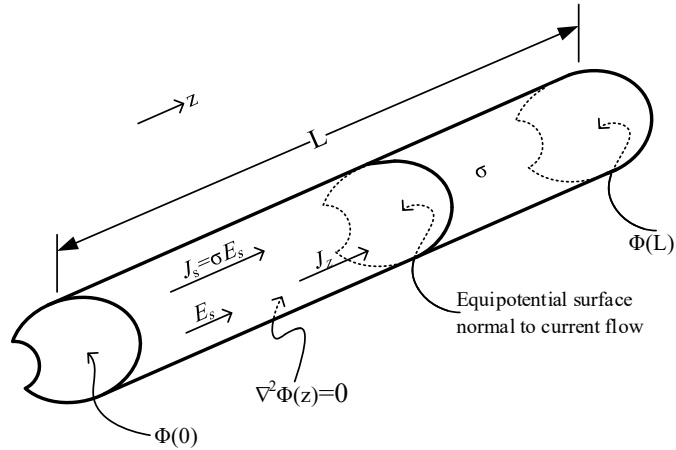

Figure 3.3. Straight conductor of arbitrary cross section and of length L with equipotential surfaces normal to direction of current flow. Redrawn from [92].

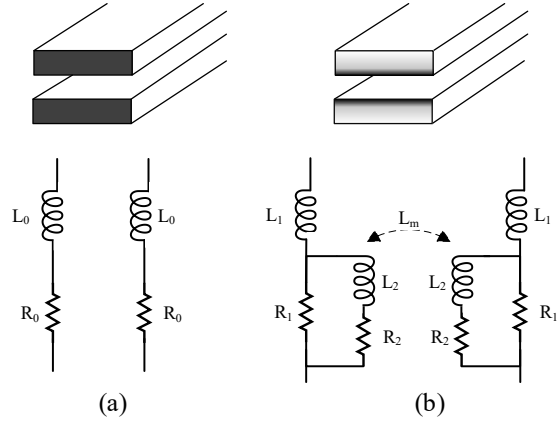

(a)          (b)

Figure 3.4. Equivalent circuit model for coils surrounded by air in (a) Early forward method and (b) Improved forward method. Each coil consists of two conductors. The circuit elements are explained in the text.

Knowing that $J_{sk}$ is constant, to determine its magnitude, one can integrate both sides of (3.12) over the $k$-th current-carrying conductor cross-section and use (3.11), as follows:

$$J_{sk} = \frac{I_k}{S_k} + j\omega \frac{\sigma_k}{S_k} \iint_{R_k} A_z(x,y) \, ds. \tag{3.20}$$

Substituting (3.20) in (3.11) gives:

$$J_{zk}(x,y) = \frac{I_k}{S_k} + j\omega \frac{\sigma_k}{S_k} \iint_{R_k} A_z \, ds - j\omega \sigma_k A_z(x,y) \tag{3.21}$$

where $\sigma_k$ is the conductivity of the $k$-th current-carrying conductor. As can be seen, (3.21) has a substantial discrepancy with (3.13). By substituting (3.21) into (3.10), the governing equation in $\Omega_S$ becomes:

$$\frac{1}{\mu_0} \nabla^2 A_z(x,y) - j\omega \sigma_k A_z(x,y) + j\omega \frac{\sigma_k}{S_k} \iint_{R_k} A_z(x,y) \, ds = -\frac{I_k}{S_k} \quad \text{in} \quad \Omega_S. \tag{3.22}$$

The equation for the regions $\Omega_N$ and $\Omega_C$ is the same as (3.15). Thus, (3.15) and (3.22) form the forward problem formulation in 2D MIT based on the improved method. To the best of our knowledge, this is a new formulation for modeling of MIT forward problem. This formulation considers skin and proximity effects in the source (coil) region.

The discretized FE equivalents of (3.15) and (3.22) are obtained from the following equation [30]:

$$\left[ \frac{1}{\mu_0} \mathcal{S} + j\omega\sigma(\mathcal{T} - \mathcal{Q}\mathcal{P}^{-1}\mathcal{Q}^T) \right] \mathcal{A} = \mathcal{Q}\mathcal{P}^{-1} I \tag{3.23}$$

where the rectangular matrix $\mathcal{Q}$ and diagonal matrix $\mathcal{P}$ are explained in [91].

The improved method does not require adding new nodes to the problem. Consequently, the size of the final FE system of equations is the same for both early and improved methods. Generally, the improved method does not considerably increase the computational complexity of the problem.

Figure 3.4(b) illustrates the equivalent circuit model for coils surrounded by air in the improved method. The inductance $L_1$ is that caused by the magnetic flux outside each conductor. The inductance $L_2$ and resistance $R_2$ are connected in parallel to $R_1$ to represent the skin effect [93]. The mutual inductance $L_m$ represents the proximity effect between the two conductors.

### 3.1.3 Boundary condition

The MIT forward problem is a boundary-value problem which can only be solved if accompanied by appropriate interface and boundary conditions. At the interface of two different material, the continuity of normal component of $\vec{B}$ and tangential component of $\vec{H}$ dictates [94]:

$$A_{z_1} = A_{z_2} \quad (3.24)$$

$$\frac{1}{\mu_1}(\nabla A_{z_1} \cdot \vec{n}) = \frac{1}{\mu_2}(\nabla A_{z_2} \cdot \vec{n}) \quad (3.25)$$

where $A_{z_1}$ and $A_{z_2}$ are MVPs at adjacent points on the interface between material 1 and 2, $\mu_1$ and $\mu_2$ are the permeability of material 1 and 2, respectively, and $\vec{n}$ is the outward unite vector from material 1. Boundary conditions, i.e., conditions at the extremities of the problem, are obtained by extending the interface conditions. Appendix 5 shows how (3.24) and (3.25) are extracted.

### 3.1.4 Induced voltages in sensing coils

The output of the forward problem is the induced voltages in sensing coils. The induced voltage $V$, for a single-turn coil, is computed by integrating around the sensing coil as follows:

$$V = -j\omega \iint_{S_{in}} \vec{B} \cdot d\vec{s} \quad (3.26)$$

where $S_{in}$ is the internal region bounded by the sensing coil. Using the definition of MVP ($\vec{B} = \nabla \times \vec{A}$) and Stokes' theorem ($\iint_{S_{in}} (\nabla \times \vec{A}) \cdot d\vec{s} = \oint_C \vec{A} \cdot d\vec{\ell}$), the induced voltage becomes:

$$V = -j\omega \oint_C \vec{A} \cdot d\vec{\ell} \quad (3.27)$$

where $C$ is a closed contour bounding the internal area of the sensing coil. Equation (3.27) in the 2D case can be rewritten as follows:

$$V = j\omega \oint_C A_z \, \vec{a}_z \cdot d\vec{\ell} \quad (3.28)$$

For a 2D coil, only the straight segments parallel to the $z$-axis contribute in (3.28). Consequently, the induced voltage becomes [95]:

$$V = j\omega l (A_{z,\mathbf{p}} - A_{z,\mathbf{q}}) \quad (3.29)$$

where $A_{z,\mathbf{p}}$ and $A_{z,\mathbf{q}}$ are the MVP of segments parallel to the $z$-axis located at $\mathbf{p}$ and $\mathbf{q}$ in the $x-y$ plane, respectively. The length of the segments is $l$.

As can be seen from (3.29), the induced voltage is complex-valued. Figure 3.5 shows a typical phasor diagram of the induced voltage in an MIT system where $I$, $V_0$, $V_1$ and $\Delta V$ are the phasors of excitation current, induced voltage when the imaging region is empty (background voltage), induced voltage when a conductivity distribution is placed in the imaging region, and the difference voltage $V_1 - V_0$, respectively. The voltages $V_0$, $\Delta V$ and $V_1$ are proportional to the primary magnetic field, secondary magnetic field, and total magnetic field, respectively. As indicated in Section 2.2, in the low conductivity applications of MIT, including biomedical applications, the real part of the induced voltage mainly changes with conductivities inside the imaging region, and the changes in the imaginary part of the induced voltage are negligible [44], [45]. It is worthwhile mentioning that in the high conductivity applications, the imaginary part of the induced voltage $Im\{V_1\}$ mainly changes with conductivities inside the imaging region, and the changes in the real part of the induced voltage are not considerable [48]. Since this thesis focuses on the biomedical applications of MIT, the real part of the induced voltage is obtained in the following sections.

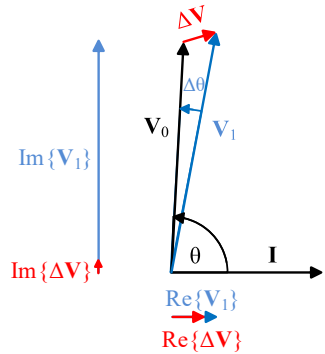

Figure 3.5. Phasor diagram of induced voltage in a typical MIT system in low conductivity application. The vectors are explained in the text.

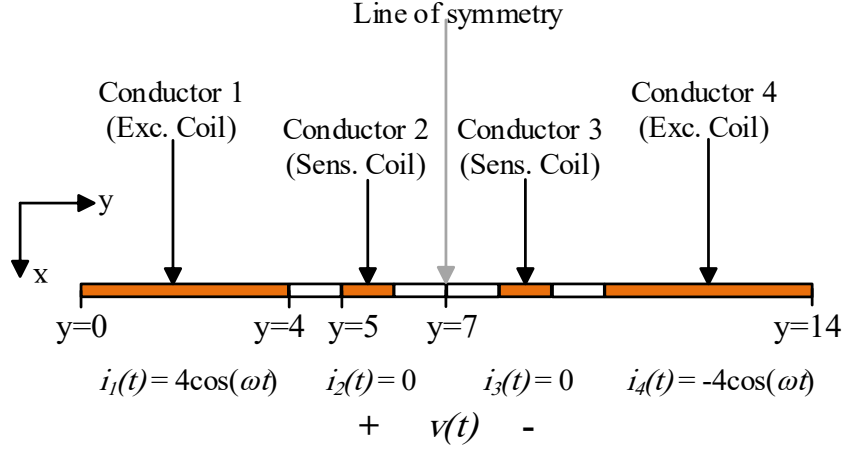

Figure 3.6. Test assembly: 1D symmetric geometry contains four conductors (two coils) surrounded by infinitely permeable iron. Note that the half of the region of interest employed in analytical solution. All dimensions are in mm.

## 3.2 Numerical experiments

### 3.2.1 Test problem: validation of the improved method

To validate the FE method implementation of the improved method, a one-dimensional (1D) test problem has been designed. Obtaining an analytical solution for 2D problems considering skin and proximity effects is an open issue. An analytical-numerical solution for the skin and proximity effects in two parallel round conductors was presented in [96]. The test assembly shown in Figure 3.6 corresponds to a 1D problem of two thin slot-embedded coils. The coils are surrounded by permeable iron with $\mu = \infty$. This problem models a one-channel MIT system in which the excitation coil is comprised of conductors 1 and 4 and the sensing coil contains conductors 2 and 3. The physical properties of the assembly are identical to those in [87]. Due to the symmetry of the problem, half of the region of interest is employed to calculate the analytical solution. The currents in the four conductors are given by $i_1(t) = 4\cos(\omega t)$ A, $i_2(t) = 0$ A, $i_3(t) = 0$ A, and $i_4(t) = -4\cos(\omega t)$ A. The conductivity and permeability of all conductors are taken $\sigma = 1$ S/m and $\mu = 1$ H/m. The analytical solution for the test problem by directly solving Maxwell's equations will be as follows (see Appendix 3):

$$A_z(y) = \begin{cases} C_1\left(e^{y\sqrt{j\omega}} + e^{-y\sqrt{j\omega}} - 0.002\right) + C_2 & 0 \leq y \leq 0.004 \\ -4y + 0.016 + C_3 & 0.004 \leq y \leq 0.005 \\ C_4\left(e^{y\sqrt{j\omega}} - e^{0.005\sqrt{j\omega}}\right) - C_5\left(e^{-y\sqrt{j\omega}} - e^{-0.005\sqrt{j\omega}}\right) + C_6 & 0.005 \leq y \leq 0.006 \\ -4y + 0.028 & 0.006 \leq y \leq 0.007 \end{cases}$$

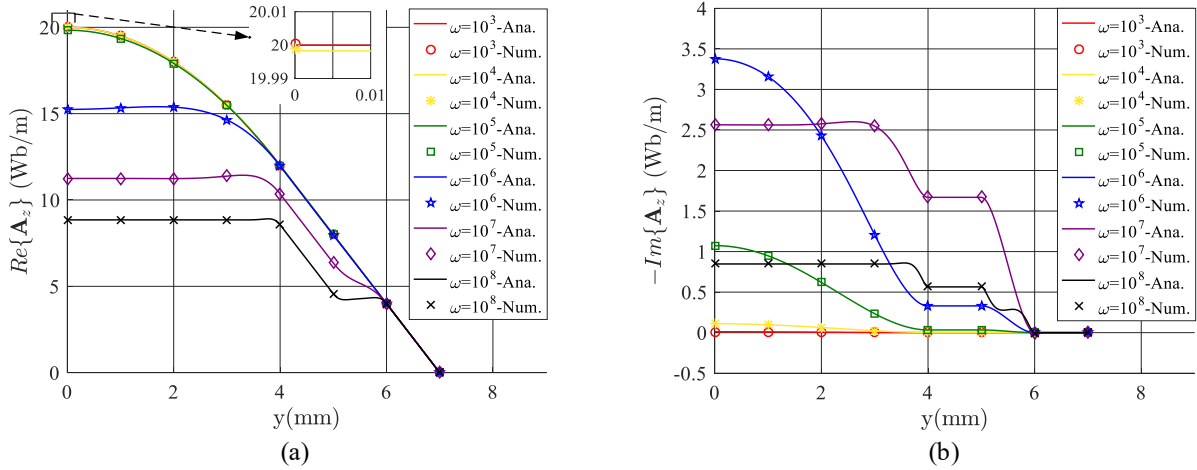

Figure 3.7. Test problem: (a) Real part and (b) Imaginary part of MVP distribution along with *y*-axis obtained from analytical solution of 1D problem (solid line) and 2D FE numerical solution of the improved method (markers) for different frequencies. The behavior of the imaginary part of MVP is explained in the text.

where

$$C_6 = 0.004 + 8\left(e^{0.001\sqrt{j\omega}} - 1\right)/\left(\sqrt{j\omega}\left(e^{0.001\sqrt{j\omega}} + 1\right)\right),$$

$$C_5 = -4e^{0.006\sqrt{j\omega}}/\left(\sqrt{j\omega}\left(e^{0.001\sqrt{j\omega}} + 1\right)\right),$$

$$C_4 = -4e^{-0.005\sqrt{j\omega}}/\left(\sqrt{j\omega}\left(e^{0.001\sqrt{j\omega}} + 1\right)\right),$$

$$C_3 = C_6 + 0.004,$$

$$C_2 = C_3 - C_1\left(e^{0.004\sqrt{j\omega}} + e^{-0.004\sqrt{j\omega}} - 0.002\right),$$

$$C_1 = -4/\left(\sqrt{j\omega}\left(e^{0.004\sqrt{j\omega}} - e^{-0.004\sqrt{j\omega}}\right)\right).$$

Due to the special boundary conditions of this test problem, the solution of a 2D problem, specified on the region $0 \leq x \leq 1$ mm and $0 \leq y \leq 7$ mm, is very close to the solution of the 1D test problem. The 2D problem was solved numerically by using the FE method implementation of (3.22).

Figure 3.7 compares MVP distribution along the *y*-axis obtained from the analytical solution of 1D problem (solid line) and 2D FE method implementation of the improved method (markers) for different frequencies. As can be seen, by increasing the frequency, the real part of $A_z$ decreases. However, the imaginary part of $A_z$ is zero at $\omega = 10^3$ rad/s (low frequency) and reaches a maximum at $\omega = 10^6$ rad/s and $y = 0$, and after that, it starts a decreasing behavior. Since the early method for modeling coils is frequency independent, the value of $A_z$ obtained by this method is constant for all frequencies as red lines in Figure 3.7. The phasor of the induce voltage $v(t)$ is obtained from (3.29). There is a good agreement between analytical and numerical results.

### 3.2.2 2D MIT problem: Example I (circular phantom)

In this subsection, we consider a simple circular phantom, model a 2D MIT system using both early and improved methods, and investigate their performance. All simulations of this thesis were executed on a core i5 2.6 GHz laptop with 8 GB RAM.

#### 3.2.2.1 Modeling set-up

Figure 3.8 shows the cross-sectional view of the 2D MIT system, which has 16 air-core coils used for both excitation and sensing. The coils are rectangular-shaped and arranged in a circular ring surrounding the object to be imaged. The imaging conducting region has a radius of $R_1$=110 mm, which is suitable for most industrial and medical applications. The size of this region is similar to the model used in [14] for ICH detection. The coils were made using 5 mm width and 500 μm thickness copper strips which are similar to strip coils in magnetic resonance imaging [97]. Sequential activation of coils using a sinusoidal alternating current of 20 A amplitude excites the imaging region. In accordance with the previous studies on the biomedical MIT, we chose three frequencies, including 100 kHz, 4 MHz, and 10 MHz, for our simulations. Using higher frequencies is not allowed since the quasi-static approximation is no longer valid, which results in a more dominant wave-propagation phase delay [24]. After each excitation using one of the coils, the induced signal in the remaining coils (except those previously used for excitation) was recorded. In this way, the independent measurements are gathered. They are labeled from 1 to 120. The measurement number 1 states Coil 1 is excited and Coil 2 is measured, measurement number 16 states Coil 2 is excited and Coil 3 is measured, and so forth. The homogeneous Dirichlet boundary condition was imposed at a radius of 250 mm.

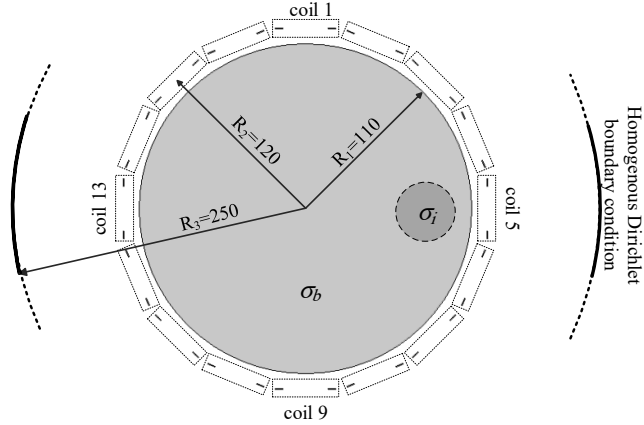

Figure 3.8. Example I: Simulation arrangement for the 2D MIT problem. The homogeneous Dirichlet boundary condition is imposed on a full circle with the radius of 250 mm (the boundary circle was partially drawn for saving space). $\sigma_i$ and $\sigma_b$ are the inclusion and background conductivities (dimensions in mm).

In Example I, to demonstrate the conductivity contrasts in the imaging region, as shown in Figure 3.8, a circular inclusion with the conductivity of $\sigma_i$ is placed in background with the conductivity of $\sigma_b$. The inclusion with a radius of 20 mm was centered at (80, 0) mm. The background also had a radius of 110 mm. Experiments were performed for four different conductivities $\sigma_i = 0, 2, 4,$ and 6 S/m and $\sigma_b = 0$ S/m. These values for electrical conductivity are approximately in the range of conductivities that appeared in the low conductivity applications of MIT.

The number of triangular elements and nodes were 84090 and 42204, respectively. We solved the forward problem based on the early and improved formulations using the FE method and calculated the induced voltages using (3.29). As explained in Section 3.2.3, in the biomedical applications of MIT, only the real part of the induced voltages is informative, and the imaginary part is negligible. Thus, we use the real part of induced voltages to compare the improved method with the early one in all experiments.

### 3.2.2.2 Voltages induced by the total magnetic field

Figure 3.9 shows the real part of induced voltages by the total magnetic field for all independent measurements. The left and right columns show respectively the real part of the induced voltage for the early, $Re\{V_1^E\}$, and improved, $Re\{V_1^I\}$, method at different frequencies. As can be seen, $Re\{V_1^E\}$ is proportional to the square of excitation frequency which is consistent with the previous studies [24]. By changing the inclusion conductivity, the induced voltages are changed distinguishably as well. However, for $Re\{V_1^I\}$ the situation is entirely different. The induced voltage is no longer proportional to the square of the excitation frequency. In addition, the change in the inclusion conductivities partially appears in the voltage induced at 4 MHz, and it is completely distinguishable at 10 MHz. A closer look at the signals obtained by the improved method at $\omega = 100$ kHz reveals that their peaks correspond to the sensor coils located around the excitation coils. This is due to the fact that the coils conductivity in the improved method was set to $5.8 \times 10^7$ S/m (conductivity of copper), whereas in the early method, the coils conductivity has no effect on the results since $\kappa = 0$ in (3.16) for coils regions. That means in addition to the conductivities to be imaged; the coils conductivities influence the induced voltages. In addition, changes in the induced voltage patterns at 4 MHz are started in sensing coils which are close to the inhomogeneity. In Figure 3.9(f), the induced voltage patterns at 10 MHz for the improved method are more similar to those of the early one. The reason will be discussed in Section 3.3.2.6.

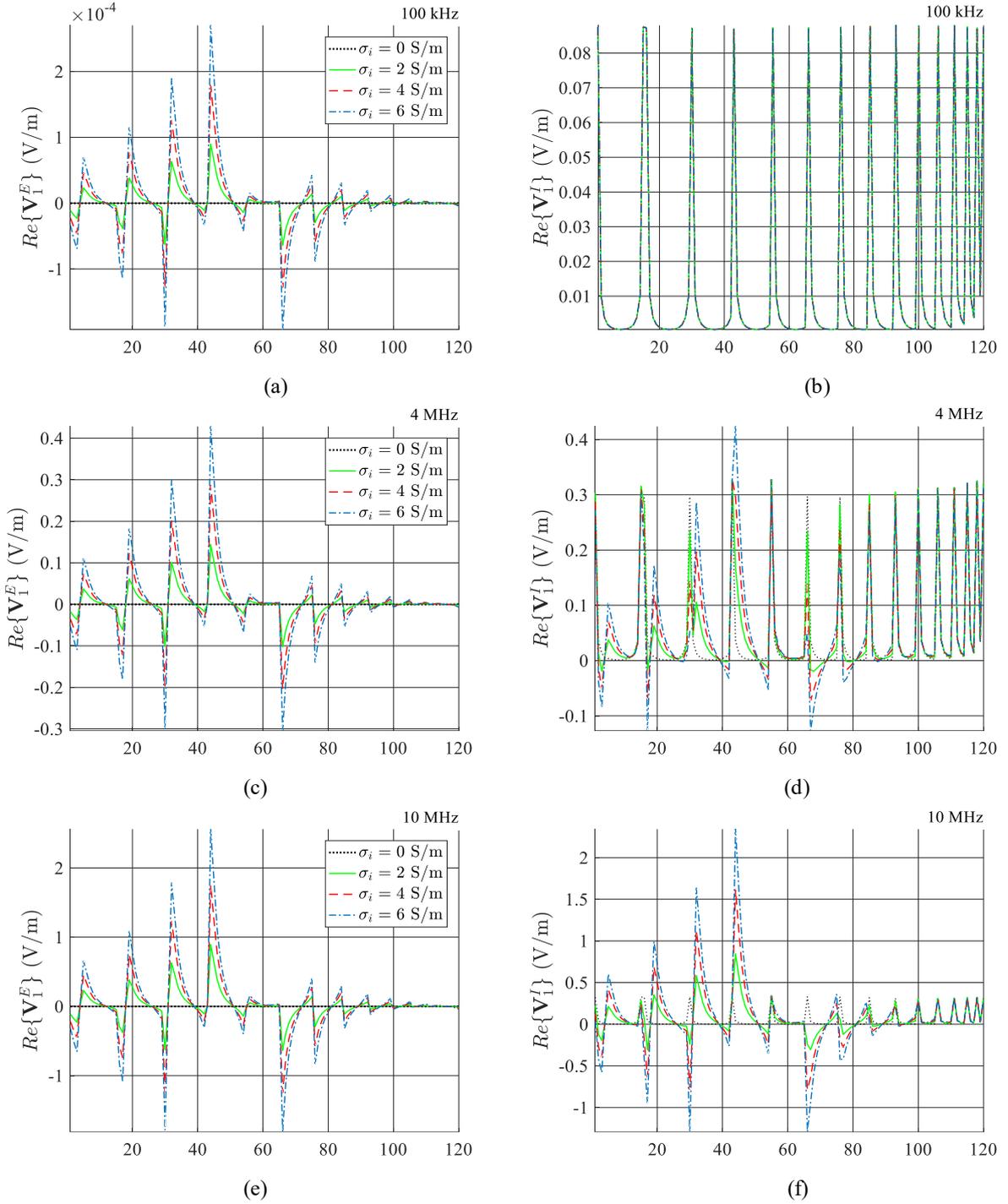

Figure 3.9. Example I: Real part of induced voltages due to the total magnetic field for (left column) the early method, $Re\{V_1^E\}$, and (right column) the improved method, $Re\{V_1^I\}$. The first to third row corresponds to excitation frequency of 100 kHz, 4MHz, and 10 MHz, respectively. Experiment was performed for $\sigma_i = 0$ S/m (black dotted line), $\sigma_i = 2$ S/m (green solid line), $\sigma_i = 4$ S/m (red dashed line), and $\sigma_i = 6$ S/m (blue dash-dotted line). The background conductivity $\sigma_b$ was zero. The horizontal axis is the measurement number.

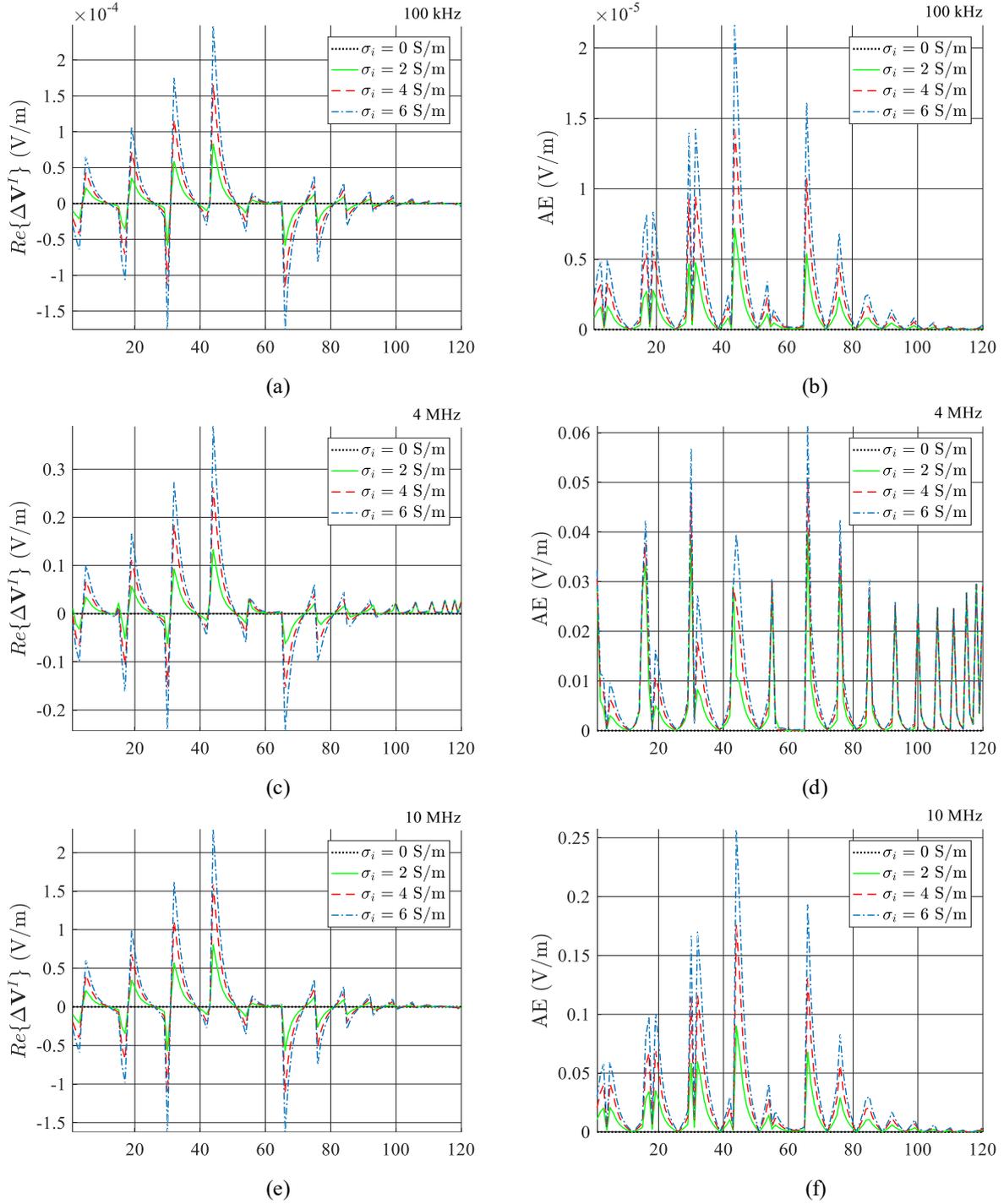

Figure 3.10. Example I: (left column) Real part of induced voltages due to the secondary magnetic field in the improved method, $Re\{\Delta V^I\}$, at (a) 100 kHz, (c) 4 MHz, and (e) 10 MHz. (right column) Absolute error between the induced voltages due to secondary fields, $AE = |Re\{\Delta V^I\} - Re\{\Delta V^E\}|$, at (b) 100 kHz, (d) 4 MHz, and (f) 10 MHz. Experiment was performed for $\sigma_i = 0$ S/m (black dotted line), $\sigma_i = 2$ S/m (green solid line), $\sigma_i = 4$ S/m (red dashed line), and $\sigma_i = 6$ S/m (blue dash-dotted line). The background conductivity $\sigma_b$ was zero. The horizontal axis is the measurement number. Note $Re\{\Delta V^E\}$ is the same as $Re\{V_1^E\}$, given in Figure 3.9.

### 3.2.2.3 Voltages induced by the secondary magnetic field

In order to compare the early and improved methods, the induced voltage due to the secondary magnetic field is computed as $\Delta V = V_1 - V_0$ for the improved method, where $V_0$ is the induced voltage when the imaging region was air-filled. As mentioned before, in the low conductivity applications of MIT, only the real part of the induced voltages is used; therefore, we consider $Re\{\Delta V_1^I\} = Re\{V_1^I\} - Re\{V_0^I\}$.

It is worthwhile mentioning when the imaging region is air-filled, one deals with the induced voltage of the primary magnetic field. In the early method, the induced voltage is pure imaginary, but; in the improved method, this voltage is complex-valued because of considering the coils conductivity. It provides that $Re\{V_0^E\} = 0$, and the induced voltage due to the secondary magnetic field in the early method is obtained as $Re\{\Delta V^E\} = Re\{V_1^E\}$.

The left column of Figure 3.10 illustrates the real part of the induced voltage due to the secondary field obtained by the improved method, $Re\{\Delta V^I\}$, at different frequencies. As can be seen, the patterns of the induced voltages for all frequencies are almost similar to the early method, i.e., the left column of Figure 3.9. However, $Re\{\Delta V^I\}$ is slightly smaller than those of the early method. This is due to the loss of skin and proximity effects in the coils.

### 3.2.2.4 Measure of accuracy

To compare the results obtained from the early and improved methods, the absolute error between the induced voltages due to secondary fields was computed as $AE = |Re\{\Delta V^I\} - Re\{\Delta V^E\}|$. The right column of Figure 3.10 shows the absolute error at different frequencies. As can be seen, AE changes with the inclusion conductivity value inside the imaging region; thus, it is a function of conductivity value to be imaged. In addition, AE increases with the frequency, so it is a function of operating frequency. The local maximums in error are related to the sensor coils located close to the inhomogeneity when their adjacent coils are excited. The percentage of relative error was also calculated over all measurements as

$$\text{RE (\%)} = \frac{\|Re\{\Delta V^I\} - Re\{\Delta V^E\}\|_2}{\|Re\{\Delta V^I\}\|_2} \times 100 \tag{3.30}$$

where $\|\cdot\|_2$ denotes the $L^2$ norm. The percentage of relative error at 100 kHz was 8.5% for all $\sigma_i$. This error at 4 MHz was 43%, 25%, and 19 % for $\sigma_i = 2$ S/m, $\sigma_i = 4$ S/m, and $\sigma_i = 6$ S/m, respectively. At 10 MHz, the error was 11% for all $\sigma_i$. These results manifest that RE is a function of conductivity values and excitation frequency. It is worth noting that the error in the early method is increased by increasing the coil area, the number of coil turns, the conductivity value of the medium, and by bringing the coils closer together.

### 3.2.2.5 Effects of conductivity distribution of imaging region

To manifest the interaction between coils and conductivity distribution inside the imaging region, we performed the simulation for $\sigma_i = 0$ S/m and the different background conductivities $\sigma_b = 2, 4,$ and 6 S/m at 1 MHz. Figure 3.11 shows the result for $Re\{V_1^I\}$. As can be seen, the patterns of induced voltages start to change with the conductivity in the imaging region at 1 MHz, unlike the previous simulation (Figure 3.9(d)), which occurred at 4 MHz. In this case, at the frequency 100 kHz, similar to the case in Figure 3.9(b), the changes of conductivities inside the imaging region are not seen in $Re\{V_1^I\}$. This shows that the value and distribution of the conductivities to be imaged influence the skin and proximity effects in addition to the excitation frequency. It should be noted that the patterns of $Re\{V_1^E\}$ are different from those which were shown in Figure 3.9.

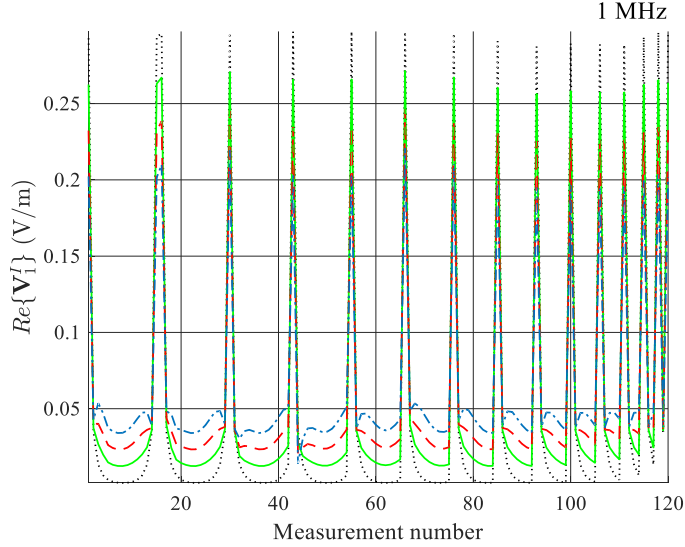

Figure 3.11. Example I: Real part of induced voltages for the improved method, $Re\{V_1^I\}$, at 1 MHz for background conductivity of $\sigma_b = 0$ S/m (black dotted line), $\sigma_b = 2$ S/m (green solid line), $\sigma_b = 4$ S/m (red dashed line), and $\sigma_b = 6$ S/m (blue dash-dotted line). The inclusion conductivity $\sigma_i$ was zero.

#### 3.2.2.6 Analysis of results obtained for Example I

As seen in Section 3.3.2.2, the induced voltage pattern due to the total magnetic field at 10 MHz for the improved method was more similar to that of the early method whereas the patterns of the two methods were different at 100 kHz and 4 MHz. To find the reason, it should be explored when (3.22) will be equal to (3.14). It occurs under three conditions: *i*) coil conductivity is zero, *ii*) the angular frequency is zero, or *iii*) the expression $A_z(x,y) = \frac{1}{S_k}\iint_{R_k} A_z(x,y)\,ds$ is held. The last condition is held when the MVP is almost constant on the cross-sectional area of the $k$-th conductor. Observation of $A_z$ on the cross-section of coils in Section 3.3.2.2 reveals that the last condition is met for the MVP at 10 MHz. It is notable at this frequency the penetration depth is $\delta = \sqrt{2/\omega\sigma\mu} \approx 20$ µm, which is very smaller than the dimensions of the cross-sectional area $S_k$ (5 mm × 0.5 mm).

As seen in Section 3.3.2.3, the patterns of the induced voltages due to the secondary magnetic field by the two methods for all frequencies were almost similar. At first glance, it may seem that using gradiometer [98] or state-difference imaging [13] techniques to obtain induced voltages by the secondary magnetic field can compensate for the error due to neglecting of skin and proximity effects in coils. In consequence, the early method can be used favorably as the MIT forward model. In order to show the importance of the improved method, we displayed $Re\{\Delta V^E\} - Re\{\Delta V^I\}$ at 10 MHz with the inclusion conductivity $\sigma_i = 6$ S/m in the blue solid-line in Figure 3.12. Then, on the same figure, the real part of the induced voltage due to the secondary field, $Re\{\Delta V^I\}$, was plotted at 10 MHz using three different small inclusion conductivity values. As shown in Figure 3.12, the values of inclusion conductivity were $\sigma_i = 0.3$ S/m (green dashed line), $\sigma_i = 0.5$ S/m (red dotted line), and $\sigma_i = 0.9$ S/m (purple dash-dotted line). Only measurement numbers 40 to 60 have been plotted to show the difference clearly. As can be seen from Figure 3.12, the induced voltage obtained from the improved method, $Re\{\Delta V^I\}$, at 10 MHz with a 0.5 S/m inclusion (red dotted-line) is close to the error between the results of the early and improved methods at 10 MHz with a

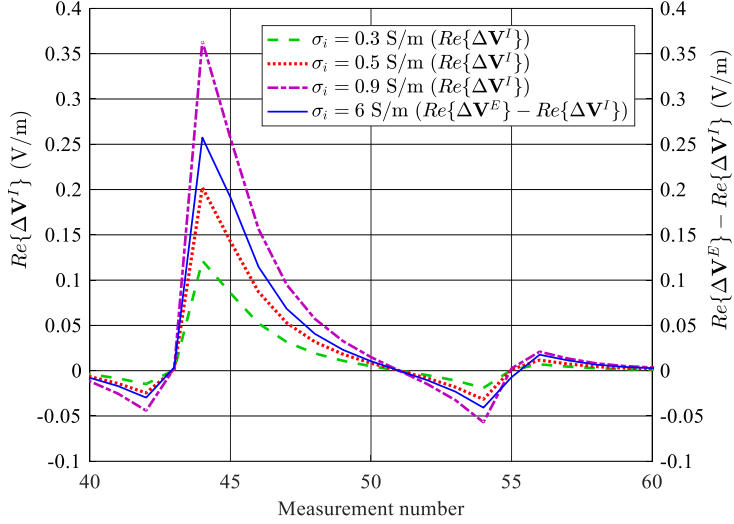

Figure 3.12. Example I: $Re\{\Delta V^I\}$ at 10 MHz with a (green dashed line) $\sigma_i$=0.3 S/m, (red dotted line) $\sigma_i$=0.5 S/m, and (purple dash-dotted line) $\sigma_i$=0.9 S/m. (blue solid line) Error between the induced voltages due to secondary fields, $Re\{\Delta V^E\} - Re\{\Delta V^I\}$, at 10 MHz with $\sigma_i$=6 S/m. The background conductivity $\sigma_b$ was zero. Only measurement numbers 40 to 60 have been plotted to clearly show the difference.

6 S/m inclusion. In this case, the skin and proximity effects produce equivalent to more than 0.5 S/m negative contrast and make a meaningful difference in terms of conductivity contrast of the improved method. In other words, the early method has almost predicted the induced voltages for a 6.5 S/m inclusion when a 6 S/m inclusion was actually placed in the imaging space. It is well-known from inverse problems that small errors in the forward computations can result in considerable errors in the image reconstruction. As noted in Sections 3.3.2.4 and 3.3.2.5, the error in the early method is increased by the coil area, the number of coil turns, the conductivity value of the medium, and by bringing the coils closer together.

In this example, we assumed the same coil has been used for excitation and sensing, and it has only one turn and a minimal cross-section. New low-conductivity MIT designs are now tending to use much higher excitation currents to amplify the secondary magnetic field and improve performance in noise reduction [29]. This requires a large coil cross section or alternatively can be achieved by increasing the coil's turns. Furthermore, the error will be more considerable in MIT systems which use separate coils for excitation and sensing or utilize parallel excitation mode.

The computation time for the early and improved methods was 88.84 s and 91.3 s, respectively, in Example I. For the same mesh, the improved method generally has the same order of computational complexity as the early one. However, for the improved method, the finite elements inside and around source conductors should be refined sufficiently in order to obtain a more accurate result. It demands higher computational resources and takes a longer time. For the early method, no such refinement is needed.

It is worthwhile noting that a similar procedure can be done for a high conductivity MIT example. As shown before, the error of the early method originates from ignoring skin and proximity effects, which occur in coils. If conducting coils are similar in both low and high conductivity MIT applications, the most important factor in determining error will be the operating frequency. Since in the high conductivity MIT applications, lower frequencies are used, the error would be less considerable compared to the low conductivity applications, in which higher frequencies are applied. It is notable that, in the imaging region, considering skin effect in conductive sub-regions is less considerable for low conductivity MIT compared to high conductivity applications [59].

### 3.2.3 2D MIT problem: Example II (human head phantom)

In this subsection, we consider a 2D human head phantom and use the MIT system modeled in the previous section to investigate the performance of early and improved methods in a biomedical application.

#### 3.2.3.1 Modeling set-up

Figure 3.13(a) shows an axial cross-section of a realistic head model freely available in [99]. Similar 2D models have been used in [14], [47], [71] for realistic simulation of MIT. The normal head model contained five tissue types including, scalp, skull, CSF, grey matter (GM), and white matter (WM). In order to simulate a hemorrhage, a disk-shaped lesion with a radius of 18 mm was embedded in the model, as shown in Figure 3.13(b). The conductivities of brain tissues in different frequencies were sourced from [100], as indicated in Table 3.1. The conductivity of blood was taken for hemorrhage. Figure 3.13(c) shows the human head phantom inside MIT setup. The modeled MIT system is exactly the same as the setup describe in Section 3.3.2.1. The number of nodes and elements which has been used for meshing the head phantom with ICH were 85818 and 171162, respectively.

For this example, we chose two frequencies, including 1 MHz and 10 MHz, for our simulations. We performed the simulation procedure similar to Example I. The obtained results were the same as previous. Thus, we did not report those results here. Instead, we consider two new simulation studies. In the first simulation, we study the feasibility of the detection of ICH by MIT. Then, we perform a simulation to see whether the frequency-difference imaging technique can compensate for ignoring the skin and proximity effects inside MIT coils.

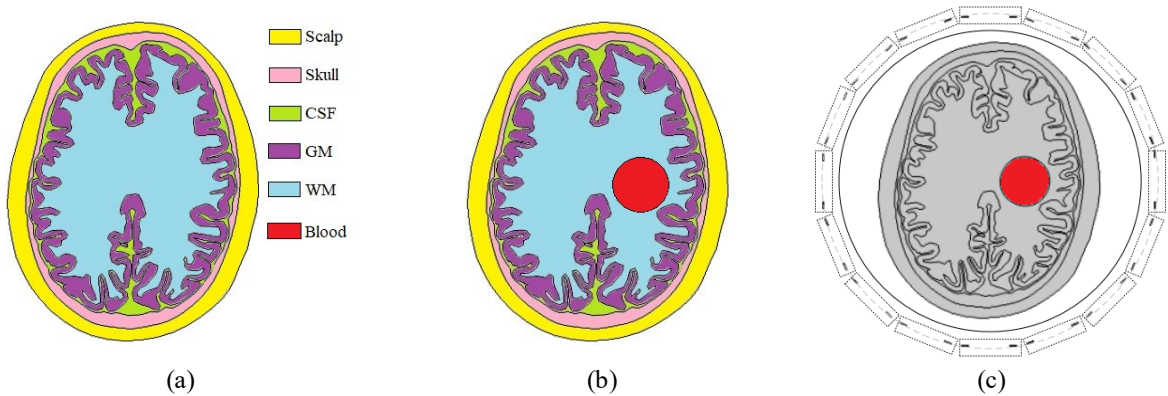

(a) (b) (c)

Figure 3.13. Example II: 2D human head phantom. (a) Different part of normal head model, (b) Embedded ICH in the head model with radius of 18 mm (c) Head model inside MIT setup.

Table 3.1 Conductivities of brain tissues at 1 MHz and 10 MHz assigned to each tissue type in the head model derived from [100]

| Tissue | Conductivity (S/m) | |
|---|---|---|
| | 1 MHz | 10 MHz |
| Scalp | 0.503 | 0.617 |
| Skull | 0.0574 | 0.0829 |
| CSF | 2 | 2 |
| GM | 0.163 | 0.292 |
| WM | 0.102 | 0.158 |
| Blood | 0.822 | 1.10 |

### 3.2.3.2 Detection of ICH

Figure 3.14(a) and Figure 3.14(c) depict the difference between the induced voltages from the normal head model, $Re\{V_1^{Normal}\}$, and the head model containing the hemorrhage, $Re\{V_1^{ICH}\}$, at frequencies of 1 MHz and 10 MHz, respectively. The red dashed line and the blue dash-dotted line illustrate the results for the early and improved forward methods, respectively. As can be seen, ICH can be resolved in both signals predicted by the early and improved forward methods. It is noteworthy that here we subtract the induced voltages by normal head model from the head model with ICH. Thus, it results in the state-difference imaging technique, and the induced voltages by the primary magnetic field will be canceled automatically. The pattern of induced signals obtained by the early and improved methods are almost similar. However, similar to results obtained in section 3.3.2.3, the signals predicted by the improved method are slightly smaller than those of the early method. This is due to the loss of skin and proximity effects in the coils.

Similar to Section 3.3.2.4, the absolute error between the induced voltages obtained from the early and improved methods was computed. Figure 3.14(b) and Figure 3.14(d) show the absolute error at frequencies of 1 MHz and 10 MHz, respectively. The pattern of AEs is almost similar to induced signals, and their amplitude is approximately 0.1 of the original ones. As can be seen, AE is a function of operating frequency. The relative error also was calculated by (3.30). The percentage of relative error at 1MHz and 10 MHz were 10.6 % and 15.3 %, respectively. It indicated that RE is a function of frequency, as well.

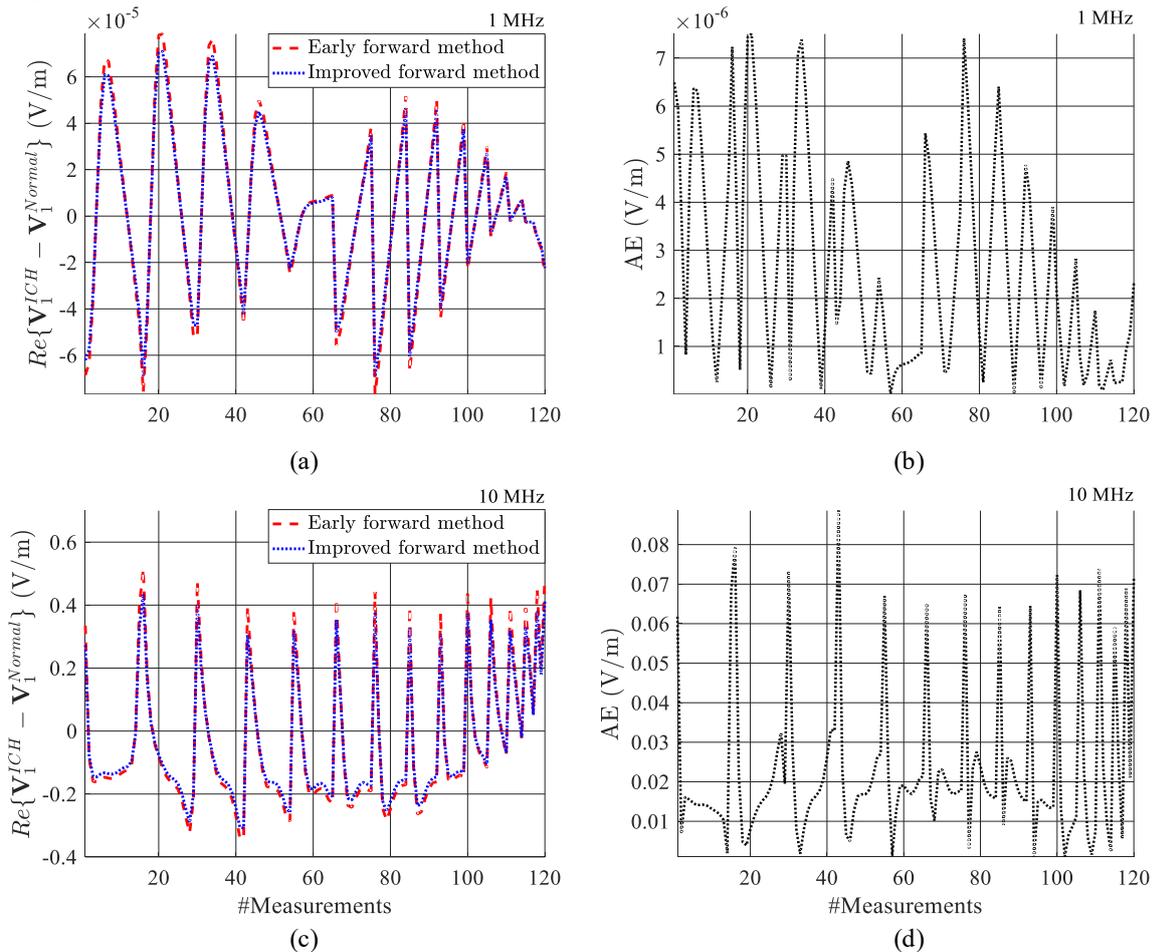

Figure 3.14. Example II: Difference between the induced voltages from the normal head model, $Re\{V_1^{Normal}\}$, and the head model with ICH, $Re\{V_1^{ICH}\}$, at (a) 1 MHz and (c) 10 MHz. Absolute error (AE) between the induced signals obtained from the early and improved methods at (b) 1MHz and (d) 10 MHz.

### 3.2.3.3 Frequency-difference imaging

The results obtained in the previous section were based on the state-difference imaging technique. As shown for Example I, the state-difference imaging technique can partially compensate for ignoring the skin and proximity effects inside MIT coils. However, in cerebral hemorrhage imaging application, the state-difference approach will not be suitable; since no before-lesion reference data set will typically be available. To overcome this limitation, frequency- difference approach has been proposed in previous studies [11], [47]. This approach, which is based on the frequency-dependent conductivity of biological tissues (as shown in Table 3.1), can reconstruct the conductivity variation between two excitation frequencies at the same time. In this approach, the difference between the real parts of induced voltages at a higher frequency $f_H$ and the real part of induced voltages at a lower frequency $f_L$ multiplied by $(f_H/f_L)^2$ has been considered as data to be used in image reconstruction. It was reported that the frequency-difference approach could reduce systematic errors in MIT image reconstruction [42]. Excitation frequencies $f_L = 1$ MHz and $f_H = 10$ MHz were chosen so as to encompass the dielectric dispersion of blood centered on about 7 MHz [100]. The difference between blood conductivity in these two frequencies is more than all other surrounding tissue in the brain.

We can obtain the induced signals in the frequency-difference technique by two approaches. In the first approach, we can feed the real part of voltages induced by the total magnetic field into this technique, i.e., $Re\{V_1^{ICH}(f_H)\} - (f_H/f_L)^2 \times Re\{V_1^{ICH}(f_L)\}$. In the second approach, we can consider the real part of voltages induced by the secondary magnetic field in the frequency-difference technique, i.e., $Re\{\Delta V^{ICH}(f_H)\} - (f_H/f_L)^2 \times Re\{\Delta V^{ICH}(f_L)\}$. Figure 3.15(a) shows the result for the first approach obtained by the early and improved methods, which are indicated by the red dashed and blue dash-dotted lines, respectively. Figure 3.15(b) shows the absolute error between the two methods. As can be seen, the error and discrepancy between the two methods are too high. This is similar to results obtained in Section 3.3.2.2 for Example I in frequencies below 4 MHz. Taking a closer look at the signals obtained by the improved method reveals that their peaks correspond to the sensor coils located around the excitation coils. This is due to the fact that the coils conductivity in the improved method was set to $5.8 \times 10^7$ S/m, whereas in the early method, the coils conductivity has no effect on the results since $\kappa = 0$ in (3.16) for coils regions. That means in addition to the conductivities to be imaged; the coils conductivities influence the induced voltages. The relative error percentage in this approach was 99% which means using the frequency-difference technique with voltages induced by the total magnetic field cannot compensate for ignoring the skin and proximity effects in coils.

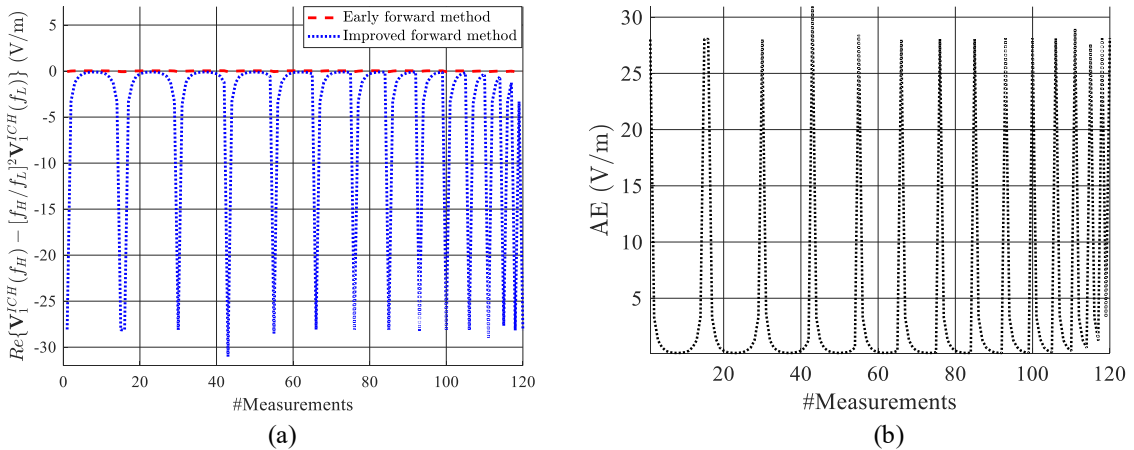

Figure 3.15. Example II: First approach for the frequency-difference technique. (a) $Re\{V_1^{Stroke}(f_H)\} - (f_H/f_L)^2 \times Re\{V_1^{Stroke}(f_L)\}$ obtained by the early and improved forward method where $f_L = 1$ MHz and $f_H = 10$ MHz. (b) The absolute error between the results of early and improved methods.

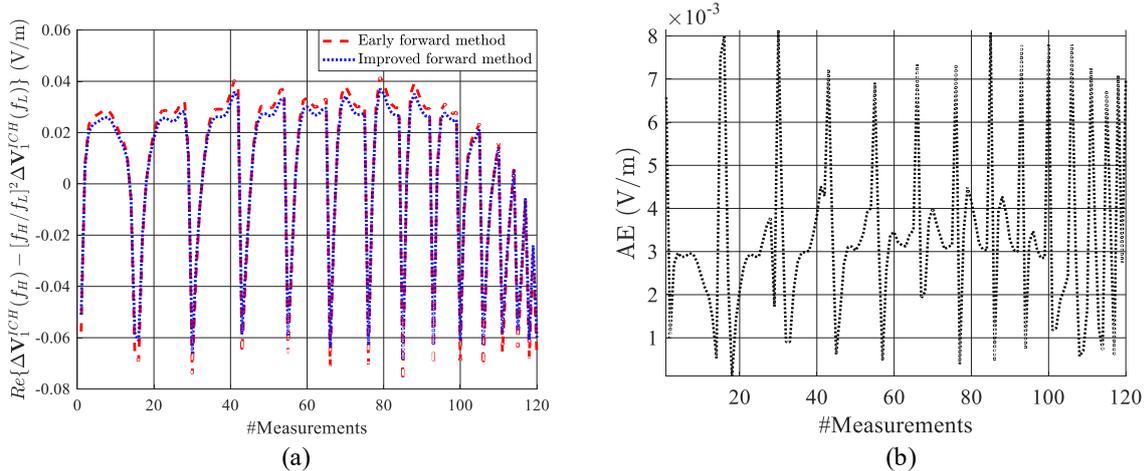

Figure 3.16. Example II: Second approach for the frequency-difference technique. (a) $Re\{\Delta V^{ICH}(f_H)\} - (f_H/f_L)^2 \times Re\{\Delta V^{ICH}(f_L)\}$, obtained by the early and improved forward method where $f_L = 1$ MHz and $f_H = 10$ MHz. (b) The absolute error between the results of early and improved methods.

Figure 3.16(a) shows the result for the second approach in which the voltages induced by the secondary magnetic field are used in the frequency-difference technique. Red dashed and blue dash-dotted lines indicate the results obtained by the early and improved methods, respectively. As can be seen, the patterns of the induced signals are almost similar to each other. However, the induced signals predicted by the improved forward method are slightly smaller than those of the early method. This is due to the loss of skin and proximity effects in the coils. Similar to results obtained in section 3.3.2.3, using voltages induced by the secondary field can partially compensate for ignoring the skin and proximity effects in MIT coils. Figure 3.16(b) depicts the absolute error between the results obtained by the early and improved methods. Again, the local maximums in error are related to the sensor coils located next to the excitation coils. The relative error percentage for this approach was 12% which shows a significant improvement compared to the first approach.

## 3.3 Conclusions

In this chapter, an improved method based on Maxwell's equations was developed to model 2D MIT forward problems. This method considers skin and proximity effects inside the exciter and sensor coils. The FE method implementation of the improved method was validated by a simple single-channel MIT problem with an analytical solution. Comparison of the early and improved methods through simulating a 16-coil 2D MIT system with a synthetic circular phantom in Example I revealed that there was a meaningful discrepancy in terms of achievable conductivity contrast from the improved one. For this example, simulations were done at 100 kHz, 4 MHz, and 10 MHz. The absolute and relative errors were obtained for induced voltages due to the secondary field. The results manifested the relative error ranging from 8.5% to 43% was a function of operating frequency and conductivity value and distribution inside the imaging region. A 2D head model was used in Example II for imaging cerebral hemorrhage application. For this example, simulations were done at 1 MHz and 10 MHz. The results revealed that induced signals predicted by both early and improved methods could detect an ICH.

Ignoring skin and proximity effects in coils appears as a numerical noise in the modeling of the MIT forward problem at higher frequencies which are desirable for biomedical MIT applications. This noise can be partially compensated by the state-difference imaging, as shown in Example I, or by the frequency-difference technique, as shown in Example II, to reconstruct the relative conductivity values. However, as will be shown in Chapter 5, to reconstruct the absolute conductivity values, it is crucial to have a highly accurate forward model which can be fitted as well as possible to measured data in the reconstruction process. A small error in the forward problem may seem negligible, whereas it can implicate considerable errors when it is entered in the process of the inverse problem for image reconstruction. Results of this chapter manifested that the improved method considering skin and proximity effect in MIT coils can significantly impress the accuracy of computation of induced voltages in biomedical MIT applications.

# Chapter 4: MIT Forward Problem II

Besides numerical methods introduced for solving MIT forward problem in Chapter 2, we can employ alternative approaches, as well. One of these approaches is mesh-less or mesh-free methods which have widely been used in the solid mechanics for the numerical solution of boundary value problems [101], [102]. These methods approximate an unknown function based on a set of nodes distributed in the problem domain, and no elements or usual connection between nodes to define elements (element connectivity) are needed to construct the discrete equations. This feature is considerably advantageous to solve problems in complex geometries like the irregular geometry of the human head in which generation of a mesh is a time-consuming and challenging task [31]. Furthermore, in some applications like imaging of lung ventilation by EIT [32], the shape and size of the boundary and internal regions of the object under study are subject to frequent deformations. Since the mesh-free methods require only nodal data and the element connectivity is not needed, they are more accurate compared to the mesh-based method like FEM in problems with moving boundaries and objects inside the imaging region [103].

Numerous mesh-free methods have been presented to solve the boundary value problems, such as the element-free Galerkin (EFG) method [104], wavelet Galerkin method [105], and mesh-less local Petrov-Galerkin [106]. In this thesis, we choose the EFG method to solve the MIT forward problem for the first time. This method has been employed to solve the EIT forward problem in several studies [8], [103], [104]. EFG has also been used for the computation of static and quasi-static electromagnetic fields, especially in non-destructive testing applications [107]–[110].

The EFG basis functions do not satisfy the Kronecker delta criterion [111], [112] (see more details in Section 4.2.1). Therefore, they cannot be used to construct an interpolant; but rather approximate a function, and the name approximant is used for them [113]. This property makes the imposition of essential boundary conditions difficult compared to FEM. It is possibly the main drawback of the EFG method. To address the difficulty, several ways have been proposed, including the use of Lagrange multipliers [114], substitution method [115], interpolating shape function [115], and coupling with finite element method [8]. The most efficient and accurate method is the last one [113]. In this thesis, we use this method which results in the hybrid FE-EFG method for modeling MIT forward problem. In fact, we use the EFG method for modeling the imaging region while employing the finite element method to model the region surrounding the imaging region. In this way, the region including complex geometries and/or moving objects are modeled by the EFG method; and the region including coils and external boundary (on which the essential boundary condition is imposed) are modeled by FEM. Another advantage of this approach is that modeling the improved forward problem in the coils regions (Equation (3.22)) using the EFG method is no longer needed. It is noteworthy that it can be done by the method presented in [116] for solving integro-differential equations.

## 4.1 Element free Galerkin (EFG) method

The EFG method was first presented by Belytschko et al. [114] using the extension of the diffuse element method (DEM). This method belongs to mesh-less approaches, which utilize a set of nodes distributed in the solution domain to construct the discrete equations. EFG method is a proper method for the numerical solution of partial differential equations. The method utilizes the moving least square (MLS) approximants to approximate the solution of equations. In fact, the MLS approximation is the integral part of the EFG method and generates the shape functions employed in EFG. In the following, we briefly introduce the procedure of generating EFG shape functions using the MLS approximations.

### 4.1.1 Moving least square approximation

The MLS approximation is widely used to construct shape functions for mesh-less methods. This approach was first proposed by mathematicians for curve fitting, multidimensional surface construction, and regression. Generally, MLS approximation relies on three parts: 1) a compact support weight function; 2) a polynomial basis, and; 3) a set of position-dependent coefficients. The weight function is associated with each node and is nonzero only over a small subdomain around a particular node. This subdomain is called the support domain [113]. The support of the weight function defines a node's domain of influence. The domain of influence of a particular node is a subdomain over which that particular node contributes to the approximation of the solution.

In MLS approximation, a function $u(\mathbf{x})$ in $\Omega \in \mathcal{R}^2$ is approximated by $u^h(\mathbf{x})$ as follows:

$$u^h(\mathbf{x}) = \sum_{j=0}^{m_{po}} p_j(\mathbf{x}) a_j(\mathbf{x}) \equiv \mathbf{p}^T(\mathbf{x})\mathbf{a}(\mathbf{x}) \tag{4.1}$$

where $\mathbf{x} = (x, y)$ is an approximation point, $p_j(\mathbf{x})$ is a monomial basis function of order $j$, $a_j(\mathbf{x})$ is a position-dependent coefficient, $\mathbf{p}(\mathbf{x})$ and $\mathbf{a}(\mathbf{x})$ are column matrices containing $p_j(\mathbf{x})$ and $a_j(\mathbf{x})$, respectively, and $m_{po}$ indicates the maximum order in $\mathbf{p}(\mathbf{x})$. In 2D, for a linear basis $\mathbf{p}(\mathbf{x})$ is given by $[1, x, y]^T$.

Assume that the domain $\Omega$ is discretized by $N^{EFG}$ nodes. The set of nodes whose domain of influence includes the approximation point is called the support nodes of $\mathbf{x}$ and are indicated by $\mathbf{x}_i$. Now, we can use (4.1) to calculate the approximated values of the function at these nodes:

$$u^h(\mathbf{x}, \mathbf{x}_i) = \mathbf{p}^T(\mathbf{x}_i)\mathbf{a}(\mathbf{x}), \quad i = 1, 2, \dots, n_{sn} \tag{4.2}$$

where $n_{sn}$ is the number of support nodes for $\mathbf{x}$. Then, a functional of weighted residuals is then constructed using the approximated values of the function and the nodal parameters, $u_i = u(\mathbf{x}_i)$:

$$J = \sum_{I=1}^{n_{sn}} W(\mathbf{x} - \mathbf{x}_i) \underbrace{[u^h(\mathbf{x}, \mathbf{x}_i) - u(\mathbf{x}_i)]^2}_{residual} = \sum_{i=1}^{n_{sn}} W_i(\mathbf{x})[\mathbf{p}^T(\mathbf{x}_i)\mathbf{a}(\mathbf{x}) - u_i]^2 \quad (4.3)$$

where $W(\mathbf{x} - \mathbf{x}_i) = W_i(\mathbf{x})$ is a weight function. In the MLS approximation, at an arbitrary point $\mathbf{x}$, $\mathbf{a}(\mathbf{x})$ is chosen to minimize the weighted residual (4.3). It results in [113]:

$$\mathbf{A}(\mathbf{x})\mathbf{a}(\mathbf{x}) = \mathbf{B}(\mathbf{x})\mathbf{u} \quad (4.4)$$

where

$$\mathbf{A}(\mathbf{x}) = \sum_{i=1}^{n_{sn}} W_i(\mathbf{x})\mathbf{p}(\mathbf{x}_i)\mathbf{p}^T(\mathbf{x}_i) \quad (4.5)$$

$$\mathbf{B}(\mathbf{x}) = [\mathbf{B}_1 \; \mathbf{B}_2 \; ... \; \mathbf{B}_n] \\ \mathbf{B}_i = W_i(\mathbf{x})\mathbf{p}(\mathbf{x}_i) \quad (4.6)$$

$$\mathbf{u} = [u_1 \; u_2 \; ... \; u_n]^T \quad (4.7)$$

Assuming that $\mathbf{A}(\mathbf{x})$ is invertible, (4.4) can then be solved for $\mathbf{a}(\mathbf{x})$:

$$\mathbf{a}(\mathbf{x}) = \mathbf{A}^{-1}(\mathbf{x})\mathbf{B}(\mathbf{x})\mathbf{u} \quad (4.8)$$

Substituting (4.8) in (4.2) leads to

$$u^h(\mathbf{x}) = \sum_{i=1}^{n_{sn}} \sum_{j=0}^{m_{po}} p_j(\mathbf{x})(\mathbf{A}^{-1}(\mathbf{x})\mathbf{B}(\mathbf{x}))_{ji} u_i \quad (4.9)$$

or

$$u^h(\mathbf{x}) = \sum_{i=1}^{n_{sn}} \phi_i(\mathbf{x}) u_i \quad (4.10)$$

where the MLS shape function $\phi_i$ is defined by

$$\phi_i(\mathbf{x}) = \sum_{j=0}^{m} p_j(\mathbf{x})(\mathbf{A}^{-1}(\mathbf{x})\mathbf{B}(\mathbf{x}))_{ji} u_i = \mathbf{p}^T(\mathbf{x})\mathbf{A}^{-1}(\mathbf{x})\mathbf{B}_i(\mathbf{x}) \quad (4.11)$$

Note that the shape function does not satisfy the Kronecker delta property: $\phi_i(\mathbf{x}_j) \neq \delta_{ij}$; therefore $u^h(\mathbf{x}_i) \neq u_i$, and they cannot be used to construct an interpolant, but rather approximates of a function.

To obtain the discrete system of equations for the EFG method, we require to determine the spatial derivative of the MLS basis functions. The derivative is obtained as follows:

$$\phi_{i,x} = (\mathbf{p}^T \mathbf{A}^{-1} \mathbf{B}_i)_{,x} \\ = (\mathbf{p}^T)_{,x} \mathbf{A}^{-1} \mathbf{B}_i + \mathbf{p}^T (\mathbf{A}^{-1})_{,x} \mathbf{B}_i + \mathbf{p}^T \mathbf{A}^{-1} (\mathbf{B}_i)_{,x} \quad (4.12)$$

where $(\cdot)_{,x}$ denotes derivative with regard to $\mathbf{x}$. To obtain $(\mathbf{A}^{-1})_{,x}$, one can use the following method:

$$\begin{aligned} \mathbf{A}\mathbf{A}^{-1} &= \mathbf{I} \\ (\mathbf{A}\mathbf{A}^{-1})_{,x} &= \mathbf{0} \\ \mathbf{A}_{,x}\mathbf{A}^{-1} + \mathbf{A}(\mathbf{A}^{-1})_{,x} &= \mathbf{0} \\ (\mathbf{A}^{-1})_{,x} &= -\mathbf{A}^{-1}(\mathbf{A})_{,x}\mathbf{A}^{-1} \end{aligned} \quad (4.13)$$

where

$$(\mathbf{A})_{,x} = \sum_{i=1}^{n_{sn}} (W_i(\mathbf{x}))_{,x} \mathbf{p}(\mathbf{x}_i)\mathbf{p}^T(\mathbf{x}_i) \quad (4.14)$$

In addition, $(\mathbf{B}_i)_{,x}$ is obtained as follows:

$$\mathbf{B}_{i,x} = \big(W_i(\mathbf{x})\big)_{,x}\mathbf{p}(x_i) \tag{4.15}$$

### 4.1.2 Weight function

Various weight functions are used for MLS approximation which mostly are bell-shaped [111]. In this thesis, we use the cubic spline weight function, which is a popular choice. The function in 1D space is defined as follows [113]:

$$W(r) = \begin{cases} \dfrac{2}{3} - 4r^2 + 4r^3 & \text{for } r \leq \dfrac{1}{2} \\ \dfrac{4}{3} - 4r + 4r^2 - \dfrac{4}{3}r^3 & \text{for } \dfrac{1}{2} < r \leq 1 \\ 0 & \text{for } r > 1 \end{cases} \tag{4.16}$$

where $r = |x - x_i|/d_{mi}$, $d_{mi} = d_{max}c_i$, $d_{max}$ is a scaling factor, and $c_i$ is the difference between node $x_i$ and its nearest neighbor. $d_{max}$ is chosen such that matrix $\mathbf{A}$ is non-singular. If nodes are distributed uniformly, then $c_i$ can be chosen as the distance between nodes.

To generate 2D weight functions, we define the tensor product of the 1D weight function at any given point as follows [113]:

$$W(\mathbf{r}) = W(r_x) \cdot W(r_y) = W_x \cdot W_y \tag{4.17}$$

where

$$W(r_x) = \begin{cases} \dfrac{2}{3} - 4r_x^2 + 4r_x^3 & \text{for } r_x \leq \dfrac{1}{2} \\ \dfrac{4}{3} - 4r_x + 4r_x^2 - \dfrac{4}{3}r_x^3 & \text{for } \dfrac{1}{2} < r_x \leq 1 \quad r_x = \dfrac{|x - x_i|}{d_{mx}} \\ 0 & \text{for } r_x > 1 \end{cases} \tag{4.18a}$$

$$W(r_y) = \begin{cases} \dfrac{2}{3} - 4r_y^2 + 4r_y^3 & \text{for } r_y \leq \dfrac{1}{2} \\ \dfrac{4}{3} - 4r_y + 4r_y^2 - \dfrac{4}{3}r_y^3 & \text{for } \dfrac{1}{2} < r_y \leq 1 \quad r_y = \dfrac{|y - y_i|}{d_{my}} \\ 0 & \text{for } r_y > 1 \end{cases} \tag{4.18b}$$

and

$$d_{mx} = d_{max}\, c_{xi} \tag{4.19a}$$
$$d_{my} = d_{max}\, c_{yi} \tag{4.19b}$$

$c_{xi}$ and $c_{yi}$ are determined at a particular node by searching for enough neighbor nodes to make $\mathbf{A}$ matrix non-singular everywhere in the domain, and thus invertible. This is necessary to compute the shape function. If the nodes are uniformly distributed, the values $c_{xi}$ and $c_{yi}$ are equal to the distance between nodes in the $x$ and $y$ directions, respectively.

The derivative of the weight function is also required. It can be obtained as follows [113]:

$$(W)_{,x} = \dfrac{dW_x}{dx} \cdot W_y \tag{4.20a}$$

$$(W)_{,y} = \dfrac{dW_y}{dy} \cdot W_x \tag{4.20b}$$

### 4.1.3 Extraction of discrete system of equations

As mentioned earlier, we use the EFG method to solve the forward problem inside conducting region $\Omega_C$ and use FEM to solve the forward problem in other regions ($\Omega_N$ and $\Omega_S$ in the improved forward method). As introduced in Chapter 3, the governing equation in $\Omega_C$ is:

$$-\frac{1}{\mu_0}\nabla^2 A_Z + j\omega\sigma A_Z = 0 \quad in \;\; \Omega_C \tag{4.21}$$

which is simplified as follows:

$$-\frac{1}{\mu_0}\frac{\partial^2 A_Z}{\partial x^2} - \frac{1}{\mu}\frac{\partial^2 A_Z}{\partial y^2} + j\omega\sigma A_Z = 0 \quad in \;\; \Omega_C \tag{4.22}$$

By obtaining the weak form of (4.22) and using the Galerkin method, one can obtain the following system of equations to approximate unknown $A_Z$ at an approximation point $x_q$ using its support nodes [117]:

$$\mathcal{K}_{n_{sn} \times n_{sn}} \hat{\mathcal{A}}_{n_{sn} \times 1} = 0 \quad in \;\; \Omega_C \tag{4.23}$$

where

$$\mathcal{K}_{i,j} = \int_{S(x_q)} \left\{ \frac{1}{\mu}\left[\left(\frac{d\phi_i}{dx}\right)\left(\frac{d\phi_j}{dx}\right) + \left(\frac{d\phi_i}{dy}\right)\left(\frac{d\phi_j}{dy}\right)\right] + j\omega\sigma\phi_i\phi_j \right\} \quad i,j = 1, \dots, n_{sn} \tag{4.24}$$

and $\hat{\mathcal{A}}$ is a column matrix containing nodal parameters of support nodes of the approximation point $x_g$. The hat mark indicates that the matrix contains the nodal parameter and not the node potential. The domain $S(x_q)$ is S-th support domain around $x_q$. To evaluate the integral in (4.24), we use the Gaussian quadrature rule. Therefore, it is necessary to define integration cells over the problem domain. These cells should have a sufficient number of quadrature points to result in nonsingular (4.23). In this thesis, the quadrature points are considered as the approximation points.

Equation (4.23) is assembled by integrating over the whole $\Omega_C$ domain (considering all quadrature points and their support nodes) and the following system of equations will be obtained:

$$\mathcal{K}^{EFG} \hat{\mathcal{A}}^{EFG} = \mathbf{0} \tag{4.25}$$

where $\mathcal{K}^{EFG}$ is an $N^{EFG} \times N^{EFG}$ square matrix, $\hat{\mathcal{A}}^{EFG}$ is an $N^{EFG} \times 1$ column matrix containing the nodal parameters, and $N^{EFG}$ is the total number of EFG nodes. Equation (4.25) is only defined over $\Omega_C$ and to obtain a complete system of equations for the forward problem, it is coupled with the FEM system of equations in the next section.

It is noteworthy that although the heart of the EFG method is an element-free method, as it was observed, in order to evaluate the integral (4.24) by numerical integration methods, it is required to use a background mesh. Therefore, the EFG method is not really a pure element-free method, and its name is not so fitting. Probably the name was chosen for it in contrast to the finite element method [116].

## 4.2 Hybrid FE-EFG method

By doing FE matrix assembly procedures for (3.23), one can find a system of equations as follows:

$$\mathcal{K}^{FE} \mathcal{A}^{FE} = \mathcal{F} \tag{4.26}$$

where $\mathcal{K}^{FE}$ and $\mathcal{F}$ are obtained by assembling the left- and right-hand side of (3.23) for the improved forward method, respectively, and $\mathcal{A}^{FE}$ contains the phasors of all node potentials. The size of $\mathcal{K}^{FE}, \mathcal{F}$, and $\mathcal{A}^{FE}$ are $N^{FE} \times N^{FE}$, $N^{FE} \times 1$, and $N^{FE} \times 1$, respectively, and $N^{FE}$ is the total number of FE nodes. It is noteworthy that the boundary conditions are embedded in (4.26). To couple FE and EFG methods, the following continuity conditions have to be enforced on the interface boundary $\Gamma^{int}$ between two regions $\Omega^{FE}$ and $\Omega^{EFG}$ (see Figure 4.1) [117]:

$$A_z^{FE}(\mathbf{x}) = A_z^{EFG}(\mathbf{x}) \quad \mathbf{x} \in \Gamma^{int} \tag{4.27}$$

$$\frac{1}{\mu_{FE}} \nabla A_z^{FE}(\mathbf{x}) \cdot \vec{n} \bigg|_{\Gamma^{int}} = \frac{1}{\mu_{EFG}} \nabla A_z^{EFG}(\mathbf{x}) \cdot \vec{n} \bigg|_{\Gamma^{int}} \tag{4.28}$$

where $\vec{n} = \vec{n}^{FE} = -\vec{n}^{EFG}$ is outward unit vector which is normal to the $\Gamma^{int}$, and $\mu_{FE}$ and $\mu_{EFG}$ are the permeability in FE and EFG regions, respectively. To couple (4.25) and (4.26) and impose (4.27) and (4.28) as well, we use the Lagrange multiplier method [117]. In this way, one can consider the following system of equations [117]:

$$\begin{bmatrix} \mathcal{K}^{FE} & 0 & \mathcal{H}^{FE} \\ 0 & \mathcal{K}^{EFG} & \mathcal{H}^{EFG} \\ (\mathcal{H}^{FE})^T & (\mathcal{H}^{EFG})^T & 0 \end{bmatrix} \begin{bmatrix} \mathcal{A}^{FE} \\ \mathcal{A}^{EFG} \\ \eta \end{bmatrix} = \begin{bmatrix} \mathcal{F} \\ 0 \\ 0 \end{bmatrix} \tag{4.29}$$

where

$$(\mathcal{H}^{FE})^T = \begin{bmatrix} 0 & \cdots & -1|_{1,i} & 0 \\ \vdots & \vdots & \ddots & \vdots \\ 0 & \cdots & -1|_{N_s,i} & 0 \end{bmatrix}_{N_s \times N^{FE}} \tag{4.30}$$

$$(\mathcal{H}^{EFG})^T = \begin{bmatrix} \phi_1(\mathbf{x}_{L1}) & \phi_2(\mathbf{x}_{L1}) & \cdots & \phi_N(\mathbf{x}_{L1}) \\ \phi_1(\mathbf{x}_{L2}) & \phi_2(\mathbf{x}_{L2}) & \cdots & \phi_N(\mathbf{x}_{L2}) \\ \vdots & \vdots & \ddots & \vdots \\ \phi_1(\mathbf{x}_{LN_s}) & \phi_2(\mathbf{x}_{LN_s}) & \cdots & \phi_N(\mathbf{x}_{LN_s}) \end{bmatrix}_{N_s \times N^{EFG}} \tag{4.31}$$

Here, $N_s$ is the number of common nodes on the interface $\Gamma^{int}$, $i$ is the global node number in the FE region, and $\eta$ is a column matrix containing Lagrange multipliers.

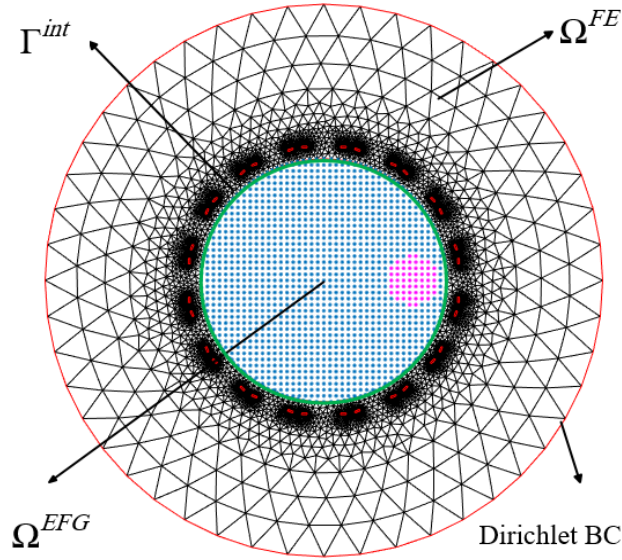

Figure 4.1 Domain handled by EFG and FE regions for a 16-coil MIT system. The domain meshed by triangular elements is $\Omega^{FE}$. The domain is discretized by nodes is $\Omega^{EFG}$. The circular domain indicated by magenta nodes is a target conductivity object placed in a background conductivity. The green circle shows the interface boundary $\Gamma^{int}$. Coils are indicated by red rectangles around the imaging region. The Dirichlet boundary condition is applied on the outer circle shown by red.

It is noteworthy that in contrast to FEM, which naturally handles the discontinuities in material properties (for the electromagnetic problems, these properties are conductivity, permittivity, and permeability), the EFG method, due to the high-order continuity of its shape functions, requires special techniques to treat the material discontinuities [115]. One of the methods to treat the discontinuities in the material properties is the use of the Lagrange multiplier [115]. For our equation, (4.22), the discontinuity in the permeability $\mu$ should be considered for implementing the EFG method. However, this property is continuous over all problem domains in our study since the permeability of most biological tissues are the same as the free space permeability, $\mu_0$.

## 4.3 Numerical experiments

In this section, we consider a simple circular phantom as shown in Figure 3.8, model a 2D MIT system using the improved forward method, solve the forward problem using both FE and FE-EFG methods, and compare their performance.

### 4.3.1 Modeling set-up

The modeling set-up is exactly the same as the one presented in Section 3.3.2.1. Figure 3.8 shows the cross-sectional view of the 2D MIT system. The imaging conducting region has a radius of $R_1$=110 mm, which has been modeled by the EFG method in the hybrid numerical method. The coils were made using 5 mm width and 500 μm thickness copper strips which have been modeled by the FE method in the hybrid approach. Sequential activation of coils using a sinusoidal alternating current of 20 A amplitude excites the imaging region. We choose 10 MHz excitation frequency, which is a typical frequency in biomedical MIT applications. Only the independent measurements are gathered. They are labeled from 1 to 120. The homogeneous Dirichlet boundary condition was imposed at a radius of 250 mm. The space outside of the imaging region is modeled by the FE method in the hybrid approach (see Figure 4.1).

Similar to Chapter 3, to demonstrate the conductivity contrasts in the imaging region, a circular inclusion with the conductivity of $\sigma_i$ is placed in background with the conductivity of $\sigma_b$. The inclusion with a radius of 20 mm was centered at (80, 0) mm. The background also had a radius of 110 mm. Experiments were performed for four different conductivities $\sigma_i = 0$, 2, 4, and 6 S/m and $\sigma_b = 0$ S/m.

Table 4.1 indicates mesh statistics for FE and FE-EFG methods. The number of nodes and elements are chosen so that the run-time for both methods is almost the same. We solved the forward problem based on the improved formulation using both FE and hybrid FE-EFG methods and calculated the induced voltages using (3.29). In addition, as indicated in Table 4.1, we have used an extremely fine mesh to solve the forward problem by the FE method as a

Table 4.1 Mesh statistics for FE, hybrid FE-EFG, and ground truth in each region.

| Mesh statistics<br>Method | Imaging region ($\Omega_C$) | | Outside imaging region ($\Omega_S + \Omega_N$) | |
|---|---|---|---|---|
| | # Nodes | # Elements | # Nodes | # Elements |
| FE | 6302 | 12648 | 12317 | 24500 |
| FE-EFG | 1308 | N/A | 12105 | 23874 |
| Ground truth (FE with extremely fine mesh) | 34135 | 68296 | 49916 | 99280 |

N/A: not applicable.

ground truth. The result obtained by the FE and FE-EFG methods will be compared with the ground truth.

To evaluate (4.24) using Gaussian quadrature rules, a regular square mesh with 1156 nodes and 1089 cells was considered. In each integration cell, 4×4 Gauss quadrature points were used to evaluate (4.24). For the EFG method, linear basis functions with the cubic spline weight function and a $d_{max}$ value of 1.7 was considered.

As explained in Chapter 3, in biomedical applications of MIT, only the real part of the induced voltages is informative, and the imaginary part is negligible. Thus, we use the real part of induced voltages to report the result of numerical solution of the forward problem in all following experiments.

In addition to results will be presented in this section, we have evaluated the hybrid FE-EFG method in comparison to an analytical solution in Appendix 6.

### 4.3.2 Induced voltages by the total magnetic field

Figure 4.2(a) and Figure 4.2(b) show the real part of voltages induced by the total magnetic field for all independent measurements calculated by FE and FE-EFG methods, respectively. The voltages were obtained based on the improved forward method introduced in Chapter 3, $Re\{V_1^I\}$. As can be seen, by changing the inclusion conductivity, the induced voltages calculated by both numerical methods are changed distinguishably as well. These figures are

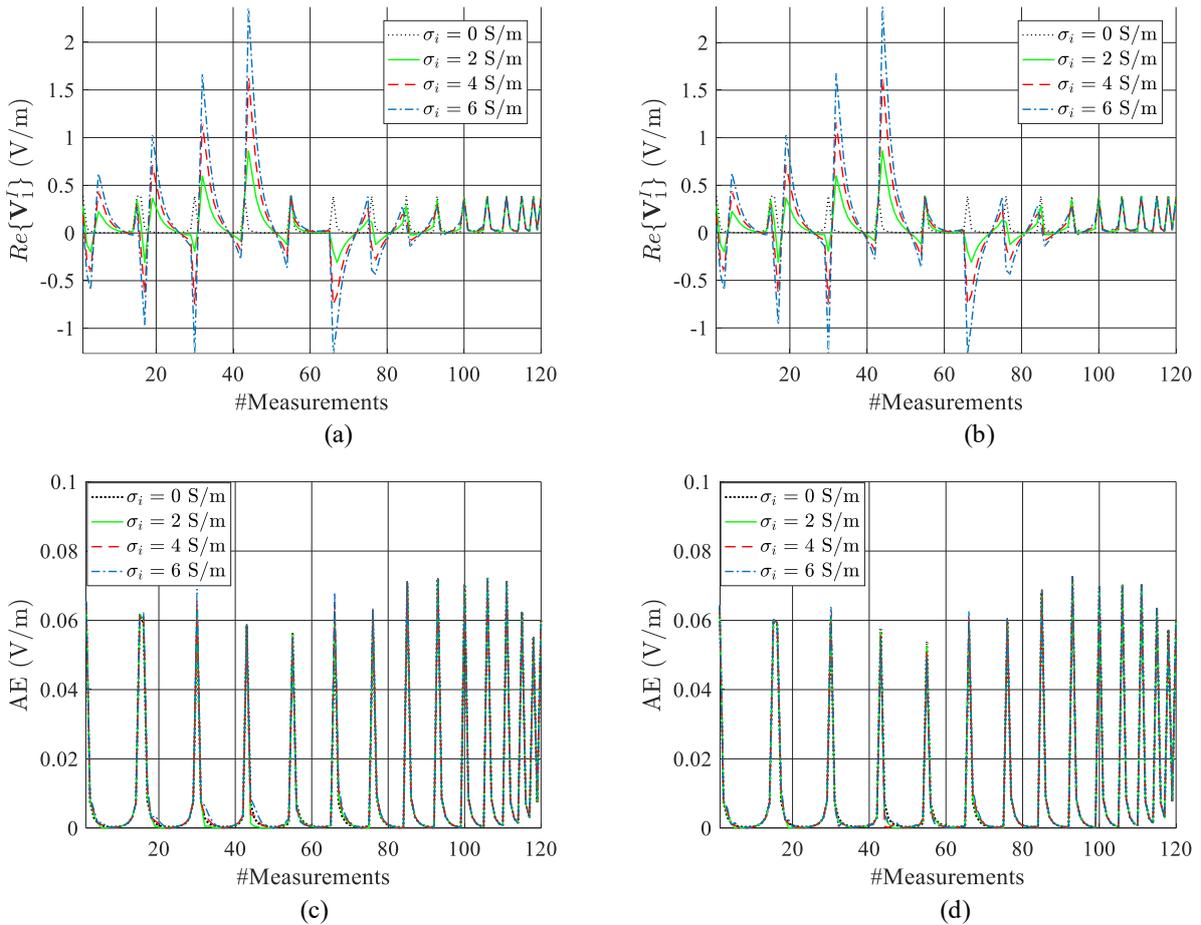

Figure 4.2. Real part of induced voltages due to the total magnetic field, $Re\{V_1^I\}$, calculated by (a) FE method and (b) FE-EFG method. Absolute error of (c) FE method and (d) FE-EFG method compared to the ground truth, $AE = |Re\{V_{1,GT}^I\} - Re\{V_1^I\}|$. Experiment was performed for $\sigma_i = 0$ S/m (black dotted line), $\sigma_i = 2$ S/m (green solid line), $\sigma_i = 4$ S/m (red dashed line), and $\sigma_i = 6$ S/m (blue dash-dotted line). The background conductivity $\sigma_b$ was zero.

similar to Figure 3.9(f) presented in Chapter 3.

To evaluate the results obtained from FE and FE-EFG methods, we compare the voltage induced by these methods by those calculated by the ground truth. For this purpose, we calculated the absolute error as AE = $\left|Re\{V_{1,GT}^I\} - Re\{V_1^I\}\right|$ where $Re\{V_{1,GT}^I\}$ is the real part of voltages induced by the total magnetic field and calculated using the ground truth mesh and $Re\{V_1^I\}$ is the ones calculated using each numerical method. Figure 4.2(c) and Figure 4.2(d) show the absolute error for FE and FE-EFG methods, respectively. As can be seen, the absolute error is almost the same for both methods; however, for the FE-EFG method, it is slightly smaller.

Using (3.30), we also calculated the relative error of the voltages induced by the total magnetic field for each numerical method in comparison with the ground truth. The percentage of relative error for the FE method was 13%, 7.3%, and 5.1% for $\sigma_i = 2$ S/m, $\sigma_i = 4$ S/m, and $\sigma_i = 6$ S/m, respectively. The percentage of relative error for the FE-EFG method was 12.8%, 7.1%, and 4.9% for $\sigma_i = 2$ S/m, $\sigma_i = 4$ S/m, and $\sigma_i = 6$ S/m, respectively. As can be seen, the result of the FE-EFG method is more accurate. This is while the run-time for both numerical methods was 16 seconds.

### 4.3.3 Induced voltages by the secondary magnetic field

Figure 4.3(a) and Figure 4.3(b) illustrate the real part of the induced voltage due to the secondary magnetic field calculated by FE and FE-EFG methods, respectively. The voltages were obtained based on the improved forward method introduced in Chapter 3, $Re\{\Delta V^I\}$. As can be seen, by changing the inclusion conductivity, the induced voltages calculated by both numerical methods are changed distinguishably as well. The patterns of the induced voltages for all target values are almost similar to Figure 4.2(a) and Figure 4.2(b). The reason was explained in Chapter 3; but briefly, it is due to small values for the background voltages at 10 MHz.

To compare FE and FE-EFG methods for induced voltages by the secondary magnetic field, we again compute the absolute error as AE = $\left|Re\{\Delta V_{1,GT}^I\} - Re\{\Delta V_1^I\}\right|$ where $Re\{\Delta V_{1,GT}^I\}$ is the real part of voltages induced by the secondary magnetic field and calculated using the ground truth mesh and $Re\{\Delta V_1^I\}$ is the ones calculated using each numerical method. Figure 4.3(c) and Figure 4.3(d) show the absolute error for FE and FE-EFG methods, respectively. As can be seen, the absolute error is almost the same for both methods; however, for the FE-EFG method is slightly smaller.

Again using (3.30), we calculated the relative error of the voltages induced by the secondary magnetic field for each numerical method in comparison with the ground truth. The relative error percentage for the FE method was 0.55%, 0.54%, and 0.53 % for $\sigma_i = 2$ S/m, $\sigma_i = 4$ S/m, and $\sigma_i = 6$ S/m, respectively. The relative error percentage for the FE-EFG method was 0.35 % for all target values. This manifests that the FE-EFG method is more accurate compared to the FE method. This is while the run-time for both numerical methods was 16 seconds.

## 4.4 Conclusions

In this chapter, for the first time, a hybrid FE-EFG numerical method was developed to solve the 2D MIT improved forward problem, which is based on Maxwell's equations. In this proposed numerical method, the EFG method was employed to solve the problem in the imaging region, and the FE method was used to solve the problem in coil and air around imaging region. It was shown that mesh-less methods are more fruitful in regions which is susceptible to change in the shape and position of the objects [118]. Therefore, we chose the EFG method for solving the problem in the imaging region. On the other hand, the FE method was employed for the constant part of the problem. In this way, we benefited from the advantages of both numerical methods.

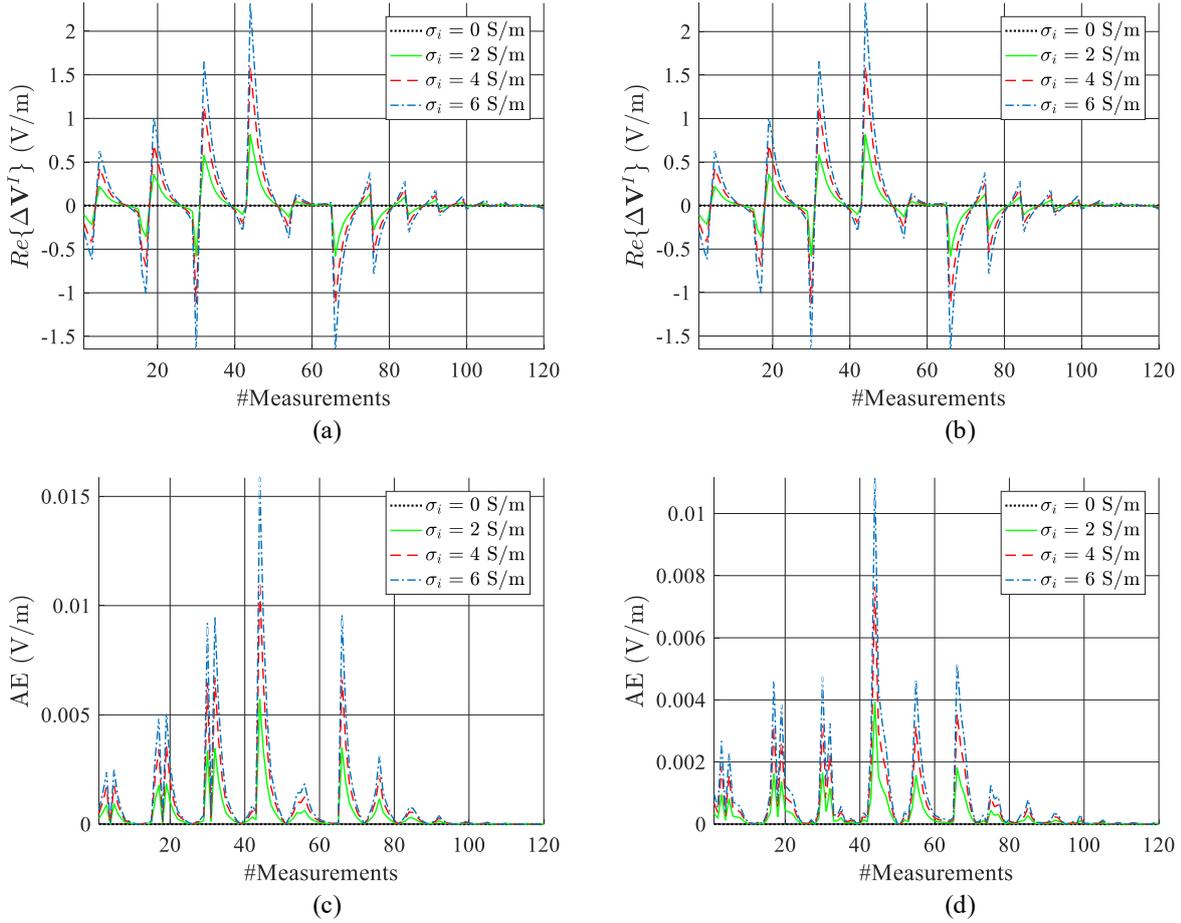

Figure 4.3. Real part of induced voltages due to the secondary magnetic field, $Re\{\Delta V^I\}$, calculated by (a) FE method and (b) FE-EFG method. Absolute error of (c) FE method and (d) FE-EFG method compared to the ground truth, $AE = |Re\{\Delta V^I_{1,GT}\} - Re\{\Delta V^I_1\}|$. Experiment was performed for $\sigma_i = 0$ S/m (black dotted line), $\sigma_i = 2$ S/m (green solid line), $\sigma_i = 4$ S/m (red dashed line), and $\sigma_i = 6$ S/m (blue dash-dotted line). The background conductivity $\sigma_b$ was zero.

Comparison of FE and FE-EFG methods through solving the forward problem for a 16-coil 2D MIT system with a synthetic circular phantom revealed almost the same accuracy for both methods in the same run-time. It is noteworthy that the obtained result for the FE-EFG method was based on a uniform distribution of the node in the imaging region. It is while for the FE method using uniform mesh results in the worse results [117]. The main drawback of the EFG method is imposing the essential boundry condition. In this chapter, we addressed this difficulty by coupling EFG and FE methods. In addition, choosing parameter $d_{max}$ which directly impresses on the accuracy of the result is another diffuctly for EFG method. Results of this chapter manifested that the FE-EFG method holds promise for solving MIT forward problem more accurately in comparison to the FE method.

# Chapter 5: MIT Inverse Problem I

MIT inverse problem aims to reconstruct images of conductivity distribution inside the imaging region. As reviewed in Chapter 2, various algorithms have been developed for biomedical MIT image reconstruction [16], [61], [74]. Single-step linear reconstruction methods such as Tikhonov regularization [16] and truncated singular value decomposition [74] are fast and easy, but they are only able to reconstruct qualitative images. Using iterative linear algorithms like the Landweber method [46] improves the resolution, especially in high conductivity applications. However, they still produce qualitative images and are appropriate for difference imaging. When the objective in the MIT inverse problem is the reconstruction of quantitative images, the nonlinear approaches such as the GN algorithm, Levenberg Marquardt method [78], and Powell's Dog Leg method [78] are applied. However, these nonlinear approaches require extensive computation costs in each iteration. In this chapter, we introduce the regularized GN algorithm to solve the biomedical MIT inverse problem based on the improved forward method.

## 5.1 Regularized Gauss-Newton algorithm

As indicated in Section 2.2.2, the real part of induced voltages has been used in the biomedical MIT. Thus, in the following sections, the real part of induced voltages has been considered to obtain the GN algorithm equations for updating the conductivity distribution.

Considering the regularized GN algorithm, a solution of the MIT inverse problem can be attained by minimizing the least-squares objective function $\phi$ given by [41]:

$$\phi(\boldsymbol{\sigma}) = \arg\min_{\boldsymbol{\sigma}} \left\{ \frac{1}{2} (\mathbf{V}_M - \mathbf{V}_F(\boldsymbol{\sigma}))^T (\mathbf{V}_M - \mathbf{V}_F(\boldsymbol{\sigma})) + \frac{1}{2} \lambda \boldsymbol{\sigma}^T \mathbf{R}^T \mathbf{R} \boldsymbol{\sigma} \right\} \qquad (5.1)$$

where $\boldsymbol{\sigma} \in \mathcal{R}^n$ is the conductivity column matrix, $\mathbf{V}_M \in \mathcal{R}^m$ denotes the real-valued column matrix which contains the real part of measured voltages, $\mathbf{V}_F(\boldsymbol{\sigma}): \mathcal{R}^n \to \mathcal{R}^m$ is the real-valued column matrix obtained from the forward solver, and the numbers $n$ and $m$ represent the number of image pixels and independent measurements, respectively.

Due to the diffuse nature of induced eddy currents, the MIT inverse problem is severely ill-posed. Therefore, the last term on the right-hand side of (5.1) is added to regularize the problem according to the Tikhonov regularization method. The matrix $\mathbf{R} \in \mathcal{R}^{m \times n}$ is a regularization matrix and $\lambda$ is a regularization parameter. The regularization matrix could be an identity matrix or a difference operator between neighboring conductivities. The parameter $\lambda$ should be chosen optimally to provide a balance between the stability and accuracy of the inverse problem solution.

By expanding (5.1), one can obtain

$$\phi(\boldsymbol{\sigma}) = \frac{1}{2}\left(\mathbf{V}_M^T\mathbf{V}_M - \mathbf{V}_M^T\mathbf{V}_F(\boldsymbol{\sigma}) - \mathbf{V}_F(\boldsymbol{\sigma})^T\mathbf{V}_M + \mathbf{V}_F(\boldsymbol{\sigma})^T\mathbf{V}_F(\boldsymbol{\sigma})\right) + \frac{1}{2}\lambda\boldsymbol{\sigma}^T\mathbf{R}^T\mathbf{R}\boldsymbol{\sigma} \qquad (5.2)$$

To find a candidate value of $\boldsymbol{\sigma}$ that minimizes $\phi(\boldsymbol{\sigma})$, we differentiate (5.2) with respect to $\boldsymbol{\sigma}$ (see the footnote[1] relation)

$$\left[\frac{\partial \phi}{\partial \boldsymbol{\sigma}}\right]_{n\times 1} = \boldsymbol{\phi}'(\boldsymbol{\sigma}) = \frac{1}{2}\left(-\left(\mathbf{V}_F'(\boldsymbol{\sigma})\right)^T\mathbf{V}_M - \left(\mathbf{V}_F'(\boldsymbol{\sigma})\right)^T\mathbf{V}_M + \left(\mathbf{V}_F'(\boldsymbol{\sigma})\right)^T\mathbf{V}_F + \left(\mathbf{V}_F'(\boldsymbol{\sigma})\right)^T\mathbf{V}_F\right)$$
$$+ \lambda\mathbf{R}^T\mathbf{R}\boldsymbol{\sigma} \qquad (5.3)$$

By simplifying (5.3), we have

$$\boldsymbol{\phi}'(\boldsymbol{\sigma}) = -\left(\mathbf{V}_F'(\boldsymbol{\sigma})\right)^T\left(\mathbf{V}_M - \mathbf{V}_F(\boldsymbol{\sigma})\right) + \lambda\mathbf{R}^T\mathbf{R}\boldsymbol{\sigma} = \mathbf{J}^T\left(\mathbf{V}_F(\boldsymbol{\sigma}) - \mathbf{V}_M\right) + \lambda\mathbf{R}^T\mathbf{R}\boldsymbol{\sigma} \qquad (5.4)$$

where

$$\mathbf{J} = \mathbf{V}_F'(\boldsymbol{\sigma}) = \frac{\partial \mathbf{V}_F(\boldsymbol{\sigma})}{\partial \boldsymbol{\sigma}} \qquad (5.5)$$

is the Jacobian matrix with the dimension of $m \times n$. Since $\boldsymbol{\phi}'(\boldsymbol{\sigma})$ is a nonlinear function of $\boldsymbol{\sigma}$, one can take a Taylor series expansion around $\boldsymbol{\sigma}_k$:

$$\boldsymbol{\phi}'(\boldsymbol{\sigma}) = \boldsymbol{\phi}'(\boldsymbol{\sigma}_k) + \boldsymbol{\phi}''(\boldsymbol{\sigma}_k)\mathbf{d}_k + \mathcal{O}(\|\mathbf{d}_k\|^2) \qquad (5.6)$$

where $\mathbf{d}_k = \boldsymbol{\sigma} - \boldsymbol{\sigma}_k$, $k$ is the iteration number, and $\mathcal{O}(\|\mathbf{d}_k\|^2)$ is higher-order terms. The term $\boldsymbol{\phi}''$ is called the modified Hessian matrix, given by

$$\mathbf{H}(\boldsymbol{\sigma}) = \boldsymbol{\phi}''(\boldsymbol{\sigma}) = (\mathbf{V}_F'(\boldsymbol{\sigma}))^T(\mathbf{V}_F'(\boldsymbol{\sigma}) + \mathbf{0}) + (\mathbf{V}_F''(\boldsymbol{\sigma}))^T\left(\mathbf{I} \otimes (\mathbf{V}_F(\boldsymbol{\sigma}) - \mathbf{V}_M)\right) + \lambda\mathbf{R}^T\mathbf{R}$$
$$= \mathbf{J}^T\mathbf{J} + \sum_{j=1}^{k} V_{F_j}''(\boldsymbol{\sigma})\left(V_{F_j}(\boldsymbol{\sigma}) - V_{M_j}\right) + \lambda\mathbf{R}^T\mathbf{R} \qquad (5.7)$$

where $\otimes$ is the Kronecker matrix product. The dimension of $\mathbf{H}$ is $n \times n$. The modified term returns to $\lambda\mathbf{R}^T\mathbf{R}$ which presents due to the regularization. The term $\mathbf{V}_F''(\boldsymbol{\sigma})$ in (5.7) is difficult to calculate explicitly, but often it is negligible relative to $\mathbf{J}^T\mathbf{J}$ [119]. Therefore, we may approximate the modified Hessian matrix by:

$$\mathbf{H} \approx \mathbf{J}^T\mathbf{J} + \lambda\mathbf{R}^T\mathbf{R} \qquad (5.8)$$

By keeping the linear terms of (5.6) and setting it equal to zero, one can obtain:

$$-\boldsymbol{\phi}'(\boldsymbol{\sigma}_k) = \mathbf{H}_k\mathbf{d}_k \qquad (5.9)$$

where $\mathbf{H}_k$ is Hessian matrix computed for $\boldsymbol{\sigma}_k$. Substituting (5.4) and (5.8) into (5.9), assuming $\mathbf{H}$ is invertible, solving for $\mathbf{d}$, we obtain:

$$\mathbf{d}_k = (\mathbf{J}_k^T\mathbf{J}_k + \lambda\mathbf{R}^T\mathbf{R})^{-1}\left[\mathbf{J}_k^T(\mathbf{V}_M - \mathbf{V}_{F_k})\right] - \lambda\mathbf{R}^T\mathbf{R}\boldsymbol{\sigma}_k \qquad (5.10)$$

By adding the update term $\mathbf{d}_k$ to the current iteration $\boldsymbol{\sigma}_k$, the conductivity distribution at $(k+1)$-th iteration will be as follows:

---

[1] $\frac{\partial \mathbf{a}^T f(\mathbf{p})}{\partial \mathbf{p}} = \frac{\partial f(\mathbf{p})^T \mathbf{a}}{\partial \mathbf{p}} = (f'(\mathbf{p}))^T \mathbf{a}$

$$\boldsymbol{\sigma}_{k+1} = \boldsymbol{\sigma}_k + (\mathbf{J}_k^T \mathbf{J}_k + \lambda \mathbf{R}^T \mathbf{R})^{-1} \left[ \mathbf{J}_k^T (\mathbf{V}_M - \mathbf{V}_{F_k}) \right] - \lambda \mathbf{R}^T \mathbf{R} \boldsymbol{\sigma}_k \tag{5.11}$$

## 5.2 Jacobian matrix calculation

The Jacobian matrix, as is computed by the forward problem, is employed to solve the inverse problem. Elements of the matrix specify the sensitivity of simulated voltages to the conductivity of image pixels. The Jacobian matrix is generally calculated using two different techniques in electrical tomography [119]: the sensitivity technique and the standard technique. The former is based on Geselowitz reciprocity theorem [120] and is widely used in MIT studies [16], [41], [44]. This method is mainly an independent formulation regardless of the method used in solving the forward problem. In the latter, the Jacobian is directly calculated by the rigorous numerical differentiation of the discretized governing equation in the forward problem with respect to the electrical conductivity. This method has widely been used in EIT studies with the complete electrode model [8], [121], whereas it has only been used in one MIT study [122]. In [122], the standard technique was called the direct method.

Since the sensitivity technique assumes that total current density is a constant space function in coil regions, it cannot be applied to the improved forward method. As a result, we introduce the standard technique to calculate the Jacobian matrix for the improved forward method. In the following sections, the methods of calculation of the Jacobian matrix are presented in detail.

### 5.2.1 Sensitivity method

In Figure 5.1(a), assume the applied current $I_j$ to the port j induces the electric field distribution $\vec{E}_j(\mathbf{x})$ inside the domain $\Omega$. The induced voltage $U_i$ can be measured from port i. In Figure 5.1(b), suppose that the applied current $I_i$ to the port i induces the electric field distribution $\vec{E}_i(\mathbf{x})$ inside the domain $\Omega$. The induced voltage $U_j$ can be measured from port j. One can write Ohm's law for Figure 5.1(a) as follows:

$$\vec{J} = \sigma \vec{E}_j \tag{5.12}$$

where $\vec{J}$ is the current density distribution inside $\Omega$. Taking the inner product of two sides of (5.12) with $\vec{E}_i$ and then integrating over $\Omega$ result in:

$$\int_\Omega \vec{J} \cdot \vec{E}_i \, d\Omega = \int_\Omega \sigma \vec{E}_j \cdot \vec{E}_i \, d\Omega \tag{5.13}$$

Left hand sided of (5.13) can be rewritten as follows [75]:

$$\int_\Omega \vec{J} \cdot \vec{E}_i \, d\Omega = \int_\Omega \vec{J} ds \cdot \vec{E}_i \, d\ell \tag{5.14}$$

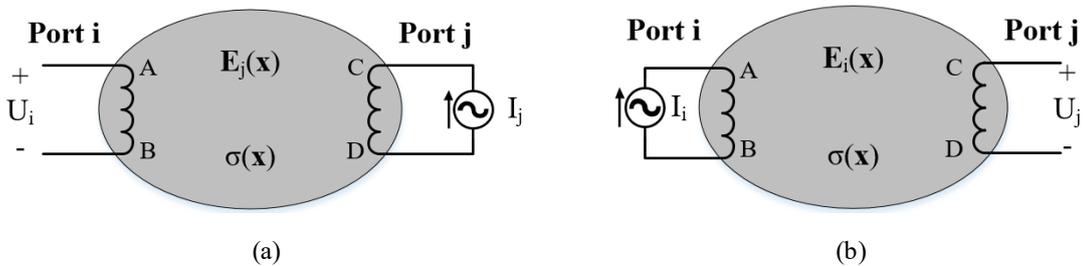

(a)                                        (b)

Figure 5.1 (a) Domain $\Omega$ with conductivity distribution of $\sigma(\mathbf{x})$. The applied current $I_j$ to the port j induces the electric field distribution $\vec{E}_j(\mathbf{x})$ inside the domain. The potential difference $U_i$ can be measured between points A and B. (b) The applied current $I_i$ is to the port i induces the electric field distribution $\vec{E}_i(\mathbf{x})$ inside the domain. The potential differenc $U_j$ can be measured between points C and D.

Assume the input current density to the port j is constant, $\vec{J}_0$. The normal component of $\vec{J}$ is zero except for input and output points of the current. Consequently, the current density at the current input point ($\mathbf{x}_1$) and the current output point ($\mathbf{x}_2$) are defined as follows:

$$\vec{J} = \vec{J}_0 \delta(\mathbf{x} - \mathbf{x}_1) \quad \text{at the current input point}$$
$$\vec{J} = \vec{J}_0 \delta(\mathbf{x} - \mathbf{x}_2) \quad \text{at the current output point}$$

Substituting above current density in (5.14) gives [75]:

$$\int_\Omega \vec{J} ds \cdot \vec{E}_i \, d\ell = \int_\Omega \vec{J}_0 \delta(\mathbf{x} - \mathbf{x}_1) ds \cdot \vec{E}_i(\mathbf{x}) \, d\ell - \int_\Omega \vec{J}_0 \delta(\mathbf{x} - \mathbf{x}_2) ds \cdot \vec{E}_i(\mathbf{x}) \, d\ell$$
$$= \int_\Omega \vec{J}_0 \delta(\mathbf{x} - \mathbf{x}_1) ds \cdot \vec{E}_i(\mathbf{x}_1) \, d\ell - \int_\Omega \vec{J}_0 \delta(\mathbf{x} - \mathbf{x}_2) ds \cdot \vec{E}_i(\mathbf{x}_2) \, d\ell \quad (5.15)$$
$$= \big(u_i(\mathbf{x}_1) - u_i(\mathbf{x}_2)\big) I_j = U_{CD} I_j$$

Substituting (5.15) in (5.13) results in:

$$U_{CD} I_j = \int_\Omega \sigma \vec{E}_j \cdot \vec{E}_i \, d\Omega \qquad (5.16)$$

For Figure 5.1(b), a similar procedure can be followed. The result would be:

$$U_{CD} I_j = U_{AB} I_i = \int_\Omega \sigma \vec{E}_j \cdot \vec{E}_i \, d\Omega \qquad (5.17)$$

Based on the reciprocity theorem, the trans-impedance seen from port i, $U_{AB}/I_j$, is equal to the transimpedance seen from port j, $U_{CD}/I_i$,

$$Z_T = \frac{U_{CD}}{I_i} = \frac{U_{AB}}{I_j} \qquad (5.18)$$

If $I_i = I_j = I_c$ then from (5.17) and (5.18), one can obtain:

$$U_{CD} = U_{AB} = \frac{1}{I_c} \int_\Omega \sigma \vec{E}_j \cdot \vec{E}_i \, d\Omega \qquad (5.19)$$

If $\vec{A} - \vec{A}$ formulation (see Appendix 2) is employed for the forward problem, then $\vec{E} = -j\omega\vec{A}$ where $\vec{A}$ and $\omega$ are the MVP and angular frequency, respectively. Substituting this relation into (5.19) gives

$$U_{CD} = U_{AB} = -\frac{\omega^2}{I_c} \int_\Omega \sigma \vec{A}_j \cdot \vec{A}_i \, d\Omega \qquad (5.20)$$

Assuming that $\Omega$ is discretized to $N_{FE}$ finite elements, (5.20) can be written as follows:

$$U_{CD} = U_{AB} = -\frac{\omega^2}{I_c} \sum_{e=1}^{N_{FE}} \sigma_e \int_{\Omega_e} \vec{A}_{e,j} \cdot \vec{A}_{e,i} \, d\Omega \qquad (5.21)$$

where $\sigma_e$ and $\Omega_e$ is the conductivity and the interior of $e$-th element, respectively, and $\vec{A}_{e,j}$ and $\vec{A}_{e,i}$ are MVP inside $e$-th element when port j and port i are excited, respectively. The sensitivity of the measurement i,j to a change in the conductivity in the $e$-th element, $\partial V_{ij}/\partial \sigma_e$, is then,

$$S^e = \frac{\partial U_{CD}}{\partial \sigma_e} = \frac{\partial U_{AB}}{\partial \sigma_e} = \frac{\partial V_{ij}}{\partial \sigma_e} = -\frac{\omega^2}{I_c} \int_{\Omega_e} \vec{A}_{e,j} \cdot \vec{A}_{e,i} \, d\Omega \qquad (5.22)$$

In 2D case, $\vec{A} = A_z \vec{a}_z$, and (5.22) would be as follows:

$$\frac{\partial V_{ij}}{\partial \sigma_e} = -\frac{\omega^2}{I_c} \int_{R_e} A_{z\,e,j} A_{z\,e,i} \, d\Omega \qquad (5.22)$$

where $R_e$ is the cross-sectional region of the $e$-th image pixel. If the first order triangular elements are used in the FE method, $A_z$ inside each element can be approximated as $A_z(x, y) \cong \mathcal{N}_e(x, y)\,\mathcal{A}_e$ where $\mathcal{A}_e = [A_1 \quad A_2 \quad A_3]^T$ contains the nodes' potential of the element and $\mathcal{N}_e$ is a matrix containing corresponding shape functions. Then, for each image pixel with the conductivity $\sigma_e$, (5.22) in discrete form is:

$$\frac{\partial V_{ij}}{\partial \sigma_e} = -\frac{\omega^2}{I_c} \mathcal{A}_{e,i} \left( \int_{R_e} \mathcal{N}_e \, \mathcal{N}_e^T \, d\Omega \right) \mathcal{A}_{e,j}^T \tag{5.23}$$

The integral of multiplication of basis functions in (5.23) is calculated as follows:

$$\int_{R_e} \mathcal{N}_e \, \mathcal{N}_e^T \, d\Omega = \int_{R_e} \begin{bmatrix} N_e^1 \\ N_e^2 \\ N_e^3 \end{bmatrix} [N_e^1 \quad N_e^2 \quad N_e^3] \, d\Omega$$

$$= \int_{R_e} \begin{bmatrix} (N_e^1)^2 & N_e^1 N_e^2 & N_e^1 N_e^3 \\ N_2^e N_e^1 & (N_e^2)^2 & N_e^2 N_e^3 \\ N_e^3 N_e^1 & N_e^3 N_e^2 & (N_e^3)^2 \end{bmatrix} d\Omega \tag{5.24}$$

To compute (5.24), we use the following relation from the FE method:

$$\int_{R_e} (N_e^1)^a \, (N_2^e)^b (N_e^3)^c dx^2 = \frac{a!\, b!\, c!}{(a + b + c + 2)!} 2\Delta_e \tag{5.25}$$

where $\Delta_e$ is the area of $e$-th element. Using (5.25), one can simplify (5.24) as follows:

$$\int_{R_e} \mathcal{N}_e \, \mathcal{N}_e^T \, d\Omega = \begin{bmatrix} \frac{1}{6} & \frac{1}{12} & \frac{1}{12} \\ \frac{1}{12} & \frac{1}{6} & \frac{1}{12} \\ \frac{1}{12} & \frac{1}{12} & \frac{1}{6} \end{bmatrix} \Delta_e = \mathcal{M}_e \tag{5.26}$$

Thus, the equation for calculation of the Jacobian matrix elements with sensitivity method in 2D MIT will be as follows:

$$\frac{\partial V_{ij}}{\partial \sigma_e} = -\frac{\omega^2}{I_c} \mathcal{A}_{e,i} \, \mathcal{M}_e \, \mathcal{A}_{e,j}^T \tag{5.27}$$

### 5.2.2 Standard method

As seen from (3.29) in Chapter 3, the induced voltage $V$ is a function of magnetic vector potential $A_z$. Thus, to obtain $\partial V/\partial \sigma$, the calculation of $\partial A_z/\partial \sigma$ is required. To get this term in the discrete form, we start with FE equations of the forward problem. After doing the FE matrix assembly procedures for (3.16) or (3.23), one can find a system of equations as follows:

$$\mathcal{K}\mathcal{A} = \mathcal{F} \tag{5.28}$$

where $\mathcal{K}_{N \times N}$ and $\mathcal{F}_{N \times 1}$ are obtained by assembling the left- and right-hand side of (3.16) and (3.23) for the early and improved forward method, respectively, $\mathcal{A}_{N \times 1}$ contains the phasors of all node potentials, and $N$ is the total number of FE nodes.

By differentiation of (5.28) with respect to $\sigma_e$, one can obtain:

$$\mathcal{K} \frac{\partial \mathcal{A}}{\partial \sigma_e} + \frac{\partial \mathcal{K}}{\partial \sigma_e} \mathcal{A} = 0 \tag{5.29}$$

or

$$\mathcal{K} \frac{\partial \mathcal{A}}{\partial \sigma_e} = -\frac{\partial \mathcal{K}}{\partial \sigma_e} \mathcal{A}. \tag{5.30}$$

The term $\partial\mathcal{F}/\partial\sigma_e$ is zero as the source currents is not dependent upon $\sigma_e$. The Gaussian elimination method can be used to solve the resulting linear system of equations (5.30) for $\partial\mathcal{A}/\partial\sigma_e$. In (5.30), $\mathcal{K}$ and $\mathcal{A}$ are known from solving the forward problem. In our application, each element in FEM is a triangle; therefore each element has three nodes. In this case, the matrix $\partial\mathcal{K}/\partial\sigma_e$ in (5.30) has at most nine nonzero elements, no matter how large the dimension of $\mathcal{K}$ is. It means that the matrix is very sparse.

By considering the discrete form of (3.29) and derivating with respect to $\sigma_e$, the Jacobian matrix elements in the standard technique can be written as follows:

$$\frac{\partial V_{ij}}{\partial \sigma_e} = j\omega l \left( \left[\frac{\partial \mathcal{A}_{ij}}{\partial \sigma_e}\right]_p - \left[\frac{\partial \mathcal{A}_{ij}}{\partial \sigma_e}\right]_q \right) \quad (5.31)$$

where $V_{ij}$ and $\mathcal{A}_{ij}$ are the induced voltage and the column matrix containing the phasors of all node potentials when the $j$-th coil is excited and $i$-th coil is measured, respectively, and $\partial\mathcal{A}_{ij}/\partial\sigma_e$ is obtained by solving (5.30). Here, $\left[\partial\mathcal{A}_{ij}/\partial\sigma_e\right]_p$ and $\left[\partial\mathcal{A}_{ij}/\partial\sigma_e\right]_q$ indicate the $p$-th and $q$-th element of $\partial\mathcal{A}_{ij}/\partial\sigma_e$, corresponding to points **p** and **q** in (3.29), respectively.

Elements of the Jacobian matrix obtained by (5.27) or (5.31) are complex values. In biomedical MIT, the real part of those enters in the computations of the inverse problem.

## 5.3 Adaptive method for choosing the regularization parameter

The solution of the inverse problem (5.1) is very sensitive to the regularization parameter $\lambda$, and thus the accurate choice of $\lambda$ is crucial. There are several methods to determine the regularization parameters for inverse algorithms. For single-step inverse algorithms, several methods such as L-curves [47] or Morozov-criterion [54] have been used. However, in this thesis, we uses the GN algorithm, which is a nonlinear iterative algorithm and the mentioned methods are not applicable. There are a few studies in MIT literature to address the choice of the regularization parameter for nonlinear algorithms. In [41], the regularization parameter has been chosen empirically in which the parameter is changed by trial and error to obtain the best image. Inspired by [123], an adaptive method has been proposed in [78] to choose the regularization parameter in the MIT inverse problem. In this thesis, we employ this method for choosing the regularization parameter. In the following, we briefly explain the procedure of the adaptive method.

Let consider the Taylor expansion of the objective function $\phi(\sigma)$, defined by (5.2), around the current iteration $\sigma_k$ as follows:

$$\phi(\sigma) = \phi(\sigma_k) + \mathbf{d}_k^T \phi'(\sigma_k) + \frac{1}{2}\mathbf{d}_k^T \mathbf{H}_k \mathbf{d}_k + \mathcal{O}(\|\mathbf{d}_k\|^3) \quad (5.32)$$

where $\mathcal{O}(\|\mathbf{d}_k\|^3)$ indicates higher-order terms. Let consider the first three terms of (5.32) as a quadratic model L(**d**) which approximates $\phi(\sigma)$ in the neighborhood of $\sigma_k$:

$$\mathrm{L}(\mathbf{d}_k) = \phi(\sigma_k) + \mathbf{d}_k^T \phi'(\sigma_k) + \frac{1}{2}\mathbf{d}_k^T \mathbf{H}_k \mathbf{d}_k \quad (5.33)$$

The approximation error is controlled by $\mathcal{O}(\|\mathbf{d}_k\|^3)$ which means that the model L is a good approximation if the update term $\mathbf{d}_k$ is small. The regularization parameter $\lambda$ can be chosen in such a way that the error in model L is minimized. In the adaptive method, $\lambda$ is updated in each iteration. Thus, we replace it with $\lambda_k$.

To update $\lambda_k$ a gain factor $\rho$ is defined which measures the error in the model L compared to the objective function $\phi$. When the update term $\mathbf{d}_k$ is calculated by (5.10), $\rho$ is evaluated as follows:

$$\rho = \frac{\phi(\pmb{\sigma}_k) - \phi(\pmb{\sigma}_k + \mathbf{d}_k)}{L(\mathbf{0}) - L(\mathbf{d}_k)} \tag{5.34}$$

where the numerator and denominator show the actual and predicted reduction in the objective function, respectively. By using (5.33), we expand the denominator:

$$L(\mathbf{0}) - L(\mathbf{d}_k) = -\mathbf{d}_k^T \phi'(\pmb{\sigma}_k) - \frac{1}{2}\mathbf{d}_k^T \mathbf{H}_k \mathbf{d}_k \tag{5.35}$$

By using (5.9), one can replace $\mathbf{H}_k \mathbf{d}_k$ with $-\phi'(\pmb{\sigma}_k)$. Now, (5.35) will be as follows:

$$L(\mathbf{0}) - L(\mathbf{d}_k) = -\frac{1}{2}\mathbf{d}_k^T \phi'(\pmb{\sigma}_k) \tag{5.36}$$

If the Hessian matrix $\mathbf{H}_k$ is positive definite[1], then $\mathbf{d}_k^T \mathbf{H}_k \mathbf{d}_k = -\mathbf{d}_k^T \phi'(\pmb{\sigma}_k) > 0$. It means the denominator of (5.34) is always positive and its sign is equal to the sign of the numerator. Based on the value of the gain factor $\rho$, three cases can be defined:

i. $\rho$ is negative or zero. It means the numerator is negative and consequently the descending condition $\phi(\pmb{\sigma}_k + \mathbf{d}_k) < \phi(\pmb{\sigma}_k)$ is not fulfilled. As a result, the update term $\mathbf{d}_k$ is rejected, $\lambda_k$ is increased to constrain $\mathbf{d}_k$ to be smaller, and the iteration is repeated.
ii. $\rho$ is posotive and smaller than 1. It means that the model L is a poor approximation. As a result, the update term $\mathbf{d}_k$ is accepted whereas $\lambda_k$ should increase in the next iteration.
iii. $\rho$ is close to 1. It means that the model L is a good approximation. As a result, the update term $\mathbf{d}_k$ is accepted and $\lambda_k$ is decreased in the next iteration.

Based on the above definitions, the following strategy is adapted to update $\pmb{\sigma}_k$ and $\lambda_k$:

$$\begin{aligned}&\text{if } \rho > 0 \\ &\quad \lambda_{k+1} = \lambda_k \times \max(0.5, 1 - (2\rho - 1)^3); \quad \eta = 2; \\ &\quad \pmb{\sigma}_{k+1} = \pmb{\sigma}_k + \mathbf{d}_k; \\ &\text{else} \\ &\quad \lambda_{k+1} = \lambda_k \times \eta; \quad \eta = \eta \times 2; \\ &\quad \pmb{\sigma}_{k+1} = \pmb{\sigma}_k;\end{aligned} \tag{5.37}$$

## 5.4 Full algorithm

Figure 5.2 shows a pseudo-code for the regularized GN algorithm presented in this chapter. This pseudo-code is adapted based on the work presented in [78] with minor modifications. In iterative algorithms, choosing a good initial guess $\pmb{\sigma}_0$ is very important to the fast convergence. Here, the initial guess is computed by a single step Tikhonov method as:

$$\pmb{\sigma}_0 = -\left(\mathbf{J}_h^T \mathbf{J}_h + \lambda_0 \mathbf{R}^T \mathbf{R}\right)^{-1} \mathbf{J}_h^T \mathbf{V}_M \tag{5.38}$$

where $\mathbf{J}_h$ is the Jacobian matrix calculated for a given homogeneous conductivity value $\sigma_h$. To find $\lambda_0$, the maximum of diagonal elements of $\mathbf{J}_h^T \mathbf{J}_h$ is obtained and multiplied by a coefficient $\tau$ [78]. The coefficient is chosen by the user so that the algorithm becomes stable. In addition, based on the prior information about the minimum ($\nu$) and maximum ($\xi$) values of the conductivity, a constraint is applied to the conductivity elements within $\pmb{\sigma}_{new}$. Here, $\pmb{\sigma}_{new}$ is a

---
[1] It means that $\mathbf{H}$ is non-singular and for any nonzero vector $\mathbf{z}$: $\mathbf{z}^T \mathbf{H} \mathbf{z} > 0$.

**Problem**: $\arg\min_{\sigma} \{\phi(\sigma)\}$; $\phi(\sigma) = \frac{1}{2}(V_M - V_F(\sigma))^T(V_M - V_F(\sigma)) + \frac{1}{2}\lambda\sigma^T R^T R\sigma$

**Inputs**: Simulated measurements $V_M$; Coefficient $\tau$; Coefficient $\eta$; Minimum conductivity value $\nu$; Maximum conductivity value $\xi$; Maximum number of iterations $K$;

1- Select the initial regularization parameter
$\lambda_0 = \tau \times \max(\text{diag}[J_h^T J_h])$; where $J_h$ is Jacobian matrix calculated for a given homogeneous conductivity value $\sigma_h$;

2- Find $\sigma_0$ using single step Tikhonov reconstruction method: $\sigma_0 = -(J_h^T J_h + \lambda_0 R^T R)^{-1} J_h^T V_M$;

3- Evaluate the objective function: $\phi(\sigma_0) = \frac{1}{2}(V_M - V_F(\sigma_0))^T(V_M - V_F(\sigma_0)) + \frac{1}{2}\lambda_0 \sigma^T R^T R\sigma_0$;

4- Compute the gradient: $g = \phi'(\sigma_0) = J_0^T(V_F(\sigma_0) - V_M) + \lambda_0 R^T R\sigma_0$;

5- Calculate: $A = J_0^T J_0$;

6- Set $\lambda_k = \lambda_0$; $\sigma_k = \sigma_0$; $k = 0$; *exit*=**false**;

7- Start algorithm:
    **while** $k < K$ and *exit* $\neq$ **true**
        Compute the Hessian matrix: $H = A + \lambda_k R^T R$
        Calculate the update term: $d_k = -gH$
        Update the conductivity distribution: $\sigma_{new} = \sigma_k + d_k$;
        Constrain the conductivity distribution: $i = 1:n \mid \sigma_{new}[i] = \begin{cases} \max(\nu, \sigma_{new}[i]) \\ \min(\xi, \sigma_{new}[i]) \end{cases}$
        Evaluate the objective function:
        $\phi(\sigma_{new}) = \frac{1}{2}(V_M - V_F(\sigma_{new}))^T(V_M - V_F(\sigma_{new})) + \frac{1}{2}\lambda_k \sigma_{new}^T R^T R\sigma_{new}$;

        Calculate the gain ratio: $\rho = \frac{\phi(\sigma_k) - \phi(\sigma_{new})}{L(0) - L(d)}$;

        **if** $\rho > 0$
            $\sigma_{k+1} = \sigma_{new}$;
            $\lambda_{k+1} = \lambda_k \times \max(0.5, 1 - (2\rho - 1)^3)$; $\eta = 2$;
            $g = J_{k+1}^T(V_F(\sigma_{k+1}) - V_M) + \lambda R^T R\sigma_{k+1}$;
            $A = J_{k+1}^T J_{k+1}$;
        **else**
            $\sigma_{k+1} = \sigma_k$;
            $\lambda_{k+1} = \lambda_k \times \eta$; $\eta = \eta \times 2$
            **if** $\eta > 32$ *exit* = **true**; **end**
        **end**

        $k = k + 1$
    **end**

Figure 5.2 A pseudo code for the regularized GN algorithm with adaptive method of choosing regularization parameter

column matrix containing conductivity values which it should be decided that accepted or rejected as $\sigma_{k+1}$ based on the value of $\rho$.

## 5.5 Numerical experiments

In this section, a 2D MIT problem is modeled and the results of numerical implementation of the inverse problem using both early and improved forward methods, introduced in Chapter 3, are presented and compared. All simulations were executed on a core i5 2.6 GHz laptop with 8 GB of RAM.

### 5.5.1 Modeling set-up

Figure 5.3(a) shows the cross-sectional view of the 2D MIT system, including eight air-core coils used for both excitation and sensing. The coils are rectangular-shaped and arranged in a circular ring surrounding the imaging region. The imaging region has a radius of 7 cm. Sequential activation of coils using a sinusoidal alternating current of 1 A amplitude excites the imaging region. The previous studies on the low conductivity MIT used excitation frequency in the range of 0.1-13 MHz [7], [53], [98]. Accordingly, we chose 10 MHz for our simulations. Using higher frequencies is desirable but not allowed because the magneto-quasi-static approximation is no longer valid, and it results in more dominant wave-propagation phase delay [24]. After each excitation coil was activated, the induced voltages in the remaining coils (except those previously used for excitation) were measured. The homogeneous Dirichlet boundary condition was imposed at a radius of 10 cm.

In order to evaluate and compare the reconstruction results using the early and improved forward methods, two examples have been considered. In Example I, a circular inclusion with the conductivity of $\sigma_t = 10$ S/m is placed in background with the conductivity of $\sigma_b = 2$ S/m. In Example I.A, as shown in Figure 5.4(a), the inclusion with a radius of 1.5 cm was centered at (-5, 0) cm, and in Example I.B, as shown in Figure 5.5(a), the inclusion with a radius of 1 cm was centered at (-5.5, 0) cm. In Example II, two circular inclusions with the conductivity of $\sigma_{t1} = 5$ S/m and $\sigma_{t2} = 10$ S/m are placed in background with the conductivity of $\sigma_b = 2$ S/m. In Example II.A, as shown in Figure 5.6(a), $\sigma_{t1}$ and $\sigma_{t2}$ with a radius of 1.5 cm were centered at (-3.5, 3.5) cm and (-3.5, -3.5) cm, respectively. In Example II.B, as shown in Figure 5.7(a), $\sigma_{t1}$ and $\sigma_{t2}$ with a radius of 1 cm were centered at (-3.9, 3.9) cm and (-3.9, -3.9) cm, respectively. The theoretical limit given for the minimum detectable inhomogeneity radius for the modeled system is $r_{min} = 1.3$ cm. This limit is obtained from $r_{min} = R/\sqrt{m}$, where $R$ and $m$ are the radius of imaging region and the number of independent measurements, respectively [35]. Since the radius 1 cm of the small target object is a little smaller than the limit, we placed the center of the small target object closer to the boundary.

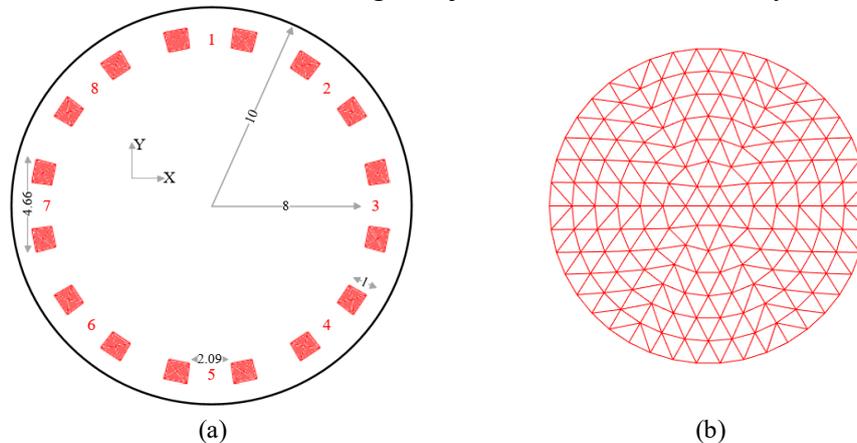

Figure 5.3. (a) Coils arrangement and cross section view of the MIT problem model. Dimension is in cm. (b) The mesh including 294 triangular pixels used for solving the inverse problem.

The forward problem is solved by the FE method based on the early and improved forward methods. The overall number of triangular elements and nodes in the FE model was 768 and 409, respectively. The inverse problem was solved by the regularized GN algorithm. As shown in Figure 5.3(b), the mesh including 294 uniform triangular pixels was used to solve the inverse problem. As illustrated, pixels have almost the same size. In addition, to avoid an inverse crime, the simulated measured data has been produced by solving the improved forward method on a very fine mesh with about $10^5$ triangular elements and $5 \times 10^4$ nodes. The value for coefficient $\eta$, the minimum conductivity $\nu$, the maximum conductivity $\xi$, and the maximum number of iterations $K$ were set to 2, $10^{-4}$ S/m, 20 S/m, and 30, respectively, for all following simulations.

In addition to results will be presented in this section, the reconstruction results for a 16-coil MIT system will be presented in Appendix 8.

### 5.5.2 Performance parameters

Performance parameters are required to evaluate reconstructed images. The principal parameter to evaluate the performance of imaging systems is the point spread function (PSF) [124]. The PSF is the response of the imaging system to a small circular perturbation scanning the entire medium (see Appendix 7). In electrical impedance tomography, several figures of merit based on the PSF concept have been defined [125], [126]. Here, some of these figures of merit are used to compare the quality of the reconstructed image in Example I. We use four performance parameters (PPs): conductivity contrast (CC), resolution (RES), position error (PE), and relative error (RE). To define CC, RES, and PE, a threshold is applied to the reconstructed image as follows:

$$[\boldsymbol{\sigma}^t]_i = \begin{cases} 1 & if \ [\boldsymbol{\sigma}]_i > \sigma_{thr} \\ 0 & \text{otherwise.} \end{cases} \quad (5.39)$$

where $[\boldsymbol{\sigma}]_i$ and $[\boldsymbol{\sigma}^t]_i$ are the $i$-th image pixel and the $i$-th thresholded amplitude image pixel, respectively. In the binary column matrix $\boldsymbol{\sigma}^t$, the non-zero elements correspond to image pixels whose conductivity value exceeds the threshold $\sigma_{thr}$. The threshold value provides a trade-off to distinguish between the visually important effects and background in the reconstructed image.

- *Conductivity contrast (CC)* measures the ratio between the conductivity of the reconstructed target object to that of its surrounding background [125]. The target object and background conductivity values are determined based on the thresholded amplitude set of the reconstructed image. Then, the average of pixels' conductivities labeled as the background ($\sigma_b$) and the target object ($\sigma_t$) are calculated and CC is obtained as $\sigma_t/\sigma_b$.
- *Resolution (RES)* is calculated as [126]:

$$\text{RES} = \sqrt{A^t/A^0} \quad (5.40)$$

where $A^t = \sum_k [\boldsymbol{\sigma}^t]_k$ is the number of pixels greater than $\sigma_{thr}$ and $A^0$ is the area (in pixels) of the entire imaging region.
- *Position error (PE)* shows the position discrepancy between the centroid of the target object in the reconstructed image and the simulated medium. PE is defined by [126]:

$$\text{PE} = r_t - r_h \quad (5.41)$$

where $r_t$ and $r_h$ are the radial position of the centroid of the actual target and reconstructed target, respectively. It is desired that PE is small and shows low variability for targets at different radial positions.

- *Relative error* for reconstructed conductivity image at $k$-th iteration is calculated as:

$$\text{RE}_k(\%) = \frac{\|\boldsymbol{\sigma}^{true} - \boldsymbol{\sigma}_k\|_2}{\|\boldsymbol{\sigma}^{true}\|_2} \qquad (5.42)$$

where $\|\cdot\|_2$ denotes $L^2$ norm and $\boldsymbol{\sigma}^{true}$ is a column matrix contains true conductivity distribution.

### 5.5.3 Example I: One target object

In this example, the imaging region includes one target object with a radius of 1.5 cm in Example I.A and with a radius of 1 cm in Example I.B. Figure 5.4(b)-(d) illustrate the reconstructed conductivity images by using the early forward problem for Example I.A. The homogeneous conductivity value $\sigma_h$ and the coefficient $\tau$ were 1 S/m and 3, respectively. The voltages induced by the secondary and total magnetic fields (refer to Chapter 3) based on the

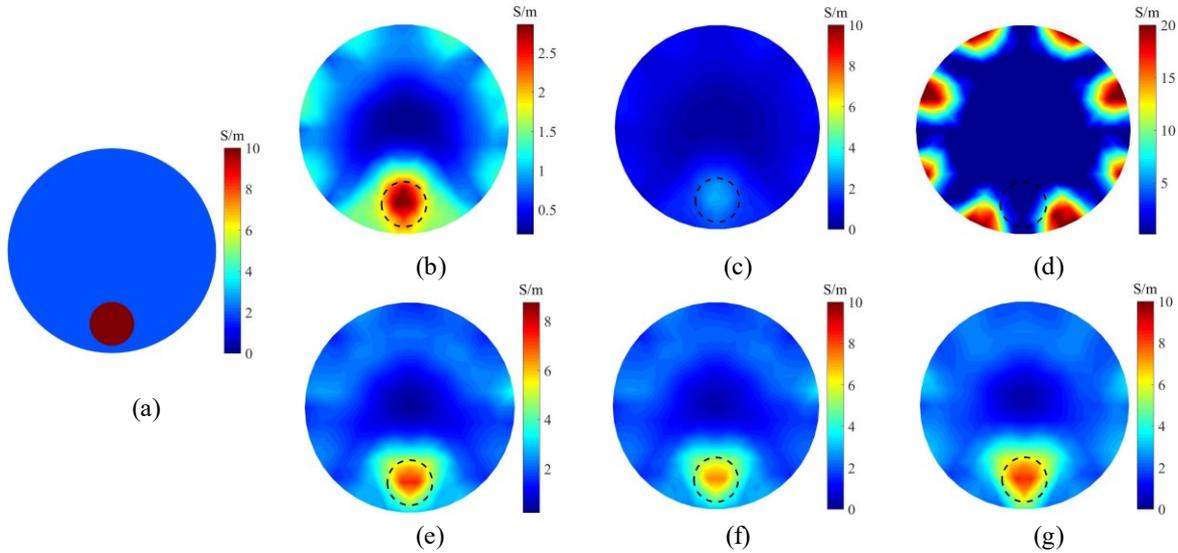

Figure 5.4 Example I.A: Imaging region contains one target object with radius of 1.5 cm. (a) True conductivity distribution. Reconstructed conductivity images using (b) the early forward method and the secondary field (colorbar is scaled to minimum and maximum reconstructed values), (c) the early forward method and the secondary field (colorbar is scaled to [0 10] S/m), (d) the early forward method and the total field, (e) the improved forward method and the secondary field (colorbar is scaled to minimum and maximum reconstructed values), (f) the improved forward method and the secondary field (colorbar is scaled to [0 10] S/m), and (g) the improved forward method and the total field. The homogeneous conductivity value $\sigma_h$ the coefficient $\tau$ were 1 S/m and 3, respectively. The target object and background conductivity were $\sigma_t = 10$ S/m and $\sigma_b = 2$ S/m, respectively.

Table 5.1. Example I.A: Performance parameters (PPs) computed for different cases of forward method, Jacobian matrix calculation technique, and magnetic field used for computation of induced voltage. The parameter $K$ indicates the iteration number for each case. PPs are explained in text.

| Case | | | PPs | | | | | | | |
|---|---|---|---|---|---|---|---|---|---|---|
| Forward method | Jacobian matrix calculation technique | Magnetic field | $\sigma_t$ (S/m) 10* | $\sigma_b$ (S/m) 2* | CC - 5* | RES - 0.2* | PE (cm) 0* | $K$ - - | $RE_K$ (%) - | Run-time (s) - |
| Early | Sensitivity | Secondary field | 3.00 | 0.92 | 3.26 | 0.14 | 0.22 | 29 | 68 | 15 |
| Early | Standard | Secondary field | 3.00 | 0.92 | 3.26 | 0.14 | 0.22 | 29 | 68 | 26 |
| Improved | Standard | Secondary field | 7.61 | 2.00 | 3.81 | 0.17 | 0.15 | 27 | 37 | 36 |
| Improved | Standard | Total field | 8.14 | 2.20 | 3.70 | 0.18 | 0.08 | 21 | 36 | 31 |

* Ideal value

early forward method are indicated by $\Delta V^E$ and $V^E$, respectively. In Figure 5.4(b) and Figure 5.4(c), $\Delta V^E$ have been used for reconstruction, and two different colorbar scales have been applied to display the results. In Figure 5.4(b), the colorbar is scaled to the minimum and maximum value of estimated conductivity
values and, in Figure 5.4(c), it is scaled to [0  10] S/m. As can be seen, using the voltages induced by the secondary field can partially compensate for the impact of ignoring skin and proximity effects in the early forward method while it sacrifices the conductivity contrast in the reconstructed image. In Figure 5.4(d), $V^E$ has been used for reconstruction. As can be seen, when the voltages induced by the total magnetic field are computed by the early forward method, conductivity distribution is not meaningfully reconstructed. It means that ignoring skin and proximity effects in coils in the forward problem implicates considerable errors in the reconstructed image, as explained in Chapter 3.

Figure 5.4(e)-(g) illustrate the reconstructed conductivity images by using the improved forward problem for Example I.A. The homogeneous conductivity value $\sigma_h$ and the coefficient $\tau$ were 1 S/m and 3, respectively. The voltages induced by the secondary and total magnetic fields based on the improved forward method are indicated by $\Delta V^I$ and $V^I$, respectively. In Figure 5.4(e) and Figure 5.4(f), $\Delta V^I$ have been used for reconstruction and two different colorbar scales have been applied to display the results. In Figure 5.4(e), the colorbar is scaled to the minimum and maximum value of estimated conductivity values and, in Figure 5.4(f), it is scaled to [0  10] S/m. In Figure 5.4(g), $V^I$ has been used for reconstruction. As can be seen, when the improved forward method is applied, using voltages induced by both total and secondary magnetic fields can detect the target object. However, it seems that using $V^I$ results in a visually better-reconstructed image.

Table 5.1 indicates PPs obtained for Example I.A. Since using $V^E$ in the inverse problem could not meaningfully reconstruct the conductivity distribution, PPs are indeterminable. Thus, they are not reported in Table 5.1. The parameter $K$ indicates the iteration number for each

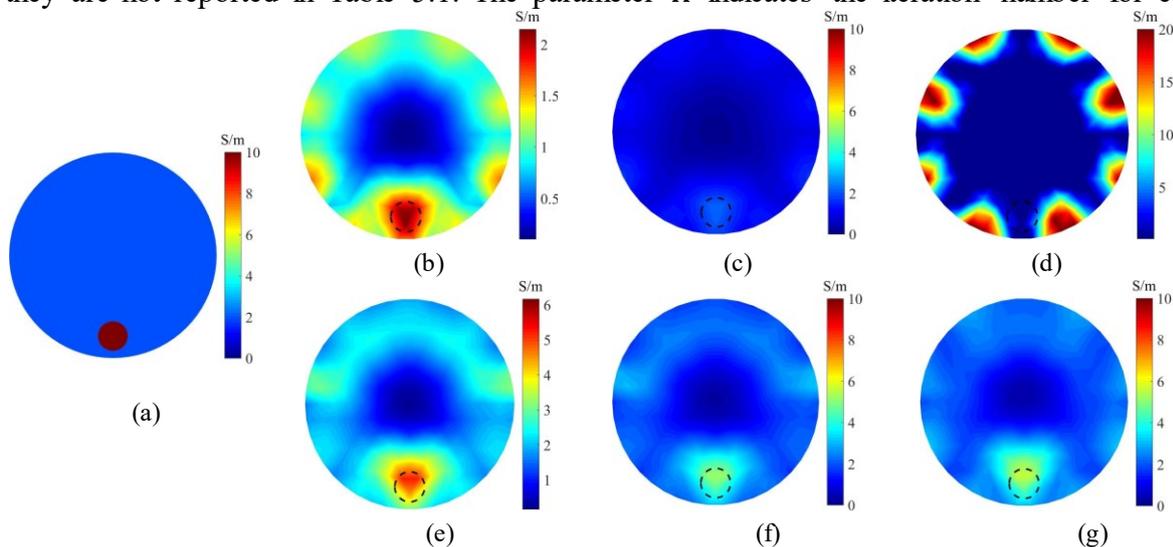

Figure 5.5 Example I.B: Imaging region contains one target object with radius of 1 cm. (a) True conductivity distribution. Reconstructed conductivity images using (b) the early forward method and the secondary field (colorbar is scaled to minimum and maximum reconstructed values), (c) the early forward method and the secondary field (colorbar is scaled to [0 10] S/m), (d) the early forward method and the total field, (e) the improved forward method and the secondary field (colorbar is scaled to minimum and maximum reconstructed values), (f) the improved forward method and the secondary field (colorbar is scaled to [0 10] S/m), and (g) the improved forward method and the total field. The homogeneous conductivity value $\sigma_h$ was 1 S/m. For reconstructions by the early forward method, the coefficient $\tau$ was 4. For reconstructions by the improved forward method, the coefficient $\tau$ was 3 and 2.2 when the voltages induced by the secondary and total fields are used, respectively. The target object and background conductivity were $\sigma_t = 10$ S/m and $\sigma_b = 2$ S/m,

Table 5.2 Example II.A: Performance parameters (PPs) computed for different cases of forward method, Jacobian matrix calculation technique, and magnetic field used for computation of induced voltage. The parameter $K$ indicates the iteration number for each case. PPs are explained in text.

| Case | | | PPs | | | | | | | |
|---|---|---|---|---|---|---|---|---|---|---|
| Forward method | Jacobian matrix calculation technique | Magnetic field | $\sigma_t$ (S/m) 10* | $\sigma_b$ (S/m) 2* | CC - 5* | RES - 0.14* | PE (cm) 0* | $K$ - - | $RE_K$ (%) - | Run-time (s) - |
| Early | Sensitivity | Secondary field | 2.00 | 0.81 | 2.47 | 0.19 | 0.10 | 18 | 64 | 9 |
| | Standard | | 2.00 | 0.81 | 2.47 | 0.19 | 0.10 | 18 | 64 | 20 |
| Improved | Standard | Secondary field | 5.13 | 1.89 | 2.71 | 0.18 | 0.57 | 23 | 35 | 32 |
| Improved | Standard | Total field | 5.81 | 2.00 | 2.91 | 0.16 | 0.65 | 22 | 35 | 29 |

* Ideal value

case. As can be seen, using $V^I$ results in the best performance except for CC. For the early forward method, we tested both sensitivity and standard techniques for Jacobian matrix calculation. The reconstructed images were the same. However, the runtime was different. As expected, the standard technique was more time-consuming.

Figure 5.5(b)-(d) illustrate the reconstructed conductivity images by using the early forward problem for Example I.B. The homogeneous conductivity value $\sigma_h$ and the coefficient $\tau$ were 1 S/m and 4, respectively. In Figure 5.5(b) and Figure 5.5(c), $\Delta V^E$ have been used for reconstruction and two different colorbar scales have been applied to display the results. In Figure 5.5(b), the colorbar is scaled to the minimum and maximum value of estimated conductivity values and, in Figure 5.5(c), it is scaled to [0 10] S/m. Similar to Example I.A, using the voltages induced by the secondary field can partially compensate for the impact of ignoring skin and proximity effects in the early forward method while it sacrifices the conductivity contrast in reconstructed image. However, in Figure 5.5(c) compared to Figure 5.4(c), the target object was barely detected. It means that when the target object becomes smaller, the compensatory effect of using the secondary field becomes less. In Figure 5.5(d), $V^E$ has been used for reconstruction. As can be seen, when the voltages induced by the total magnetic field are computed by the early forward method, conductivity distribution is not meaningfully reconstructed.

Figure 5.5(e)-(g) illustrate the reconstructed conductivity images by using the improved forward problem for Example I.B. The homogeneous conductivity value $\sigma_h$ was 1 S/m. The coefficient $\tau$ was 3 and 2.2 when the voltages induced by the secondary and total fields are used, respectively. Figure 5.5(e) and Figure 5.5(f), $\Delta V^E$ have been used for reconstruction and two different colorbar scales have been applied to display the results. In Figure 5.5(e), the colorbar is scaled to the minimum and maximum value of estimated conductivity values and, in Figure 5.5(f), it is scaled to [0 10] S/m. In Figure 5.5(g), $V^I$ has been used for reconstruction. Similar to Example I.A, when the improved forward method is applied, using voltages induced by both total and secondary magnetic fields can detect the target object.

Table 5.2 indicates PPs obtained for Example I.B. Similar to Example I.A, PPs are not reported for Figure 5.5(d). As can be seen, using $V^I$ results in the best performance except for PE. As seen from Fig 3(g), the center of the reconstructed target object slightly moved towards the origin. Table 5.2 shows the reconstructed $\sigma_t$ and $\sigma_b$ were 2 and 0.89, respectively, when $\Delta V^E$ was used, respectively. Consequently, in Figure 5.5(c), the target object cannot be distinguished according to the given colorbar scale. Similar to Example I.A, the standard technique was more time-consuming.

### 5.5.4 Example II: Two target objects

In this example, the imaging region includes two target objects with the target conductivities $\sigma_{t1} = 5$ S/m (upper target) and $\sigma_{t2} = 10$ S/m (lower target). The radius of targets is 1.5 cm in Example II.A and 1 cm in Example II.B.

Figure 5.6 illustrates the reconstructed conductivity images for Example II.A. The homogeneous conductivity value $\sigma_h$ and the coefficient $\tau$ were 1 S/m and 3, respectively. In Figure 5.6(b)-(d), $\Delta V^E$, $\Delta V^I$, and $V^I$ have been used for reconstruction, respectively. Similar to the previous examples, when the voltages induced by the total magnetic field are computed by the early forward method, conductivity distribution is not meaningfully reconstructed, and the corresponding image is not shown in Figure 5.6. As shown in Figure 5.6(b), using $\Delta V^E$ in this example partially detects the target object with higher conductivity value and the target with the lower conductivity cannot be distinguished. When the improved forward method is applied, using voltages induced by both total and secondary magnetic fields can detect both target objects. However, using $\Delta V^I$ results in lower contrast in the reconstructed images. The relative error percentage at the final iteration was $RE_{30}$=67%, $RE_{30}$=35%, and $RE_{27}$=32% when $\Delta V^E$, $\Delta V^I$, and $V^I$ were used, respectively.

Figure 5.7 illustrates the reconstructed conductivity images for Example II.B. The homogeneous conductivity value $\sigma_h$ was 1 S/m. The coefficient $\tau$ was 4 and 3 for the early and improved forward methods, respectively. In Figure 5.7(b)-(d), $\Delta V^E$, $\Delta V^I$, and $V^I$ have been used for reconstruction, respectively. Similar to the previous examples, when the voltages induced by the total magnetic field are computed by the early forward method, conductivity distribution is not meaningfully reconstructed and the related image is not shown in Figure 5.7. As shown in Figure 5.7(b), when $\Delta V^E$ is used the target objects cannot be distinguished in the [0 10] colorbar scale. When the improved forward method is applied, using voltages induced by both total and secondary magnetic fields can detect both target objects. However, using $\Delta V^I$ results in lower contrast in the reconstructed images. The relative error percentage at the final iteration was $RE_{22}=67\%$, $RE_{26}=38\%$, and $RE_{19}=36\%$ when $\Delta V^E$, $\Delta V^I$, and $V^I$ were used, respectively.

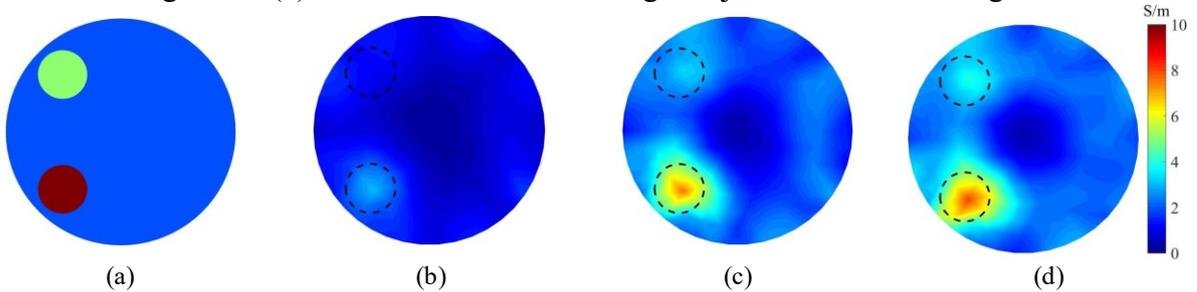

(a)        (b)        (c)        (d)

Figure 5.6 Example II.A: Imaging region contains two target objects with radius of 1.5 cm. (a) True conductivity distribution. Reconstructed conductivity images using (b) the early forward method and the secondary field, (c) the improved forward method and the secondary field, and (d) the improved forward method and the total field. The homogeneous conductivity value $\sigma_h$ and the coefficient $\tau$ were 1 S/m and 3, respectively. The target object conductivities were $\sigma_{t1} = 5$ S/m (upper target) and $\sigma_{t2} = 10$ S/m (lower target) and the background conductivity was $\sigma_h = 2$ S/m.

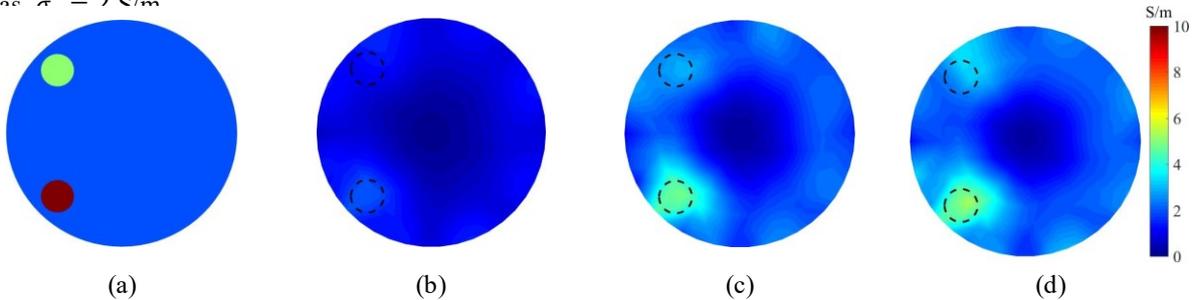

(a)        (b)        (c)        (d)

Figure 5.7 Example II.B: Imaging region contains two target objects with radius of 1 cm (a) True conductivity distribution. Reconstructed conductivity images using (b) the early forward method and the secondary field, (c) the improved forward method and the secondary field, and (d) the improved forward method and the total field. The homogeneous conductivity value $\sigma_h$ was 1 S/m. The coefficient $\tau$ was 4 and 3 for the early and improved forward methods, respectively. The target object conductivities were $\sigma_{t1} = 5$ S/m (upper target) and $\sigma_{t2} = 10$ S/m (lower target) and the background conductivity was $\sigma_h = 2$ S/m.

### 5.5.5 Noise study

In this subsection, we study the robustness of the reconstruction algorithm against the noise when $\Delta V^E$, $\Delta V^I$, and $V^I$ are used. For this purpose, we chose Example I.A in which the target object was detected when $\Delta V^E$, $\Delta V^I$, and $V^I$ were used. We added complex white Gaussian noise to the simulated measured voltages and considered 40 dB [72], 30 dB [41], and 20 dB [73] SNR. For each SNR, we repeated the experiment 50 times. To evaluate the performance of the reconstruction, we used the thresholded amplitude conductivity image obtained by (18). For SNR= 40 dB, the target object was detected in all 50 thresholded images when $\Delta V^E$, $\Delta V^I$, or $V^I$ was used. When SNR decreased to 30 dB, using $\Delta V^E$ or $\Delta V^I$ in GN algorithm resulted in detection of target objects in 45 thresholded images (out of 50) while using $V^I$ resulted in detection of target objects in all thresholded images. By decreasing the SNR to 20 dB, using $\Delta V^E$, $\Delta V^I$, and $V^I$ resulted in detection of target objects in 24, 29, and 40 thresholded images, respectively.

### 5.5.6 Discussion

As seen in Example I and Example II, when $V^E$, induced voltages obtained from the total field and computed by the early forward method, is used in the inverse problem, the conductivity distribution is not reconstructed meaningfully. It manifests that ignoring skin and proximity effects inside MIT coils in the forward problem implicates considerable errors in the reconstructed image. As mentioned in Chapter III, using gradiometer or state-difference imaging techniques to obtain voltages induced by the secondary magnetic field, $\Delta V^E$, can partially compensate error due to neglecting skin and proximity effects in coils. However, as seen in Example I.B and Example II.B, when target objects become smaller, it is hard to distinguish them in the reconstructed images. In addition, in Example II.A where large target objects are placed in the imaging region, using $\Delta V^E$ partially reconstructs the target object with higher conductivity, and the target with lower conductivity remains unresolvable.

It is noteworthy that the reconstructed conductivity values using the voltages induced by the secondary magnetic field, $\Delta V^E$ and $\Delta V^I$, are lower compared to the true conductivities. In other words, using the secondary magnetic field data causes the conductivity values to be underestimated. Consequently, using the secondary magnetic field data to compensate for error due to neglecting skin and proximity effects in coils, will be at the cost of producing qualitative images. As seen from numerical experiments, to reconstruct the absolute conductivity values, it is necessary to use the total field data and considering skin and proximity effects inside MIT coils.

Here, we are not dealing with a linear system in which the superposition principle can be applied. Removing the primary field compensates partially for the error caused by ignoring the skin and proximity effects in coils, but not completely. In fact, by placing the target object in the imaging region, losses caused by the skin and proximity effects in coils change compared to when the imaging region is empty (primary field). This change is due to the interaction between coils and the conductivities to be imaged, as shown in Chapter 3. Consequently, removing the primary field from the total field cannot completely compensate for ignoring the skin and proximity effects in coils. In addition, as seen in Section 5.6.5, using the secondary data has another drawback. The reconstruction procedure based on the secondary field data has less robustness against the noise.

As expected and seen from the simulation results, the standard technique for calculation of the Jacobian matrix is more computationally demanding compared to the sensitivity one. In this work, we observed both standard and sensitivity techniques had the same performance in terms of the reconstructed conductivity for the early forward method. However, in [122], it has been shown that the standard technique is more accurate in some situations.

## 5.6 Conclusions

In this chapter, numerical conductivity image reconstruction based on the improved forward method was developed for 2D biomedical MIT. As discussed in Chapter 3, the improved forward method is based on Maxwell's equations and considers skin and proximity effects inside the exciter and sensor coils. By modeling an 8-coil 2D MIT system through two different numerical experiments, the importance of employing the improved forward method in the MIT conductivity reconstruction was investigated. Results of this chapter manifested that the error due to neglecting the skin and proximity effects can be partially compensated by the difference imaging; however, it will be at the cost of producing qualitative images. Furthermore, to reconstruct the absolute conductivity values in the low conductivity MIT applications, including biomedical MIT, it is crucial to use the improved forward method and voltages induced by the total magnetic field.

# Chapter 6: MIT Inverse Problem II

Conductivity image reconstruction in MIT so far reported in the literature can be divided into two categories of applications: low-conductivity and high-conductivity MIT applications [1]. Cerebral hemorrhage imaging [16][17] and conductive fluid imaging [98] are examples of the former, and molten metal flow visualization [127] and monitoring of steel solidification [128] are examples of the latter (see Table 6.1 for more information). As illustrated in Chapter 2, in the low-conductivity category, the voltage which is in phase with the excitation current and is equal to the real part in the phasor representation, is used for conductivity reconstruction. Furthermore, in the high-conductivity category, the voltage which is in quadrature phase with the excitation current and is equal to the imaginary part in the phasor representation, has been used in [48], [129], [130]. In [9] and [131], both real and imaginary parts of the induced voltage have been employed while permeability, permittivity, or both together with conductivity have been used in the forward model, and they have been estimated in the inverse problem. Some MIT research groups have used other forms of the induced voltage [11], [14], [29], [38]–[40], [46], [47]. Some of them have used the quadrature component (imaginary part) of an induced voltage ratio [38]–[40], and some other groups have used the phase shift of the induced voltage in low-conductivity applications, especially in biomedical applications [11], [14], [29], [46], [47]. As clarified in 2.2.2, these forms fall in the first above category, that is, utilizing the imaginary part of an induced voltage ratio and the phase shift of the induced voltage are equivalent to using the in-phase component (real part) of the induced voltage.

There are some situations in which both real and imaginary parts of the induced voltage contain information about conductivity. For example, when conductivity image reconstruction is needed in the mid-range conductivity applications or when imaging regions include vastly different conductivity values, the information of both real and imaginary parts of the induced voltage can be exploited to provide a conductivity image with better quality. Consequently, there is a need for developing of a technique which uses complex-valued data for MIT conductivity imaging. This chapter focuses on the development of a regularized GN algorithm which uses complex-valued data to reconstruct MIT conductivity images. Up to now, the conductivity image reconstruction techniques for the mid-range conductivity MIT have not been well established, and the necessity of using both real and imaginary parts of the induced

Table 6.1 Frequency and conductivity ranges in various MIT applications

| Conductivity value category | Typical Values $f$ | Typical Values $\sigma$ (S/m) | Explanation |
|---|---|---|---|
| Low | 1-10 MHz | $0.1 - 10$ | Medical and industrial applications [17], [98] |
| Mid-range | 0.05- 1 MHz | $10 - 10^4$ | Applications for imaging regions with carbon fibers [135] or biomedical conductive polymers [33] |
| High | 5-100 kHz | $10^4 - 10^7$ | Industrial applications [128] |

voltage in MIT conductivity image reconstruction has not been explained in the literature.

## 6.1 Induced voltage in a single channel MIT

As can be seen from the forward problem equations presented in Chapter 3, the magnetic vector potential $A_z$ is a complex function of conductivity and frequency. Thus, based on (3.29), the induced voltage is also a complex function of conductivity and frequency. For a single-channel MIT system (see Figure 2.1), when the imaging region includes an inhomogeneity, the induced voltage $V_1$ can be given as (see Figure 2.3):

$$V_1 = |V_0|\psi(\sigma, \omega, Q) \tag{6.1}$$

where $|V_0|$ is the magnitude of induced voltage when the imaging region of the system is empty, and it is so-called background voltage, and $\psi$ is a nonlinear complex-valued function of the distribution of conductivity $\sigma$, the system geometry $Q$, and the angular frequency $\omega$. As noted in Chapter 2, $V_0$ and $V_1$ are voltages induced by the primary and total magnetic field, respectively. The complex function $\psi$ can be written in terms of its real and imaginary components as:

$$\psi = \psi_{\text{Re}}(\sigma, \omega, Q) + j\psi_{\text{Im}}(\sigma, \omega, Q) \tag{6.2}$$

Table 6.1 shows frequency and conductivity ranges in various MIT applications. In low-conductivity applications, such as cerebral hemorrhage imaging, a frequency in the MHz range is usually applied to obtain the desired SNR for the induced voltage. In this case, the real part of the induced voltage mainly includes information on conductivity, and the imaginary part can be considered as a constant [41], [44], [45], [129]. Therefore, (6.1) can be written as:

$$V_1 \cong |V_0|(\psi_{\text{Re}}(\sigma, \omega, Q) + jC_1) \tag{6.3}$$

where $C_1$ is a constant.

In the high-conductivity MIT applications, such as molten metal casting process visualization, the imaginary part of induced voltages has been used for conductivity image reconstruction [48], [129], [130], [132] since they have larger values and can improve SNR [132]. In other words, in these applications, the imaginary part of the induced voltage mainly contains information on conductivity, and the real part can be considered as a constant. In consequence, (6.1) can be written as:

$$V_1 \cong |V_0|\big(C_2 + j\psi_{\text{Im}}(\sigma, \omega, Q)\big) \tag{6.4}$$

where $C_2$ is a constant.

In low- and high-conductivity applications, respectively, $\psi_{\text{Re}}(\sigma, \omega, Q)$ and $\psi_{\text{Im}}(\sigma, \omega, Q)$ predominantly contribute to the induced voltage. However, there are potential applications where both $\psi_{\text{Re}}$ and $\psi_{\text{Im}}$ can significantly contribute to the induced voltage, such as imaging of regions with carbon fibers [133] or biomedical conductive polymers [33]. Assuming both the real and imaginary parts of induced voltage consist of information on conductivity, both the real and imaginary parts must be used for conductivity reconstruction. Substituting (6.2) into (6.1) yields the following equation:

$$V_1 = |V_0|(\psi_{\text{Re}}(\sigma, \omega, Q) + j\psi_{\text{Im}}(\sigma, \omega, Q)) \tag{6.5}$$

This technique, i.e., using complex voltages to reconstruct real-valued conductivity, should work for all MIT applications regardless of the conductivity values.

## 6.2 Techniques for conductivity reconstruction in MIT

Based on the previous section, we present two techniques for MIT conductivity reconstruction: early and proposed techniques. In the following, these techniques are introduced.

### 6.2.1 Early technique

In the early technique depending on the conductivity value of the region of interest, two types of data have been used.

*A. Real part data*

In the inverse problems of the low-conductivity MIT, in-phase components of the induced voltages, which are equal to the real parts in phasor representation, are predominant with respect to the quadrature components, which are equivalent to the imaginary parts in the phasor representation. Similarly, the real parts of the calculated complex voltages in the forward solver are predominant with respect to the imaginary parts. Consequently, the researchers have used the real-valued column matrix $\mathbf{V}_M$, obtained from induced voltages, and the real-valued column matrix $\mathbf{V}_F$ obtained from simulated complex-valued voltages in the forward solver. Therefore, in the early technique with real part data, the real-valued data for the measured data and the simulated data are obtained from $\mathbf{V}_M = \text{Re}\{\overline{\mathbf{V}}_M\}$ and $\mathbf{V}_F = \text{Re}\{\overline{\mathbf{V}}_F\}$, respectively, where $\overline{\mathbf{V}}_M$ and $\overline{\mathbf{V}}_F$ are complex-valued column matrices of induced and calculated voltages, respectively. In this chapter, we use bar mark to discriminate between real-valued and complex-valued matrices.

*B. Imaginary part data*

In the inverse problems of high-conductivity MIT, the imaginary parts of the induced voltages are predominant with respect to the real parts. Similarly, the imaginary parts of the simulated complex-valued voltages in the forward solver are predominant with respect to the real parts. Therefore, in the early technique with imaginary part data, the real-valued data for the measured data and the simulated data are obtained from $\mathbf{V}_M = \text{Im}\{\overline{\mathbf{V}}_M\}$ and $\mathbf{V}_F = \text{Im}\{\overline{\mathbf{V}}_F\}$, respectively.

The regularized GN algorithm introduced in Chapter 5 are used to reconstruct conductivity distribution by the early technique with real and imaginary part data.

### 6.2.2 Proposed technique

In this section, we adapt the regularized GN algorithm to implement the proposed technique, which uses complex-valued voltages. Based on this technique, a solution of the MIT inverse problem is attained by minimizing the least-squares objective function $\phi_p$ given by:

$$\phi_p(\boldsymbol{\sigma}) = \arg\min_{\boldsymbol{\sigma}} \left\{ \frac{1}{2}(\overline{\mathbf{V}}_M - \overline{\mathbf{V}}_F(\boldsymbol{\sigma}))^{\text{H}}(\overline{\mathbf{V}}_M - \overline{\mathbf{V}}_F(\boldsymbol{\sigma})) + \frac{1}{2}\lambda \boldsymbol{\sigma}^{\text{T}}\mathbf{R}^{\text{T}}\mathbf{R}\boldsymbol{\sigma} \right\} \tag{6.1}$$

where superscript H stands for conjugate transpose, which is so-called Hermitian transpose. In (6.1), $\boldsymbol{\sigma} \in \mathcal{R}^n$ is the conductivity column matrix, $\overline{\mathbf{V}}_M \in \mathcal{C}^m$ denotes the complex-valued column matrix which contains the measured voltages, $\overline{\mathbf{V}}_F(\boldsymbol{\sigma}): \mathcal{C}^n \to \mathcal{C}^m$ is the complex-valued column matrix obtained from the forward solver, and the numbers $n$ and $m$ represent the number of image pixels and independent measurements, respectively. The matrix $\mathbf{R} \in \mathcal{R}^{m \times n}$ is a regularization matrix and $\lambda$ is a regularization parameter.

It is noteworthy that both $\overline{\mathbf{V}}_M$ and $\overline{\mathbf{V}}_F$ are complex-valued column matrices in (6.1) whereas $\phi_p$ still remain real like $\phi$ in (5.1). By expanding (6.1), one obtains

$$\phi_p(\boldsymbol{\sigma}) = \frac{1}{2}(\overline{\mathbf{V}}_M^H \overline{\mathbf{V}}_M - \overline{\mathbf{V}}_M^H \overline{\mathbf{V}}_F - \overline{\mathbf{V}}_F^H \overline{\mathbf{V}}_M + \overline{\mathbf{V}}_F^H \overline{\mathbf{V}}_F) + \frac{1}{2} \lambda \boldsymbol{\sigma}^T \mathbf{R}^T \mathbf{R} \boldsymbol{\sigma}$$
$$= \frac{1}{2}((\overline{\mathbf{V}}_M^*)^T \overline{\mathbf{V}}_M - (\overline{\mathbf{V}}_M^*)^T \overline{\mathbf{V}}_F - (\overline{\mathbf{V}}_F^*)^T \overline{\mathbf{V}}_M + (\overline{\mathbf{V}}_F^*)^T \overline{\mathbf{V}}_F) + \frac{1}{2} \lambda \boldsymbol{\sigma}^T \mathbf{R}^T \mathbf{R} \boldsymbol{\sigma}$$
(6.2)

To find a candidate value of $\boldsymbol{\sigma}$ that minimizes $\phi_p$, we differentiate (6.2) with respect to $\boldsymbol{\sigma}$:

$$\left[\frac{\partial \phi_p}{\partial \boldsymbol{\sigma}}\right]_{n \times 1} = \boldsymbol{\phi}'_p(\boldsymbol{\sigma}) = \frac{1}{2}\left(-(\overline{\mathbf{V}}_F')^T \overline{\mathbf{V}}_M^* - (\overline{\mathbf{V}}_F^{*'})^T \overline{\mathbf{V}}_M + (\overline{\mathbf{V}}_F^{*'})^T \overline{\mathbf{V}}_F + (\overline{\mathbf{V}}_F')^T \overline{\mathbf{V}}_F^*\right) + \lambda \mathbf{R}^T \mathbf{R} \boldsymbol{\sigma}$$
(6.3)

By simplifying (6.3), one can write:

$$\boldsymbol{\phi}'_p = \frac{1}{2}\left((\overline{\mathbf{V}}_F^{*'})^T (\overline{\mathbf{V}}_F - \overline{\mathbf{V}}_M) + (\overline{\mathbf{V}}_F')^T (\overline{\mathbf{V}}_F^* - \overline{\mathbf{V}}_M^*)\right) + \lambda \mathbf{R}^T \mathbf{R} \boldsymbol{\sigma}$$
$$= \frac{1}{2}\left((\overline{\mathbf{V}}_F^{*'})^T (\overline{\mathbf{V}}_F - \overline{\mathbf{V}}_M) + \left[(\overline{\mathbf{V}}_F^{*'})^T (\overline{\mathbf{V}}_F - \overline{\mathbf{V}}_M)\right]^*\right) + \lambda \mathbf{R}^T \mathbf{R} \boldsymbol{\sigma}$$
$$= \mathrm{Re}\{(\overline{\mathbf{V}}_F')^H (\overline{\mathbf{V}}_F - \overline{\mathbf{V}}_M)\} + \lambda \mathbf{R}^T \mathbf{R} \boldsymbol{\sigma}$$
$$= \mathrm{Re}\{\overline{\mathbf{J}}^H (\overline{\mathbf{V}}_F - \overline{\mathbf{V}}_M)\} + \lambda \mathbf{R}^T \mathbf{R} \boldsymbol{\sigma}$$
(6.4)

Here, $\overline{\mathbf{J}}$ is the complex-valued Jacobian matrix and is given by

$$\overline{\mathbf{J}} = \overline{\mathbf{V}}'_F = \mathrm{Re}\{\overline{\mathbf{V}}'_F\} + j\,\mathrm{Im}\{\overline{\mathbf{V}}'_F\} = \frac{\partial \mathrm{Re}\{\overline{\mathbf{V}}'_F\}}{\partial \boldsymbol{\sigma}} + j\frac{\partial \mathrm{Im}\{\overline{\mathbf{V}}'_F\}}{\partial \boldsymbol{\sigma}}$$
(6.5)

By taking a Taylor series expansion of $\boldsymbol{\phi}'_p(\boldsymbol{\sigma})$ around $\boldsymbol{\sigma}_k$, keeping the linear terms, and setting equal to $\mathbf{0}$ to minimize $\phi_p$:

$$\boldsymbol{\phi}'_p(\boldsymbol{\sigma}) = \boldsymbol{\phi}'_p(\boldsymbol{\sigma}_k) + \boldsymbol{\phi}''_p(\boldsymbol{\sigma}_k) \mathbf{d}_k = \mathbf{0}$$
(6.6)

where $\mathbf{d}_k = \boldsymbol{\sigma} - \boldsymbol{\sigma}_k$ and $k$ is the iteration number. The term $\boldsymbol{\phi}''_p$ is called the modified Hessian matrix, given by

$$\mathbf{H}_p = \boldsymbol{\phi}''_p = \frac{\partial}{\partial \boldsymbol{\sigma}}(\mathrm{Re}\{(\overline{\mathbf{V}}'_F)^H (\overline{\mathbf{V}}_F - \overline{\mathbf{V}}_M)\} + \lambda \mathbf{R}^T \mathbf{R} \boldsymbol{\sigma})$$
$$= \mathrm{Re}\left\{\frac{\partial}{\partial \boldsymbol{\sigma}}\left((\overline{\mathbf{V}}'_F)^H (\overline{\mathbf{V}}_F - \overline{\mathbf{V}}_M)\right)\right\} + \lambda \mathbf{R}^T \mathbf{R}$$
$$= \mathrm{Re}\{(\overline{\mathbf{V}}'_F)^H (\overline{\mathbf{V}}'_F + 0) + (\overline{\mathbf{V}}''_F)^H (\mathbf{I} \otimes (\overline{\mathbf{V}}_F - \overline{\mathbf{V}}_M))\} + \lambda \mathbf{R}^T \mathbf{R}$$
$$= \mathrm{Re}\{\overline{\mathbf{J}}^H \overline{\mathbf{J}} + (\overline{\mathbf{V}}''_F)^H (\mathbf{I} \otimes (\overline{\mathbf{V}}_F - \overline{\mathbf{V}}_M))\} + \lambda \mathbf{R}^T \mathbf{R}$$
(6.7)

The term $\overline{\mathbf{V}}''_F$ in (6.7) is difficult to calculate explicitly, but often it is negligible relative to $\overline{\mathbf{J}}^H \overline{\mathbf{J}}$. Therefore, we may approximate the modified Hessian matrix by:

$$\mathbf{H}_p = \mathrm{Re}\{\overline{\mathbf{J}}^H \overline{\mathbf{J}}\} + \lambda \mathbf{R}^T \mathbf{R}$$
(6.8)

Substituting (6.4) and (6.8) into (6.6), assuming $\mathbf{H}_p$ is invertible, solving for $\mathbf{d}_k$, we obtain:

$$\boldsymbol{\sigma}_{k+1} = \boldsymbol{\sigma}_k + \left[ \text{Re}\{\bar{\mathbf{J}}_k^H \bar{\mathbf{J}}_k\} + \lambda \mathbf{R}^T\mathbf{R} \right]^{-1} \left[ \text{Re}\{\bar{\mathbf{J}}_k^H(\bar{\mathbf{V}}_M - \bar{\mathbf{V}}_{F_k})\} - \lambda \mathbf{R}^T\mathbf{R}\boldsymbol{\sigma}_k \right] \quad (6.9)$$

Considering $\bar{\mathbf{J}}_k = \text{Re}\{\bar{\mathbf{J}}_k\} + j\text{Im}\{\bar{\mathbf{J}}_k\}$ and $\bar{\mathbf{J}}_k^H = (\text{Re}\{\bar{\mathbf{J}}_k\})^T - j(\text{Im}\{\bar{\mathbf{J}}_k\})^T$, one obtains

$$\bar{\mathbf{J}}_k^H \bar{\mathbf{J}}_k = ((\text{Re}\{\bar{\mathbf{J}}_k\})^T - j(\text{Im}\{\bar{\mathbf{J}}_k\})^T)(\text{Re}\{\bar{\mathbf{J}}_k\} + j\text{Im}\{\bar{\mathbf{J}}_k\}) \quad (6.10)$$

Equation (6.10) can be rewritten as

$$\bar{\mathbf{J}}_k^H \bar{\mathbf{J}}_k = (\text{Re}\{\bar{\mathbf{J}}_k\})^T\text{Re}\{\bar{\mathbf{J}}_k\} + (\text{Im}\{\bar{\mathbf{J}}_k\})^T\text{Im}\{\bar{\mathbf{J}}_k\}$$
$$+ j((\text{Re}\{\bar{\mathbf{J}}_k\})^T\text{Im}\{\bar{\mathbf{J}}_k\} - (\text{Im}\{\bar{\mathbf{J}}_k\})^T\text{Re}\{\bar{\mathbf{J}}_k\})) \quad (6.11)$$

Then, the real part of $\bar{\mathbf{J}}_k^H \bar{\mathbf{J}}_k$ becomes

$$\text{Re}\{\bar{\mathbf{J}}_k^H \bar{\mathbf{J}}_k\} = (\text{Re}\{\bar{\mathbf{J}}_k\})^T\text{Re}\{\bar{\mathbf{J}}_k\} + (\text{Im}\{\bar{\mathbf{J}}_k\})^T\text{Im}\{\bar{\mathbf{J}}_k\} \quad (6.12)$$

which can be written in the following form

$$\text{Re}\{\bar{\mathbf{J}}_k^H \bar{\mathbf{J}}_k\} = [\text{Re}\{\bar{\mathbf{J}}_k\} \quad \text{Im}\{\bar{\mathbf{J}}_k\}] \begin{bmatrix} \text{Re}\{\bar{\mathbf{J}}_k\} \\ \text{Im}\{\bar{\mathbf{J}}_k\} \end{bmatrix} = \mathbf{J}_{p_k}^T \mathbf{J}_{p_k} \quad (6.13)$$

where the column matrix $\mathbf{J}_{p_k}$ is $\begin{bmatrix} \text{Re}\{\bar{\mathbf{J}}_k\} \\ \text{Im}\{\bar{\mathbf{J}}_k\} \end{bmatrix}$, which is the Jacobian matrix at the iteration $k$ in the proposed technique. Therefore, (6.9) can then be simplified to

$$\boldsymbol{\sigma}_{k+1} = \boldsymbol{\sigma}_k + \left[ \mathbf{J}_{p_k}^T \mathbf{J}_{p_k} + \lambda \mathbf{R}^T\mathbf{R} \right]^{-1} \left[ \mathbf{J}_{p_k}^T \left( \mathbf{V}_{pM} - \mathbf{V}_{pF_k} \right) - \lambda \mathbf{R}^T\mathbf{R}\boldsymbol{\sigma}_k \right] \quad (6.14)$$

where the real-valued column matrices $\mathbf{V}_{pM}$ and $\mathbf{V}_{pF_k}$ are $\begin{bmatrix} \text{Re}\{\bar{\mathbf{V}}_M\} \\ \text{Im}\{\bar{\mathbf{V}}_M\} \end{bmatrix}$ and $\begin{bmatrix} \text{Re}\{\bar{\mathbf{V}}_{F_k}\} \\ \text{Im}\{\bar{\mathbf{V}}_{F_k}\} \end{bmatrix}$, respectively.

When the adaptive method introduced in Section 5.4 is used to determine the regularization parameter, (6.14) is rewritten as follows"

$$\boldsymbol{\sigma}_{k+1} = \boldsymbol{\sigma}_k + \left[ \mathbf{J}_{p_k}^T \mathbf{J}_{p_k} + \lambda_k \mathbf{R}^T\mathbf{R} \right]^{-1} \left[ \mathbf{J}_{p_k}^T \left( \mathbf{V}_{pM} - \mathbf{V}_{pF_k} \right) - \lambda_k \mathbf{R}^T\mathbf{R}\boldsymbol{\sigma}_k \right] \quad (6.15)$$

It is noteworthy that (6.15) can be used for updating the conductivity in the early technique providing that, depending on conductivity value to be imaged, the real part or imaginary part of $\mathbf{J}_{p_k}$, $\mathbf{V}_{pM}$, and $\mathbf{V}_{pF_k}$ is employed.

In [9] and [131], the complex-valued Jacobian matrices have been used to reconstruct the permeability, permittivity, or both together with conductivity. However, in these references, the Jacobian matrices have not been derived and have different structures from the one which was presented here.

## 6.3 Numerical experiments

This section contains five numerical 2D experiments to demonstrate the feasibility of the proposed technique. First, the simulation set-up will be presented. Four PPs are introduced in Section 6.4.2 and used to evaluate and compare the early and proposed techniques in the first three examples, which include a one-target object imaging region. In the fourth example, a two-target object example will be studied, and finally, a noise study example will be presented. All simulations were executed on a core i5 2.6 GHz laptop with 8 GB of RAM.

### 6.3.1 Modeling set-up

A model of a 2D MIT system is considered to investigate the performance of two techniques. This system is exactly the same as the one employed in Section 5.6.1 (see Figure 5.3(a)). The forward problem was solved by the FE method and the improved forward method. The numbers of triangular elements and nodes in the FE model were 3072 and 1585, respectively. In addition, the simulated measured voltages have been produced by a very fine mesh with 49152 triangular elements and 24769 nodes. All meshes were generated by the TOAST toolbox [134]. Dirichlet boundary condition was imposed at a radius of 10 cm. The inverse problem was solved by the regularized GN algorithm based on the early and proposed techniques. Triangular pixels were used for solving the inverse problem (see Figure 5.3(b)). The regularization parameter was chosen based on the adaptive method introduced in Section 5.4. The value for coefficient $\eta$ and minimum conductivity $\nu$ were set to 2 and $10^{-4}$ S/m, respectively, for all following simulations.

### 6.3.2 Performance parameters

In order to evaluate and compare the early and proposed techniques, we use conductivity contrast (CC), resolution (RES), and position error (PE), introduced in Section 5.6.2. In addition to these performance parameters, another PP is introduced and employed in this chapter; shape deformation (SD). SD measures the extent to which reconstructed images faithfully represent the shape of the target image. It is the fraction of the reconstructed thresholded amplitude set (Equation (5.39)) which does not fit within a circle of an equal area [126]:

$$\text{SD} = \sum_{i \notin C}[\boldsymbol{\sigma}^t]_i \Big/ \sum_{i}[\boldsymbol{\sigma}^t]_i \qquad (6.16)$$

where $[\boldsymbol{\sigma}^t]_i$ is the $i$-th thresholded amplitude image pixel and $C$ is a circle centered at the centroid of the $\boldsymbol{\sigma}^t$ with an area equivalent to $A_t$ (in pixels). The desired SD is low and uniform over the whole imaging region.

### 6.3.3 Example I: One-target object with low-conductivity

In the first numerical experiment, as shown in Figure 6.2(a), it is assumed the imaging region includes a conductive hexagonal bar (target object) with $\sigma_t = 6$ S/m. This hexagon can be simply approximated by a circle with the diameter $r = 2$ cm and the center (0, -4) cm. The diameter is a little smaller than the approximate limit given for minimum detectable inhomogeneity diameter [35] (see Section 5.6.1). The background conductivity is $\sigma_b = 0.5$ S/m. A sinusoidal alternating current with a driving frequency of 10 MHz and current amplitude of 1 A was applied to the excitation coils. When the true conductivity distribution presents in the imaging region, Figure 6.1 shows the real (in blue) and imaginary (in red) parts of the voltages induced by the secondary magnetic field, $\Delta\mathbf{V}_M$, as a function of measurement number. As shown in this figure and as noted in Section 6.2, in the low-conductivity application, the real part of the induced voltage mainly changes with conductivities inside the imaging region, and the changes in the imaginary part of the induced voltage are not considerable.

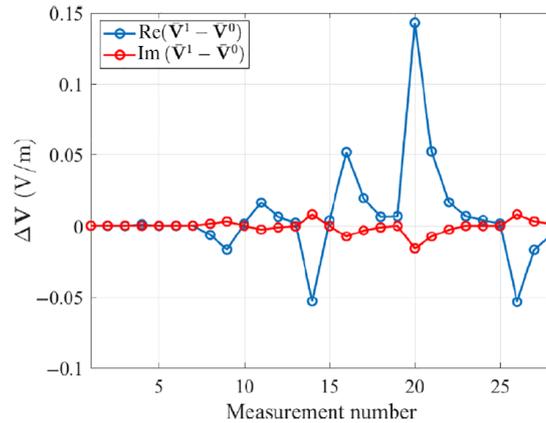

Figure 6.1 Example I: real (in blue) and imaginary (in red) parts of the voltages induced by the secondary magnetic field, $\Delta \mathbf{V}_M$, as a function of measurement number when the true conductivity distribution presents in the imaging region.

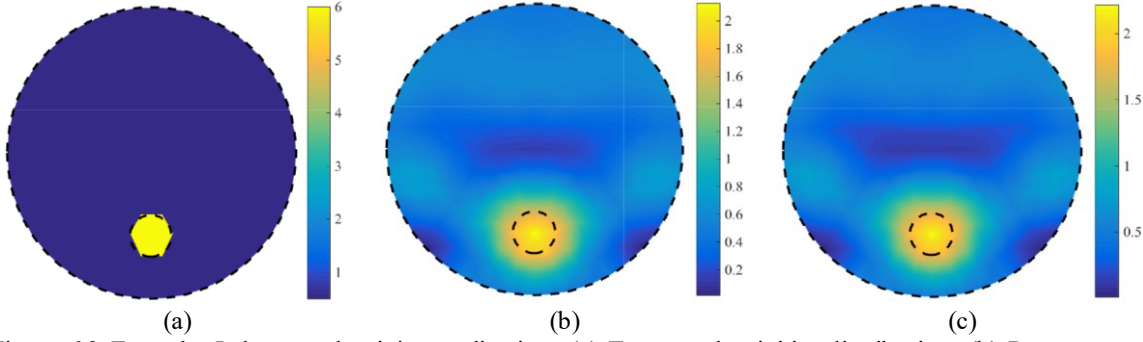

(a)                                 (b)                               (c)

Figure 6.2 Example I: low-conductivity application; (a) True conductivities distribution; (b) Reconstructed image by the early technique using real part of data; and (c) Reconstructed image by the proposed technique. The maximum number of iterations, the homogeneous conductivity value $\sigma_h$, and the coefficient $\tau$ were 20, 10 S/m, and 50, respectively, for both techniques. The regularization parameter at iteration 20-th was $4.2 \times 10^{-9}$ and $3.4 \times 10^{-9}$ for the early and proposed techniques, respectively. The excitation frequency was 10 MHz. The target object and background conductivity were $\sigma_t = 6$ S/m and $\sigma_b = 0.5$ S/m, respectively.

Table 6.2 Example I: Performance parameters (PPs)

| PPs | Early technique | Proposed technique |
|---|---|---|
| PE (cm) | -0.03 | -0.03 |
| RES | 0.04 | 0.04 |
| SD | 0.54 | 0.50 |
| CC | 1.39 | 1.48 |
| Run time (s) | 42 | 43 |

With regard to conductivity values, the real part of induced voltages has been used in the early technique. The maximum number of iterations, the homogeneous conductivity value $\sigma_h$, and the coefficient $\tau$ were 20, 10 S/m, and 50, respectively, for both techniques (see Section 5.5). Figure 6.2(b)-(c) illustrate the reconstructed conductivity images by the early and proposed techniques, respectively. The results indicate that inhomogeneity is detected by both techniques and Figure 6.2(c) manifests the proposed technique works well for low-conductivity image reconstruction.

PPs obtained for each technique are given in Table 6.2. This table shows that (a) the SD in the proposed technique is 0.04 cm smaller than that of in the early technique, and (b) an increase of 0.09 in CC is achieved by the proposed method. The rest of PPs were the same in both techniques. In general, comparing the PPs in Table 6.2 manifests techniques have the same performance in this example.

### 6.3.4 Example II: One-target object with high-conductivity

In the second simulation, as shown in Figure 6.4(a), it is assumed that a conductive hexagonal bar with $\sigma_t = 10^7$ S/m is centered at (0, -4) cm. This hexagon can be simply approximated by a circle with a diameter $r = 2$ cm. This conductivity value is close to the conductivity of metals. In this case, the background conductivity is zero. The excitation frequency was 50 kHz. When the true conductivity distribution presents in the imaging region, Figure 6.3 shows the real (in blue) and imaginary (in red) parts of the voltages induced by the secondary magnetic field, $\Delta \mathbf{V}_M$, as a function of measurement number. As shown in this figure and as noted in Section 6.2, in the high-conductivity application, the imaginary part of the induced voltage mainly changes with conductivities inside the imaging region, and the changes in the real part of the induced voltage are not considerable.

With regard to conductivity values, the imaginary part of induced voltages has been used in the early technique. The maximum number of iterations, the homogeneous conductivity value $\sigma_h$, and the coefficient $\tau$ were 30, $10^4$ S/m, and 500, respectively, for both techniques. Figure 6.4(b)-(c) depict the reconstructed conductivity images obtained from the early and proposed techniques, respectively. The results indicate that inhomogeneity is detected by both techniques. Furthermore, Figure 6.4(c) manifests the proposed technique works better for high-conductivity image reconstruction.

PPs obtained for each technique are given in Table 6.3. This table shows that (a) the absolute value of PE in the proposed technique is 0.09 cm smaller than that of in the early technique, (b) the RES in the proposed method is 0.04 lesser than that of in the early method, (c) a reduction of 0.3 in SD is attained by the proposed method, and (d) an increase of 15857 in CC is achieved by the proposed method. In general, comparing the PPs in Table 6.3 manifests the proposed technique has superior performance compared to the early technique in this example.

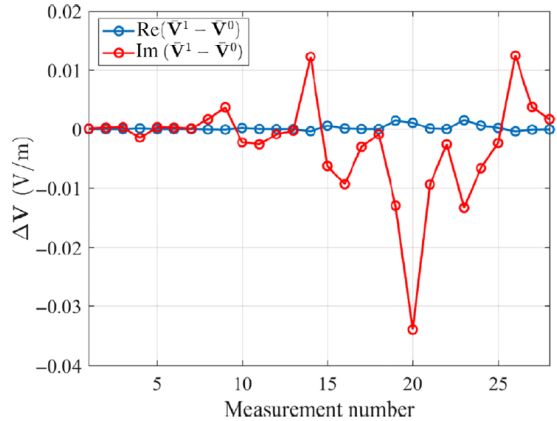

Figure 6.3 Example II: real (in blue) and imaginary (in red) parts of the voltages induced by the secondary magnetic field, $\Delta \mathbf{V}_M$, as a function of measurement number when the true conductivity distribution presents in the imaging region.

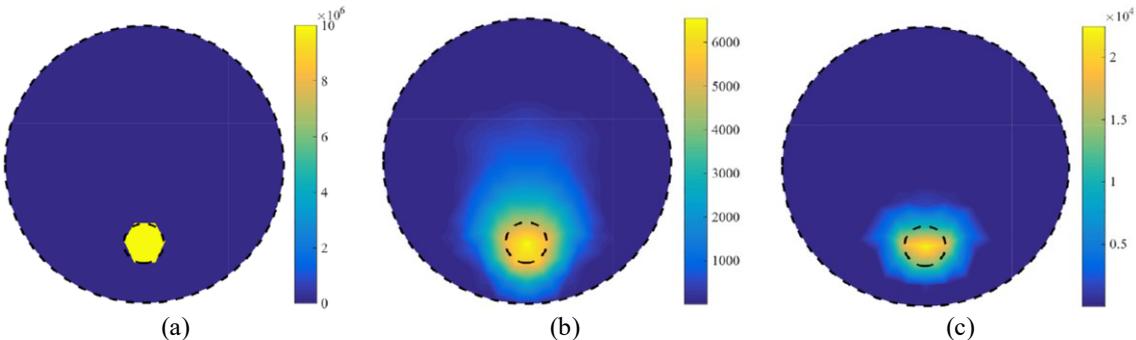

Figure 6.4 Example II: high-conductivity application; (a) True conductivities distribution; (b) Reconstructed image by the early technique using real part of data; and (c) Reconstructed image by the proposed technique. The maximum number of iterations, the homogeneous conductivity value $\sigma_h$, and the coefficient $\tau$ were 30, $10^4$ S/m, and 500, respectively, for both techniques. The regularization parameter at iteration 30-th was $3 \times 10^{-14}$ and $5 \times 10^{-15}$ for the early and proposed techniques, respectively. The excitation frequency was 50 kHz. The target object and background conductivity were $\sigma_t = 10^7$ S/m and $\sigma_b = 0$ S/m, respectively.

Table 6.3 Example I: Performance parameters (PPs)

| PPs | Early technique | Proposed technique |
|---|---|---|
| PE (cm) | - 0.17 | - 0.08 |
| RES | 0.07 | 0.03 |
| SD | 0.70 | 0.40 |
| CC | 4938 | 20795 |
| Run time (s) | 63 | 63 |

### 6.3.5 Example III: One-target object with mid-range conductivity

In the third experiment, it is assumed that a conductive hexagonal bar with $\sigma_t = 1000$ S/m moves from the periphery toward the center of the imaging region. Its center is placed at (0, -6), (0, -4), and (0, -2) cm as shown in Figure 6.6(a). The background conductivity is $\sigma_b = 100$ S/m. The conductivity value of inhomogeneity and background were chosen based on the conductivity value of carbon fiber [135] and biomedical conductive polymers [33], which fall into the mid-range conductivity category. The excitation frequency was 500 kHz. When the true conductivity distribution presents in the imaging region, Figure 6.5(a)-(c) show the real (in blue) and imaginary (in red) parts of the voltages induced by the secondary magnetic field, $\Delta\mathbf{V}_M$, as a function of measurement number for the target object with centers at (0, -6), (0, -4), and (0, -2), respectively. As shown in these figures and as noted in Section 6.2, in the mid-range conductivity application, both real and imaginary parts of the induced voltage change with conductivities inside the imaging region and are considerable.

The maximum number of iterations was set to 30. The homogeneous conductivity value $\sigma_h$ was 10, $10^4$, and 700 S/m for the early technique with the real part of data, the early technique with the imaginary part of data, and the proposed technique, respectively. The coefficient $\tau$ was 50, 500, and 100 for the early technique with the real part of data, the early technique with the imaginary part of data, and the proposed technique, respectively. Corresponding to true images with the target objects centered at (0, -6), (0, -4), and (0, -2) cm, the regularization parameter at iteration 30-th, $\lambda_{30}$, was obtained as follows. In the early technique with real part data, $\lambda_{30}$ was $2 \times 10^{-11}$, $1.8 \times 10^{-11}$, and $1.8 \times 10^{-11}$, respectively. In the early technique with imaginary part data, $\lambda_{30}$ was $1.7 \times 10^{-11}$, $1 \times 10^{-11}$, and $1 \times 10^{-11}$, respectively. In the proposed technique, $\lambda_{30}$ was $2.8 \times 10^{-11}$, $4.8 \times 10^{-11}$, and $4.8 \times 10^{-11}$, respectively.

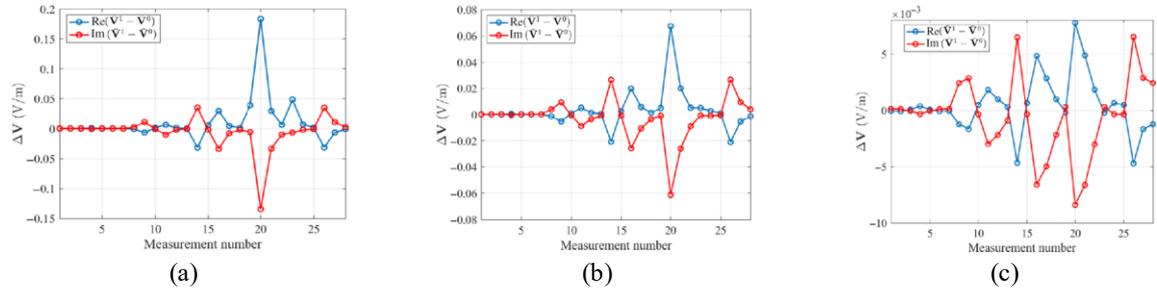

Figure 6.5 Example III: real (in blue) and imaginary (in red) parts of the voltages induced by the secondary magnetic field, $\Delta\mathbf{V}_M$, as a function of measurement number for the target object with centers at (a) (0, -6), (b) (0, -4), and (c) (0, -2); when the true conductivity distribution presents in the imaging region.

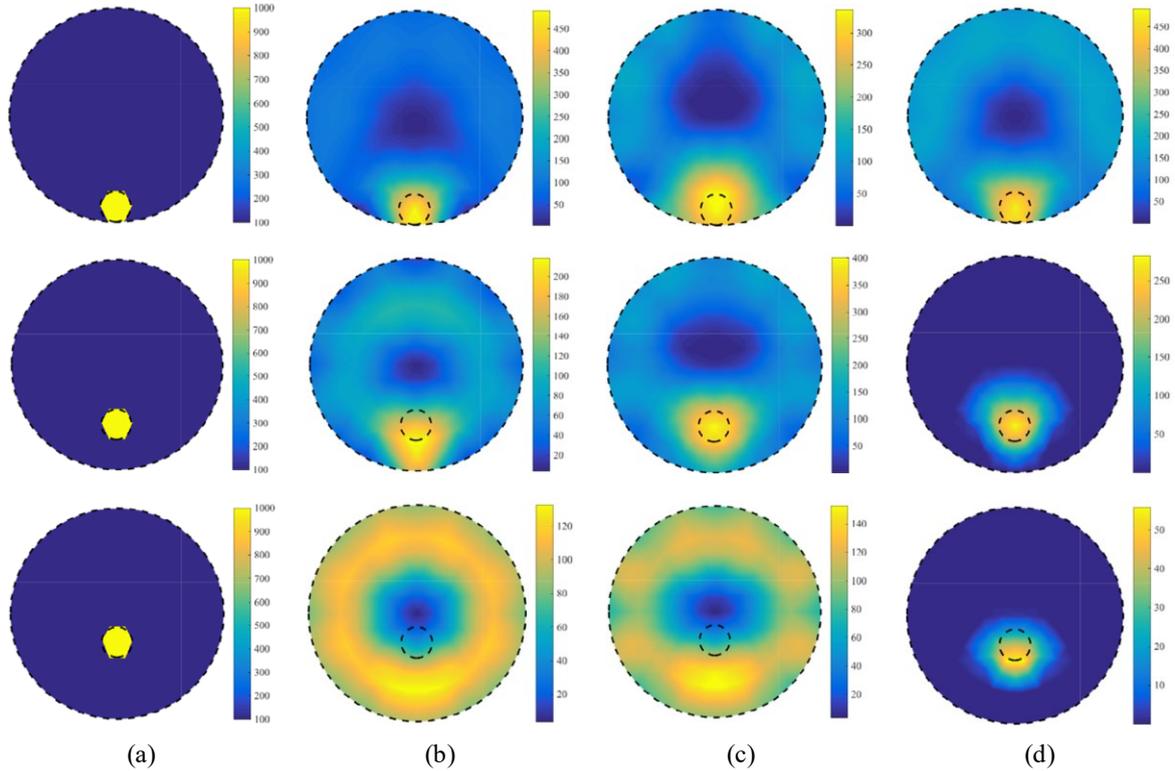

(a) (b) (c) (d)

Figure 6.6 Example III: mid-range conductivity application; (a) True conductivities distribution for the target object with centers at (0, -6), (0, -4), and (0, -2) cm in the first to third row, respectively; (b Reconstructed images by the early technique using the real part of data, $\sigma_h$ = 10 S/m and $\tau$ = 50; (c) Reconstructed images by the early technique using the imaginary part of data, $\sigma_h$ = $10^4$ S/m and $\tau$ = 500; (d) Reconstructed images by the proposed technique using the complex-valued, $\sigma_h$ = 700 S/m and $\tau$ = 100. The excitation frequency was 500 kHz. The target object and background conductivity were $\sigma_t$ = 1000 S/m and $\sigma_h$ = 100 S/m.

Table 6.4 Example III: Performance parameters (PPs)

| PPs | Target object position | Early technique | | Proposed technique |
|---|---|---|---|---|
| | | Re | Im | |
| PE (cm) | (0, -6) | 0.49 | 1.04 | 0.69 |
| | (0, -4) | -0.68 | 0.01 | 0.14 |
| | (0, -2) | a | a | -0.87 |
| RES | (0, -6) | 0.06 | 0.13 | 0.08 |
| | (0, -4) | 0.11 | 0.14 | 0.05 |
| | (0, -2) | a | a | 0.03 |
| SD | (0, -6) | 0.68 | 0.84 | 0.76 |
| | (0, -4) | 0.81 | 0.85 | 0.62 |
| | (0, -2) | a | a | 0.40 |
| CC | (0, -6) | 288 | 188 | 228 |
| | (0, -4) | 106 | 198 | 211 |
| | (0, -2) | a | a | 50 |

Figure 6.6(b)-(d) illustrate the reconstructed conductivity images by the early technique with the real part of data, the early technique with the imaginary part of data, and the proposed technique for corresponding target object position, respectively. As shown in Figure 6.7(d), the proposed technique can reconstruct conductivity images for all target object positions. Whereas the early technique cannot reconstruct images for the target object near the center. This is an evidence for our assumption which there are applications where both the real and imaginary

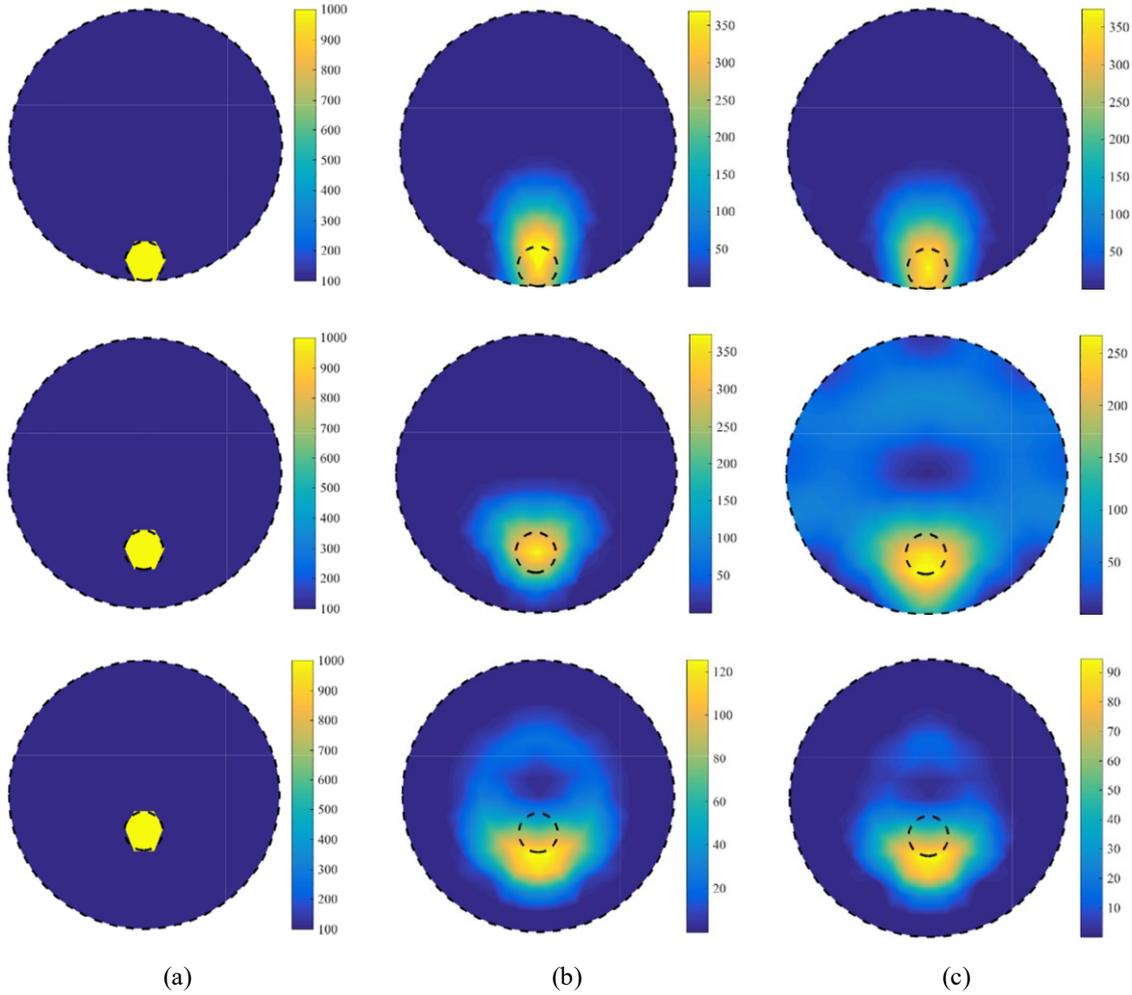

(a)                (b)                (c)

Figure 6.7 Example III: Reconstructed images by the proposed technique with other values for $\sigma_h$ and $\tau$; (a) True conductivities distribution for the target object with centers at (0, -6), (0, -4), and (0, -2) cm in the first to third row, respectively; (b) Reconstructed images with $\sigma_h = 10$ S/m and $\tau = 50$; and (c) Reconstructed images with $\sigma_h = 10^4$ and $\tau = 500$. The maximum number of iterations was 30. The excitation frequency was 500 kHz. The target object and background conductivity were $\sigma_t = 1000$ S/m and $\sigma_b = 100$ S/m, respectively.

parts of induced voltages consist of information on conductivity, and both the real and imaginary parts are necessary to be used for conductivity image reconstruction.

Table 6.4 shows the PPs for both techniques. The character "a" in Table 6.4 means that when the target object is close to the center, the algorithm based on the early technique fails to converge. Note that the corresponding images in Figure 6.6 show that the target object is not detectable for this position. As a result, the PPs are indeterminable for this position. For other positions of the target object, the PPs of the proposed technique are in the range of the early one.

It is worthwhile mentioning that the proposed technique was also able to reconstruct images by the values of $\sigma_h$ and $\tau$ which have been used for the early technique. Figure 6.7(b) shows the reconstructed images by the proposed method with $\sigma_h = 10$ S/m and $\tau = 50$ for the target object with centers at (0, -6), (0, -4), and (0, -2) cm in the first to third row, respectively. Figure 6.7(c) shows the reconstructed images by the proposed method with $\sigma_h = 10^4$ S/m and $\tau = 500$ for each target object position. It is notable that the early technique using the real part of data and the early technique using the imaginary part of data with $\sigma_h = 700$ S/m and $\tau = 100$ were not able to reconstruct images for the target objects with centers at (0, -2) and (0, -4). This result implies that the proposed technique is less sensitive to these two inputs than the early one.

### 6.3.6 Example IV: Two-target objects with different high-conductivities

In this example, we used two-target objects with vastly different conductivities in a high-conductivity application. The excitation frequency was 50 kHz. The conductivities of lower and upper target objects were $\sigma_{t1} = 10^7$ and $\sigma_{t2} = 10^5$ S/m, respectively. The centers of lower and upper target objects were placed at (0, -4) and (0, 4) cm, respectively. The maximum number of iterations, the homogeneous conductivity value $\sigma_h$, and the coefficient $\tau$ were 30, $10^4$ S/m, and 500, respectively, for both techniques.

Figure 6.8 shows the results of this numerical experiment. As shown in Figure 6.8(b), the early technique could not reconstruct two separate target objects, while as shown in Figure 6.8(c), the proposed technique could reconstruct two clearly separate target objects. A reason for this result can be explained by singular values of the Hessian matrices. Figure 6.9 shows the first 70 normalized singular values of the Hessian matrices for the proposed technique, calculated for the true conductivity distribution, on a logarithmic scale. The roughly linear decay of the first 28 (number of independent measurements) singular values for the early technique (solid blue line) and the first 56 singular values for the proposed technique (dotted red line) show that the problems are severely ill-posed and require regularization. Since the singular values jump in a less singular value number when the problem is solved with the early technique; thus, the problem is more ill-posed with the early technique. Results imply that there is more information in complex-valued data for conductivity image reconstruction. Thus, the proposed technique can be useful in MIT systems which include target objectives with wide conductivity ranges, for example, the MIT system used in [136].

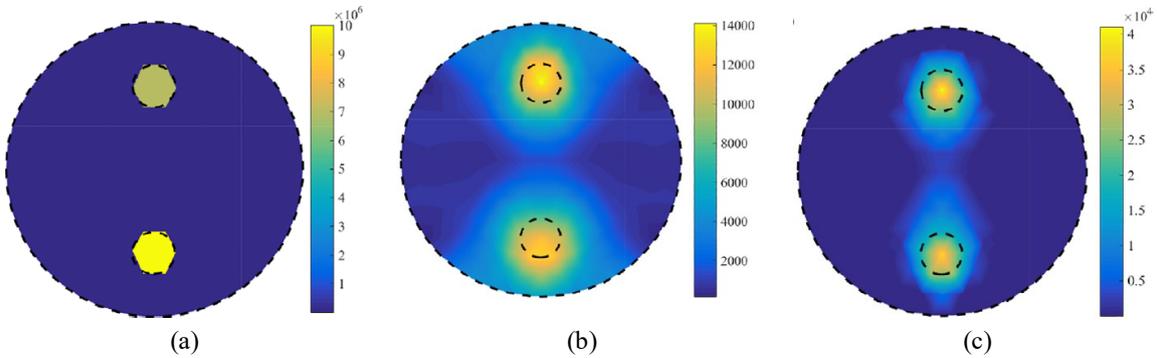

(a)           (b)           (c)

Figure 6.8 Example IV: two-target objects with different high-conductivities; (a) True conductivities distribution; (b) reconstructed image by the early technique using imaginary part of data; and (c) Reconstructed image by the proposed technique. The maximum number of iterations, the homogeneous conductivity value $\sigma_h$, and the coefficient $\tau$ were 30, $10^4$ S/m, and 500, respectively, for both techniques. The regularization parameter at iteration 30-th was $1.6 \times 10^{-14}$ and $2.4 \times 10^{-15}$ for early and proposed techniques, respectively. The excitation frequency was 50 kHz. The target objects and background conductivity were $\sigma_{t1} = 10^7$ S/m, $\sigma_{t2} = 10^5$ S/m, and $\sigma_b = 0$ S/m, respectively.

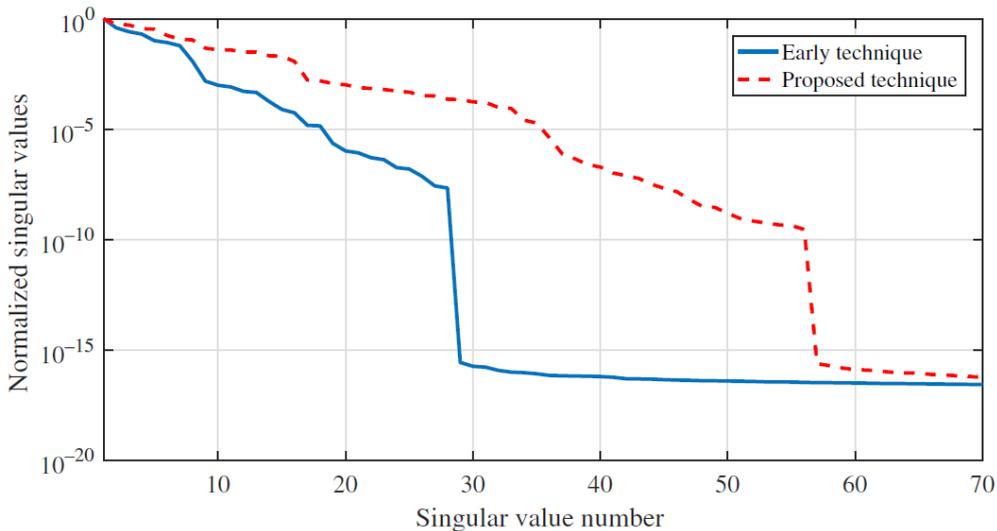

Figure 6.9 Example IV: singular values of Hessian matrices for the early technique (solid blue line) and the proposed technique (dotted red line).

### 6.3.7 Example V: Noise study in the mid-range conductivity application

In this example, we evaluate the noise robustness of the proposed technique compared to the early one. For this purpose, we chose the case given in Example III where the center of the target object was at (0, -4) cm. In this case, the proposed technique and the early technique with both real and imaginary data types have detected the target object. We added 0.3% complex white Gaussian noise to the simulated measured voltages and used them for both techniques. We repeated the experiment 50 times while setting a seed value for the noise before starting the reparation. After applying threshold ($\boldsymbol{\sigma}^t$ image), the early technique with the real part of data detected the target object 10 times, the early technique with the imaginary part of data detected the target object 5 times, and the proposed technique detected the target object in all 50 images.

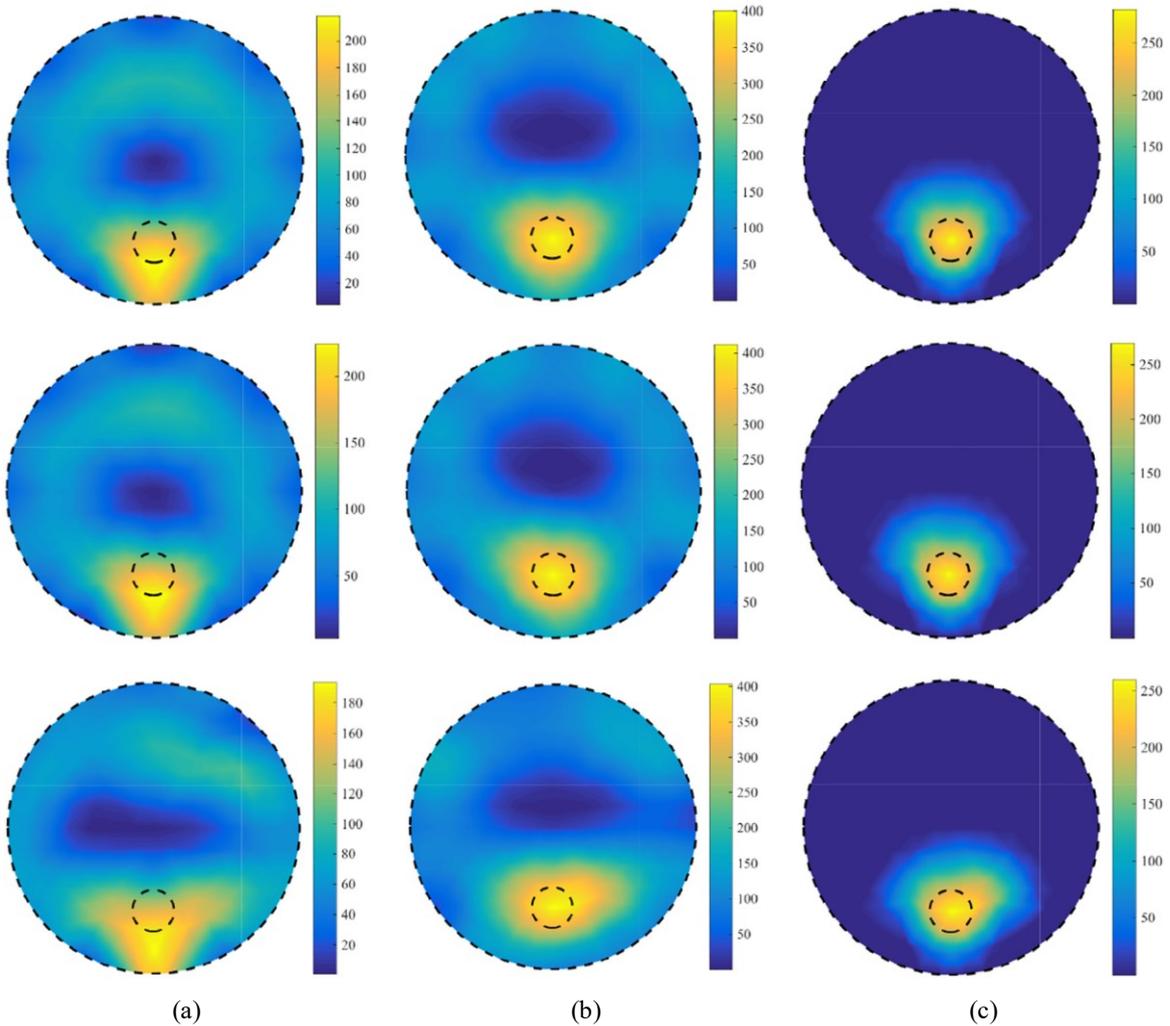

(a) (b) (c)

Figure 6.10 Example V: Noise study for 0%, 0.3%, and 1% noise levels corresponding to the first to third rows from the top, respectively; (a) Reconstructed images by the early technique using real part of data, $\sigma_h = 10$ S/m and $\tau = 50$; (b) Reconstructed images by the early technique using imaginary part of data, $\sigma_h = 10^4$ S/m and $\tau = 500$; and (c) Reconstructed images by the proposed technique, $\sigma_h = 700$ S/m and $\tau = 100$. The maximum number of iterations was 30. The regularization parameter at iteration 30-th was $1.8 \times 10^{-11}$ in the early technique with real part of data, $1 \times 10^{-11}$ in the early technique with imaginary part of data, and $4.8 \times 10^{-11}$ in the proposed technique. The excitation frequency was 500 kHz. The target object and background conductivity were $\sigma_t = 1000$ S/m and $\sigma_b = 100$ S/m, respectively. True conductivities distribution was the case given in Example III where the center of the target object was at (0, -4) cm.

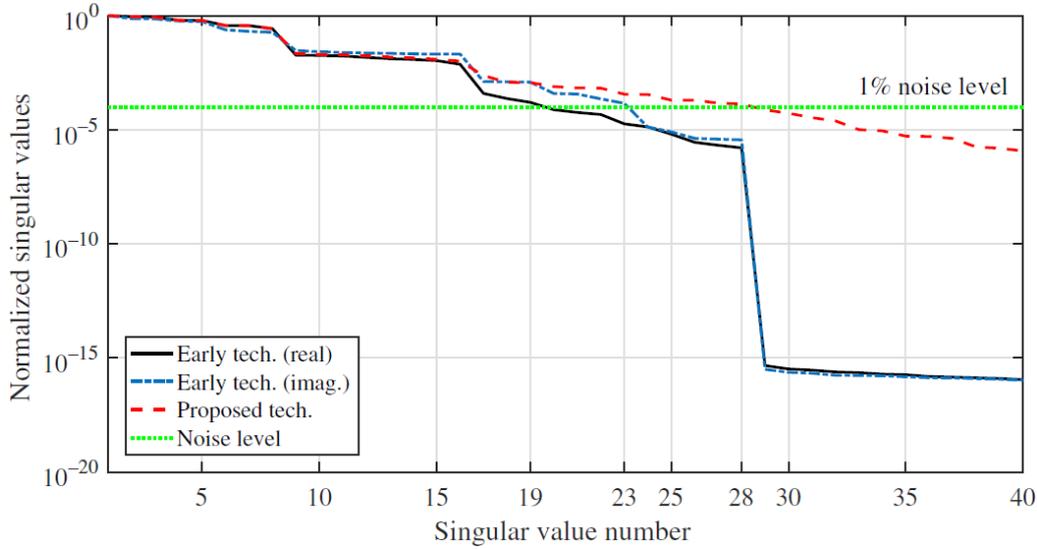

Figure 6.11 Example V: singular values of Hessian matrices for the early and proposed techniques and 1%

Figure 6.10 shows the reconstructed images by the early and proposed techniques for the noise levels 0%, 0.3%, and 1%. The number of iterations, the homogeneous conductivity value $\sigma_h$, and the coefficient $\tau$ were identical to those given in Example III. As can be seen, the proposed method is more robust against noise. Based on the Picard criteria, only the singular values above the noise level will contribute to the inverse problem solution [137]. Figure 6.11 shows that for a 1% noise level, the first 28 singular values in the proposed technique are involved in image reconstruction, and the others are under the noise level, whereas the first 19 and 23 singular values in the early technique with real and imaginary parts of induced voltage, respectively, are relevant to reconstruction. This implies that there is more information in complex-valued data, and the proposed technique is more stable and has more robustness against the noise. This is another potential advantage that the proposed technique might bring.

## 6.4 Conclusions

In this chapter, a technique for MIT conductivity reconstruction was proposed. The technique uses both the real and imaginary parts of the induced voltages to reconstruct MIT conductivity images. Using the regularized GN algorithm based on complex-valued data for the first time in MIT studies, equations were derived and implemented for updating the conductivity, regularization parameter, and the Jacobian matrix.

By modeling a 2D MIT system through five different numerical experiments, the performance of the proposed technique was investigated. Four PPs were defined to evaluate the obtained results. Based on the results of the first numerical experiment, which included a one-target object with low conductivity, the early and proposed techniques had almost the same performance. Based on the results of the second numerical experiment, which included a one-target object with high conductivity, the proposed technique had superior performance compared to the early technique. Results of the third numerical experiment, which included a one-target object with mid-range conductivity, showed that when the target object was close to the center of the imaging region, the proposed technique could yield PPs, but the early technique could not. For other positions of the target object, the PPs of the proposed technique were in the range of the early one. The results obtained in the fourth numerical experiment indicated that when the imaging region included vastly different conductivity values, there was more information in complex-valued data compared to real data. This result was obtained by singular value decomposition of the Hessian matrix. In the fifth numerical experiment, results demonstrated that for a given noise level, the proposed technique had greater singular values. Consequently, the proposed technique had more robustness against noise and showed more stable behavior. In the numerical experiments, the regularization parameter was obtained based on the adaptive method, which uses the two inputs homogeneous conductivity value ($\sigma_h$) and multiple coefficient ($\tau$). Results showed that the proposed technique is less sensitive to these two inputs than the early technique. The outcomes of this chapter confirm the necessity of using both real and imaginary parts of the induced voltage on conductivity image reconstruction in MIT. In addition, the proposed technique can be used in MIT systems which include target objectives with a wide conductivity range. As a result, this technique will help further focus on the development and improvement of MIT systems.

# Chapter 7: Conclusions and Future Work

In this thesis, improved forward problem modeling in MIT for biomedical applications was investigated through numerical simulations. This chapter summarizes the results carried out from this thesis and discusses the main conclusions drawn from the study. It also suggests some potential future works in the continuation of this study.

## 7.1 Conclusions

This thesis has been tried to cover four main objectives:
1- Presenting an improved forward model for the biomedical MIT in which skin and proximity effects were incorporated in exciter and sensor coils.
2- Solving the biomedical MIT forward problem by a hybrid mesh-based and mesh-free method.
3- Investigating the importance of the improved forward model for conductivity image reconstruction.
4- Developing a technique for mid-range conductivity image reconstruction.

For the first objective, an improved method based on Maxwell's equations was developed to model 2D MIT forward problems in Chapter 3. This method considers skin and proximity effects inside the coils. Comparison of the improved method with the early one through numerical simulations revealed that there was a meaningful discrepancy in terms of achievable conductivity contrast from the improved one. The results manifested the error of the early method was a function of operating frequency, and conductivity value and distribution inside the imaging region. Ignoring skin and proximity effects in coils appears as a numerical noise in the modeling of the MIT forward problem at higher frequencies which are desirable for biomedical MIT applications. This noise can be partially compensated by the state-difference imaging or the frequency-difference technique to reconstruct the relative conductivity values. Results manifested that the improved method, which considers skin and proximity effect in MIT coils, can significantly impress the accuracy of computation of induced voltages in the biomedical MIT applications.

For the second objective, a hybrid FE-EFG numerical method was developed to solve the 2D MIT improved forward problem in Chapter 4. In this proposed numerical method, the EFG method was employed to solve the problem in the imaging region, and the FE method was used to solve the problem in coil and air around imaging regions. In this way, we benefited from the advantages of both numerical methods. Comparison of FE and FE-EFG methods showed almost the same accuracy for both methods in the same run-time.

For the third objective, numerical conductivity image reconstruction based on the improved forward method was developed for 2D biomedical MIT in Chapter 5. The regularized GN algorithm was adapted to the improved forward method. Results obtained from solving the inverse problem manifested that the error due to neglecting the skin and proximity effects can be partially compensated by the difference imaging; however, it will be at the cost of producing qualitative images. Furthermore, to reconstruct the absolute conductivity values in biomedical applications of MIT, it is crucial to use the improved forward method and voltages induced by the total magnetic field. A small error in the forward problem may seem negligible, whereas it can implicate considerable errors when it is entered in the process of the inverse problem for image reconstruction.

For the fourth objective, a new technique for MIT conductivity reconstruction was proposed. The technique uses both the real and imaginary parts of the induced voltages to reconstruct MIT conductivity images. The regularized GN algorithm was adapted to this technique. Based on the obtained results from the numerical experiments, the proposed technique had almost the superior performance over the early one for all ranges of the conductivity distribution, especially for the mid-range conductivity. The outcomes confirm the necessity of using both real and imaginary parts of the induced voltage on conductivity image reconstruction in MIT.

From work carried out in this thesis and current literature, it seems MIT has a long way to go before it can successfully be introduced as a clinical imaging modality. However, improvements approached in this thesis will help further focus on the development and progress of MIT systems.

## 7.2 Future Work

As mentioned before, the most interesting biomedical application of MIT is ICH imaging. However, it seems that MIT is still young for this application and needs further efforts to address the research gaps. This thesis was contributed to this aim, and good signs of progress have been achieved. In the following, potential research areas which could be considered as future work to improve MIT imaging modality are outlined.

- **Follow up research for the improved forward method:**
1- In this thesis, we showed the importance of the improved forward method by 2D simulations. As a vital research area, it is necessary to develop the improved forward model to three-dimensional case and use human head models to approach more realistic situations.

2- In this thesis, the improved forward method was validated by an analytical solution. The experimental verification of the improved forward model is of high interest.

3- The developed improved forward model takes only the real conductivity. Complex conductivity (conductivity and permittivity) can be considered in the improved forward method straightforwardly by replacing the real conductivity with complex conductivity for the imaging region equation. Furthermore, the conductivities have been assumed to be isotropic in the improved forward method. The method can be extended to consider anisotropic conductivities, as well.

- **Follow up research for the FE-EFG forward solver:**
4- In this thesis, the hybrid FE-EFG method was employed for solving the improved forward method. Another option can be solving the improved forward model with the pure EFG method using the approach presented in [116]. The obtained result for the FE-EFG method in this thesis was based on a uniform distribution of the node in the imaging region. It would be interesting to see the performance of the EFG method for solving the biomedical MIT forward problem with arbitrarily distributed nodes.

5- The realistic human head model incorporates irregular outer and inner shapes, and it is vital to be modeled accurately. Any errors in modeling the geometry can lead to a wrong solution of the inverse problem. One interesting research area would be the evaluation of the efficacy of the 3D EFG method for solving the forward problem in the realistic head model compared to FEM.

- **Follow up research for the inverse problem:**
6- The reconstruction results presented in this thesis were based on synthetic data. Using experimental data to reconstruct conductivity distribution is of high interest.

7- Based on the current literature, it seems that biomedical MIT in the current state is more valuable in monitoring the development of ICH instead of detecting or identifying the stroke type. One option to increase the quality of the reconstructed conductivity for this application may be using an individualized head model obtained by MRI or CT. In this way, the geometry of the problem will be known, and only the conductivity of each region should be estimated.

8- In this thesis, the importance of using complex voltage to reconstruct conductivity was shown numerically. The technique was implemented by the regularized GN algorithm. It can be extended to other reconstruction algorithms, as well. Experimental evaluation of the superiority of this technique for various conductivity distribution values would be interesting.

- **Further follow up research:**
To increase the chances for MIT to become a medical imaging modality, further researches are required. Some suggestions are provided below based on the author's view:

1- It seems that the MIT systems in the current form are more sensitive to near the border regions than to the center of the imaging region. Thus, it is more successful to image large peripheral ICH. An interesting suggestion to increase the sensitivity of MIT to the central region would be parallel excitation which has been used in industrial systems [84], [138]. However, it must be guaranteed that the power absorbed in the tissue from the electromagnetic fields is within safety limits.

2- It would be valuable to develop a software package for the numerical study of MIT, something like EIDORS (electrical impedance and optical reconstruction software) [139]. Current commercial software packages for solving the electromagnetic fields are not suitable for solving the forward problem, and there is no option for solving the inverse problem inside them. As a consequence, developing a dedicated software package would pave the way for further progress in MIT.

3- MIT inherently suffers from poor spatial resolution. However, it should be noted that there is no conventional imaging modality for high resolution and high contrast imaging of conductivity. It is while high resolution and contrast information about the conductivity distribution inside biological tissues, obtained noninvasively, is of high interest for a wide range of medical applications. The poor resolution of MIT originates from insufficient independent measurement. To overcome the difficulty, a recently developed hybrid approach combining MIT with MR imaging is proposed [140]. The MR scanner provides information about the interior magnetic field induced by the current field induced by the MIT experiment. This interior information has the potential for a vast improvement of the resolution and contrast in conductivity reconstruction. However, in this way, some advantages of MIT, like being fast and low-cost would be sacrificed.

# Appendix 1: Extraction of equation (2.1)

In this appendix, we present the extraction of (2.1). Figure A1.1 shows a simple model of two coaxial coils with a target domain which known as a single channel MIT system in the literature [34]. By passing an alternating current from the excitation coil placed around the target object, an alternating magnetic field is generated which is called the primary magnetic field, $B_0$ (the magnitude of field). This generates another magnetic field inside the conductive target object by inducing eddy currents, this field is called the secondary magnetic field, $\Delta B$. The total magnetic field, $B_1$, which is the sum of the primary and secondary fields, can be measured by the sensing coil. To explore the relation between the electrical properties of the tissue and the induced field in the sensing coil, let us consider Figure A1.1. In this figure, a disk (as a biological tissue) with a radius of r and thickness of t is coaxially placed between two small coils with a radius of b. The distance between the center of two coils is 2a. The conductivity and relative permittivity of the disk are $\sigma$ and $\varepsilon_r$, respectively, and it is non-magnetic ($\mu_r = 1$). The following assumptions are presumed:

1- $\delta \gg t$; where $\delta$ is skin depth and defined as follows:

$$\delta = \sqrt{\frac{2}{\omega\mu\sigma}} \tag{A1.1}$$

Here, $\omega$, $\mu$, and $\sigma$ are the angular frequency, the permeability of disk, and the conductivity of disk, respectively. Assumption 1 means that the attenuation produced by the disk in the excitation field is negligible (weakly coupled field approximation).

2- $R \gg b$. It means that coils are considered as magnetic dipoles. A magnetic dipole is characterized by the magnetic moment vector $\vec{m}$ as follows:

$$\vec{m} = I\pi b^2\, \vec{a}_z \tag{A1.2}$$

where I and b are current amplitude and radius of the coil. In fact, the magnetic dipole moment is equal to the product of current amplitude and area of the coil. Its direction is obtained by the right hand rule.

First, we only consider the conductivity of the disk and then enter the permittivity. We obtain the magnetic flux density at a distant point of a small circular loop (excitation coil) of a radius of b that carries current I. In other words, we want to evaluate $\vec{B}$ at a distant point in the spherical coordinate system (see Figure A1.2). We select the center of the loop at the origin. The source coordinates are primed, and the field evaluation point coordinates are unprimed.

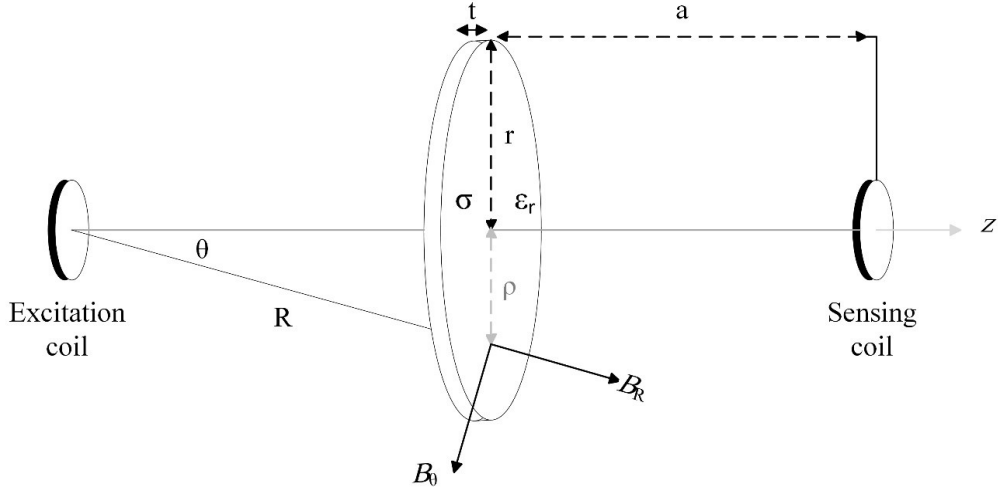

Figure A1.1. A single channel MIT system. The magnetic field generated by the excitation coil has been shown in a conductive disk at the spherical coordinate. The radius and thickness of the disk are $r$ and $t$, respectively. It is coaxially placed between the excitation and sensing coils. The distance between the centers of coils is 2a. The conductivity and relative permittivity of disk is $\sigma$ and $\varepsilon_r$, respectively, and it is assumed to be non-magnetic ($\mu_r = 1$). It is presumed that the penetration depth of excitation field is greater than t. Redrawn from [34].

The magnetic vector potential (MVP) produced by the loop is obtained as follows [141]:

$$\vec{A}(R, \theta, \varphi) = \frac{\mu_0 I}{4\pi} \oint_{C'} \frac{d\vec{\ell'}}{R_1}, \quad d\vec{\ell'} = \vec{a}_{\varphi'}\, b\, d\varphi' \tag{A1.3}$$

Here $C'$ is a closed contour surrounding the loop. It is obvious that, due to symmetry, the magnetic field is independent of the angle $\varphi$ of the field point. For simplicity, we consider point $P(R, \theta, \pi/2)$ in the $y - z$ plane to evaluate the field. It should be noted that $\vec{a}_\varphi$ at $P$ is equal to $-\vec{a}_x$ whereas $\vec{a}_{\varphi'}$ in $d\vec{\ell'}$ is

$$d\vec{\ell'} = \vec{a}_{\varphi'}\, b\, d\varphi' = (-\vec{a}_x \sin\varphi' + \vec{a}_y \cos\varphi') b\, d\varphi' \tag{A1.4}$$

For every $Id\vec{\ell'}$, there is another symmetrically located differential current element in the other side of the $y$-axis that will contribute an equal amount to $\vec{A}$ in $-\vec{a}_x$ direction but will cancel the contribution of $Id\vec{\ell'}$ in the $\vec{a}_y$ direction. In other words, the field is symmetric with regard to the $y$-axis. By substituting (A1.4) to (A1.3), one obtains:

$$\vec{A} = -\vec{a}_x \frac{\mu_0 I}{4\pi} \int_0^{2\pi} \frac{b\, \sin\varphi'}{R_1}\, d\varphi' \rightarrow \vec{A} = \vec{a}_\varphi \frac{\mu_0 Ib}{2\pi} \int_{-\frac{\pi}{2}}^{\frac{\pi}{2}} \frac{\sin\varphi'}{R_1}\, d\varphi' \tag{A1.5}$$

Using the law of cosines in $OPP'$ triangle ($O$ origin, $P$ field evaluation point, $P'$ current element point), we have:

$$R_1^2 = R^2 + b^2 - 2bR\cos\psi \tag{A1.6}$$

where $R\cos\psi$ is the projection of R on the radius $OP'$ and is the same as $R\sin\theta\sin\varphi'$ (Projection of $R$ on the $y$-axis, $OP''$, and then projection on the radius $OP'$):

$$R_1^2 = R^2 + b^2 - 2bR\sin\theta\sin\varphi'$$

$$\frac{1}{R_1} = \frac{1}{R}\left(1 + \frac{b^2}{R^2} - \frac{2b}{R}\sin\theta\sin\varphi'\right)^{-\frac{1}{2}} \tag{A1.7}$$

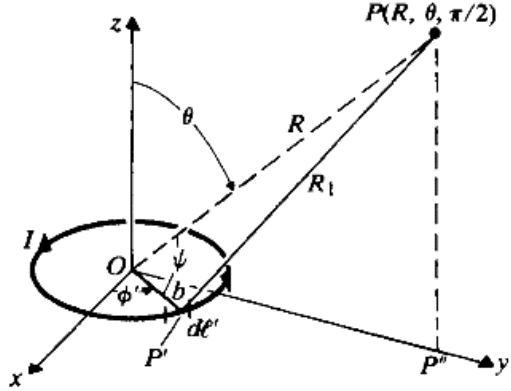

Figure A1.2. A small circular loop carrying current $I$ [141].

Since $R \gg b$, thus $R^2 \gg b^2$ and $\frac{b^2}{R^2} \ll 1$. This term can be neglected in (A1.7):

$$\frac{1}{R_1} \cong \frac{1}{R}\left(1 - \frac{2b}{R}\sin\theta \sin\varphi'\right)^{-\frac{1}{2}} \tag{A1.8}$$

Using Taylor series expansion of (A1.8), one obtains:

$$\frac{1}{R_1} \cong \frac{1}{R}\left(1 + \frac{b}{R}\sin\theta \sin\varphi'\right) \tag{A1.9}$$

Substituting (A1.9) in (A1.5) results in:

$$\vec{A} = \vec{a}_\varphi \frac{\mu_0 Ib}{2\pi R} \int_{-\frac{\pi}{2}}^{\frac{\pi}{2}} \left(1 + \frac{b}{R}\sin\theta \sin\varphi'\right) \sin\varphi' \, d\varphi' \tag{A1.10}$$

Evaluation of the integral (A1.10) gives:

$$\vec{A} = \vec{a}_\varphi \frac{\mu_0 Ib^2}{4R^2} \sin\theta \tag{A1.11}$$

Using the definition of MVP, $\vec{B} = \nabla \times \vec{A}$, we obtain:

$$\vec{B} = \frac{\mu_0 Ib^2}{4R^3} (\vec{a}_R 2\cos\theta + \vec{a}_\theta \sin\theta) \tag{A1.12}$$

In (A1.12), we can substitute $m/\pi$ for $Ib^2$:

$$\vec{B} = \frac{\mu_0 m}{4\pi R^3} (\vec{a}_R 2\cos\theta + \vec{a}_\theta \sin\theta) \tag{A1.13}$$

where m is the magnitude of the magnetic dipole moment vector. The magnetic field of a magnetic dipole can be resolved into radial and tangential components, $\vec{B}_R$ and $\vec{B}_\theta$, as follows:

$$\vec{B}_R = \frac{\mu_0 m}{4\pi R^3} 2\cos\theta \, \vec{a}_R \tag{A1.14}$$

$$\vec{B}_\theta = \frac{\mu_0 m}{4\pi R^3} \sin\theta \, \vec{a}_\theta \tag{A1.15}$$

As seen from Figure A1.1, $\cos\theta = a/R$ and $\sin\theta = \rho/R$. The magnitude of the field along the z-axis on the disk shown in Figure A1.1 is obtained as follows:

$$B_z = B_R \cos\theta - B_\theta \sin\theta = \frac{\mu_0 m}{4\pi R^3}(2\cos^2\theta - \sin^2\theta)$$

$$= \frac{\mu_0 m}{4\pi}\left(\frac{2a^2 - \rho^2}{R^5}\right) \tag{A1.16}$$

$$= \frac{\mu_0 m}{4\pi} \frac{2a^2 - \rho^2}{(a^2 + \rho^2)^{\frac{5}{2}}}$$

The magnetic flux passing through a circular path of radius ρ and centered on the axis as follows:

$$\Phi(\rho) = \iint\limits_{0\ 0}^{2\pi\ \rho} \vec{B}_z \cdot d\vec{s} = \iint\limits_{0\ 0}^{2\pi\ \rho} B_z(\rho')\, \rho'\, d\varphi'\, d\rho' \tag{A1.17}$$

Substituting (A1.16) in (A1.17) and evaluating the integral result in:

$$\Phi(\rho) = \frac{\mu_0 m}{2} \frac{\rho^2}{(a^2 + \rho^2)^{\frac{3}{2}}} \tag{A1.18}$$

The induced voltage around this path is $-j\omega\Phi$. It also equals the line integral of the induced electric field around the closed path, $2\pi\rho E_\varphi$ where $E_\varphi$ is the $\varphi$-component of the induced electric field. Hence:

$$-j\omega\Phi = 2\pi\rho E_\varphi \;\rightarrow\; E_\varphi = \frac{-j\omega\Phi}{2\pi\rho} \tag{A1.19}$$

Substituting (A1.19) in (A1.18) gives:

$$E_\varphi = \frac{-j\omega\mu_0 m}{4\pi} \frac{\rho}{(a^2 + \rho^2)^{\frac{3}{2}}} \tag{A1.20}$$

The current density induced in the disk is $J_\varphi = \sigma E_\varphi$. Thus:

$$J_\varphi = \frac{-j\omega\sigma\mu_0 m}{4\pi} \frac{\rho}{(a^2 + \rho^2)^{\frac{3}{2}}} = \frac{-jm}{2\pi\delta^2} \frac{\rho}{(a^2 + \rho^2)^{\frac{3}{2}}} \tag{A1.21}$$

where $\delta$ is the skin depth introduced in (A1.1). A thin annulus of the disk between ρ and ρ + dρ, carries a current dI, where,

$$dI = J_\varphi\, t\, d\rho = \frac{-jmt}{2\pi\delta^2} \frac{\rho}{(a^2 + \rho^2)^{\frac{3}{2}}}\, d\rho \tag{A1.22}$$

From Biot-Savart's law, the magnitude of the magnetic flux density at a point (0, 0, z) on the axis of a circular loop carrying current $I_{cl}$ and with a radius of $k$ can be evaluated as follows [141]:

$$B_{cl} = \frac{\mu_0 I_{cl}\, k^2}{2(z^2 + k^2)^{\frac{3}{2}}} = \frac{\mu_0 m_{cl}}{2\pi(z^2 + k^2)^{\frac{3}{2}}} \tag{A1.23}$$

Using (A1.23), the magnitude of magnetic flux density caused by dI at the center of the sensing coil is obtained as follows:

$$dB = \frac{\mu_0 \rho^2\, dI}{2(a^2 + \rho^2)^{\frac{3}{2}}} = \frac{-j\mu_0 mt}{4\pi\delta^2} \frac{\rho^3\, d\rho}{(a^2 + \rho^2)^3} \tag{A1.24}$$

and the magnitude of the magnetic flux density caused by the excitation coil at the center of the sensing coil will be:

$$B_0 = \frac{\mu_0 m}{2\pi((2a)^2 + b^2)^{\frac{3}{2}}} \xrightarrow{4a^2 \gg b^2} B_0 \cong \frac{\mu_0 m}{16\pi a^3} \tag{A1.25}$$

Hence, we can obtain:

$$\frac{dB}{B_0} = \frac{-4j\, t\, a^3}{\delta^2} \frac{\rho^3\, d\rho}{(a^2 + \rho^2)^3} \tag{A1.26}$$

The relative change in the magnetic flux density due to induced current in the whole disk is found by integrating (A1.26) for ρ: 0 → r; which gives:

$$\frac{\Delta B}{B_0} = \frac{-j\, t\, a^3}{\delta^2}\left[\frac{1}{a^2} - \frac{a^2 + 2r^2}{(a^2 + r^2)^2}\right] \tag{A1.27}$$

where r is the disk radius.

For considering the permittivity of the disk, it is enough to change $\sigma \to \sigma + j\omega\,\varepsilon_0\varepsilon_r$ in the skin depth $\delta$ in (A1.27):

$$\frac{\Delta B}{B_0} = \frac{-j\, t\, a^3}{2}(\sigma + j\omega\varepsilon_0\varepsilon_r)\omega\mu_0\left[\frac{1}{a^2} - \frac{a^2 + 2r^2}{(a^2 + r^2)^2}\right] \tag{A1.28}$$

where $\varepsilon_0$ and $\varepsilon_r$ are the free space permittivity and the relative permittivity of the disk, respectively. Finally, we can write:

$$\frac{\Delta B}{B_0} = \frac{\Delta V}{V_0} = \frac{-j\, t\, a^3}{2}(\sigma + j\omega\varepsilon_0\varepsilon_r)\omega\mu_0\left[\frac{1}{a^2} - \frac{a^2 + 2r^2}{(a^2 + r^2)^2}\right] \tag{A1.29}$$

where $\Delta V$ and $V_0$ are the change in the voltage induced in the sensing coil due to the presence of a conductive disk and the voltage induced by the excitation coil in the sensing coil, respectively. In other words, $\Delta V$ and $V_0$ are induced by the secondary and primary magnetic fields, respectively. Equation (A1.29) can be simplified as follows:

$$\frac{\Delta B}{B_0} = \frac{\Delta V}{V_0} = Q\mu_0\omega(\omega\varepsilon_0\varepsilon_r - j\sigma) \tag{A1.30}$$

where Q is a constant coefficient related to the problem geometry:

$$Q = \left(\frac{t a^3}{2}\right)\left[\frac{1}{a^2} - \frac{a^2 + 2r^2}{(a^2 + r^2)^2}\right] \tag{A1.31}$$

In brief, passing magnetic flux density caused by the excitation coil, $B_0$, through the disk produces a voltage difference inside it. The voltage difference induces an electric field in the disk. The electric field causes to induce eddy currents in the disk. These currents generate another magnetic field with the magnetic flux density $\Delta B$. The resultant field induced a voltage in the sensing coil which causes to change of the primary voltage $V_0$. It should be noted that $\Delta V/V_0$ is independent of the coil's geometry and the excitation current because they are simplified in the numerator and the denominator of (A1.30).

One can write the voltage induced by the total field in the sensing coil as follows:

$$V_1 = V_0 + \Delta V = [(Q\omega^2\mu_0\varepsilon_0\varepsilon_r + 1) - jQ\omega\mu_0\sigma]V_0 \tag{A1.32}$$

The phase angle and the magnitude of the induced voltage can be obtained as follows:

$$\angle V_1 = \tan^{-1}\left(\frac{-Q\omega\mu_0\sigma}{Q\omega^2\mu_0\varepsilon_0\varepsilon_r + 1}\right) \tag{A1.33}$$

$$|V_1| = \sqrt{(Q\omega^2\mu_0\varepsilon_0\varepsilon_r + 1)^2 + (Q\omega\mu_0\sigma)^2}\,|V_0| \tag{A1.34}$$

To see the importance of the measuring the phase angle of the induced voltage in the biomedical MIT, consider the following example (inspired by [34]). Consider r = 4.5 cm, a = 8.75 cm, and t = 7.9 cm in Figure A1.1. The excitation frequency, the disk conductivity, and the relative permittivity of the disk are $f$=10 MHz, $\sigma = 1$, and $\varepsilon_r = 80$. Then, (A1.33) and (A1.34) gives:

$$\angle V_1 = -0.684° = -684\, m°$$
$$|V_1| = 1.0006\,|\mathbf{V_0}|$$

As seen, the change in the voltage magnitude is too small and it is hard to measure it in practice. But, the change in the phase angle may be measurable. As a result, the phase noise is a very important criteria in the biomedical MIT systems.

# Appendix 2: Various biomedical MIT forward problem formulations

In this appendix, various formulations employed for modeling the biomedical MIT forward problem are introduced. The mathematical treatment of fields is often simplified by working with potential functions instead of fields. Hence, vector field expressions are generally expressed by scalar and vector potentials. There are several formulations for the MIT forward problem in the literature which have been expressed in terms of electrical scalar potential V and magnetic vector potential $\vec{A}$. In the following, we introduce formulations employed in biomedical MIT.

## A2.1 $\vec{A} - \vec{A}, V$ formulation

$\vec{A} - \vec{A}, V$ formulation is one of the most widely used formulations in modeling MIT forward problem [41], [62]. This formulation uses magnetic vector potential $\vec{A}$ in the entire domain ($\Omega_N \cup \Omega_C$) and electrical scalar potential V in the conducting domain $\Omega_C$. Due to the solenoidal nature of the magnetic flux density $\vec{B}$ ($\nabla \cdot \vec{B} = 0$), one can define the magnetic vector potential $\vec{A}$ as follows:

$$\vec{B} = \nabla \times \vec{A} \qquad (A2.1)$$

Using (A2.1), the constitutive relations $\vec{B} = \mu_0 \vec{H}$, and Ampere's law $\nabla \times \vec{H} = \vec{J}_0$, the forward problem in $\Omega_N$ is defined as follows:

$$\frac{1}{\mu_0} \nabla \times (\nabla \times \vec{A}) = \vec{J}_0 \qquad (A2.2)$$

The current density $\vec{J}_0$ is only non-zero for the excitation coil.

Using (A2.1), the constitutive relations $\vec{B} = \mu_0 \vec{H}$, and Ampere-Maxwell's equation $\nabla \times \vec{H} = (\sigma + j\omega\varepsilon)\vec{E}$, the forward problem in $\Omega_C$ is defined as follows:

$$\frac{1}{\mu_0} \nabla \times (\nabla \times \vec{A}) = (\sigma + j\omega\varepsilon)\vec{E} \qquad (A2.3)$$

Substituting (A2.1) in the first Maxwell's equation, $\nabla \times \vec{E} = -j\omega\vec{B}$, gives:

$$\nabla \times \vec{E} = -j\omega(\nabla \times \vec{A}) \quad \rightarrow \quad \nabla \times (\vec{E} + j\omega\vec{A}) = 0 \qquad (A2.4)$$

Using the vector identity $\nabla \times (\nabla V) = 0$, one can write:

$$\vec{E} = -j\omega\vec{A} - \nabla V \tag{A2.5}$$

Substituting (A2.5) in (A2.3) results in:

$$\frac{1}{\mu_0}\nabla \times (\nabla \times \vec{A}) + (\sigma + j\omega\varepsilon)(j\omega\vec{A} + \nabla V) = 0 \tag{A2.6}$$

Taking the divergence of (A2.3) and knowing that the divergence of the curl function is zero, we obtain the following equation in $\Omega_C$ (the continuity equation):

$$\nabla \cdot \left((\sigma + j\omega\varepsilon)\vec{E}\right) = 0 \tag{A2.7}$$

Using (A2.5), one can rewrite (A2.7):

$$\nabla \cdot \left((\sigma + j\omega\varepsilon)(j\omega\vec{A} + \nabla V)\right) = 0 \tag{A2.8}$$

Hence, we have two equations (A2.6) and (A2.8) for two unknowns $\vec{A}$ and $V$ in $\Omega_C$. It is noteworthy that (A2.8) can be obtained by taking the divergence of (A2.6). It means that these equations are not independent [51].

In the literature, various boundary conditions have been considered for $\vec{A} - \vec{A}, V$ formulation. The easiest one is the far-field boundary condition which assumes that the normal component of the magnetic flux density on a far external boundary is zero, $\vec{B} \cdot \vec{n} = 0$. Here, $\vec{n}$ stands for the outward normal unit vector on the boundary. Furthermore, the normal component of the current density on the boundary of conducting region $\Omega_C$ should be zero, $\vec{n} \cdot (j\omega\vec{A} + \nabla V) = 0$ [1].

If in (A2.6) and (A2.8) the permittivity is considered to be zero, the displacement current will be ignored in the formulation.

It should be noted that so far, we only have determined the curl of $\vec{A}$ ($\vec{B} = \nabla \times \vec{A}$). Based on the Helmholtz theorem, a vector field is determined up to an additive constant if both its divergence and its curl are specified anywhere [141]. The divergence of $\vec{A}$ is as yet undetermined. The value of $\nabla \cdot \vec{A}$ can be chosen arbitrarily, without affecting on the physical problem. However, if it is not fixed, it may cause some numerical errors. Different choices for $\nabla \cdot \vec{A}$ are referred to as choices of gauge. For all possible gauge choices, field quantities are unique, but a smart choice of gauge can lead to ease of computation. The best-known gauge used in electromagnetics is extremely simple, $\nabla \cdot \vec{A} = 0$ [142]. It is known as Coulomb's gauge. Using the following vector identity,

$$\nabla \times (\nabla \times \vec{A}) = \nabla(\nabla \cdot \vec{A}) - \nabla^2\vec{A} \tag{A2.9}$$

and applying Coulomb's gauge, one can rewrite the $\vec{A} - \vec{A}, V$ formulation as follows:

$$\begin{aligned}
-\frac{1}{\mu_0}\nabla^2\vec{A} &= \vec{J}_0 && \text{in } \Omega_N \\
-\frac{1}{\mu_0}\nabla^2\vec{A} + (\sigma + j\omega\varepsilon)(j\omega\vec{A} + \nabla V) &= 0 && \text{in } \Omega_C \\
\nabla \cdot \left((\sigma + j\omega\varepsilon)(j\omega\vec{A} + \nabla V)\right) &= 0
\end{aligned} \tag{A2.10}$$

## A2.2 $\vec{A} - \vec{A}$ formulation

If the conductivity in $\Omega_C$ is constant; i.e., $\nabla\sigma = 0$, it can be shown that the term $\nabla V$ can be dropped from (A2.10) [143]. It leads to $\vec{A} - \vec{A}$ formulation [13], [15], [16], [29], [45]–[47], [57]:

$$-\frac{1}{\mu_0}\nabla^2\vec{A} = \vec{J}_0 \qquad \text{in } \Omega_N$$

$$-\frac{1}{\mu_0}\nabla^2\vec{A} + (\sigma + j\omega\varepsilon)j\omega\vec{A} = 0 \qquad \text{in } \Omega_C$$

(A2.11)

In [51] has been stated that $\vec{A} - \vec{A}$ formulation has less numerical stability compared to $\vec{A} - \vec{A}, V$ formulation.

## A2.3 $\vec{A}_r - \vec{A}_r, V$ formulation

The main drawback of $\vec{A} - \vec{A}, V$ formulation is that coils should be modeled. For example, if FEM is employed to solve the forward problem, then discretizing the coil structures into the FEM mesh model is necessary. This discretization process increases the difficulty and complexity of the mesh model; moreover, the system becomes very inflexible due to the restriction from the size and the location of the coil sensors [69]. Hence, another formulation has been employed called $\vec{A}_r - \vec{A}_r, V$ formulation [52]–[54], [69], [144]. In this formulation, $\vec{A}_r$ represents the reduced magnetic vector potential in the entire domain and V represents the electrical scalar potential only in $\Omega_C$. In this formulation, the excitation field generated by coils is computed by the Biot-Savart law, and modeling of coils is no longer required [69], [144].

To define $\vec{A}_r$ in non-conduction region $\Omega_N$, the magnetic flux density $\vec{B}$ is separated into $\vec{B}_s$ and $\vec{B}_r$. The former is caused by the source (excitation coil) and the latter is caused by the presence of the material. One can write:

$$\vec{B} = \vec{B}_s + \vec{B}_r = \mu_0\vec{H}_s + \nabla \times \vec{A}_r \tag{A2.12}$$

where $\vec{H}_s$ is the magnetic field intensity caused by the excitation coil and is obtained from the Biot-Savart integral:

$$\vec{H}_s = \frac{1}{4\pi}\int \frac{\vec{J}_0 \times \vec{r}}{|\vec{r}|^3}dV \tag{A2.13}$$

and also satisfy

$$\nabla \times \vec{H}_s = \vec{J}_0 \tag{A2.14}$$

where $\vec{J}_0$ is the current density in the excitation coil and $\vec{r}$ is the displacement vector. Substituting (A2.12) and (A2.14) in Ampere's law in $\Omega_N$ yields

$$\nabla \times \left(\frac{1}{\mu}\nabla \times \vec{A}_r\right) = \nabla \times \vec{H}_s - \nabla \times \frac{\mu_0}{\mu}\vec{H}_s \tag{A2.15}$$

Similar to the non-conducting domain, the magnetic flux density is split into two parts in $\Omega_C$:

$$\vec{B} = \vec{B}_s + \vec{B}_r = \nabla \times \vec{A}_s + \nabla \times \vec{A}_r \tag{A2.16}$$

Here $\nabla \times \vec{A}_s = \mu_0\vec{H}_s$. This assumption allows to define the electric field intensity in $\Omega_C$ as follows:

$$\vec{E} = -j\omega(\vec{A}_s + \vec{A}_r) - \nabla V \tag{A2.17}$$

Similar to $\vec{A} - \vec{A}, V$ formulation, in $\Omega_C$ we can write:

$$\nabla \times \left(\frac{1}{\mu}\nabla \times \vec{A}_r\right) + (\sigma + j\omega\varepsilon)(j\omega\vec{A}_r + \nabla V)$$
$$= -j\omega(\sigma + j\omega\varepsilon)\vec{A}_s - \nabla \times \left(\frac{\mu_0}{\mu}\vec{H}_s\right) \tag{A2.18}$$

Furthermore, based on the continuity equation, we can write in $\Omega_C$:

$$\nabla \cdot \left((\sigma + j\omega\varepsilon)(j\omega\vec{A}_r + \nabla V)\right) = -\nabla \cdot \left(j\omega(\sigma + j\omega\varepsilon)\vec{A}_s\right) \tag{A2.19}$$

If $\mu = \mu_0$ in both $\Omega_N$ and $\Omega_C$, $\vec{A}_r - \vec{A}_r, V$ formulation can be summarized as follows:

$$\frac{1}{\mu_0} \nabla \times (\nabla \times \vec{A}_r) = 0 \qquad \text{in } \Omega_N$$

$$\frac{1}{\mu_0} \nabla \times (\nabla \times \vec{A}_r) + (\sigma + j\omega\varepsilon)(j\omega\vec{A}_r + \nabla V)$$
$$= -j\omega(\sigma + j\omega\varepsilon)\vec{A}_s - \nabla \times \vec{H}_s \qquad \text{in } \Omega_C \qquad (A2.20)$$

$$\nabla \cdot \left((\sigma + j\omega\varepsilon)(j\omega\vec{A}_r + \nabla V)\right) = -\nabla \cdot \left(j\omega(\sigma + j\omega\varepsilon)\vec{A}_s\right)$$

For coupling formulations in $\Omega_N$ and $\Omega_C$, the continuity of the tangential component of the magnetic vector potential on the interface of domains should be satisfied. This enforces the continuity of the normal component of magnetic flux density. The continuity of the tangential component of the magnetic field intensity should be satisfied on the interface, as well [51]. On the far boundary, the Dirichlet boundary condition $\vec{A}_r \times \vec{n} = -\vec{A}_s \times \vec{n}$ is prescribed [19].

In $\vec{A}_r - \vec{A}_r, V$ formulation, it is very important to evaluate the Biot-Savart integral accurately for ensuring the convergence of the solution [145].

## A2.4 $\vec{A}_0 - \vec{A}_0, V$ formulation

In addition to presented formulations, another formulation has widely used in the biomedical MIT [11], [20], [55], [58], [59], [63]. This formulation is named $\vec{A}_0 - \vec{A}_0, V$ formulation and is based on the weakly coupled field approximation. This approximation assumes that the attenuation in the excitation field produced by the conductive objects within the imaging region is negligible. In this case, it can be assumed that the primary magnetic field is independent of the target object, and therefore it can be calculated in advance. It means that the magnetic vector potential is known and shown by $\vec{A}_0$. This assumption is true for biomedical MIT, in which the conductivity of the target object is low and causes the decoupling of the field equations.

Similar to (A2.5), one can write:

$$\vec{E} = -j\omega\vec{A}_0 - \nabla V \qquad (A2.21)$$

where

$$\vec{A}_0 = \frac{\mu_0}{4\pi} \int \frac{\vec{J}_0}{|\vec{r}|^3} dV \qquad (A2.22)$$

Using the Coulomb's gauge and the continuity equation in $\Omega_C$ give [146]:

$$\nabla \cdot \left((\sigma + j\omega\varepsilon)\nabla V\right) = -j\omega\vec{A}_0 \cdot \nabla(\sigma + j\omega\varepsilon) \qquad (A2.23)$$

In addition, on the boundary of the conducting domain, we have (the normal component of the current density should be zero):

$$\frac{\partial V}{\partial \vec{n}} = -j\omega\vec{A}_0 \cdot \vec{n} \qquad (A2.24)$$

By solving (A2.23) with boundary condition (A2.24), the electric scalar potential V is found in $\Omega_C$. Then, using (A2.21) and $\vec{J}_e = (\sigma + j\omega\varepsilon)\vec{E}$, one can find the induced eddy current density in in $\Omega_C$. By substituting $\vec{J}_e$ for $\vec{J}_0$ in (A2.22), we can compute the magnetic vector potential at the sensor coils [146].

# Appendix 3: Analytical solution of the test problem in Section 3.3.1

In this appendix, we provide the analytical solution of the test problem presented in Section 3.3.1. The test assembly shown in Figure A3.1(a) corresponds to a 1D problem of two thin slot-embedded coils. The coils are surrounded by permeable iron with $\mu = \infty$. This problem models a one-channel MIT system in which the excitation coil is comprised of conductors 1 and 4 and the sensing coil contains conductors 2 and 3. The physical properties of the assembly are identical to those in [87]. Due to the symmetry of the problem, half of the region of interest is employed to calculate the analytical solution (Figure A3.1(b)). The currents in the four conductors are given by $i_1(t) = 4\cos(\omega t)$ A, $i_2(t) = 0$ A, $i_3(t) = 0$ A, and $i_4(t) = -4\cos(\omega t)$ A. The conductivity and permeability of all conductors are taken $\sigma = 1$ S/m and $\mu = 1$ H/m.

The governing differential equation in terms of the magnetic field intensity $\vec{H}$ is given by

$$\nabla \times (\nabla \times \vec{H}) + j\omega\mu\sigma\vec{H} = 0 \tag{A3.1}$$

By assuming that $\vec{H}$ has only one component in the $x$ direction, $H_x$, (A3.1) is simplified as

$$-\frac{\partial^2 H_x}{\partial y^2} + j\omega\mu\sigma H_x = 0 \tag{A3.2}$$

With regard to Figure A3.1(b), the boundary conditions for (A3.2) is given as follows:

$$\begin{aligned} H_x|_{y=0} &= 0 \\ H_x|_{y=4} &= -4 \\ H_x|_{y=5} &= -4 \\ H_x|_{y=6} &= -4 \\ H_x|_{y=7} &= -4 \end{aligned} \tag{A3.3}$$

which are obtained by evaluation $\oint \vec{H} \cdot d\vec{l} = I$. By replacing the physical properties of the problem in (A3.2), we have:

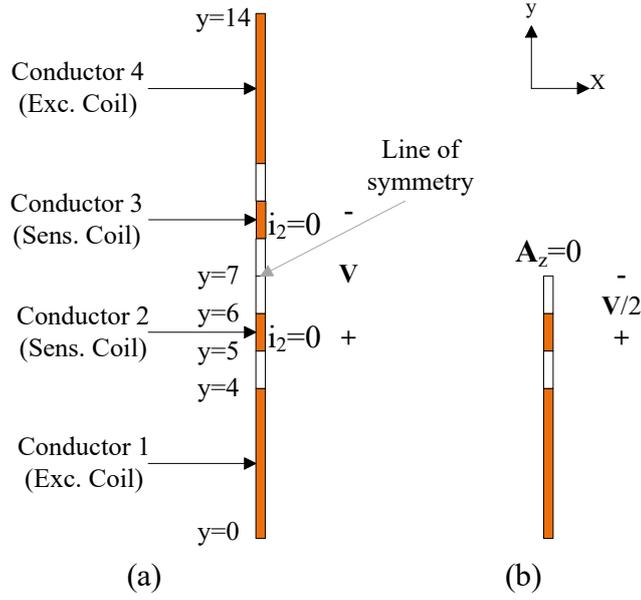

Figure A3.1. Test assembly: (a) 1D symmetric geometry contains four conductors (two coils) surrounded by infinitely permeable iron, (b) Half of the region of interest. All dimensions are in mm.

$$\begin{cases} -\dfrac{\partial^2 H_x}{\partial y^2} + j\omega H_x = 0 & 0 \leq y \leq 0.004 \\ -\dfrac{\partial^2 H_x}{\partial y^2} = 0 & 0.004 \leq y \leq 0.005 \\ -\dfrac{\partial^2 H_x}{\partial y^2} + j\omega H_x = 0 & 0.005 \leq y \leq 0.006 \\ -\dfrac{\partial^2 H_x}{\partial y^2} = 0 & 0.006 \leq y \leq 0.007 \end{cases} \quad (A3.4)$$

By solving (A3.4) and applying boundary conditions (A3.3), $H_x$ can be computed as follows:

$$H_x = \begin{cases} -4\dfrac{e^{y\sqrt{j\omega}} - e^{-y\sqrt{j\omega}}}{e^{4\sqrt{j\omega}} - e^{-4\sqrt{j\omega}}} & 0 \leq y \leq 0.004 \\ -4 & 0.004 \leq y \leq 0.005 \\ K_1 e^{y\sqrt{j\omega}} + K_2 e^{-y\sqrt{j\omega}} & 0.005 \leq y \leq 0.006 \\ -4 & 0.006 \leq y \leq 0.007 \end{cases} \quad (A3.5)$$

where $K_1 = (0.027 - j0.031)$ and $K_2 = (78.970 + j56.720)$.

For the test problem, we can obtain the magnetic vector potential as follows:

$$H_x = \dfrac{\partial A_z}{\partial y} \to A_z = \int H_x dy + C \quad (A3.6)$$

where C is a constant of integration. Using (A3.6), $A_z$ is obtained as follows:

$$A_z(y) = \begin{cases} C_1 \left(e^{y\sqrt{j\omega}} + e^{-y\sqrt{j\omega}} - 0.002\right) + C_2 & 0 \leq y \leq 0.004 \\ -4y + 0.016 + C_3 & 0.004 \leq y \leq 0.005 \\ C_4 \left(e^{y\sqrt{j\omega}} - e^{0.005\sqrt{j\omega}}\right) - C_5 \left(e^{-y\sqrt{j\omega}} - e^{-0.005\sqrt{j\omega}}\right) + C_6 & 0.005 \leq y \leq 0.006 \\ -4y + 0.028 & 0.006 \leq y \leq 0.007 \end{cases} \quad (A3.7)$$

The boundary conditions for $A_z$ can be defined as follows:

$$\left.\dfrac{\partial A_z}{\partial y}\right|_{y=0} = 0 \quad (A3.8)$$

$$A_z|_{y=4^-} = A_z|_{y=4^+} \; ; \quad \left.\frac{\partial A_z}{\partial y}\right|_{y=4^-} = \left.\frac{\partial A_z}{\partial y}\right|_{y=4^+}$$

$$A_z|_{y=5^-} = A_z|_{y=5^+} \; ; \quad \left.\frac{\partial A_z}{\partial y}\right|_{y=5^-} = \left.\frac{\partial A_z}{\partial y}\right|_{y=5^+}$$

$$A_z|_{y=6^-} = A_z|_{y=6^+} \; ; \quad \left.\frac{\partial A_z}{\partial y}\right|_{y=6^-} = \left.\frac{\partial A_z}{\partial y}\right|_{y=6^+}$$

$$A_z|_{y=7^-} = A_z|_{y=7^+} \; ; \quad \left.\frac{\partial A_z}{\partial y}\right|_{y=7^-} = \left.\frac{\partial A_z}{\partial y}\right|_{y=7^+}$$

Using (A3.8), the constants in (A3.7) are obtained as follows:

$$C_6 = 0.004 + 8\left(e^{0.001\sqrt{j\omega}} - 1\right)\Big/\left(\sqrt{j\omega}\left(e^{0.001\sqrt{j\omega}} + 1\right)\right),$$

$$C_5 = -4e^{0.006\sqrt{j\omega}}\Big/\left(\sqrt{j\omega}\left(e^{0.001\sqrt{j\omega}} + 1\right)\right),$$

$$C_4 = -4e^{-0.005\sqrt{j\omega}}\Big/\left(\sqrt{j\omega}\left(e^{0.001\sqrt{j\omega}} + 1\right)\right),$$

$$C_3 = C_6 + 0.004,$$

$$C_2 = C_3 - C_1\left(e^{0.004\sqrt{j\omega}} + e^{-0.004\sqrt{j\omega}} - 0.002\right),$$

$$C_1 = -4\Big/\left(\sqrt{j\omega}\left(e^{0.004\sqrt{j\omega}} - e^{-0.004\sqrt{j\omega}}\right)\right).$$

# Appendix 4: Biomedical MIT systems

This appendix provides an overview of the history of biomedical MIT. The progress in industrial MIT systems was initially faster in which the conductivity materials to be imaged are high (>$10^6$ S/m) and produce large induced signals. In addition, inductive sensors were very popular in the metallurgical industry. In contrast, the development of biomedical MIT systems has been more problematic because the conductivities to be imaged (<10 S/m) are many orders of magnitude lower than those of metals and provide considerably weaker signals. In the last two decades, the developing of biomedical MIT systems has attracted the attention of researchers. However, the systems have not yet been used in clinical applications. In the following, we review the details of biomedical MIT systems presented in the literature.

The first biomedical MIT system was reported in 1993 by Al-Zeibak and Sunders [65]. The operating frequency of the system was 2 MHz. The system was equipped with two coaxial coils, and the object to be imaged was mounted on a turntable so that its orientation between coils in the horizontal plane could be varied. The results showed that the system could clearly distinguish the conductivities corresponding to fat-free and fat tissues. The system also was able to image the geometry of simple objects. However, results were questioned by Griffiths et al. [34] in 1999 using an analytical solution.

In 1999, Griffiths et al. [34] developed a single-channel MIT system for biomedical applications. The operating frequency of the system was 10 MHz. A phase-sensitive detector was employed to measure the phase angle of induced signals. By displacement and rotating the saline solutions with conductivities ranging from 0.001- 6 S/m in the space between coils, tomographic images was obtained. The experimental results were close to that predicted theoretically. In this study, using an analytical solution, it was predicted that a 1 S/m change in the medium conductivity causes less than 0.1% change in amplitude and 0.7° change in phase angle of the induced voltage. Then, they concluded the suitability of the measuring phase angle for the biomedical MIT systems.

In 2000, Korjenevsky et al. [66] developed for the first time a multi-channel biomedical MIT system. The system was equipped with 16 exciter and sensor coils. In the measurement circuitry part of the system, the phase angle of induced voltages was directly measured. The excitation frequency was 20 MHz which was mixed down to 20 kHz to facilitate the measuring of the phase angle of the receiving signal. Using this system, images of a saline bottle with various conductivities were reconstructed. Figure A4.1 shows an image of this system.

In 2001, for the first time, Scharfetter et al. [147] developed a multi-frequency biomedical MIT system which was equipped with planar gradiometer coils. The gradiometer consists of two coils placed next to each other in such a way that the gradient of the field is measured. The induced voltage in an ideal gradiometer is zero when there is no target object in the imaging region since the same magnetic flux passes from both coils. The operating frequency ranges were 20-370 kHz. The system consisted of an excitation coil, a reference coil, and a small planer gradiometer. By rotating the target object in the excitation field, the data required for conductivity reconstruction was gathered. This system was therefore very slow and very sensitive to drifts and movement artifacts. In 2008, to eliminate these disadvantages, this research group developed a new system in which sixteen transceiver units were arranged in two groups in the form of an upper and a lower ring with 8 units per ring [148]. Each transceiver unit consisted of a shielded solenoid excitation coil and a receiving planar gradiometer. In this way, they increased the imaging speed significantly. The operating frequency range for the new system was from 50 kHz to 1.5 MHz.

In 2003, Watson et al. [38] developed an MIT system named "Glamorgan system mark 1" which in terms of geometry was similar to the system presented in [66] but working at a lower frequency of 10 MHz. In this study, two types of data extraction techniques were compared: (1) direct-phase measurement and (2) measurement of the in-phase and quadrature components of the signal with a vector voltmeter. The system had 16 coils wrapped on a perspex coil former of 5 cm diameter. Each coil former had a 2-turn transmitter coil and a 2-turn receiver coil. A cylindrical aluminum screen with a diameter of 35 cm and a height of 25 cm encircled the coil array. The transmitter and receiver circuits were housed in 16 separate metal enclosures attached to the outside of the screen. The received signal was downconverted to 10 kHz, and a digital lock-in amplifier extracted the in-phase and quadrature components of the received signal. The phase noise of 27 m° was reported for this system. Both techniques of data extraction produced similar results for phase noise, with an apparent small advantage for the direct-phase technique. In 2008, this system was improved for reconstruction conductivity images from samples with low conductivities (<10 S/m) like multiphase flow in pipelines and biological tissue. The new system was named "Cardiff Mk2" system, and its phase noise was 17 m°.

In 2008, Vauhkonen et al. [68] in Philips research center developed a 16 channel MIT system for imaging of conductivities less than 5 S/m. The system was named "Philips research system", and its operating frequency was 10 MHz. The system was capable of a parallel readout of 16 receiver channels. The parallel measurements were carried out using high-quality audio sampling devices. Compared to the sequentially working systems, the Philips research system improved the measurement speed by a factor of 16 due to enabling the parallel readout. The system was able to reconstruct both difference and static images. In [149], 8 m° phase noise was reported for this system.

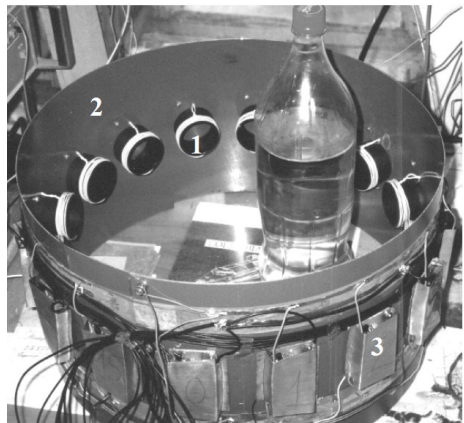

Figure A4.1 An image of the developed biomedical MIT system in [66]. The system consist the following parts: 1- exciter and sensor coils, 2- electromagnetic shield, and 3- transceiver modules. A saline bottle was placed in the imaging space.

In 2009, Xu et al. [10] proposed a new MIT system for human brain imaging. This system was different from the previous MIT systems in two ways: (1) coils were arrayed on a hemisphere surrounding the imaging region, which was similar to a helmet. (2) A single cancellation sensor was employed to detect the phase drift. As shown by Figure A4.2, the first feature causes that the imaging region encircles the human head. The system was composed of 15 measurement sensors and a cancellation sensor. Axial gradiometers were used to cancel the primary field. Printed circuit board (PCB) coils with 10 and 30 turns were used for excitation and sensor coils, respectively. The operating frequency of this system was 120 kHz which increased to 1 MHz in the subsequent study in 2010 [150]. Griffith et al. [151] also proposed a similar hemispherical MIT helmet coil array for imaging cerebral hemorrhage in 2010.

In 2012, Caeiros et al. [152] replaced the typical coils' setup with one named the twin coils setup technique which could increase the number of independent measurements. An excitation coil was placed at the center of a circular setup, and four pairs of sensing coils were placed aroud it by the same angular spacing. Each pair of sensing coils was positioned on diametrically opposing sides of the circle surrounding the imaging space. The pair had the same distance from the excitation coil, and its voltage was measured differentially. Thus, it worked as a gradiometer. The target object was placed over a plate which can be rotated and moved vertically. In this way, a larger number of independent projections could be attained, and consequently, the ill-posedness of the inverse problem reduced. The prototype structure was made of a PVC material to be electrically, and magnetically inerted and any metallic object was removed from within a 1 m radius of the excitation coil. The operating frequency was 1.5 MHz. The number of measurements obtained by this system was 936.

In 2012, Wei and Soleimani [69] developed an MIT system with an operating frequency of 10 MHz for prospective biomedical applications. All signal driving, switching, and data acquisition tasks were accomplished by National Instrument products which eased the system design whilst providing satisfactory performance. As shown by Figure A4.3, the system utilized 8 excitation coils and 8 sensing coils. The coils were attached in adjacent positions and evenly spread around a 23 cm diameter perspex cylinder. The coils were arranged in such a way that each excitation coil was positioned between two sensing coils and vice versa. The reported phase noise was 4 m°. In this study, images from the saline tank were reconstructed.

In 2012, Trakic et al. [40] investigated the feasibility of rotating a single transceiver coil to image the low conductivity samples like biological tissues. The system was capable of emulating an array of 200 transceiver coils by time-division multiplexing. The operating frequency was 5 MHz. The single transceiver coil was rotated around imaging space by a stepper motor. Authors claimed that in this way, the number of independent measurements, and in consequence, the quality of reconstructed images would increase.

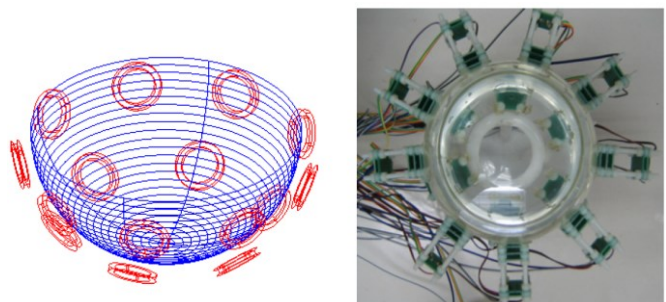

Figure A4.2 (Left) Schematic diagram of the sensor arrangement used in [10]. (Right) Top view of the sensor

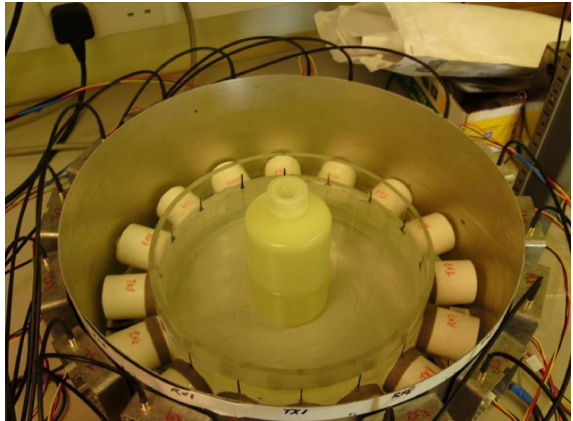

Figure A4.3 Coils array arrangement for developed system in [69].

In 2015, Luo and Jiang [153] proposed using the Helmholtz coil in MIT for imaging cerebral hemorrhage. The Helmholtz coil is composed of two coaxial coils with equal radius. Their distance is equal to the radius of coils, and the amplitude and direction of the excitation current are the same through the two coils. Unlike spiral coils which were widely used in MIT, the Helmholtz coil could generate a uniform excitation field, which caused to have a uniform map for the sensitivity of induced voltage to the conductivity. This uniform excitation field provided the possibility of using the linear back-projection algorithm. The main drawback of this system was the small imaging space. The experimental results showed that this system could reconstruct the conductivity of a phantom which mimicked a cerebral hemorrhage, but the target object was enlarged compared with the actual one.

In 2016, Marmugi and Renzoni [23] proposed a novel instrument based on optical atomic magnetometers (OAM) to measure the secondary field in MIT for imaging the heart's conductivity. OAM sensitivity can be up to $10^7$ times larger than that of a standard pick-up coil of the same size below 50 MHz. They can reach a sensitivity comparable to the performances of super-conductive quantum interference devices (SQUIDs), whereas they operate at room temperature and are unshielded in contrast to SQUIDs. They also have an extreme potential for miniaturization. It is a promising study that difficulties caused by the small secondary magnetic field in the biomedical MIT can be overcome by using OAM.

In 2021, Tan et al. [154] developed a portable, flexible, and extensible MIT system with high-phase detection precision MIT system for low-conductivity medium imaging. The proposed system was designed to be modular. The sensor modules were used as both the excitation and the sensing, and the detected signals were directly demodulated in the modules. The system was not required to other instruments for the signal generation and demodulation. The reported phase noise for the system was 6.8 m°. Here, we named it Tianjin University system.

In Appendix 1, it was shown that in biomedical applications of MIT, it is important to measure the phase change in the sensing coils. Figure A4.4 shows the trend of biomedical MIT systems' progresses from the phase measurement precision point of view during several years. As can be seen, the phase noise of practical MIT systems is decreasing over time. It can be inferred that as research on biomedical MIT continues, clinical applications will be available

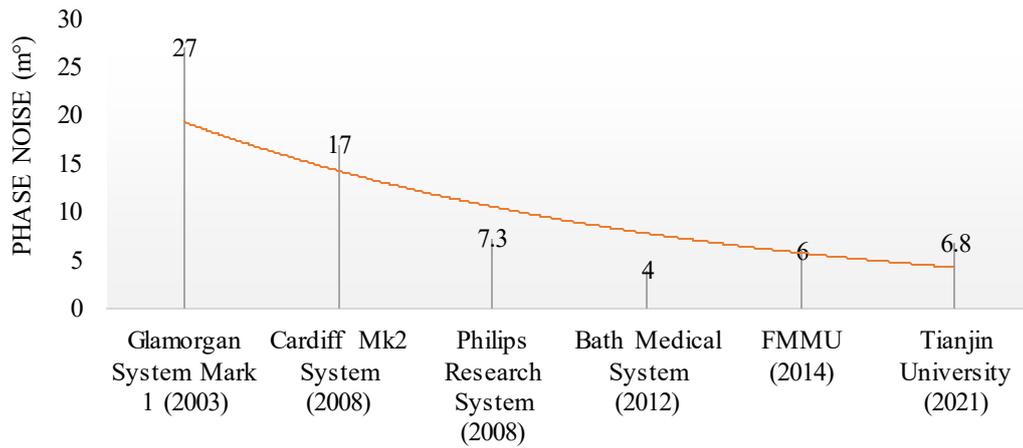

Figure A4.4 The phase noise versus the year of development for some biomedical MIT systems. The dashed-line shows the trend of MIT systems progress. The information about FMMU system was presented in [13], [155].

in the near future. Furthermore, using OAMs can possibly accelerate this process.

# Appendix 5: Interface and boundary conditions in 2D MIT

Boundary-value problems can only be solved if accompanied by interface conditions where the physical properties of the media changes and by boundary conditions applicable at the boundary of the problem. In eddy current problems, usually the continuity of normal component of $\vec{B}$ and the continuity of tangential component of $\vec{H}$ are considered [94]. In this thesis, we have formulated the 2D MIT forward problem in terms of the MVP. Finding the potential still implies solving a boundary-value problem which must be accompanied by interface and boundary conditions. In this appendix, we obtain the appropriate interface and boundary conditions on MVP which are dictated by the continuity of normal component of $\vec{B}$ and the continuity of tangential component of $\vec{H}$. It is noteworthy that the continuity of normal component of $\vec{B}$ is equivalent to the continuity of tangential component of $\vec{E}$ [141].

## A5.1 Continuity of normal component of $\vec{B}$

Figure A5.1 shows an interface between materials 1 and 2. A closed small path parallels the interface within material 1 and returns in material 2. The length and width of the path are $\Delta w$ and $\Delta h$, respectively. The magnetic flux that threads through the closed path is given by:

$$\varphi = \int \vec{B} \cdot \mathrm{d}s = \int (\nabla \times \vec{A}) \cdot \mathrm{d}s = \oint \vec{A} \cdot \mathrm{d}l \qquad (A5.1)$$

If $\Delta h$ approaches zero, the flux enclosed by the closed path therefore vanishes. This can only happen if the tangential component of $\vec{A}$ has the same value on both sides of the interface [142]:

$$\begin{aligned}\nabla h \rightarrow 0 \quad &\Rightarrow \varphi \rightarrow 0 \Rightarrow \oint \vec{A} \cdot \mathrm{d}l = 0 \\ &\overrightarrow{\Delta w} \cdot (\vec{A}_2 - \vec{A}_1) = 0 \\ &\vec{n} \times (\vec{A}_2 - \vec{A}_1) = 0\end{aligned} \qquad (A5.2)$$

where $\vec{n}$ is outward unite vector from material 1. As seen from Figure A5.1, in 2D problem, only tangential component of $\vec{A}$ is available. It means the equality of MVP at adjacent points on the material interface:

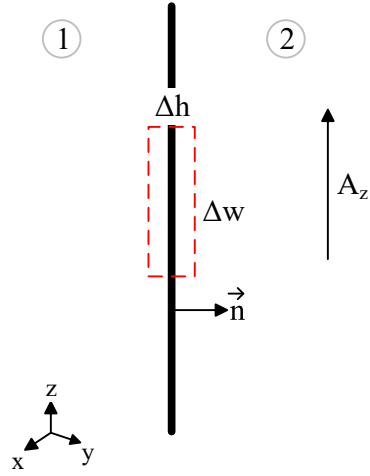

Figure A5.1 Interface conditions at a surface separating two different materials.

$$A_{z_1} = A_{z_2} \tag{A5.3}$$

On the other hand, the continuity of normal component of $\vec{B}$ on the interface says:

$$\vec{B} \cdot \vec{n} = 0 \tag{A5.4}$$

Substituting $\vec{B} = \nabla \times \vec{A}$ in (A5.4) results in:

$$(\nabla \times \vec{A}) \cdot \vec{n} = 0 \implies \nabla \cdot (\vec{A} \times \vec{n}) = 0 \tag{A5.5}$$

which again means that the tangential component of $\vec{A}$ should has the same value on both sides of the interface. In 2D, it would be $A_{z_1} = A_{z_2}$ on the interface of materials. Thus, the equality of MVP at adjacent points on material interfaces guarantees the continuity of normal component of $\vec{B}$ in 2D MIT problem.

## A5.2 Continuity of normal tangential of $\vec{H}$

Figure A5.2 shows an interface between materials 1 and 2. A closed small path parallels the interface within material 1 and returns in material 2. The length and width of the path are $\Delta w$ and $\Delta h$, respectively. The current passes through the closed path is given by:

$$I = \oint \vec{H} \cdot dl = \oint \left(\frac{1}{\mu} \nabla \times \vec{A}\right) \cdot dl \tag{A5.6}$$

For 2D MIT problem, $\vec{A} = A_z(x,y)\vec{a}_z$ and

$$\nabla \times \vec{A} = \begin{vmatrix} \vec{a}_x & \vec{a}_y & \vec{a}_z \\ \frac{\partial}{\partial x} & \frac{\partial}{\partial y} & \frac{\partial}{\partial z} \\ 0 & 0 & A_z \end{vmatrix} = \vec{a}_x \frac{\partial A_z}{\partial y} - \vec{a}_y \frac{\partial A_z}{\partial x} = \begin{vmatrix} \vec{a}_x & \vec{a}_y & \vec{a}_z \\ \frac{\partial A_z}{\partial x} & \frac{\partial A_z}{\partial y} & 0 \\ 0 & 0 & 1 \end{vmatrix} = \nabla A_z \times \vec{a}_z \tag{A5.7}$$

Substituting (A5.7) in (A5.6) gives:

$$I = \oint (\nabla A_z \times \vec{a}_z) \cdot dl \tag{A5.8}$$

Letting $\Delta h$ approaches zeroes, we have:

$$J_{sn} \Delta w = \oint (\nabla A_z \times \vec{a}_z) \cdot dl \tag{A5.9}$$

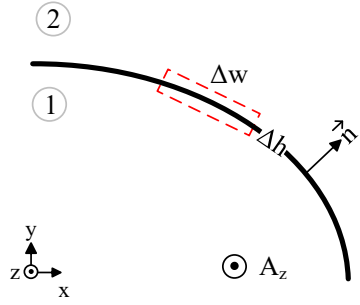

Figure A5.2 Interface conditions at a surface separating two different materials.

where $J_{sn}$ is the surface current density on the interface normal to the closed path. In Figure A5.2, the positive direction of $J_{sn}$ for the chosen path is out of the paper. By expanding (A5.9), we have:

$$J_{sn}\Delta w = \overrightarrow{\Delta w} \cdot \left[\left(\frac{1}{\mu_2}\nabla A_{z_2} \times \vec{a}_z\right) - \left(\frac{1}{\mu_1}\nabla A_{z_1} \times \vec{a}_z\right)\right]$$
$$= \Delta w\, \vec{n} \times \left[\left(\frac{1}{\mu_2}\nabla A_{z_2} \times \vec{a}_z\right) - \left(\frac{1}{\mu_1}\nabla A_{z_1} \times \vec{a}_z\right)\right] \quad (A5.10)$$

For media with finite conductivity $J_{sn}$ is equal to zero [141]:

$$\vec{n} \times \left(\frac{1}{\mu_2}\nabla A_{z_2} \times \vec{a}_z\right) = \vec{n} \times \left(\frac{1}{\mu_1}\nabla A_{z_1} \times \vec{a}_z\right) \quad (A5.11)$$

Using the bac-cab rule [141], we have:

$$\frac{1}{\mu_2}\left[\nabla A_{z_2}\underbrace{(\vec{a}_z \cdot \vec{n})}_{=0} - \vec{a}_z(\nabla A_{z_2}\cdot \vec{n})\right] = \frac{1}{\mu_1}\left[\nabla A_{z_1}\underbrace{(\vec{a}_z \cdot \vec{n})}_{=0} - \vec{a}_z(\nabla A_{z_1}\cdot \vec{n})\right] \quad (A5.12)$$

which is simplified to:

$$\frac{1}{\mu_2}(\nabla A_{z_2} \cdot \vec{n}) = \frac{1}{\mu_1}(\nabla A_{z_1} \cdot \vec{n}) \quad (A5.13)$$

On the other hand, the continuity of tangential component of $\vec{H}$ on the interface of media with finite conductivity says:

$$\vec{H} \times \vec{n} = 0 \quad (A5.14)$$

Substituting $\vec{H} = \mu \nabla \times \vec{A}$ in (A5.14) results in:

$$\left(\frac{1}{\mu}\nabla \times \vec{A}\right) \times \vec{n} = 0 \quad (A5.15)$$

For 2D MIT, (A5.15) would be:

$$\left(\frac{1}{\mu}\nabla A_z \times \vec{a}_z\right) \times \vec{n} = 0 \quad (A5.16)$$

which again gives (A5.13). Thus, the boundary condition of continuous tangential $\vec{H}$ requires:

$$\frac{1}{\mu_1}\frac{\partial A_{z1}}{\partial n} = \frac{1}{\mu_2}\frac{\partial A_{z2}}{\partial n} \quad (A5.17)$$

It is noteworthy that boundary conditions, i.e., conditions at the extremities of the boundary-value problem, are obtained by extending the interface conditions.

# Appendix 6: Analytical solution for a 2D MIT problem

In this appendix, we present an analytical solution for a simple problem to evaluate FE and FE-EFG numerical methods.

Let's consider a simple example of a circular target object with a radius $r_1$ inside a circular imaging region with a radius $r_2$, as shown in Figure A6.1. One may picture the imaging region as the inside of an infinitely long cylinder whose outside extends to infinity. For the sake of simplicity, we consider the target object at the center of the imaging region. The target object and background conductivity are $\sigma_1$ and $\sigma_2$, respectively. The excitation field is generated by a current density distribution of $J = J_0 \sin \varphi$ (in A/m). This can be realized by current strips installed vertically at radius $r_3$ such that $J$ flows perpendicular to the page. This problem was designed by inspiration of the one presented in [89].

One can rewrite Eq. (3.15) in the polar coordinate for each region of Figure A6.1 as follows:

$$\frac{\partial^2 A_k}{\partial r^2} + \frac{1}{r}\frac{\partial A_k}{\partial r} + \frac{1}{r^2}\frac{\partial^2 A_k}{\partial \varphi^2} = j\alpha_k^2 A_k \quad \text{for} \quad k = 1, 2, 3, 4 \tag{A6.1}$$

where $A_k$ is the magnetic vector potential in the z-direction at k-th layer and $\alpha_k = \sqrt{\omega \sigma_k \mu}$.

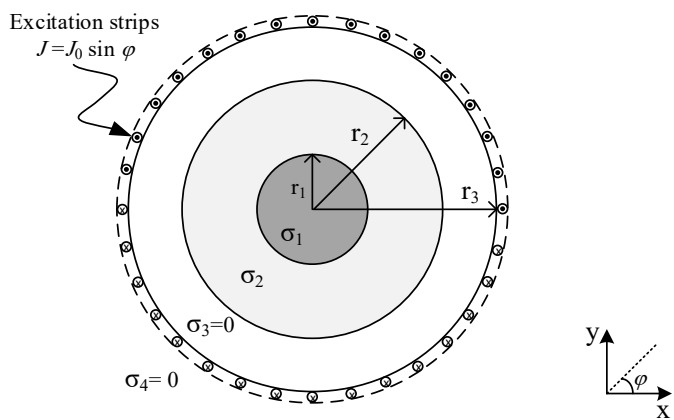

Figure A6.1 2D analytical problem assembly containing a target object ($\sigma_1$) within a background conductivity ($\sigma_2$) in a parallel magnetic field. The problem has four layers $k = 1, 2, 3,$ and $4$.

Using the separation of variables method, one can consider $A_k(r,\varphi) = R_k(r)\Phi_k(\varphi)$ and then the above equation becomes:

$$\frac{r^2 R_k'' + r R_k' - j\alpha_k^2 R_k}{R_k} = \frac{-\Phi_k''}{\Phi_k} \quad \text{for} \quad k = 1, 2, 3, 4 \tag{A6.2}$$

The solution to the right hand side of (A6.2) must be periodic, so the differential equation is

$$\frac{-\Phi_k''}{\Phi_k} = n^2 \tag{A6.3}$$

which has solutions

$$\Phi_k(\varphi) = B_n \cos(n\varphi) + C_n \sin(n\varphi) \tag{A6.4}$$

where $n = 0, 1, 2, \ldots$ and $B_n$ and $C_n$ are unknown coefficients. Since the excitation field is generated by $J = J_0 \sin\varphi$, in (A6.4) only $C_n \sin(n\varphi)$ remains.

Substituting (A6.3) in (A6.2) gives:

$$\frac{r^2 R_k'' + r R_k' - j\alpha_k^2 R_k}{R_k} = n^2 \tag{A6.5}$$

which has solutions

$$R_k(r) = D_n I_n(\alpha_k r) + E_n K_n(\alpha_k r) \tag{A6.6}$$

where $I_n$ and $K_n$ are modified Bessel functions of the first and second kinds, respectively, and $D_n$ and $E_n$ are unknown coefficients. Since $A_k(r,\varphi)$ must be finite, (A6.6) for each layer would be as follows:

$K_n(\alpha_k r)|_{r=0} = \infty \Rightarrow$

$$R_1(r) = D_n I_n(\alpha_1 r) \qquad k = 1$$

$$R_2(r) = D_n I_n(\alpha_2 r) + E_n K_n(\alpha_2 r) \qquad k = 2$$

$I_n(\alpha_3 r)|_{\alpha_3=0} = r^n$ and $K_n(\alpha_3 r)|_{\alpha_3=0} = r^{-n} \Rightarrow$

$$R_3(r) = D_n r^n + E_n r^{-n} \qquad k = 3$$

$r \to \infty \Rightarrow r^n \to \infty \Rightarrow$

$$R_4(r) = E_n r^{-n} \qquad k = 4$$

Combining solutions gives the general solution for each region as follows:

$$\begin{aligned} A_1(r,\varphi) &= \sum_{n=1}^{\infty} a_n I_n(\alpha_1 r) \sin(n\varphi) & 0 \leq r < r_1 \\ A_2(r,\varphi) &= \sum_{n=1}^{\infty} [b_n I_n(\alpha_2 r) \\ &\quad + c_n K_n(\alpha_2 r)] \sin(n\varphi) & r_1 \leq r < r_2 \\ A_3(r,\varphi) &= \sum_{n=1}^{\infty} [d_n r^n + e_n r^{-n}] \sin(n\varphi) & r_2 \leq r < r_3 \\ A_4(r,\varphi) &= \sum_{n=1}^{\infty} f_n r^{-n} \sin(n\varphi) & r > r_3 \end{aligned} \tag{A6.7}$$

where $a_n$, $b_n$, $c_n$, $d_n$, $e_n$, and $f_n$ are constant coefficients to be determined from the boundary conditions Since the excitation filed is generated by $\sin\varphi$, (A6.7) exists only for $n = 1$. The boundary conditions that have to be fulfilled are:

$$A_k|_{r=r_k} = A_{k+1}|_{r=r_k} \qquad \text{for} \quad k = 1,2,3 \tag{A6.8a}$$

$$\frac{1}{\mu}\frac{\partial A_k}{\partial r}\bigg|_{r=r_k} = \frac{1}{\mu}\frac{\partial A_{k+1}}{\partial r}\bigg|_{r=r_k} \qquad \text{for} \quad k = 1,2 \tag{A6.8b}$$

$$\frac{\partial A_3}{\partial r}\bigg|_{r=r_3} - \frac{\partial A_4}{\partial r}\bigg|_{r=r_3} = \mu J_0 \sin\varphi \tag{A6.8c}$$

Applying (A6.8a) to (A6.7) and considering $n = 1$ give:

$$a_1 I_1(\alpha_1 r_1) - b_1 I_1(\alpha_2 r_1) - c_1 K_1(\alpha_2 r_1) = 0$$
$$b_1 I_1(\alpha_2 r_2) + c_1 K_1(\alpha_2 r_2) - d_1 r_2 - e_1 r_2^{-1} = 0$$
$$d_1 r_3 + e_1 r_3^{-1} - f_1 r_3^{-1} = 0$$

Applying (A6.8b) to (A6.7) and considering $n = 1$ give:

$$a_1\left[\alpha_1 I_2(\alpha_1 r_1) + \frac{I_1(\alpha_1 r_1)}{r_1}\right] - b_1\left[\alpha_2 I_2(\alpha_2 r_1) + \frac{I_1(\alpha_2 r_1)}{r_1}\right]$$
$$- c_1\left[\frac{K_1(\alpha_2 r_1)}{r_1} - \alpha_2 K_2(\alpha_2 r_1)\right] = 0$$

$$b_1\left[\alpha_2 I_2(\alpha_2 r_2) + \frac{I_1(\alpha_2 r_2)}{r_2}\right] + c_1\left[\frac{K_1(\alpha_2 r_2)}{r_2} - \alpha_2 K_2(\alpha_2 r_2)\right]$$
$$- d_1 + e_1 r_2^{-2} = 0$$

Applying (A6.8c) to (A6.7) and considering $n = 1$ give:

$$d_1 - e_1 r_3^{-2} + f_1 r_3^{-2} = \mu J_0$$

So, we have six equations and six unknowns which constitute a system of equations as follows:

$$\begin{bmatrix} I_1(\alpha_1 r_1) & -I_1(\alpha_2 r_1) & -K_1(\alpha_2 r_1) & 0 & 0 & 0 \\ 0 & I_1(\alpha_2 r_2) & K_1(\alpha_2 r_2) & -r_2 & -r_2^{-1} & 0 \\ 0 & 0 & 0 & 1 & r_3^{-1} & -r_3^{-1} \\ \left[\alpha_1 I_2(\alpha_1 r_1) + \frac{I_1(\alpha_1 r_1)}{r_1}\right] & -\left[\alpha_2 I_2(\alpha_2 r_1) + \frac{I_1(\alpha_2 r_1)}{r_1}\right] & -\left[\frac{K_1(\alpha_2 r_1)}{r_1} - \alpha_2 K_2(\alpha_2 r_1)\right] & 0 & 0 & 0 \\ 0 & \left[\alpha_2 I_2(\alpha_2 r_2) + \frac{I_1(\alpha_2 r_2)}{r_2}\right] & \left[\frac{K_1(\alpha_2 r_2)}{r_2} - \alpha_2 K_2(\alpha_2 r_2)\right] & -1 & r_2^{-2} & 0 \\ 0 & 0 & 0 & 1 & -r_3^{-2} & r_3^{-2} \end{bmatrix}\begin{bmatrix} a_1 \\ b_1 \\ c_1 \\ d_1 \\ e_1 \\ f_1 \end{bmatrix} = \begin{bmatrix} 0 \\ 0 \\ 0 \\ 0 \\ 0 \\ \mu J_0 \end{bmatrix}$$

Considering $\sigma_1 = 2$ S/m, $\sigma_2 = 1$ S/m, $r_1 = 40$ mm, $r_2 = 100$ mm, and $r_3 = 150$ mm gives:

$$a_1 = 5.23 \times 10^{-08} - j8.39e \times 10^{-08}$$
$$b_1 = 7.77 \times 10^{-08} - j1.17 \times 10^{-07}$$
$$c_1 = 7.75 \times 10^{-11} - j1.16 \times 10^{-10}$$
$$d_1 = 6.28 \times 10^{-07}$$
$$e_1 = -8.6e \times 10^{-11} - j6.24 \times 10^{-10}$$
$$f_1 = 1.41 \times 10^{-08} - j6.24 \times 10^{-10}$$

In the next step, the problem will be solved numerically by FE and FE-EFG methods. The excitation filed in the analytical problem is generated by surface current densities which is unrealistic. For this reason, instead of using excitation strips as shown in Figure A6.1, we use

Table A6.1 Mesh statistics for FE and hybrid FE-EFG methods in each region.

| Mesh statistics<br>Method | Imaging region $(k = 1,2)$ | | Outside imaging region $(k = 3)$ | |
|---|---|---|---|---|
| | # Nodes | # Elements | # Nodes | # Elements |
| FE | 630 | 1328 | 406 | 653 |
| FE-EFG | 766 | N/A | 401 | 644 |

N/A: not applicable.

the MVPs calculated by the analytical solution at radius of 120 mm. Then, imposed these MVPs as the Dirichlet boundary conditions to numerical methods.

Table A6.1 indicates mesh statistics used in FE and FE-EFG methods. The number of nodes and elements are chosen so that the run-time for both methods is almost the same. The run-time for both numerical methods was 1.5 seconds.

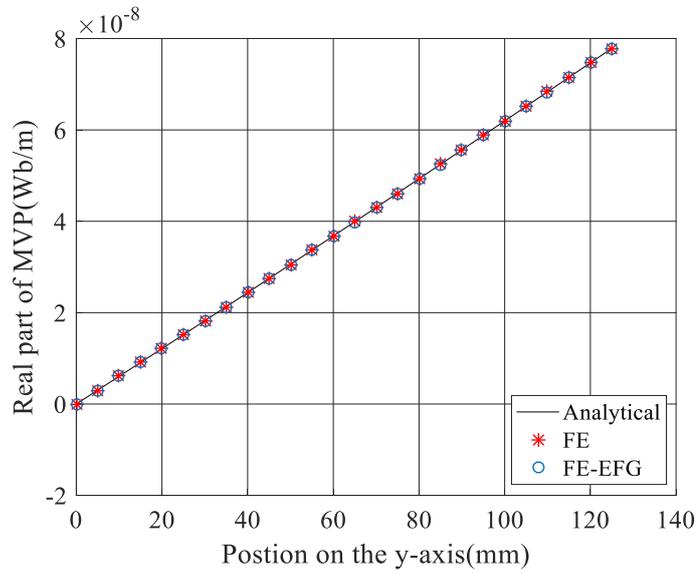

(a)

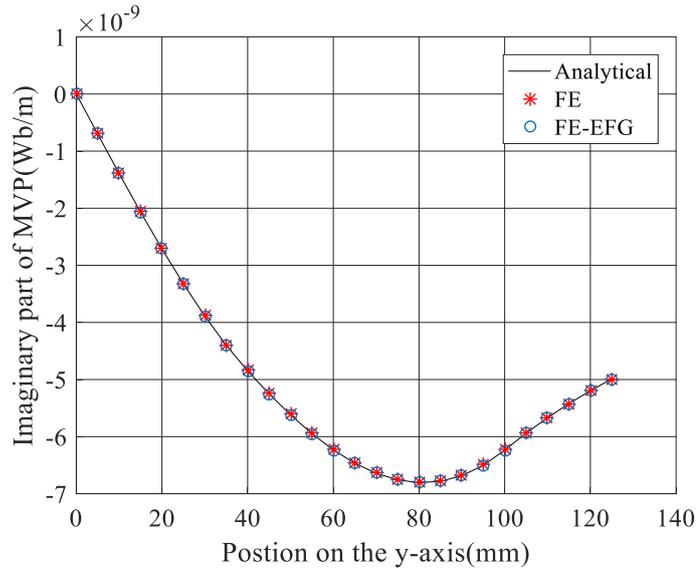

(b)

Figure A6.2 (a) the real and (b) the imaginary part of MVPs calculated by (black line) analytical solution, (red stars) FE method, and (blue circles) FE-EFG method for points located on *y*-axis, from (0, 0) to (0, 125) mm (step by 5 mm).

To evaluate (4.24) using Gaussian quadrature rules, a regular square mesh with 625 nodes and 576 cells was considered. In each integration cell, 4×4 Gauss quadrature points were used to evaluate (4.24). For the EFG method, linear basis functions with the cubic spline weight function and a $d_{max}$ value of 1.7 was considered.

Figure A6.2 compares MVPs obtained from numerical solutions to those of obtained by the analytical solution. We calculated the percentage of the relative error for FE and EFG methods for points located on $y$-axis, from (0, 0) to (0, 125) mm (step by 5 mm). The relative error for the real and imaginary parts of MVPs obtained by FE method was 0.001% and 0.190%, respectively. The relative error for the real and imaginary parts of MVPs obtained by FE-EFG method was 0.073% and 0.073%, respectively. It is noteworthy that in the low-conductivity MIT the imaginary part of MVP is important. Because it is proportional to the real part of the induced voltage.

# Appendix 7: Note on PSF and FWHM in MIT

Point spread function (PSF) is the impulse response of a medical imaging system which measures how the medical imaging system spreads the image of a point impulse [124]. Measuring the size of PSF is the most powerful scheme to evaluate the medical imaging system performance. However, in imaging modalities which require the domain to be discretized, e.g., MIT, the discreteness of the domain forces application of a 'small' circular contrast rather than a point impulse to be able to measure the PSF [126], [156]–[159]. Thus, in the electrical tomography literature, PSF is somehow the pulse response of the system and is my calculated as a response to a small circular conductivity contrast [126]. The 'small' can indicate a diameter of less than 5% of the imaging region diameter [126].

The spatial resolution of an imaging system quantifies how close two contrasts must be before the reconstruction will blur them into one form, and is a function of the point spread function of a small contrast [156]. The most direct metric of the spatial resolution is the full width at half maximum (FWHM) of PSF. This is the (full) width of the PSF at one-half its maximum value [124]. Since it is a linear measurement, the shape of the PSF is important [160]. For circular PSFs, the FWHM is the same whichever angle the measurement is made at. In the case of MIT and EIT images, however, some image reconstruction methods cause the PSFs to become elongated ellipsoid regions towards the edge of the image. In this case the resolution is different depending on whether the slice through the PSF is made in the radial direction (i.e. on a line from the image center through the PSF) or in a direction perpendicular to the radius [160]. Thus, FWHM will not cope so well with any irregular PSF shapes and in the electrical tomography is therefore not sufficient for measuring the spatial resolution across the full range of image reconstruction method.

An alternative to the linear values obtained from measuring the FWHM is to measure the area of the PSF. The resolution may then be expressed as the ratio of the area of the PSF divided by the area of the entire image. Since this is a two-dimensional measurement, it cannot be directly compared with the linear FWHM values, so to convert back to a linear form, the square-root of the area ratio should be employed [160]. A convenient method for calculating the area of the PSF is to count the number pixels with a value above a given threshold [126]. This leads to a performance parameter called RES which is widely used in the electrical tomography studies [126], [157]–[164]. We have used this parameter in this thesis as presented in (5.40).

It is noteworthy that a desired RES should be uniform and small over the entire imaging region. However, its uniformity is much more important than its smallness. Since, MIT is understood to be a low resolution modality and distinguish nearby targets is less important [126].

# Appendix 8: Reconstruction results for a 16-coil MIT system

In this appendix the reconstruction results for a 16-coil MIT system have been presented. The modeling set-up is exactly the same as the one presented in Section 3.3.2.1. Figure A8.1(a) shows the cross-sectional view of the 2D MIT system.

The forward problem is solved by the FE method based on the early and improved forward methods. The overall number of triangular elements and nodes in the FE model was 8200 and 4117, respectively. The inverse problem was solved by the regularized GN algorithm. As shown in Figure A8.1(b), the mesh including 546 uniform triangular pixels was used to solve the inverse problem. As illustrated, pixels have almost the same size. In addition, to avoid an inverse crime, the simulated measured data has been produced by solving the improved forward method on a very fine mesh with about $8.4 \times 10^4$ triangular elements and $4.2 \times 10^4$ nodes.

The value for coefficient $\eta$, the minimum conductivity $\nu$, the maximum conductivity $\xi$, and the maximum number of iterations $K$ were set to 2, $10^{-4}$ S/m, 12 S/m, and 30, respectively, for all following simulations.

In order to evaluate and compare the reconstruction results using the early and improved

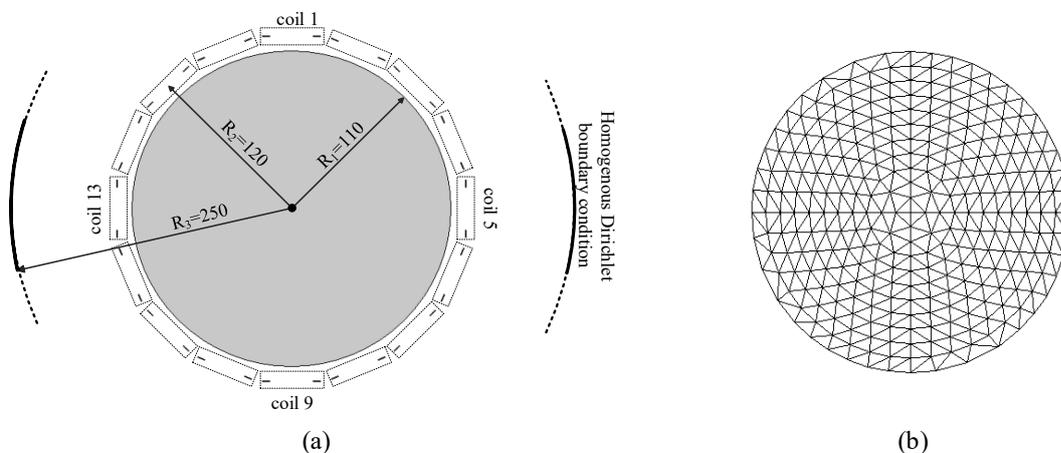

(a)          (b)

Figure A8.1 (a) Coils arrangement and cross section view of the MIT problem model. The homogeneous Dirichlet boundary condition is imposed on a full circle with the radius of 250 mm (the boundary circle was partially drawn for saving space). Dimension is in mm. (b) The mesh including 546 triangular pixels used for solving the inverse problem.

forward methods, three examples have been considered. In Example I, as shown in Figure A8.2(a), a circular inclusion with a radius of 20 mm was centered at (-45, 60) mm. The conductivity of inclusion and background were $\sigma_t = 10$ S/m and $\sigma_b = 2$ S/m, respectively. The theoretical limit given for the minimum detectable inhomogeneity radius for the modeled system is $r_{min} \approx 11$ mm. In Example II, as shown in Figure A8.4(a), two circular inclusions $\sigma_{t1}$ and $\sigma_{t2}$ with a radius of 20 mm were centered at (-45, 60) mm and (45, 60) mm, respectively. The conductivity of inclusions and background was of $\sigma_{t1} = 10$ S/m, $\sigma_{t2} = 5$ S/m, and $\sigma_b = 2$ S/m, respectively. In Example III, as shown in Figure A8.5(a), a ring with an inner radius 30 mm and an outer radius 60 mm was centered at (0, -30) mm and two circular inclusions $\sigma_{t1}$ and $\sigma_{t2}$ with a radius of 20 mm were centered at (-45, 60) mm and (45, 60) mm, respectively. The conductivity of ring and circles were 10 S/m and the conductivity of background was 2 S/m. This phantom was designed inspired from Austria scatterer phantom [165].

## A8.1 Example I: One target object

Figure A8.2(b)-(c) illustrate the reconstructed conductivity images by using the early forward problem for Example I. The homogeneous conductivity value $\sigma_h$ and the coefficient $\tau$ were 0.5 S/m and 3.5, respectively. The voltages induced by the secondary and total magnetic fields based on the early forward method are indicated by $\Delta V^E$ and $V^E$, respectively. As can be seen from Figure A8.2(b), using the voltages induced by the secondary field can compensate for the impact of ignoring skin and proximity effects in the early forward method. In Figure A8.2(c), $V^E$ has been used for reconstruction. As seen, when the voltages induced by the total magnetic field are computed by the early forward method, the target object cannot be resolved. It means that ignoring skin and proximity effects in coils in the forward problem implicates considerable errors in the reconstructed image.

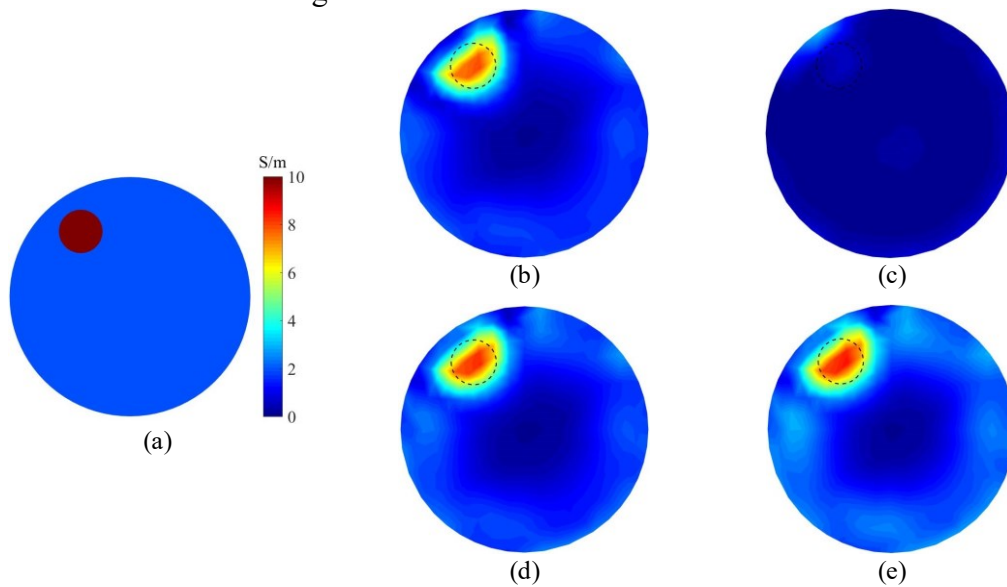

Figure A8.2 Example I: Imaging region contains one target object with radius of 20 mm. (a) True conductivity distribution. Reconstructed conductivity images using (b) the early forward method and the secondary field, (c) the early forward method and the total field, (d) the improved forward method and the secondary field, and (e) the improved forward method and the total field. The homogeneous conductivity value $\sigma_h$ the coefficient $\tau$ were 0.5 S/m and 3.5, respectively. The target object and background conductivity were $\sigma_t = 10$ S/m and $\sigma_b = 2$ S/m, respectively.

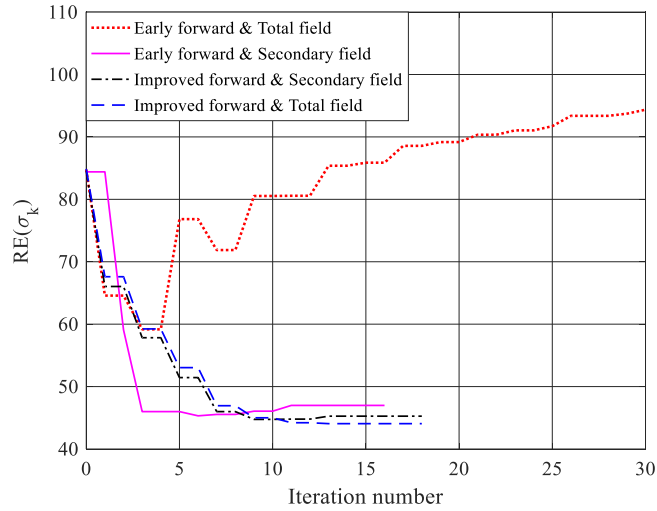

Figure A8.3 Example I: Relative error versus iteration number for each case.

Table A8.1 Example I: Performance parameters (PPs) computed for different cases of forward method, Jacobian matrix calculation technique, and magnetic field used for computation of induced voltage. The parameter $K$ indicates the iteration number for each case. PPs are explained in section 5.5.2.

| Case | | | PPs | | | | | | |
|---|---|---|---|---|---|---|---|---|---|
| Forward method | Jacobian matrix calculation technique | Magnetic field | $\sigma_t$ (S/m) $10^*$ | $\sigma_b$ (S/m) $2^*$ | CC - $5^*$ | RES - $0.2^*$ | PE (mm) $0^*$ | $RE_K$ (%) - | Run-time (min) - |
| Early | Sensitivity | Secondary field | 7.5 | 1.3 | 5.8 | 0.2 | -1 | 68 | 1.1 |
| | Standard | | 7.5 | 1.3 | 5.8 | 0.2 | -1 | 68 | 6.9 |
| Improved | Standard | Secondary field | 7.8 | 1.4 | 5.7 | 0.2 | -0.3 | 37 | 7.8 |
| Improved | Standard | Total field | 7.9 | 1.7 | 4.6 | 0.2 | -0.1 | 36 | 7.8 |

\* Ideal value

Figure A8.2(d)-(e) illustrate the reconstructed conductivity images by using the improved forward problem for Example I. The homogeneous conductivity value $\sigma_h$ and the coefficient $\tau$ were 0.5 S/m and 3.5, respectively. The voltages induced by the secondary and total magnetic fields based on the improved forward method are indicated by $\Delta V^I$ and $V^I$, respectively. In Figure A8.2(d) and Figure A8.2(e), $\Delta V^I$ and $V^I$ have been used for reconstruction, respectively. As can be seen, when the improved forward method is applied, using voltages induced by both total and secondary magnetic fields can detect the target object.

Figure A8.3 shows the relative error of the reconstructed conductivity in each iteration for the various cases in Example I. When $V^E$ is used for reconstruction, the algorithm does not converge. However, when $\Delta V^E$, $\Delta V^I$ or $V^I$ is applied, the algorithm converges. The minimum error was obtained when $V^I$ was employed.

Table A8.1 indicates PPs obtained for Example I. Since using $V^E$ in inverse problem could not meaningfully reconstruct the conductivity distribution, PPs are indeterminable. Thus, they are not reported in Table 5.2. The parameter $K$ indicates the iteration number for each case. As can be seen, using $V^I$ results in the best performance. For the early forward method, we tested both sensitivity and standard techniques for Jacobian matrix calculation. The reconstructed images were the same. However, the runtime was different. As expected, the standard

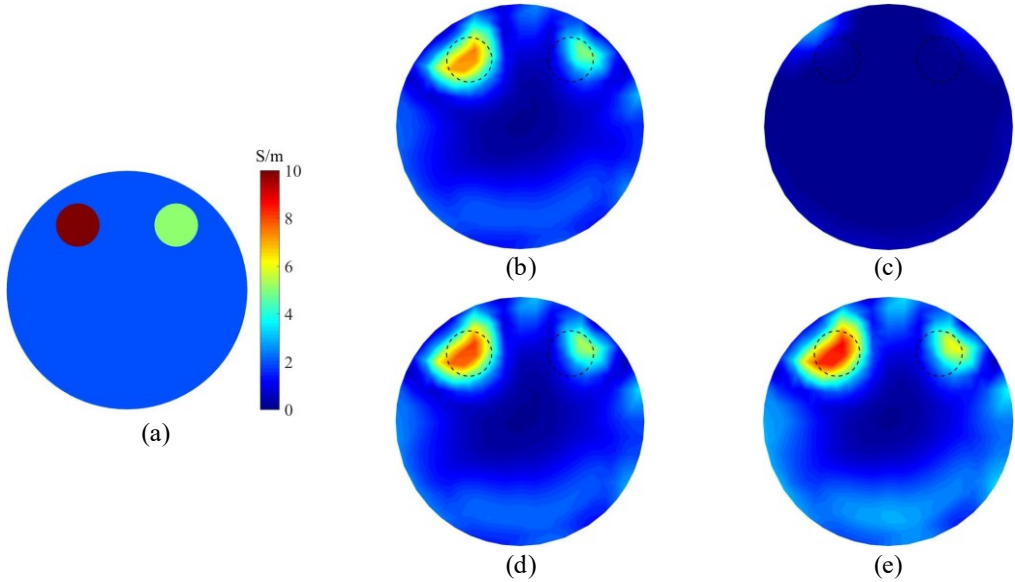

Figure A8.4 Example II: Imaging region contains two target objects with radius of 20 mm. (a) True conductivity distribution. Reconstructed conductivity images using (b) the early forward method and the secondary field, (c) the early forward method and the total field, (d) the improved forward method and the secondary field, and (e) the improved forward method and the total field. The homogeneous conductivity value $\sigma_h$ and the coefficient $\tau$ were 0.5 S/m and 3.5, respectively. The target object conductivities were $\sigma_{t1} = 10$ S/m (left target) and $\sigma_{t2} = 5$ S/m (right target) and the background conductivity was $\sigma_b = 2$ S/m.

technique was more time-consuming.

## A8.2 Example II: Two target objects

Figure A8.4(b)-(c) illustrate the reconstructed conductivity images by using the early forward problem for Example II. The homogeneous conductivity value $\sigma_h$ and the coefficient $\tau$ were 0.5 S/m and 3.5, respectively. As can be seen from Figure A8.4(b), using the voltages induced by the secondary field can compensate for the impact of ignoring skin and proximity effects in the early forward method. In Figure A8.4(c), $V^E$ has been used for reconstruction. As seen, when the voltages induced by the total magnetic field are computed by the early forward method, target objects cannot be resolved. It means that ignoring skin and proximity effects in coils in the forward problem implicates considerable errors in the reconstructed image

Figure A8.4(d)-(e) illustrate the reconstructed conductivity images by using the improved forward problem for Example II. The homogeneous conductivity value $\sigma_h$ and the coefficient $\tau$ were 0.5 S/m and 3.5, respectively. In Figure A8.4(d) and Figure A8.4(e), $\Delta V^I$ and $V^I$ have been used for reconstruction, respectively. As can be seen, when the improved forward method is applied, using voltages induced by both total and secondary magnetic fields can detect target objects.

## A8.3 Example III: Phantom similar to Austria scatterer

Figure A8.5(b)-(c) illustrate the reconstructed conductivity images by using the early forward problem for Example III. The homogeneous conductivity value $\sigma_h$ and the coefficient $\tau$ were 0.5 S/m and 4, respectively. In Figure A8.5(c), $V^E$ has been used for reconstruction. As seen, when the voltages induced by the total magnetic field are computed by the early forward method, target objects cannot be resolved. Again, it means that ignoring skin and proximity effects in coils in the forward problem implicates considerable errors in the reconstructed image. Figure A8.5(d)-(e) illustrate the reconstructed conductivity images by using the

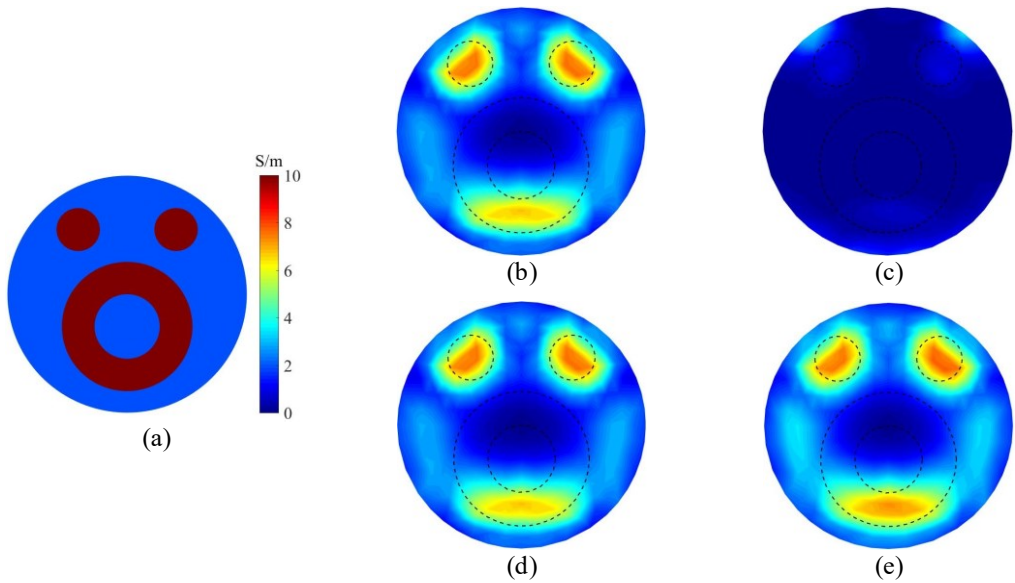

Figure A8.5 Example III: Imaging region contains two target objects with radius of 20 mm and a ring with thickness of 30 mm. (a) True conductivity distribution. Reconstructed conductivity images using (b) the early forward method and the secondary field, (c) the early forward method and the total field, (d) the improved forward method and the secondary field, and (e) the improved forward method and the total field. The homogeneous conductivity value $\sigma_h$ and the coefficient $\tau$ were 0.5 S/m and 4, respectively. The target object conductivities were 10 S/m and the background conductivity was $\sigma_b = 2$ S/m.

improved forward problem for Example III. The homogeneous conductivity value $\sigma_h$ and the coefficient $\tau$ were 0.5 S/m and 3.5, respectively.

When $\Delta V^E$ (Figure A8.5(b)), $\Delta V^I$ (Figure A8.5(d)) or $V^I$ (Figure A8.5(e)) is applied, the almost similar result was obtained: the circular object were detected and only the bottom of the ring was resolved. Again, it says that using $\Delta V^E$ can compensate for the impact of ignoring skin and proximity effects in the early forward method.

# Appendix 9: Publications

1- S. Alidoust, R. Jafari, **H. Yazdanian**, "Design and development of a wideband excitation unit for medical and industrial magnetic induction tomography ", in *ICBEM2016*, Tehran, Iran, 2016 (In Persian).

2- **H. Yazdanian** and R. Jafari, "Improvement on conductivity image reconstruction in magnetic induction tomography", In *MeMeA 2018*, Rome, Italy, 2018, pp. 1-5.

3- **H. Yazdanian**, and R. Jafari, "Comparing and improving techniques for conductivity image reconstruction in MIT: A simulation study", *International Journal of Numerical Modeling: Electronic Networks, Devices and Fields*, vol. 34, no.1, p. e2794, 2020.

4- **H. Yazdanian**, R. Jafari, and H. Moghaddam, "Solution of 2D MIT Forward Problem by Considering Skin and Proximity Effects in Coils", *IEEE Transactions on Computational Imaging*, vol. 7, pp.22-31, 2020.

5- **H. Yazdanian**, R. Jafari, and H. Moghaddam, "An Improved Technique for Solving the MIT Inverse Problem by Considering Skin and Proximity Effects in Coils", Manuscript submitted for publication (revised).

# References

[1] H. Y. Wei and M. Soleimani, "Electromagnetic tomography for medical and industrial applications: Challenges and opportunities [Point of View]," *Proceedings of the IEEE*, vol. 101, no. 3, pp. 559–565, 2013.

[2] R. H. Bayford, "Bioimpedance tomography (electrical impedance tomography)," *Annual Review of Biomedical Engineering*, vol.8, pp. 63-91, 2006.

[3] W. Q. Yang and L. Peng, "Image reconstruction algorithms for electrical capacitance tomography," *Measurement science and technology*, vol. 14, no. 1, pp. R1, 2003.

[4] L. Ma and M. Soleimani, "Magnetic induction tomography methods and applications: A review," *Measurement Science and Technology*, vol. 28, no. 7, pp. 072001, 2017.

[5] "ELISA 800VIT." [Online]. Available: http://www.swisstom.com/en/products/elisa-800vit-2. [Accessed: 07-Sep-2021].

[6] "PulmoVista 500." [Online]. Available: https://www.draeger.com/en_uk/Products/PulmoVista-500. [Accessed: 07-Sep-2021].

[7] L. Ma, A. Hunt, and M. Soleimani, "Experimental evaluation of conductive flow imaging using magnetic induction tomography," *International Journal of Multiphase Flow*, vol. 72, pp. 198-209, 2015.

[8] M. Hadinia, R. Jafari, and M. Soleimani, "EIT image reconstruction based on a hybrid FE-EFG forward method and the complete-electrode model," *Physiological Measurement*, vol. 37, no. 6, pp. 863–878, 2016.

[9] M. Soleimani, "Simultaneous reconstruction of permeability and conductivity in magnetic induction tomography," *Journal of Electromagnetic Waves and Applications*, vol. 23, no. 5, pp. 785–798, 2009.

[10] Z. Xu, H. Luo, W. He, C. He, X. Song, and Z. Zahng, "A multi-channel magnetic induction tomography measurement system for human brain model imaging," *Physiological Measurement*, vol. 30, no. 6, pp. S175, 2009.

[11] M. Zolgharni, H. Griffiths, and P. D. Ledger, "Frequency-difference MIT imaging of cerebral haemorrhage with a hemispherical coil array: Numerical Modeling ," *Physiological Measurement*, vol. 31, no. 8, pp. S111, 2010.

[12] Y. Chen *et al.*, "Imaging hemorrhagic stroke with magnetic induction tomography: Realistic simulation and evaluation," *Physiological Measurement*, vol. 31, no. 6, pp. 809–827, 2010.

[13] R. Liu, Y. Li, F. Fu, F. You, X. Shi, and X. Dong, "Time-difference imaging of magnetic induction tomography in a three-layer brain physical phantom," *Measurement science

*and technology*, vol. 25, no. 6, pp. 065402, 2014.

[14] Z. Xiao, C. Tan, and F. Dong, "Multi-frequency difference method for intracranial hemorrhage detection by magnetic induction tomography," *Physiological Measurement*, vol. 39, no. 5, pp. 055006, 2018.

[15] R. Chen et al., "A novel algorithm for high-resolution magnetic induction tomography based on stacked auto-encoder for biological tissue imaging," *IEEE Access*, vol. 7, pp. 185597-185606, 2019.

[16] Z. Xiao, C. Tan, and F. Dong, "3-D hemorrhage imaging by cambered magnetic induction tomography," *IEEE Transactions on Instrumentation and Measurement*, vol. 68, no. 7, pp. 2460–2468, 2019.

[17] Y. Chen, C. Tan, and F. Dong, "Combined planar magnetic induction tomography for local detection of intracranial hemorrhage," *IEEE Transactions on Instrumentation and Measurement*, vol. 70, pp. 1-11, 2020.

[18] Y. Lv and H. Luo, "A new method of haemorrhagic stroke detection via deep magnetic induction tomography," *Frontiers in Neuroscience*, vol. 15, p. 495, 2021.

[19] L. Ke, W. Zu, Q. Du, J. Chen, and X. Ding, "A bio-impedance quantitative method based on magnetic induction tomography for intracranial hematoma," *Medical & biological engineering & computing*, vol. 58, pp. 857–869, 2020.

[20] D. Gürsoy and H. Scharfetter, "Feasibility of lung imaging using magnetic induction tomography," in *IFMBE Proceedings*, Munich, Germany, 2009, pp. 525-528.

[21] D. Gürsoy and H. Scharfetter, "Magnetic induction pneumography: a planar coil system for continuous monitoring of lung function via contactless measurements," *Journal of Electrical Bioimpedance*, vol. 1, no. 1, pp.56-62, 2010.

[22] J. R. Feldkamp, "Single-coil magnetic induction tomographic three-dimensional imaging," *Journal of Medical Imaging*, vol.2, no. 1, pp. 013502, 2015.

[23] L. Marmugi and F. Renzoni, "Optical magnetic induction tomography of the heart," *Scientific reports*, vol. 6, no.1, pp. 1-8, 2016.

[24] Z. Sharon, R. Moshe, and A. Shimon, "Contactless bio-impedance monitoring technique for brain cryosurgery in a 3D head model," *Annals of biomedical engineering*, vol. 33, no. 5, pp. 616–625, 2005.

[25] World Stroke Organization, "Global stroke fact sheet," 2020. [Online]. Available: https://www.world-stroke.org/assets/downloads/WSO_Global_Stroke_Fact_Sheet.pdf. [Accessed: 07-Sep-2021].

[26] Y. V. Kalkonde, S. Alladi, S. Kaul, and V. Hachinski, "Stroke prevention strategies in the developing world," *Stroke*, vol. 49, no. 12, pp. 3092-3097, 2018.

[27] L. R. Caplan, *Caplan's Stroke: A Clinical Approach,* 5th ed., Philadelphia, PA, USA: SAUNDERS, 2016.

[28] E. C. Jauch et al., "Guidelines for the early management of patients with acute ischemic stroke: A guideline for healthcare professionals from the American Heart Association/American Stroke Association," *Stroke*, vol. 44, no. 3, pp. 870-974, 2013.

[29] M. Zolgharni, P. D. Ledger, and H. Griffiths, "Forward Modeling of magnetic induction tomography: A sensitivity study for detecting haemorrhagic cerebral stroke," *Medical & biological engineering & computing*, vol. 47, no. 12, pp. 1301–1313, 2009.

[30] R. Jafari-Shapoorabadi, A. Konrad, and A. N. Sinclair, "Comparison of three formulations for eddy-current and skin effect problems," *IEEE Transactions on Magnetics*, vol. 38, no. 2 I, pp. 617–620, 2002.

[31] N. Von Ellenrieder, C. H. Muravchik, and A. Nehorai, "A meshless method for solving the EEG forward problem," *IEEE Transactions on Biomedical Engineering*, vol. 52, no. 2, 2005.

[32] B. H. Brown, "Electrical impedance tomography (EIT): A review," *Journal of Medical*

*Engineering & Technology*. vol. 27, no. 3, pp.97-108, 2003.

[33] G. Kaur, R. Adhikari, P. Cass, M. Bown, and P. Gunatillake, "Electrically conductive polymers and composites for biomedical applications," *Rsc Advances*. vol. 5, no. 47, pp. 37553-37567, 2015.

[34] H. Griffiths, W. R. Stewart, and W. Gough, "Magnetic induction tomography: A measuring system for biological tissues," *Annals of the New York Academy of Sciences*, vol. 873, no. 1, pp. 335–345, 1999.

[35] H. Griffiths, "Magnetic induction tomography," *Measurement science and technology*, vol. 12, no. 8, p. 1126, 2001.

[36] H. Scharfetter, R. Casañas, and J. Rosell, "Biological tissue characterization by magnetic induction spectroscopy (MIS): Requirements and limitations," *IEEE transactions on biomedical engineering*, vol. 50, no. 7, pp. 870-880, 2003.

[37] R. Casañas *et al.*, "Measurement of liver iron overload by magnetic induction using a planar gradiometer: Preliminary human results," *Physiological Measurement*, vol. 25, no. 1, p. 135, 2004.

[38] S. Watson, R. J. Williams, H. Griffiths, W. Gough, and A. Morris, "Magnetic induction tomography: Phase versus vector-voltmeter measurement techniques," in *Physiological Measurement*, vol. 24, no. 2, p. 555, 2003.

[39] S. Watson, R. J. Williams, W. Gough, and H. Griffiths, "A magnetic induction tomography system for samples with conductivities below 10 S m$^{-1}$," *Measurement Science and Technology*, vol. 19, no. 4, p. 045501, 2008.

[40] A. Trakic, N. Eskandarnia, B. K. Li, E. Weber, H. Wang, and S. Crozier, "Rotational magnetic induction tomography," *Measurement Science and Technology*, vol.23, no. 2, p. 025402, 2012.

[41] M. Soleimani and W. R. B. Lionheart, "Absolute conductivity reconstruction in magnetic induction tomography using a nonlinear method," *IEEE Transactions on medical imaging*, vol. 25, no. 12, pp. 1521-1530, 2006.

[42] D. Gürsoy and H. Scharfetter, "Reconstruction artefacts in magnetic induction tomography due to patient's movement during data acquisition," *Physiological Measurement*, vol. 30, no. 6, p. S165, 2009.

[43] D. Gürsoy and H. Scharfetter, "The effect of receiver coil orientations on the imaging performance of magnetic induction tomography," *Measurement Science and Technology*, vol. 20, no. 10, p. 105505, 2009.

[44] D. Gürsoy and H. Scharfetter, "Optimum receiver array design for magnetic induction tomography," *IEEE transactions on biomedical engineering*, vol. 56, no. 5, pp.1435-1441, 2009.

[45] M. Han, X. Cheng, and Y. Xue, "Comparison with reconstruction algorithms in magnetic induction tomography," *Physiological Measurement*, vol. 37, no. 5, pp. 683–697, 2016.

[46] M. Zolgharni, P. D. Ledger, D. W. Armitage, D. S. Holder, and H. Griffiths, "Imaging cerebral haemorrhage with magnetic induction tomography: Numerical modeling," *Physiological Measurement*, vol. 30, no. 6, p. S187, 2009.

[47] Z. Xiao, C. Tan, and F. Dong, "Effect of inter-tissue inductive coupling on multi-frequency imaging of intracranial hemorrhage by magnetic induction tomography," *Measurement Science and Technology*, vol. 28, no. 8, p. 084001, 2017.

[48] M. Soleimani, W. R. B. Lionheart, and A. J. Peyton, "Image reconstruction for high-contrast conductivity imaging in mutual induction tomography for industrial applications," *IEEE Transactions on Instrumentation and Measurement*, vol. 56, no. 5, pp. 2024-2032, 2007.

[49] B. Dekdouk, M. H. Pham, D. W. Armitage, C. Ktistis, M. Zolgharni, and A. J. Peyton,

"A feasibility study on the delectability of edema using magnetic induction tomography using an analytical model," in *IFMBE Proceedings*, Berlin, Germany, 2008.

[50] Z. Xu, Q. Li, and W. He, "Analytical solution for the forward problem of magnetic induction tomography with multi-layer sphere model," in *International Conference on Life System Modeling and Simulation*, Wuxi, China 2010.

[51] O. Biro, "Edge element formulations of eddy current problems," *Computer methods in applied mechanics and engineering*, vol. 169, no. 3-4, 1999.

[52] R. Merwa, K. Hollaus, B. Oszkar, and H. Scharfetter, "Detection of brain oedema using magnetic induction tomography: A feasibility study of the likely sensitivity and detectability," *Physiological Measurement*, vol. 25, no. 1, pp. 347–354, 2004.

[53] K. Hollaus, C. Magele, R. Merwa, and H. Scharfetter, "Numerical simulation of the eddy current problem in magnetic induction tomography for biomedical applications by edge elements," *IEEE Transactions on Magnetics*, vol. 40, no. 2 II, pp. 623–626, 2004.

[54] H. Scharfetter, K. Hollaus, J. Rosell-Ferrer, and R. Merwa, "Single-step 3-D image reconstruction in magnetic induction tomography: Theoretical limits of spatial resolution and contrast to noise ratio," *Annals of biomedical engineering*, vol. 34, no. 11, pp. 1786-1798, 2006.

[55] N. G. Gençer and M. N. Tek, "Electrical conductivity imaging via contactless measurements," *IEEE Transactions on Medical Imaging*, vol. 18, no. 7, pp. 617-627, 1999.

[56] B. Ülker Karbeyaz and N. G. Gençer, "Electrical conductivity imaging via contactless measurements: An experimental study," *IEEE Transactions on Medical Imaging*, vol. 22, no. 5, pp. 627-635, 2003.

[57] C. Wang, R. Liu, F. Fu, F. You, X. Shi, and X. Dong, "Image reconstruction for Magnetic Induction Tomography and Preliminary simulations on a simple head model," in *IEMBS 2007*, *Lyon, France*, 2007.

[58] A. Morris, H. Griffiths, and W. Gough, "A numerical model for magnetic induction tomographic measurements in biological tissues," *Physiological Measurement*, vol. 22, no. 1, p. 113, 2001.

[59] B. Dekdouk, W. Yin, C. Ktistis, D. W. Armitage, and A. J. Peyton, "A method to solve the forward problem in magnetic induction tomography based on the weakly coupled field approximation," *IEEE Transactions on Biomedical Engineering*, vol. 57, no. 4, pp. 914–921, 2010.

[60] A. Ramos and J. G. B. Wolff, "Numerical modeling of magnetic induction tomography using the impedance method," *Medical & biological engineering & computing*, vol. 49, no. 2, pp. 233-340, 2011.

[61] A. V. Korjenevsky and S. A. Sapetsky, "Feasibility of the backprojection method for reconstruction of low contrast perturbations in a conducting background in magnetic induction tomography," *Physiological Measurement*, vol. 38, no. 6, p. 1204, 2017.

[62] J. M. S. Caeiros and R. C. Martins, "An optimized forward problem solver for the complete characterization of the electromagnetic properties of biological tissues in magnetic induction tomography," *IEEE transactions on Magnetics*, vol. 48, no. 12, pp. 4707–4712, 2012.

[63] S. Engleder and O. Steinbach, "Boundary integral formulations for the forward problem in magnetic induction tomography," *Mathematical Methods in the Applied Sciences*, vol. 34, no. 9, pp. 1144-1156, 2011.

[64] P. De Tillieux and Y. Goussard, "Improving the computational cost of image reconstruction in biomedical magnetic induction tomography using a volume integral equation approach," *IEEE Transactions on Antennas and Propagation*, vol. 69, no. 1, pp. 366-378, 2021.

[65] S. Al-Zeibak and N. H. Saunders, "A feasibility study of in vivo electromagnetic imaging," *Physics in Medicine & Biology*, vol. 38, no. 1, p. 151, 1993.

[66] A. Korjenevsky, V. Cherepenin, and S. Sapetsky, "Magnetic induction tomography: Experimental realization," *Physiological Measurement*, vol. 21, no. 1, pp. 89–94, 2000.

[67] R. Merwa, K. Hollaus, P. Brunner, and H. Scharfetter, "Solution of the inverse problem of magnetic induction tomography (MIT)," *Physiological Measurement*, vol. 26, no. 2, p. S241, 2005.

[68] M. Vauhkonen, M. Hamsch, and C. H. Igney, "A measurement system and image reconstruction in magnetic induction tomography," *Physiological Measurement*, vol. 29, no. 6, p. S445, 2008.

[69] H. Y. Wei and M. Soleimani, "Hardware and software design for a National Instrument-based magnetic induction tomography system for prospective biomedical applications," *Physiological Measurement*, vol. 33, no. 5, pp. 863–879, 2012.

[70] Y. Mamatjan, "Imaging of hemorrhagic stroke in magnetic induction tomography: An in vitro study," *International journal of imaging systems and technology*, vol. 24, no. 2, pp. 161–166, 2014.

[71] Z. Xiao, C. Tan, and F. Dong, "Brain tissue based sensitivity matrix in hemorrhage imaging by magnetic induction tomography," in *I2MTC 2017*, Turin, Italy, 2017.

[72] R. Merwa, P. Brunner, A. Missner, K. Hollaus, and H. Scharfetter, "Solution of the inverse problem of magnetic induction tomography (MIT) with multiple objects: Analysis of detectability and statistical properties with respect to the reconstructed conducting region," *Physiological Measurement*, vol. 27, no. 5, p. S249, 2006.

[73] R. Merwa and H. Scharfetter, "Magnetic induction tomography: Evaluation of the point spread function and analysis of resolution and image distortion," *Physiological Measurement*, vol. 28, no. 7, p. S313, 2007.

[74] D. Gürsoy and H. Scharfetter, "Anisotropic conductivity tensor imaging using magnetic induction tomography," *Physiological Measurement*, vol. 31, no. 8, p. S135, 2010.

[75] B. Dekdouk, "Image reconstruction of low conductivity material distribution using magnetic induction tomography,", PhD thesis, University of Manchester, 2011.

[76] B. Dekdouk, C. Ktistis, W. Yin, D. W. Armitage, and A. J. Peyton, "The application of a priori structural information based regularization in image reconstruction in magnetic induction tomography," in *ICEBI & EIT 2010*, Florida, USA, 2010.

[77] Q. Chen, R. Liu, C. Wang, and R. Liu, "Real-time in vivo magnetic induction tomography in rabbits: A feasibility study," *Measurement Science and Technology*, vol. 32, no. 3, p. 035402, 2020.

[78] B. Dekdouk, C. Ktistis, D. W. Armitage, and A. J. Peyton, "Absolute imaging of low conductivity material distributions using nonlinear reconstruction methods in magnetic induction tomography," *Prog. Electromagn. Res.*, 2016.

[79] R. Chen, J. Huang, Y. Song, B. Li, J. Wang, and H. Wang, "Deep learning algorithms for brain disease detection with magnetic induction tomography," *Medical Physics*, vol. 48, no. 2, pp. 745-759, 2020.

[80] Z. Chen et al., "MITNet: GAN enhanced magnetic induction tomography based on complex CNN," *arXiv Prepr. arXiv2102.07911*, 2021.

[81] M. R. Yousefi, R. Jafari, and H. Abrishami Moghaddam, "Employing dual frequency phase sensitive demodulation technique to improve the accuracy of voltage measurement in magnetic induction tomography and designing a labratoary prototype," *Journal of Control*, vol. 14, no. 3, pp. 89-102, 2020.

[82] M. R. Yousefi, R. Jafari, and H. Abrishami Moghaddam, "A combined wavelet based mesh free-finite element method for solving the forward problem in magnetic induction tomography," *Iranian Journal of Biomedical Engineering*, vol. 8, no. 1, pp. 69–86, 2014,

(in Persian).

[83] M. R. Yousefi, "Finite element-meshfree modeling in magnetic induction tomography and developing a laboratory system," PhD thesis, K. N. Toosi University of Technology, 2015.

[84] Z. Liu, M. He, and H. Xiong, "Simulation study of the sensing field in electromagnetic tomography for two-phase flow measurement," *Flow Measurement and Instrumentation*, vol. 16, no. 2–3, pp. 199–204, 2005.

[85] M. Soleimani, W. R. B. Lionheart, C. H. Riedel, and O. Dössel, "Forward problem in 3D magnetic induction tomography (MIT)," in *Proceedings of 3rd World Congress on Industrial Process Tomography*, Banff, Alberta, Canada, 2003, pp. 275–280.

[86] S. Wang, S. Huang, Y. Zhang, and W. Zhao, "Multiphysics modeling of a lorentz force-based meander coil electromagnetic acoustic transducer via steady-state and transient analyses," *IEEE Sensors Journal*, vol. 16, no. 17, pp. 6641–6651, 2016.

[87] R. Jafari, "Electromagnetic acoustic transducer analysis by the finite element method," PhD thesis, University of Toronto, 2002.

[88] R. Casanova, A. Silva, and A. R. Borges, "A quantitative algorithm for parameter estimation in magnetic induction tomography," *Measurement Science and Technology*, vol. 15, no. 7, pp. 1412–1419, 2004.

[89] A. J. Peyton *et al.*, "An overview of electromagnetic inductance tomography: Description of three different systems," *Measurement Science and Technology*, vol. 7, no. 3. pp. 261–271, 1996.

[90] M. B. Özakın and S. Aksoy, "Application of magneto-quasi-static approximation in the finite difference time domain method," *IEEE Transactions on Magnetics*, vol. 52, no. 8, pp. 1–9, 2016.

[91] J. Weiss and Z. J. Csendes, "A one-step finite element method for multiconductor skin effect problems," *IEEE Transactions on Power Apparatus and Systems*, vol. PAS-101, no. 10, pp. 3796–3803, 1982.

[92] A. Konrad, "Integrodifferential finite element formulation of two-dimensional steady-state skin effect problems," *IEEE Transactions on Magnetics*, vol. 18, no. 1, pp. 284–292, 1982.

[93] H. A. Wheeler, "Formulas for the Skin Effect," *Proceedings of the IRE*, vol. 30, no. 9, pp. 412-424, 1942.

[94] L. Turner and H. C. Han, "The calculation of transient eddy-current fields using null-field integral techniques," *IEEE Transactions on Magnetics*, vol. 23, no. 2, pp. 1811-1818, 1987.

[95] H. A. Haus and J. R. Melcher, "Magnetoquasistatic fields: superposition integral and boundary," in *Electromagnetic fields and energy*, vol. 107, Englewood Cliffs, NJ, USA: Prentice Hall, 1989, ch.8, sec. 8.6, pp. 34–36.

[96] P. Jabłoński, T. Szczegielniak, D. Kusiak, and Z. Piątek, "Analytical–numerical solution for the skin and proximity effects in two parallel round conductors," *Energies*, vol. 12, no. 18, p.3584, 2019.

[97] G. Giovannetti *et al.*, "Conductor geometry and capacitor quality for performance optimization of low-frequency birdcage coils," *Concepts in Magnetic Resonance Part B: Magnetic Resonance Engineering: An Educational Journal*, vol. 20, no. 1, pp. 9–16, 2004.

[98] J. Xiang, Y. Dong, M. Zhang, and Y. Li, "Design of a magnetic induction tomography system by gradiometer coils for conductive fluid imaging," *IEEE Access*, vol. 7, pp. 56733–56744, 2019.

[99] "SimNIBS Example Dataset." [Online]. Available: www.simnibs.org. [Accessed: 25-Jan-2020].

[100] S. Gabriel, R. W. Lau, and C. Gabriel, "The dielectric properties of biological tissues: III. Parametric models for the dielectric spectrum of tissues," *Physics in medicine & biology*, vol. 41, no. 11, pp. 2271–2293, 1996.

[101] G. R. Liu, "An overview on meshfree methods: for computational solid mechanics," *International Journal of Computational Methods*, vol. 13, no. 05, p. 1630001, 2016.

[102] Y. Chen, J. D. Lee, and A. Eskandarian, *Meshless methods in solid mechanics*, vol. 9, NY, USA: Springer, 2006.

[103] V. Cutrupi, F. Ferraioli, A. Formisano, and R. Martone, "An approach to the electrical resistance tomography based on meshless methods," *IEEE transactions on magnetics*, vol. 43, no. 4, pp. 1717-1720, 2007.

[104] M. Hadinia and R. Jafari, "An element-free Galerkin forward solver for the complete-electrode model in electrical impedance tomography," *Flow Measurement and Instrumentation*, vol. 45, pp. 68–74, 2015.

[105] M. R. Yousefi, R. Jafari, and H. A. Moghaddam, "Imposing boundary and interface conditions in multi-resolution wavelet Galerkin method for numerical solution of Helmholtz problems," *Computer Methods in Applied Mechanics and Engineering*, vol. 276, pp. 67-94, 2014.

[106] G. Hu, M. Chen, W. He, and J. Zhai, "A novel forward problem solver based on meshfree method for electrical impedance tomography," *Przeglad Elektrotechniczny*, vol. 89, pp. 234-237, 2013.

[107] L. Xuan, "Meshless element-free Galerkin method in NDT applications," in *AIP Conference Proceedings*, Brunswick, Maine, USA, 2002, pp. 1960-1967.

[108] L. Xuan, Z. Zeng, B. Shanker, and L. Udpa, "Element-free Galerkin method for static and quasi-static electromagnetic field computation," *IEEE Transactions on Magnetics*, vol. 41, no. 1, pp. 12-20, 2004.

[109] O. Bottauscio, M. Chiampi, and A. Manzin, "Eddy current problems in nonlinear media by the element-free Galerkin method," *Journal of Magnetism and Magnetic Materials*, vol. 304, no. 2, pp. e823-e825, 2006.

[110] O. Bottauscio, M. Chiampi, and A. Manzin, "Element-free Galerkin method in eddy-current problems with ferromagnetic media," *IEEE Transactions on Magnetics*, vol. 42, no. 5, pp. 1577-1584, 2006.

[111] G.-R. Liu, *Meshfree methods: moving beyond the finite element method*, 2$^{nd}$ ed., Boca Raton, FL, USA: CRC press, 2009.

[112] O. Zienkiewicz, R. Taylor, and J. Z. Zhu, *The Finite Element Method: its Basis and Fundamentals*, 7$^{th}$ ed., Oxford, United Kingdom: Elsevier Butterworth-Heinemann, 2013.

[113] J. Dolbow and T. Belytschko, "An introduction to programming the meshless: Element free Galerkin method," *Archives of computational methods in engineering*, vol. 5, no. 3, pp. 207-241, 1998.

[114] T. Belytschko, Y. Y. Lu, and L. Gu, "Element-free Galerkin methods," *International journal for numerical methods in engineering*, vol. 37, no. 2, pp. 229–256, 1994.

[115] C. Hérault and Y. Maréchal, "Boundary and interface conditions in meshless methods [for EM field analysis]," *IEEE Transactions on Magnetics*, vol. 35, no. 3, pp. 1450–1453, 1999.

[116] D. Mirzaei, "Development of Moving Least Squares Based Meshless Methods," PhD thesis, Amirkabir University of Technology, 2011 (in Persian).

[117] M. Hadinia, "Image reconstruction in diffuse optical and electrical impedance tomographies by a hybrid FE-EFG method," PhD thesis, K. N. Toosi University of Technology, 2016.

[118] M. R. Yousefi, R. Jafari, and H. A. Moghaddam, "A combined wavelet-based mesh-free

method for solving the forward problem in electrical impedance tomography," *IEEE Transactions on Instrumentation and Measurement*, vol. 62, no. 10, pp. 2629-2638, 2013.

[119] T. J. Yorkey, J. G. Webster, and W. J. Tompkins, "Comparing reconstruction algorithms for electrical impedance tomography," *IEEE Transactions on Biomedical Engineering*, vol. BME-34, no. 11, pp. 843 - 852, 1987.

[120] B. Brandstätter, "Jacobian calculation for electrical impedance tomography based on the reciprocity principle," *IEEE transactions on magnetics*, vol. 39, no. 3, pp. 1309-1312, 2003.

[121] P. J. Vauhkonen, M. Vauhkonen, T. Savolainen, and J. P. Kaipio, "Three-dimensional electrical impedance tomography based on the complete electrode model," *IEEE Transactions on Biomedical Engineering*, vol. 46, no. 9, pp. 1150-1160, 1999.

[122] A. A. Roohi Noozadi, "Forward and inverse modeling of magnetic induction tomography (MIT) for biomedical application," PhD thesis, École Polytechnique de Montréal, 2017.

[123] H. B. Nielsen, "Damping parameter in Marquardt's method, " Technical report, Technical University of Denmark, 1999.

[124] J. L. Prince and J. M. Links, *Medical imaging signals and systems*, London, United Kingdom: Pearson Education, 2006.

[125] C. Li, K. An, and K. Zheng, "The levenberg-marquardt method for acousto-electric tomography on different conductivity contrast," *Applied Sciences*, vol. 10, no. 10, p. 3482, 2020.

[126] A. Adler *et al.*, "GREIT: A unified approach to 2D linear EIT reconstruction of lung images," *Physiological Measurement*, vol. 30, no. 6, p. S35, 2009.

[127] I. Muttakin, T. Wondrak, and M. Soleimani, "Magnetic induction tomography sensors for quantitative visualization of liquid metal flow shape," *IEEE Sensors Letters*, vol. 4, no. 7, pp. 1-4, 2020.

[128] M. Soleimani *et al.*, "In situ steel solidification imaging in continuous casting using magnetic induction tomography," *Measurement Science and Technology*, vol. 31, no. 6, p. 065401, 2020.

[129] M. Soleimani, W. R. B. Lionheart, A. J. Peyton, X. Ma, and S. R. Higson, "A three-dimensional inverse finite-element method applied to experimental eddy-current imaging data," *IEEE Transactions on Magnetics*, vol. 42, no. 5, pp. 1560-1567, 2006.

[130] X. Ma, A. J. Peyton, S. R. Higson, A. Lyons, and S. J. Dickinson, "Hardware and software design for an electromagnetic induction tomography (EMT) system for high contrast metal process applications," *Measurement Science and Technology*, vol. 17, no. 1, pp. 111–118, 2006.

[131] J. Caeiros, R. C. Martins, and B. Gil, "A new image reconstruction algorithm for real-time monitoring of conductivity and permeability changes in magnetic induction tomography," in *Proceedings of the Annual International Conference of the IEEE EMBS*, San Diego, CA, USA, 2012, pp. 6239–6242.

[132] X. Ma, A. J. Peyton, M. Soleimani, and W. R. B. Lionheart, "Imaging internal structure with electromagnetic induction tomography," in *Proceedings of IEEE IMTC*, Sorrento, Italy 2006, pp. 299-303.

[133] M. H. Schulze and H. Heuer, "Textural analyses of carbon fiber materials by 2D-FFT of complex images obtained by high frequency eddy current imaging (HF-ECI)," in *Proceedings of Nondestructive Characterization for Composite Materials, Aerospace Engineering, Civil Infrastructure, and Homeland Security*, San Diego, California, USA, 2012, p. 83470S.

[134] M. Schweiger and S. Arridge, "The Toast++ software suite for forward and inverse

modeling in optical tomography," *Journal of biomedical optics*, vol. 19, no. 4, p. 040801, 2014.

[135] L. Ma and M. Soleimani, "Hidden defect identification in carbon fibre reinforced polymer plates using magnetic induction tomography," *Measurement Science and Technology*, vol. 25, no. 5, p. 055404, 2014.

[136] C. S. Park *et al.*, "A portable phase-domain magnetic induction tomography transceiver with phase-band auto-tracking and frequency-sweep capabilities," *Sensors*, vol. 18, no. 11, p. 3816, 2018.

[137] P. C. Hansen, *Rank-Deficient and Discrete Ill-Posed Problems*, Philadelphia, PA, USA: SIAM publications, 1998.

[138] Z. Liu, G. Yang, N. He, and X. Tan, "Landweber iterative algorithm based on regularization in electromagnetic tomography for multiphase flow measurement," *Flow Measurement and Instrumentation*, vol. 27, pp. 53-58, 2012.

[139] A. Adler and W. R. B. Lionheart, "Uses and abuses of EIDORS: An extensible software base for EIT," *Physiological Measurement*, vol. 27, no.5, p. S25, 2006.

[140] H. H. Eroğlu, M. Sadighi, and B. M. Eyüboğlu, "Induced current magnetic resonance electrical conductivity imaging with oscillating gradients," *IEEE Transactions on Medical Imaging*, vol. 37, no. 7, pp. 1606-1617, 2018.

[141] D. K. Cheng, "Static Magnetic Fields," in *Field and Wave Electromagnetics*, Boston, MA, USA: Addison-Wesley, 1986, ch.6, sec. 6.5, pp. 239–242.

[142] P. P. Silvester and R. L. Ferrari, "Electromagnetics of finite elements," in *Finite elements for electrical engineers*, Cambridge, United Kingdom: Cambridge University Press, 1996, ch.3, sec. 3.4, 6, pp. 79–80.

[143] T. Morisue, "Magnetic vector potential and electric scalar potential in three-dimensional eddy current problem," *IEEE Transactions on Magnetics*, vol. 18, no.2, pp. 531-535, 1982.

[144] R. Merwa, K. Hollaus, B. Brandstätter, and H. Scharfetter, "Numerical solution of the general 3D eddy current problem for magnetic induction tomography (spectroscopy)," *Physiological Measurement*, vol. 24, no. 2, p. 545, 2003.

[145] A. Kameari, "Regularization on Ill-posed source terms in FEM computation using two magnetic vector potentials," *IEEE Transactions on Magnetics*, vol. 40, no.2, pp. 1310-1313, 2004.

[146] M. Zolgharni, "Magnetic induction tomography for imaging cerebral stroke," PhD thesis, Swansea University, 2010.

[147] H. Scharfetter, H. K. Lackner, and J. Rosell, "Magnetic induction tomography: Hardware for multi-frequency measurements in biological tissues," *Physiological Measurement*, vol. 22, no.1, p. 131, 2001.

[148] H. Scharfetter, A. Köstinger, and S. Issa, "Hardware for quasi-single-shot multifrequency magnetic induction tomography (MIT): The Graz Mk2 system," *Physiological Measurement*, vol. 29, no. 6, p. S431, 2008.

[149] A. L. McEwan, M. Hamsch, S. Watson, C. H. Igney, and J. Kahlert, "A comparison of two phase measurement techniques for magnetic impedance tomography," in *World Congress on Medical Physics and Biomedical Engineering*, Munich, Germany, 2009, pp. 4-6.

[150] W. He, H. Luo, Z. Xu, and J. Wang, "Multi-channel magnetic induction tomography measurement system," in *Proceedings of 3rd International Conference on Biomedical Engineering and Informatics*, Yantai, China, 2010, pp. 402-405.

[151] H. Griffiths, M. Zolgharni, P. D. Ledger, and S. Watson, "The cardiff Mk2b MIT head array: Optimising the coil configuration," in *Journal of Physics: Conference Series*, 2010.

[152] J. Caeiros, B. Gil, N. B. Brás, and R. C. Martins, "A differential high-resolution motorized multi-projection approach for an experimental magnetic induction tomography prototype," in *Proceedings of MeMeA 2012*, Budapest, Hungary, 2012, pp. 1-4.

[153] H. Luo and X. Jiang, "The magnetic induction tomography measurement system based on Helmholtz coil," in *Proceedings of 8th International Conference on Biomedical Engineering and Informatics*, Shenyang, China, 2015, pp. 29–33.

[154] C. Tan, Y. Chen, Y. Wu, Z. Xiao, and F. Dong, "A modular magnetic induction tomography system for low-conductivity medium imaging," *IEEE Transactions on Instrumentation and Measurement*, vol. 70, pp. 1-8, 2021.

[155] G. Jin *et al.*, "A special phase detector for magnetic inductive measurement of cerebral hemorrhage," *PLoS One*, vol. 9, no. 5, p. e97179, 2014.

[156] A. Adler and R. Guardo, "Electrical impedance tomography: regularized imaging and contrast detection," *IEEE Transactions on Medical Imaging*, vol. 15, no. 2, pp. 170-179, 1996.

[157] M. Yasin, S. Böhm, P. O. Gaggero, and A. Adler, "Evaluation of EIT system performance," *Physiological Measurement*, vol. 23, no. 7, pp. 851- 866.

[158] A. Javaherian, A. Movafeghi, R. Faghihi, and E. Yahaghi, "An exhaustive criterion for estimating quality of images in electrical impedance tomography with application to clinical imaging," *Journal of Visual Communication and Image Representation*, vol. 24, no. 7, pp. 773- 785, 2013.

[159] F. Li, J. F. P. J. Abascal, M. Desco, and M. Soleimani, "Total variation regularization with Split Bregman-based method in magnetic induction tomography using experimental data," *IEEE Sensors Journal*, vol. 17, no. 4, pp. 976-985, 2017.

[160] J. L. Wheeler, W. Wang, and M. Tang, "A comparison of methods for measurement of spatial resolution in two-dimensional circular EIT images," *Physiological Measurement*, vol. 23, no. 1, pp. 169-176, 2002.

[161] M. Neumayer, G. Steiner, D. Watzenig, and H. Zangl, "Spatial resolution analysis for real time applications in electrical capacitance tomography," *Nuclear Engineering and Design*, vol. 241, no. 6, pp. 1988-1983, 2011.

[162] S. Teniou, M. Meribout, K. Al-Wahedi, A. Al-Durra, and E. Al-Hosani, "A near-infrared-based magnetic induction tomography solution to improve the image reconstruction accuracy in opaque environments," *IEEE Transactions on Magnetics*, vol. 49, no. 4, pp. 1361-1366, 2013.

[163] M. Zhang, L. Ma, and M. Soleimani, "Magnetic induction tomography guided electrical capacitance tomography imaging with grounded conductors," *Measurement*, vol. 53, pp. 171-181, 2014.

[164] G. Dingley and M. Soleimani, "Multi-frequency magnetic induction tomography system and algorithm for imaging metallic objects," *Sensors*, vol. 21, no. 11, p. 3671, 2021.

[165] K. Xu, L. Wu, X. Ye, and X. Chen, "Deep learning-based inversion methods for solving inverse scattering problems with phaseless data," *IEEE Transactions on Antennas and Propagation*, vol. 68, no. 11, pp. 7457-7470, 2020.160

## چکیده:

مقطع‌نگاری القای مغناطیسی (MIT) روشِ تصویربرداری نرم-میدانِ نسبتاً جدیدی است که می‌تواند جهت تصویربرداری از توزیع داخلی ضرایب هدایت الکتریکی یک جسم، مورد استفاده قرارگیرد. ویژگی غیرتهاجمی و غیرتماسی بودن MIT، استفاده از آن را در کاربردهای بالقوه‌ی پزشکی، مانند تصویربرداری از خونریزی مغزی، مورد توجه ساخته است. در این کاربردها ضرایب هدایت الکتریکی کوچک هستند (کوچکتر از ۲ S/m) و در نتیجه سیگنال‌های اندازه‌گیری‌شده بسیار ضعیف می‌باشند. از این رو بهبود در سیستم‌های MIT جهت استفاده در کاربردهای پزشکی ضروری می‌باشد.

تصویر در MIT با حل یک مسأله‌ی معکوس بد-وضع و غیرخطی به‌دست می‌آید. در مسأله‌ی معکوس، خروجی مسأله‌ی مستقیم با داده‌های اندازه‌گیری‌شده مقایسه و بر اساس آن ضرایب هدایت الکتریکی به‌روز رسانی می‌شوند. از این رو لازم است تا حد امکان مسأله‌ی مستقیم به صورت دقیق مدل و حل شود. مدلی که در مطالعات قبلی برای مسأله‌ی مستقیم MIT استفاده‌شده، چگالی جریان سیم‌پیچ‌ها را معلوم و یکنواخت در نظر می‌گیرد. این به معنای صرف‌نظر کردن از اثرات پوستی و همجواری در سیم‌پیچ‌ها می‌باشد. در کاربردهای پزشکی MIT فرکانس تحریک در محدوده چند مگاهرتز است که در این محدوده اثرات پوستی و همجواری در سیم‌پیچ‌ها قابل توجه خواهد بود.

نوآوری اصلی این رساله، مطالعه‌ی در نظر گرفتن اثرات پوستی و همجواری در سیم‌پیچ‌های MIT دوبعدی و ارائه‌ی یک مدل بهبود یافته برای مسأله‌ی مستقیم می‌باشد. در این مدل توزیع چگالی جریان در سیم‌پیچ‌ها مجهول درنظر گرفته می‌شود. به منظور ارزیابی عملکرد مدل بهبودیافته در کاربردهای بالقوه‌ی پزشکی، یک سیستم MIT دوبعدی با ۱۶ سیم‌پیچ مدل می‌شود و از روش المان محدود (FEM) برای حل مسأله‌ی مستقیم استفاده می‌گردد. مقایسه‌ی مدل قبلی و مدل بهبودیافته نشان می‌دهد که استفاده از مدل قبلی برای کاربردهای پزشکی ممکن است منجر به بروز خطا در محاسبه‌ی سیگنال‌های القایی شود.

نوآوری دیگر مطالعه حاضر، ارائه‌ی یک روش ترکیبی المان محدود-بدون المان گالرکین (FE-EFG) جهت حل مسأله‌ی مستقیم بهبودیافته می باشد. مطالعات قبلی در زمینه MIT پزشکی از روش های عددی مبتنی بر مش مانند FEM برای حل مسأله‌ی مستقیم بهره برده‌اند. تولید مش در این روش‌ها برای هندسه‌های پیچیده مانند ناحیه‌ی سر انسان سخت و زمان‌گیر است. از سوی دیگر، روش بدون المان گالرکین (EFG) هزینه‌ی محاسباتی نسبتاً بالایی دارد. در این رساله برای اولین بار روش ترکیبی FE-EFG جهت بهره بردن از مزایای هر دو روش مورد استفاده قرار می‌گیرد. مقایسه‌ی نتایج به‌دست آمده از روش المان محدود و روش بدون المان گالرکین برای حل مسأله‌ی مستقیم نشان می‌دهد که در یک زمان اجرای یکسان، روش ترکیبی دقیق تر است.

به منظور بررسی اهمیت مدل بهبودیافته‌ی مسأله‌ی مستقیم، نیاز است که این مدل در بازسازی تصویر ضرایب هدایت الکتریکی مورد استفاده قرار گیرد. از این رو نوآوری دیگر این رساله، ارائه یک تکنیک بهبودیافته برای حل مسأله‌ی معکوس MIT با در نظر گرفتن اثرات پوستی و همجواری در سیم‌پیچ‌ها

می‌باشد. بدین منظور روش گاوس-نیوتن تنظیم شده با مدل بهبودیافته‌ی مسأله‌ی مستقیم تطبیق داده می‌شود. پارامتر تنظیم بر اساس یک روش تطبیقی انتخاب می‌گردد. بر اساس یک تکنیک استاندارد، روش جدیدی جهت محاسبه ماتریس ژاکوبین ارائه می شود که قابل استفاده برای روش مستقیم بهبود یافته است. به منظور مقایسه‌ی روش مستقیم قبلی و بهبودیافته، یک سیستم MIT دوبعدی ۸ سیم‌پیچه مدل می‌شود و تصویر ضرایب هدایت الکتریکی برای فانتوم‌های ساده بازسازی می‌گردد. نتایج نشان می‌دهد که استفاده از روش مستقیم بهبودیافته جهت بازسازی ضرایب هدایت الکتریکی مطلق ضروری به نظر می‌رسد.

نوآوری آخر در این مطالعه بر روی مقایسه و بهبود روش‌های بازسازی ضرایب هدایت الکتریکی متمرکز می‌باشد. در مطالعات گذشته، عموماً بخش حقیقی (مؤلفه‌ی هم‌فاز) ولتاژ القایی در کاربردهایی با ضرایب هدایت الکتریکی کوچک و بخش موهومی (مؤلفه‌ی متعامد) ولتاژ القایی در کاربردهایی با ضرایب هدایت الکتریکی بزرگ مورد استفاده قرار گرفته است. با این وجود، روشی برای بازسازی ضرایب هدایت الکتریکی در کاربردهایی که ضرایب در محدوده‌ی متوسط می‌باشند ارائه نشده است. در این رساله، ضرورت استفاده از هر دو بخش حقیقی و موهومی ولتاژهای القایی در MIT نشان داده می‌شود و یک تکنیک که از هر دو بخش برای بازسازی ضرایب هدایت الکتریکی استفاده می‌کند ارائه می‌گردد. نتایج نشان می‌دهد که این تکنیک در مقایسه با تکنیک قبلی، عملکرد بهتری دارد؛ به ویژه برای کاربردهایی که ضرایب هدایت الکتریکی در محدوده‌ی متوسط قرار می‌گیرند.

**واژگان کلیدی**: اثرات پوستی و همجواری، الگوریتم گاوس-نیوتن، تصویربرداری از ضرایب هدایت الکتریکی، روش اجزای محدود، روش بدون المان گالرکین، ماتریس ژاکوبین، مسأله‌ی مستقیم، مقطع نگاری القای مغناطیسی، ولتاژهای القایی مختلط.

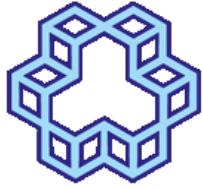

دانشگاه صنعتی خواجه نصیرالدین طوسی

دانشکده مهندسی برق

رساله‌ی دکتری

مهندسی برق- گرایش مهندسی پزشکی

# مدلسازی مسأله‌ی مستقیم بهبود یافته در مقطع نگاری القای مغناطیسی برای کاربردهای پزشکی

نگارش:

حسن یزدانیان

اساتید راهنما:

دکتر رضا جعفری

دکتر حمید ابریشمی مقدم

مهرماه ۱۴۰۰